\documentclass[reqno, 11pt]{gen-m-l}
\usepackage{latexsym, amssymb}%, showlabels}
\usepackage{hyperref}
\usepackage{amsmath}
\usepackage{amsthm}
\usepackage{epsfig}
\usepackage[mathscr]{eucal}
\numberwithin{equation}{section}
\allowdisplaybreaks[1]

\allowdisplaybreaks[1]

\newtheorem{Def}{Def.}[section]
\newtheorem{Thm}[Def]{Theorem}

\newtheorem{Lemma}[Def]{Lemma}

\newcommand{\Proof}{{\em{Proof. }}}
\newcommand{\QED}{\ \hfill $\FBox$ \\[1em]}
\newcommand{\spc}{\;\;\;\;\;\;\;\;\;\;}
\newcommand{\bra}{\mbox{$< \!\!$ \nolinebreak}}
\newcommand{\ket}{\mbox{\nolinebreak $>$}}
\newcommand{\Bra}{\langle}
\newcommand{\Ket}{\rangle}
\newfont{\Bb}{msbm10 scaled 1095}
\newcommand\C{{\mbox{\Bb C}}}
\newcommand\Z{{\mbox{\Bb Z}}}

\newcommand\R{{\mbox{\Bb R}}}
\newcommand\N{{\mbox{\Bb N}}}

\newcommand{\1}{\mbox{\rm 1 \hspace{-1.05 em} 1}}

\newcommand{\sR}{\mbox{\rm \scriptsize I \hspace{-.8 em} R}}

\newcommand{\Pdd}{\mbox{$\partial$ \hspace{-1.2 em} $/$}}
\newcommand{\slsh}{\mbox{ \hspace{-1.1 em} $/$}}
\newcommand{\Tr}{\mbox{\rm{Tr}\/}}
\newcommand{\tr}{\mbox{tr\/}}
\newcommand{\FBox}{\rule{2mm}{2.25mm}}
\newcommand{\OBox}{\raisebox{.6ex}{\fbox{}}\,}
\newcommand{\Sl}{\mbox{$\prec \!\!$ \nolinebreak}}
\newcommand{\Sr}{\mbox{\nolinebreak $\succ$}}
\newcommand{\Aslsh}{\mbox{ $\!\!A$ \hspace{-1.2 em} $/$}}

\newcommand{\Pexp}{\mbox{\rm{Pexp}}}
\newcommand{\eff}{{\mbox{\scriptsize{eff}}}}
\newcommand{\beq}{\begin{equation}}
\newcommand{\eeq}{\end{equation}}

\newcommand{\np}{n_{\mathrm{p}}}
\newcommand{\na}{n_{\mathrm{a}}}
\renewcommand{\H}{\mathscr{H}}

\makeindex

\begin{document}

\frontmatter
%\title{{{\vspace*{2cm}The Principle of the Fermionic Projector}}}
\title{{{\vspace*{2cm}The Principle of the Fermionic Projector}} \\[.5cm] \Large{Chapters 0-4}}
%\title{{{\vspace*{2cm}The Principle of the Fermionic Projector}} \\[.5cm] \Large{Chapters 5-8}}
%\title{{{\vspace*{2cm}The Principle of the \\ Fermionic Projector}} \\[.5cm] \Large{Appendices}}
\author{\Large{Felix Finster} \\[2cm]
{\normalsize{\bf{Abstract}}} \\[.3cm]
\begin{quote}
\pagestyle{empty}
\footnotesize
The  ``principle of the fermionic projector'' provides a new mathematical 
framework for the formulation of physical theories and is a promising 
approach for physics beyond the standard model. The book begins with a 
brief review of relativity, relativistic quantum mechanics and classical 
gauge theories, with the emphasis on the basic physical concepts and the 
mathematical foundations. The external field problem and Klein's paradox 
are discussed and then resolved by introducing the so-called fermionic 
projector, a global object in space-time which generalizes the notion of 
the Dirac sea. The mathematical core of the book is to give a precise 
definition of the fermionic projector and to employ methods of 
hyperbolic differential equations for its detailed analysis. The 
fermionic projector makes it possible to formulate a new type of 
variational principles in space-time. The mathematical tools for the 
analysis of the corresponding Euler-Lagrange equations are developed. A 
particular variational principle is proposed which gives rise to an 
effective interaction showing many similarities to the interactions of 
the standard model.

The main chapters of the book are easily accessible for beginning 
graduate students in mathematics or physics. Several appendices provide 
supplementary material which will be useful to the experienced researcher.
\end{quote}}
%\end{quote}}
%\address{Fakult{\"a}t f{\"u}r Mathematik, Universit{\"a}t Regensburg,
%93040 Regensburg, Germany}
%\email{Felix.Finster@mathematik.uni-regensburg.de}
%\thanks{The first author was supported in part by NSF Grant \#000000.}
%\date{November 2004}

\maketitle

\setcounter{page}{4}

\tableofcontents

\chapter*{Preface to the Second Online Edition}

\vspace*{-0.25em}

In the almost twelve years since this book was completed, the fermionic projector approach
evolved to what is known today as the theory of causal fermion systems.
There has been progress in several directions: the mathematical
setting was generalized, the mathematical methods were improved and enriched,
and the physical applications have been concretized and worked out in more detail.
The current status of the theory is presented in a coherent way in
the recent monograph~\cite{cfs}. An untechnical physical introduction is given in~\cite{dice2014}.

Due to these developments, parts of the present book are superseded by the
more recent research papers or the monograph~\cite{cfs}.
However, other parts of this book have not been developed further and are still up to date.
For some aspects not covered in~\cite{cfs}, the present book is still the best reference.
Furthermore, the present book is still of interest as being the first publication in which
the causal action principle was presented.
Indeed, comparing the presentation in the present book
to the later developments should give the reader a deeper understanding of why
certain constructions were modified and how they were improved.
In order to facilitate such a study, we now outline the developments which led from the present
book to the monograph~\cite{cfs}. In order not to change the original bibliography,
a list of references to more recent research papers is given at the end of this preface,
where numbers are used (whereas the original bibliography using letters is still
at the end of the book). Similar as in the first online edition, I took the opportunity
to correct a few typos. Also, I added a few footnotes 
beginning with ``{\textsf{Online version}:}''.
Apart from these changes, the online version coincides precisely
with the printed book in the AMS/IP series. In particular, all equation numbers
are the same.

Maybe the most important change in the mathematical setup was the
move from indefinite inner product spaces to Hilbert spaces, as we now explain in detail.
Clearly, the starting point of all my considerations was Dirac theory. On Dirac wave functions in
Minkowski space, one can introduce the two inner products
\begin{align}
( \Psi | \Phi) &= \int_{\sR^3} (\overline{\Psi} \gamma^0 \Phi)(t, \vec{x})\: d^3x \tag{1} \label{sprodMin} \\
\bra \Psi|\Phi \ket &= \int_{\sR^4} \overline{\Psi(x)} \Phi(x) \: d^4x \:. \tag{2} \label{stipMin}
\end{align}
The first inner product~\eqref{sprodMin} is positive definite and thus defines a scalar product.
For solutions of the Dirac equation, it is time independent due to current conservation,
making the solution space of the Dirac equation to a Hilbert space
(more generally, the scalar product can be computed by integrating the normal component
of the Dirac current over any Cauchy surface).
The inner product~\eqref{stipMin}, on the other hand,
is indefinite. It is well-defined and covariant even on wave functions which do not satisfy the Dirac equation,
giving rise to an indefinite inner product space (which can be given a Krein space structure).
It should be pointed out that the time integral in~\eqref{stipMin} in general diverges for solutions of the
Dirac equation, a problem which I always considered to be more of technical than of fundamental nature
(this technical problem can be resolved for example
by working as in~\eqref{56}--\eqref{58} with a $\delta$-normalization in the mass
parameter or by making use of the mass oscillation property as introduced in~\cite{infinite}).

The fermionic projector approach is based on the belief that on a microscopic scale (like the Planck scale),
space-time should not be modeled by Minkowski space but should have
a different, possibly discrete structure. Consequently, the Dirac equation in Minkowski space
should not be considered as being fundamental, but it should be replaced by equations of different type.
For such a more fundamental description, the scalar product~\eqref{sprodMin} is problematic, 
because it is not clear how the analog of an integral over a hypersurface should be defined,
and why this integral should be independent of the choice of the hypersurface.
The indefinite inner product~\eqref{stipMin}, however, can easily be generalized to 
for example a discrete space-time if one simply replaces the
integral in~\eqref{stipMin} by a sum over all space-time points.
Such considerations led me to regard the indefinite inner product~\eqref{stipMin} as being
more fundamental than~\eqref{sprodMin}. This is the reason why throughout this book,
we work preferably with indefinite inner product spaces. In particular, the structure of
``discrete space-time'' is introduced on an underlying indefinite inner product space
(see~\S\ref{psec13}).

My views changed gradually over the past few years. The first input which triggered this process
was obtained when developing the existence theory for the causal action principle.
While working on this problem in the simplest setting of a finite number of
space-time points~\cite{discrete}, it became clear that in
order to ensure the existence of minimizers, one needs to assume that the 
image of the fermionic projector~$P$ is a {\em{negative definite}} subspace
of the indefinite inner product space~$(H, \bra .|. \ket)$. The fact that~$P$ has
a definite image makes it possible to introduce a Hilbert space~$(\H, \langle .|. \rangle_\H)$
by setting~$\langle .|. \rangle_\H = -\bra .| P\, . \ket$ and dividing out the null space.
This construction, which was first given in~\cite[Section~1.2.2]{rrev}, gave
an underlying Hilbert space structure. However, at this time, the connection of the corresponding
scalar product to integrals over hypersurfaces as in~\eqref{sprodMin} remained obscure.

From the mathematical point of view, having an underlying Hilbert space structure has the major
benefit that functional analytic methods in Hilbert spaces become applicable.
When thinking about how to apply these methods, it became clear that also measure-theoretic
methods are useful. This led me to generalize the mathematical setting such as to allow
for the description of not only discrete, but also continuous space-times.
This setting was first introduced in~\cite{continuum} when working out the existence theory.
This analysis also clarified which constraints one must impose in order to obtain
a mathematically well-posed variational problem.

The constructions in~\cite{continuum} also inspired the notion of the {\em{universal measure}},
as we now outline. When working out the existence theory, it became clear that instead of using the
kernel of the fermionic projector, the causal action principle can be formulated equivalently
in terms of the local correlation operators~$F(x)$ which relate the Hilbert space scalar product
to the spin scalar product by
\[ \langle \psi | F(x) \phi \rangle_\H = -\Sl \psi(x) | \phi(x) \Sr_x \:. \]
In this formulation, the only a-priori structure of space-time is that of being a measure space~$(M, \mu)$.
The local correlation operators give rise to a mapping
\[ F \::\: M \rightarrow {\mathscr{F}} \:,\quad x \mapsto F(x) \:, \]
where~${\mathscr{F}}$ is the subset of finite rank operators on~$\H$ which are symmetric and 
(counting multiplicities) have at most~$2N$ positive and at most~$2N$ negative eigenvalues
(where~$N$ denotes the number of sectors).
Then, instead of working with the measure~$\mu$, the causal action can be expressed
in terms of the push-forward measure~$\rho = F_* \mu$, being a measure on~${\mathscr{F}}$
(defined by~$\rho(\Omega) = \mu(F^{-1}(\Omega))$).
As a consequence, it seems natural to leave out the measure space~$(M, \mu)$ and to work instead
directly with the measure~$\rho$ on~${\mathscr{F}}$, referred to as the universal measure.
We remark that working with~$(M, \mu)$ has the potential benefit that it is possible to
prescribe properties of the measure~$\rho$. In particular, if~$\mu$ is a discrete measure,
then so is~$\rho$ (for details see~\cite[Section~1.2]{continuum}). However, the analysis of the
causal action principle in~\cite{support} suggests
that minimizing measures are always discrete, even if one varies over all regular Borel measures
(which may have discrete and continuous components).
With this in mind, it seems unnecessary to arrange the discreteness of the measure~$\rho$
by starting with a discrete measure space~$(M, \mu)$.
Then the measure space~$(M, \mu)$ becomes obsolete.
These considerations led me to the conviction that one should work with the universal measure~$\rho$,
which should be varied within the class of all regular Borel measures. Working with general regular Borel measures
also has the advantage that it becomes possible to take convex combinations of universal measures,
which seems essential for getting the connection to second-quantized bosonic fields
(see the notions of decoherent replicas of space-time and of microscopic mixing of wave functions
in~\cite{qft} and~\cite{qftlimit}).

Combining all the above results led to the framework of {\em{causal fermion systems}}, where
a physical system is described by a Hilbert space~$(\H, \langle .|. \rangle_\H)$ and
the universal measure~$\rho$ on~${\mathscr{F}}$. This framework was first introduced in~\cite{rrev}.
Subsequently, the analytic, geometric and topological structures encoded in a causal fermion system
were worked out systematically; for an overview see~\cite[Chapter~1]{cfs}.

From the conceptual point of view, the setting of causal fermion systems
and the notion of the universal measure considerably changed both the
role of the causal action principle and the concept of what space-time is.
Namely, in the causal action principle in this book, one varies the fermionic projector~$P$ in
a given discrete space-time.
In the setting of causal fermion systems, however, one varies instead the universal measure~$\rho$,
being a measure on linear operators on an abstract Hilbert space.
In the latter formulation, there is no space-time to begin with. On the contrary, space-time is
introduced later as the support of the universal measure.
In this way, the causal action principle evolved from a variational principle for wave functions
in space-time to a variational principle for space-time itself as well as for all structures therein.

In order to complete the summary of the conceptual modifications, we remark that
the connection between the scalar product~$\langle .|. \rangle_\H$ and surface integrals as in~\eqref{sprodMin},
which had been obscure for quite a while, was finally clarified when working out Noether-like theorems
for causal variational principles~\cite{noether}. Namely, surface integrals now have a proper generalization to
causal fermion systems in terms of so-called {\em{surface layer integrals}}.
It was shown that the symmetry of the causal action under unitary transformations acting on~$\mathscr{F}$
gives rise to conserved charges which can be expressed by surface layer integrals.
For Dirac sea configurations, these conserved charges coincide with the surface
integrals~\eqref{sprodMin}.

Another major development concerns the description of {\em{neutrinos}}.
In order to explain how these developments came about, we first note that
in this book, neutrinos are modelled as left-handed massless Dirac particles (see~\S\ref{esec1}).
This has the benefit that the neutrinos drop out of the closed chain due to
chiral cancellations (see~\S\ref{esec21} and~\S\ref{esec22}).
When writing this book, I liked chiral cancellations, and I even regarded them as
a possible explanation for the fact that neutrinos appear only with one chirality.
As a side remark, I would like to mention that
I was never concerned about experimental observations which indicate that
neutrinos do have a rest mass, because I felt that these experiments are too indirect
for making a clear case. Namely, measurements only tell us that there are
fewer neutrinos on earth than expected from the number of neutrinos generated in fusion processes
in the sun. The conventional explanation for this seeming disappearance
of solar neutrinos is via neutrino oscillations, making it
necessary to consider massive neutrinos. However, it always seemed to me that
there could be other explanations for the lack of neutrinos on earth
(for example, a modification of the weak interaction or other, yet unknown fundamental forces),
in which case the neutrinos could well be massless.

My motivation for departing from massless neutrinos was not related to experimental evidence, but
had to do with problems of mathematical consistency. Namely, I noticed that left-handed neutrinos
do not give rise to stable minimizers of the causal action (see~\cite[Section~4.2]{cfs}).
This general result led me to incorporate right-handed neutrino components, and to
explain the fact that only the left-handed component is observed by the postulate that the
regularization breaks the chiral symmetry. This procedure cured the mathematical consistency
problems and had the desired side effect that neutrinos could have a rest mass, in agreement with
neutrino oscillations.

We now comment on other developments which are of more technical nature.
These developments were mainly triggered by minor errors or shortcomings in the present book.
First, Andreas Grotz noticed when working on his master thesis in 2007 that the
normalization conditions for the fermionic projector as given in~\eqref{eq:2a1} and~\eqref{eq:2a2}
are in general violated to higher order in perturbation theory. This error was corrected in~\cite{grotz}
by a rescaling procedure.
This construction showed that there are two different perturbation expansions: with and
without rescaling. The deeper meaning of these two expansions became clearer later when
working out different normalizations of the fermionic projector.
This study was initiated by the quest for a non-perturbative construction of the fermionic projector,
as was carried out in globally hyperbolic space-times in~\cite{finite, infinite}.
It turned out that in space-times of finite lifetime, one cannot work with the $\delta$-normalization in the mass
parameter as used in~\eqref{56}--\eqref{58} (the ``mass normalization''). Instead, a proper normalization
is obtained by using a scalar product~$(.|.)$ which is represented similar to~\eqref{sprodMin}
by an integral over a spacelike hypersurface (the ``spatial normalization'').
As worked out in detail in~\cite{norm} with a convenient contour integral method,
the causal perturbation expansion without rescaling realizes the spatial normalization condition,
whereas the rescaling procedure in~\cite{grotz} gives rise to the mass normalization.
The constructions in curved space-time in~\cite{finite, infinite} as well as the general connection between the
scalar product~$( .|. )$ and the surface layer integrals in~\cite{noether} showed that
the physically correct and mathematically consistent normalization condition is
the spatial normalization condition. With this in mind, the combinatorics of the causal perturbation
expansion in this book is indeed correct, but the resulting fermionic projector does not satisfy the mass
but the spatial normalization condition.

Clearly, the analysis of the continuum limit in Chapters~\ref{esec3}--\ref{esec5}
is superseded by the much more detailed analysis in~\cite[Chapters~3-5]{cfs}.
A major change concerns the treatment of the logarithmic singularities on the light cone,
as we now point out. In the present book, some of the contributions involving logarithms are
arranged to vanish by imposing that the regularization should satisfy the relation~\eqref{e:3C}.
I tried for quite a while to construct an example of a regularization which realizes this relation,
until I finally realized that there is no such regularization, for the following reason: \\[-0.5em]

\textsc{Lemma~I.}
There is no regularization which satisfies the condition~\eqref{e:3C}.

\Proof The linear combination of monomials~$M$ in~\eqref{e:3B} involves a factor~$T^{(1)}_{[2]}$,
which has a logarithmic pole on the light cone (see~\eqref{Tldef}, \eqref{Tadef}
and~\eqref{l:3.1}). Restricting attention to the corresponding
contribution~$\sim \log|\vec{\xi}|$, we have
\[ M \asymp -\frac{1}{16 \pi^3}\:
T^{(-1)}_{[0]}\: \overline{T^{(-1)}_{[0]}\: T^{(0)}_{[0]}}\:\log|\vec{\xi}| \:. \]
As a consequence,
\begin{align*}
(M & - \overline{M})\: \overline{T^{(0)}_{[0]}}^{-1} =
-\frac{\log|\vec{\xi}|}{16 \pi^3}\: \frac{\big| T^{(-1)}_{[0]} \big|^2}{\overline{T^{(0)}_{[0]}}}
\big( \overline{T^{(0)}_{[0]}} - T^{(0)}_{[0]} \big)  \\
&= -\frac{\log|\vec{\xi}|}{16 \pi^3}\: \bigg| \frac{T^{(-1)}_{[0]}}{T^{(0)}_{[0]}} \bigg|^2
\Big( \big| T^{(0)}_{[0]} \big|^2 - \big( T^{(0)}_{[0]} \big)^2 \Big)
= -\frac{\log|\vec{\xi}|}{8 \pi^3}\: \bigg| \frac{T^{(-1)}_{[0]}}{T^{(0)}_{[0]}} \bigg|^2
\, \Big(\text{Im} \,T^{(0)}_{[0]} \Big)^2 \leq 0 \:.
\end{align*}
Since this expression has a fixed sign, it vanishes in a weak evaluation on the light cone
only if it vanishes identically to the required degree.
According to~\eqref{l:3.1}, the function~$\text{Im}\, T^{(0)}_{[0]}$ is a regularization
of the distribution~$-i \pi \delta (\xi^2) \:\varepsilon (\xi^0)/(8 \pi^3)$ on the scale~$\varepsilon$.
Hence on the light cone it is of the order~$\varepsilon^{-1}$. This gives the claim.
\QED
This no-go result led me to reconsider the whole procedure of the continuum limit.
At the same time, I tried to avoid imposing relations between the regularization parameters,
which I never felt comfortable with because I wanted the continuum limit to work for at
least a generic class of regularizations. Resolving this important issue took
me a lot of time and effort. My considerations eventually led to the method of compensating the
logarithmic poles by a {\em{microlocal chiral transformation}}.
These construction as well as many preliminary considerations are given in~\cite[Section~3.7]{cfs}.

Finally, I would like to make a few comments on each chapter of the book.
Chapters~\ref{secintro}--\ref{secpfp} are still up to date, except for the generalizations
and modifications mentioned above. Compared to the presentation in~\cite{cfs},
I see the benefit that these chapters might be easier to read and might convey
a more intuitive picture of the underlying physical ideas.
Chapter~\ref{psec2} is still the best reference for the general derivation of the formalism of the
continuum limit. In~\cite[Chapter~2]{cfs} I merely explained the regularization effects in examples
and gave an overview of the methods and results in Chapter~\ref{psec2}, but without repeating
the detailed constructions. Chapter~\ref{esec2} is still the only reference where the
form of the causal action is motivated and derived step by step. Also, the
notion of state stability is introduced in detail, thus providing the basis for the
later analysis in~\cite{reg, vacstab}. As already mentioned above, the analysis in
Chapters~\ref{esec3}--\ref{secegg} is outdated. I recommend the reader to study
instead~\cite[Chapters~3--5]{cfs}. The Appendices are still valuable. I added a few
footnotes which point to later improvements and further developments.
%
%I hope that the reader will find these remarks helpful.
\\[1.5em]
\hspace*{1cm} \hfill Felix Finster, Regensburg, August 2016 \\[0.5cm]

%\bibliographystyle{amsplain}
%\bibliography{../../aarbeit/felix}
%\newpage

\centerline{\large{\bf{References}}}

\vspace*{.5em}

\newcommand\oldchapter{}
\let\oldchapter=\chapter
\renewcommand{\chapter}[2]{}

\providecommand{\bysame}{\leavevmode\hbox to3em{\hrulefill}\thinspace}
\providecommand{\MR}{\relax\ifhmode\unskip\space\fi MR }
% \MRhref is called by the amsart/book/proc definition of \MR.
\providecommand{\MRhref}[2]{%
  \href{http://www.ams.org/mathscinet-getitem?mr=#1}{#2}
}
\providecommand{\href}[2]{#2}

\let\chapter=\oldchapter

%!TEX root = finsterpfp.tex
\chapter*{Preface to the First Online Edition}
In the few years since the book appeared, I was frequently asked if the introductory chapters
were also available online. Also, I heard complaints that the preprints on the
arXiv on the ``principle of the fermionic projector'' were preliminary
versions which were not quite compatible with the book and with subsequent papers.
In order to improve the situation, I decided to replace my original
preprints hep-th/0001048, hep-th/0202059 and hep-th/0210121
by the corresponding chapters of the book.

I took the opportunity to correct some typos. I also added a few footnotes
beginning with ``{\textsf{Online version}:}'' 
which point out to later developments.
Apart from these changes, the present online version coincides precisely
with the book in the AMS/IP series. In particular, all equation numbers
are the same.
\\[1.5em]
\hspace*{1cm} \hfill Felix Finster, Regensburg, October 2009

\chapter*{Preface}
The basic ideas behind the ``principle of the fermionic projector'' go back to 
the years 1990-91 when I was a physics student in Heidelberg. At that time, I 
was excited about relativity and quantum mechanics, in particular about 
classical Dirac theory, but I felt uncomfortable with quantum field theory. The 
dissatisfaction with second quantized fields, which was in part simply a
beginner's natural skepticism towards an unfamiliar physical concept, was my 
motivation for thinking about possible alternatives. Today I clearly understand 
quantum field theory much better, and many of my early difficulties have 
disappeared. Nevertheless, some of my objections remain, and the idea of 
formulating physics in a unified way based on Dirac's original concept of a ``sea 
of interacting particles'' seems so natural and promising to me that I have 
pursued this idea ever since. It took quite long to get from the first 
ideas to a consistent theory, mainly because mathematical methods had to be 
developed in order to understand the ``collective behavior'' of the particles 
of the Dirac sea.

This book gives me the opportunity to present the main ideas and 
methods in a somewhat broader context, with the intention of making this area 
of mathematical physics accessible to both theoretical physicists and applied 
mathematicians. The emphasis of the main chapters is on the conceptual part,
whereas the more technical aspects are worked out in the appendices.

I am grateful to Claus Gerhardt, Joel Smoller, Shing-Tung Yau and
Eberhard Zeidler for their encouragement and support.  I would like to thank
Stefan Hoch, Niky Kamran, Johann Kronthaler, W{\"a}tzold Plaum and Joel Smoller 
for helpful comments, and Eva R{\"u}tz for the typesetting. Finally, I
am grateful to the Max Planck Institute for Mathematics in the
Sciences, Leipzig, and the Morningside Center, Beijing, for their
hospitality while I was working on the manuscript. \\[1.5em]
\hspace*{1cm} \hfill Felix Finster, Regensburg, November 2004

\mainmatter

\setcounter{chapter}{-1}
\chapter{The Principle of the Fermionic Projector -- A New Mathematical
Model of Space-Time}
\markboth{0. THE PFP -- A NEW MATHEMATICAL MODEL OF SPACE-TIME}
{0. THE PFP -- A NEW MATHEMATICAL MODEL OF SPACE-TIME}
\label{sec0}
The mathematical model of space-time has evolved in
history. In Newtonian mechanics, space is described by a Euclidean
vector space. In special relativity, space and time were combined
to Minkowski space, a vector space endowed with a scalar product
of signature $(+ \ \!\! - \ \!\! - \ \! - )$. In general
relativity, the vector space structure of space-time was given up
on the large scale and was replaced by that of a Lorentzian manifold.
The first hint that the notions of space and time should be
modified also on the microscopic scale was obtained by Planck, who
observed that the gravitational constant, Planck's constant and
the speed of light give rise to a quantity of the dimension of
length,
\[ l_P \;=\; \sqrt{\frac{\hbar \:\kappa}{c^3}} \;\approx\;
1.6 \cdot 10^{-35} \:{\mbox{m}} \;, \]
and he conjectured that for distances as tiny as this so-called Planck
length\index{Planck length},
the conventional laws of physics should no longer hold, and yet unknown physical
effects might become significant. Later, this picture was confirmed by quantum
field theory. Namely, due to the ultraviolet divergences, perturbative QFT is
well-defined only after regularization, and the regularization is then removed
using the renormalization procedure.
While renormalization ensures that the observable quantities do not depend
on the regularization, the theoretical justification for the renormalization
program lies in the assumption that the continuum theory should be valid only
down to some microscopic length scale, and it seems natural to associate this
length scale to the Planck length.

\begin{figure}[tb]
\begin{center}
\scalebox{0.22}
{\includegraphics{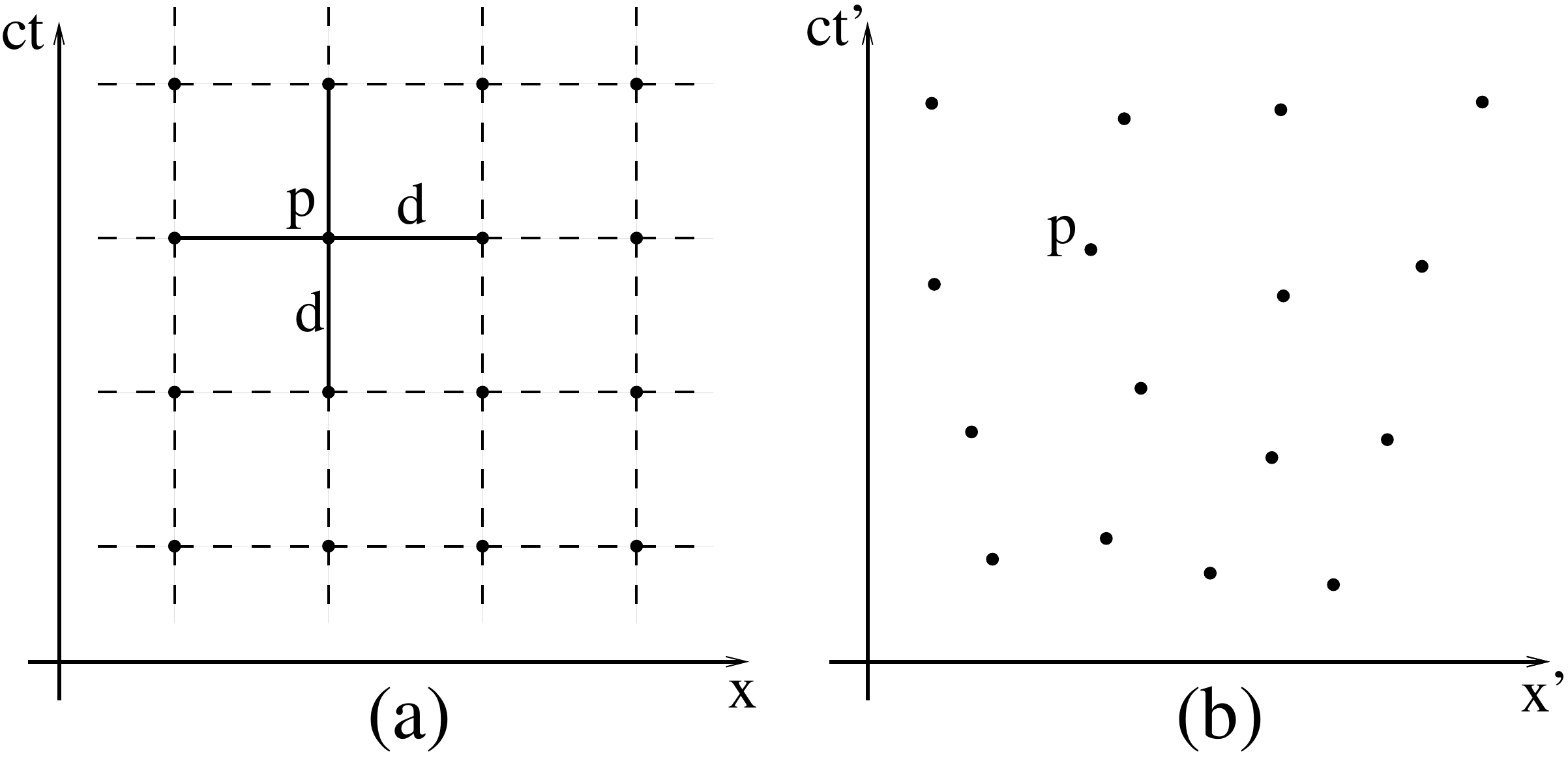}}
\caption{A lattice regularization for two different observers.}\label{fig0}
\end{center}
\end{figure}

Today most physicists agree that in order to make progress
in fundamental physics, one should go beyond the continuum field theory and try to get a better understanding of the microscopic structure of space-time.
However, giving up the usual space-time continuum leads to serious
problems, and this is one reason why there is no
consensus on what the correct mathematical model for ``Planck scale physics'' should be.
Let us illustrate the difficulties by briefly discussing
a few of the many approaches.
The simplest and maybe most natural approach is to assume that on the
Planck scale space-time is no longer a continuum but becomes in some
way ``discrete.'' This idea is for example used in lattice gauge
theories, where space-time is modeled by a four-dimensional lattice
(see Figure~\ref{fig0}(a)).
Using the specific structures of a lattice like
the nearest-neighbor relation and the lattice spacing $d$, one can set up
a quantum field theory which is ultraviolet finite~\cite{Ro}.
Lattice gauge theories are very useful for numerical
simulations~\cite{numerik}. However, they are not fully satisfying
from a conceptual point of view
because a space-time lattice is not consistent
with the equivalence principle of general relativity.
Namely, if one considers the lattice in the reference frame of an
accelerated observer (denoted in in Figure~\ref{fig0}(b) by~$(t',x')$),
the lattice points are no longer in a regular configuration.
Thus the structure of a lattice is not invariant under general
coordinate transformations and hence is not compatible with the
equivalence principle.

An alternative approach is to hold on to a space-time continuum, but
to work with objects which are spread out in space-time, thereby
avoiding the ultraviolet problems which occur for very small
distances. The most prominent example in this direction is string
theory, where physics is on a fundamental level described by
so-called strings, which are extended in space-time and are
therefore ultraviolet finite. The basic problem with such theories
is that they are formulated using the structures of an underlying
continuum space-time (like the vector space structure, the topology
or even a metric), although all observable quantities (like the
Lorentz metric, particles, fields, etc.) are to be derived from the
non-localized objects, without referring to the underlying
space-time continuum. Therefore, the structures of the underlying
``continuum background'' may seem artificial, and serious conceptual
problems arise when these background structures are not compatible
with basic physical principles (for example, a background vector
space is not compatible with the equivalence principle). For short,
one says that the theory is not background-free (for a more detailed
discussion see~\cite{Baez} and the references therein).

Thus one difficulty in finding a promising model for
Planck scale physics is that it should be background-free and
should respect the basic physical principles (like the
equivalence principle, the local gauge principle, etc.).
There are approaches which
actually meet these requirements. One is Connes' noncommutative geometry.
As pointed out by Grothendieck, there is a one-to-one correspondence
between the points of a manifold and the prime ideals of the (commutative)
algebra of functions on this manifold.  Thus the geometry of a manifold can be
recovered from an underlying algebraic structure, and this makes it possible to
extend the notions of space and time by considering more general,
noncommutative algebras (see~\cite{Co, AC2} for details and physical
applications). The other approach is quantum gravity as pursued by Ashtekar and
his school~\cite{As, Th}. The hope is that the ultraviolet divergences of
QFT should disappear as soon as gravity, quantized in a non-perturbative
way, is included.

Ultimately, a model for space-time on the Planck scale must be verified
or falsified by physical experiments. Unfortunately, experiments on the
Planck scale would require such enormously high energies that they are
at present out of reach. Indirect experiments seem possible in
principle~\cite{Ah} but have so far not been carried out.
In my opinion, one should not hope for important new experimental
input in the near future, but one should try to make due with the
experimental data which is now available. Indeed, this situation is
not as hopeless as it might appear at first sight. Namely, in present
physical models like the standard model, a
lot of information from experiments is built in empirically, like
the masses of the elementary particles, the gauge groups, the coupling
constants, etc. Therefore, one can get a connection to experiments simply by
trying to reproduce this well known empirical data. If successful,
this procedure could give strong physical evidence for a model. For
example, even if based on ad-hoc assumptions on the microscopic structure
of space-time (which cannot be verified directly), a model would be very
convincing and physically interesting if it gave the correct value for
the fine structure constant and explained e.g.\ why the strong gauge
group is $SU(3)$ or why neutrinos do not couple to the
electromagnetic field. Thus the goal of a mathematical model for space-time on
the Planck scale is to give a more fundamental explanation for the
structures and empirical parameters in the standard model. To our opinion,
only such concrete results can justify the model. Clearly, it is far
from obvious how a promising model should look, or even in which
mathematical framework it should be formulated. But at least for a
mathematician, this makes the problem only more interesting, and it seems a
challenging program to search for such models and to develop the mathematical
methods needed for their analysis.

Our point of view that the mathematical model needs justification by known
experimental data is not just a requirement which the model should fulfill at
the very end, but it also gives a few hints on how one should proceed in
order to find a
promising model. First of all, one can expect concrete results only if one
makes specific assumptions. Therefore, generalizing the notion of a Lorentzian
manifold does not seem to be sufficient, but one should make a concrete ansatz
for the microscopic structure of space-time (as it is done e.g.\ in string theory and
lattice gauge theories). Furthermore, in order to make it easier to get a
connection to well-established theories like classical field theory and quantum
mechanics, it seems a good idea to take these theories as the starting point and
to try to work as closely as possible with the objects used in these theories.
Namely, if one drops important structures of the classical theories and/or
introduces too many new structures ad hoc, it might become very difficult if
not impossible to obtain a relation to observable data.

In our model of space-time we have tried to follow the above considerations.
Our starting point is relativistic quantum mechanics and classical field theory.
We assume that space-time is discrete on the Planck scale.  But our notion of
``discrete space-time'' is much more general than a four-dimensional lattice; in particular, we do not assume any discrete
symmetries in space-time, we keep the local gauge freedom, and we also extend the
diffeomorphism invariance of a manifold in such a way that the equivalence
principle is respected in discrete space-time.
Furthermore, our model is background-free.
In contrast to string theory, we
do not introduce any new objects, but hold on to the structures already present in
classical Dirac theory.  We build in our physical ideas simply by prescribing
which of these structures we consider as being fundamental, and then carry over
these structures to discrete space-time.  In the resulting mathematical
framework, it is impossible to formulate the conventional physical equations,
and thus we propose instead new equations of different type, referred to as the
equations of discrete space-time. In a certain limiting case, the so-called
continuum limit, we get a connection to the conventional formulation of physics
in a space-time continuum. We point out that, in contrast to the
Ashtekar program, we do not work with second quantized fields.  But our
concept is that the equations of discrete space-time should also account for
the physical effects of quantized fields if one goes beyond the continuum
limit.

More specifically, we describe the physical system by the
{\em{fermionic projector}} $P(x,y)$, which can be regarded as the
projector on all occupied fermionic states of the system,
including the states of the Dirac sea.
After carrying over the fermionic projector to discrete space-time, we
can set up variational principles like our ``model variational principle''
\[ \sum_{x,y \in M} {\mathcal{L}}[P(x,y)\: P(y,x)] \;\to\; \min\:, \]
where the ``Lagrangian''~${\mathcal{L}}$ is given by
\[ {\mathcal{L}}[A] \;=\; |A^2| - \mu |A|^2 \:, \]
with~$\mu$ a Lagrangian multiplier.
Here~$|A|$ is the so-called spectral weight defined as the sum of the
absolute values of the eigenvalues of the matrix~$A$ (or, in case that~$A$
is not diagonalizable, of the zeros of its characteristic polynomial).
We study the above variational principle for a fermionic projector which in the
vacuum is the direct sum of seven identical massive sectors and one massless
left-handed sector, each of which is composed of
three Dirac seas. Analyzing the continuum limit for an
interaction via general chiral and (pseudo)scalar potentials, we find
that the sectors spontaneously form pairs, which are referred to
as blocks. The resulting so-called effective interaction can be described
by chiral potentials corresponding to the effective gauge group
\[ SU(2) \times SU(3) \times U(1)^3 \;. \]
This model has striking similarity to the standard model if the block
containing the left-handed sector is identified with the leptons and the
three other blocks with the quarks. Namely, the effective gauge fields
have the following properties.
\begin{itemize}
\item The $SU(3)$ corresponds to an unbroken gauge symmetry. The
$SU(3)$ gauge fields couple to the quarks exactly
as the strong gauge fields in the standard model.
\item The $SU(2)$ potentials are left-handed and couple to the leptons
and quarks exactly as the weak gauge potentials in the standard model.
Similar to the CKM mixing in the standard model,
the off-diagonal components of these potentials must involve a
non-trivial mixing of the generations.
The $SU(2)$ gauge symmetry is spontaneously broken.
\item The $U(1)$ of electrodynamics can be identified with an Abelian
subgroup of the effective gauge group.
\end{itemize}
The effective gauge group is larger than the gauge group of the standard
model, but this is not inconsistent because a more detailed analysis of
our variational principle should give further constraints for the
Abelian gauge potentials. Moreover, there are the following differences
to the standard model, which we derive mathematically without working
out their physical implications.
\begin{itemize}
\item The $SU(2)$ gauge field tensor $F$ must be simple in the sense
that $F=\Lambda \:s$ for a real 2-form $\Lambda$ and an $su(2)$-valued
function $s$.
\item In the lepton block, the off-diagonal $SU(2)$ gauge potentials are
associated with a new type of potential, called nil potential, which couples
to the right-handed component.
\end{itemize}
These results give a strong indication that the principle of the fermionic
projector is of physical significance.
Nevertheless, the goal of this book is not to work out our model variational
principle in all details.
Our intention is to develop the general concepts and methods from the basics,
making them easily accessible to the reader who might be interested in
applying them to other equations of discrete space-time or to related
problems.

These notes are organized as follows. In order to make the
presentation self-contained, Chapter~\ref{secintro} gives a brief
account of the mathematical and physical preliminaries.
Chapter~\ref{sec3} introduces the fermionic projector in the
continuum and provides the mathematical methods needed for its
detailed analysis. In Chapter~\ref{secpfp} we go beyond the
continuum description and introduce our mathematical model for
space-time on the Planck scale. In Chapter \ref{psec2} we develop a
mathematical formalism suitable for the analysis of the continuum
limit. In Chapter \ref{esec2} we present and discuss different
equations of discrete space-time in the vacuum, and we choose the
most promising equations as our ``model equations''. In the last
Chapters \ref{esec3}-\ref{esec5} we analyze interacting systems in
the continuum limit. The appendices contain additional material and
will be referred to from the main chapters.

% Introduction
\chapter{Preliminaries} \label{secintro}
\setcounter{equation}{0}

\section{Relativity} \label{isec1} \setcounter{equation}{0}
\setcounter{equation}{0}
In this section we briefly outline the mathematical framework
of special and general relativity (for a more detailed
introduction see~\cite{Woodhouse} and~\cite{Wald}). We always work in
{\em{normal units}} where $\hbar = c = 1$\index{units}.
In special relativity, space-time is described by
Minkowski space\index{Minkowski space} $(M, \Bra
.,. \Ket)\label{Mrangle}$, a real $4$-dimensional vector space endowed with an inner product of signature $(+ \ \!\! - \ \!\! - \ \! - )$. Thus,
choosing a pseudo-orthonormal basis $(e_i)_{i=0,\ldots,3}$ and representing the vectors of $M$ in this basis, $\xi = \sum_{i=0}^3 \xi^i e_i$, the inner product takes the form
\beq \label{minsp}
\Bra \xi, \eta \Ket \;=\; \sum_{j,k=0}^3 g_{jk}\: \xi^j\: \eta^k\:,
\eeq
where $g_{ij}$, the {\em{Minkowski metric}}, is the diagonal matrix
$g={\mbox{diag}}\,(1,-1,-1,-1)$. In what follows we usually omit the
sums using Einstein's summation convention
(i.e.\ we sum over all indices which appear twice, once as an
upper and once as a lower index). Also, we sometimes abbreviate the
Minkowski scalar product by writing $\xi \eta := \Bra \xi, \eta \Ket$
and $\xi^2 := \Bra \xi, \xi \Ket$. A pseudo-orthonormal basis
$(e_i)_{i=0,\ldots,3}$ is in physics called a {\em{reference frame}}\index{reference frame},
because the corresponding coordinate system $(x^i)$ of Minkowski space
gives the time and space coordinates for an observer in a system of
inertia. We also refer to $t := x^0$ as time and denote the spatial
coordinates by $\vec{x}=(x^1, x^2, x^3)$.

The sign of the Minkowski metric encodes the causal structure
of space-time. Namely, a vector $\xi \in M$ is said to be
\[ \left. \begin{array}{cll}
{\mbox{\em{timelike}}\index{timelike vector}} &\quad& {\mbox{if $\Bra \xi, \xi \Ket >0$}} \\
{\mbox{\em{spacelike}}\index{spacelike vector}} && {\mbox{if $\Bra \xi, \xi \Ket <0$}} \\
{\mbox{\em{null}}\index{null vector}} && {\mbox{if $\Bra \xi, \xi \Ket =0$}}\:. \end{array}
 \right\} \]
Likewise, a vector is called {\em{non-spacelike}} if it is timelike
or null. The null vectors form the double cone $L \label{E} =
\{ \xi \in M \:|\: \Bra \xi, \xi \Ket =0\}$, referred to as the
{\em{light cone}}\index{light cone}. Physically, the light cone is formed of all
rays through the origin of $M$ which propagate with the speed of
light. Similarly, the timelike vectors correspond to velocities
slower than light speed; they form the {\em{interior light
    cone}}\index{light cone!interior}
$I = \{ \xi \in M \:|\: \Bra \xi, \xi \Ket > 0 \}$.
Finally, we introduce the {\em{closed light cone}}\index{light cone!closed}
$J = \{ \xi \in M \:|\: \Bra \xi, \xi \Ket \geq 0 \}$.
The space-time trajectory of a moving object describes a curve
$q(\tau)$ in Minkowski space (with $\tau$ an arbitrary parameter).
We say that the space-time curve $q$ is timelike if the tangent vector
to $q$ is everywhere timelike. Spacelike, null, and
non-spacelike curves are defined analogously.
The usual statement of causality that no information
can travel faster than with the speed of light can then be
expressed as follows,
\begin{quote}
{\em{causality}}\index{causality}: \quad \parbox[t]{7.4cm}{
information can be transmitted only along
non-spacelike curves.} \\[-0em]
\end{quote}
The set of all points which can be joined with a given
space-time point $x$ by a non-spacelike curve is
precisely the closed light cone centered at $x$, denoted by
$J_x\label{Jx} := J - x$. It is the union of the two single cones
\begin{eqnarray*}
J_x^\lor\label{Jlorx} &=& \{ y \in M \:|\: (y-x)^2 \geq 0,\;
(y^0-x^0) \geq 0 \} \\
J_x^\land\label{Jlandx} &=& \{ y \in M \:|\: (y-x)^2 \geq 0, \;
(y^0-x^0) \leq 0 \} \:,
\end{eqnarray*}
which have the interpretation as the points in the causal future and
past of $x$, respectively. Thus we refer to $J^\lor_x$ and
$J^\land_x$ as the closed {\em{future}}\index{light cone!future} and {\em{past
light cones}}\index{light cone!past}
centered at $x$, respectively. Similarly, we also introduce the
sets $I^\lor_x\label{Ilorx}$, $I^\land_x\label{Ilandx}$ and $L^\lor_x\label{Elorx}$, $L^\land_x\label{Elandx}$.

A linear transformation of Minkowski space which leaves the form
of the Min\-kow\-ski metric~(\ref{minsp}) invariant is called a
{\em{Lorentz transformation}}\index{Lorentz transformation}.
The Lorentz transformations form
a group, the {\em{Lorentz group}}\index{Lorentz group}.
The Lorentz transformations which preserve both the time direction
and the space orientation form a subgroup of the Lorentz group,
the {\em{orthochronous proper Lorentz group}}\index{Lorentz group!orthochronous proper}.

The physical equations should be {\em{Lorentz invariant}}, meaning that
they must be formulated in Minkowski space, independent of the reference
frame. A convenient way of making Lorentz invariance manifest is to
bring the equations in tensorial form (see~\cite{Lichnerowicz} for
a good introduction). Writing out the tensor indices, we get upper
``contravariant'' and lower ``covariant'' indices, which can be
transformed into each other by contracting with the metric, e.g.\
$\xi_j = g_{jk} \xi^k$ and $\xi^k = g^{kl} \xi_l$ with
$g^{kl} = (g_{kl})^{-1}$. In order to formulate electrodynamics in
a manifestly Lorentz invariant form, one combines the electric potential
and the vector potential to a $1$-form $A=A_j dx^j$, the
{\em{electromagnetic potential}}\index{potential!electromagnetic}. The electric and magnetic
fields are then components of the electromagnetic {\em{field tensor}}
$F$ defined by
\[ F = dA \quad\qquad {\mbox{or, in components,}} \quad\qquad
F_{jk} \;=\; \partial_j A_k - \partial_k A_j\:. \]
The Maxwell equations take the form \index{Maxwell's equations}
\beq \label{maxwell}
\partial_k F^{kl} \;=\; C\, j^l\:,
\eeq
where $j$ is the so-called $4$-current and the constant $C=4 \pi e$
involves the electromagnetic coupling constant\index{electromagnetic coupling constant} (we use the sign convention where~$e>0$).
For an observer in a reference frame, the time and spatial components
$j^0$ and $\vec{\jmath}$ of the $4$-current have the interpretation as
the electric charge density and the electric current density,
respectively. Since we shall always work in the $4$-dimensional setting,
it is unambiguous to refer to $j$ simply as the {\em{electromagnetic
    current}}\index{electromagnetic current}. Now consider a point particle of mass $m$ and unit charge
$e$ in a given (=external) electromagnetic field.
Since by causality its velocity is always smaller than light speed,
its trajectory is a timelike curve $q(\tau)$. Thus we can parametrize
it by the arc length, i.e.
\[ \Bra u,u \Ket \equiv 1 \qquad {\mbox{with}} \qquad
u(\tau) := \frac{d}{d\tau} q(\tau)\:. \] In a reference frame, the
time and spatial components of the vector $m \!\cdot\! u(\tau)$ are
the energy and momentum of the particle at the space-time point
$q(\tau)$. We refer to $m u$ as the {\em{momentum}} of the particle.
The parameter $\tau$ has the interpretation as the {\em{proper
time}}\index{proper time} of an observer moving along $q$. The
equation of motion is the tensor equation $m \frac{d}{d\tau}u^j = -e
F^{jk} u_k$. Since we shall only consider particles of unit charge,
it is convenient to remove the parameter $e$ from the equation of
motion. To this end, we rescale the electromagnetic potential
according to $A \to e^{-1} A$. Then the equation of motion
simplifies to \beq \label{eom} m\: \frac{d}{d\tau} u^j \;=\;
-F^{jk}\: u_k \:, \eeq whereas the constant $C$ in the Maxwell
equations~(\ref{maxwell}) becomes $C=4 \pi e^2$.

The starting point for general relativity is the observation that
a physical process involving gravity can be understood in different
ways. Consider for example an observer at rest on earth looking
at a freely falling person (e.g.\ a diver who just jumped from a
diving board). The observer at rest argues that the earth's gravitational
force, which he can feel himself, also acts on the freely falling
person and accelerates him. The person at free fall, on the other hand,
does not feel gravity. He can take the point of view that
he himself is at rest, whereas the earth is accelerated towards him.
He then concludes that there are no gravitational fields,
and that the observer on earth merely feels the force of inertia
corresponding to his acceleration.
Einstein postulated that these two points of view should be equivalent
descriptions of the physical process. More generally, it depends on
the observer whether one has a gravitational force or an inertial
force. In other words,
\begin{quote}
{\em{equivalence principle}}\index{equivalence principle}: \quad\!\!\!\!
\parbox[t]{6.2cm}{
no physical experiment can distinguish
between gravitational and inertial forces.} \\[-0em]
\end{quote}
In mathematical language, observers correspond to
coordinate systems, and so the equivalence principle states that the
physical equations should be formulated in general (i.e.\
``curvilinear'') coordinate systems, and should in all these coordinate
systems have the same mathematical structure. This means that the
physical equations should be invariant under diffeomorphisms, and thus
space-time is to be modeled by a {\em{Lorentzian manifold}} $(M, g)$.

A Lorentzian manifold is ``locally Minkowski space'' in the sense
that at every space-time point $p \in M$, the corresponding
{\em{tangent space}}\index{tangent space} $T_pM\label{Tpm}$ is a
vector space endowed with a scalar product $\Bra .,. \Ket_p$ of
signature $(+ \ \!\! - \ \!\! - \ \! - )$. Therefore, we can
distinguish between spacelike, timelike and null tangent vectors.
Defining a non-spacelike curve $q(\tau)$ by the condition that its
tangent vector $u(\tau) \in T_{q(\tau)}M$ be everywhere
non-spacelike, our above definition of light cones and the notion of
causality immediately carry over to a Lorentzian manifold. In a
coordinate chart, the scalar product $\Bra .,. \Ket_p$ can be
represented in the form~(\ref{minsp}) where $g_{jk}\label{gjk}$ is
the so-called {\em{metric tensor}}\index{metric tensor}. In contrast
to Minkowski space, the metric tensor is not a constant matrix but
depends on the space-time point, $g_{jk} = g_{jk}(p)$. Its ten
components can be regarded as the relativistic analogue of Newton's
gravitational potential. For every $p \in M$ there are coordinate
systems in which the metric tensor coincides with the Minkowski
metric up to second order, \beq \label{gnc} g_{jk}(p) \;=\;
{\mbox{diag}}(1,-1,-1,-1)\;,\spc
\partial_j g_{kl}(p) \;=\; 0\:.
\eeq
Such {\em{Gaussian normal coordinates}}\index{Gaussian normal coordinates} correspond to the reference frame
of a ``freely falling observer'' who feels no gravitational forces. However,
it is in general impossible to arrange that also $\partial_{jk}
g_{lm}(p)=0$. This means that by going into a suitable reference frame,
the gravitational field can be transformed away locally (=in one point),
but not globally. With this in mind, a reference frame corresponding to
Gaussian normal coordinates is also called a {\em{local inertial
    frame}}\index{local inertial frame}.

The equation of motion~(\ref{eom}) and the Maxwell equations~(\ref{maxwell})
can easily be formulated on a Lorentzian manifold by the prescription
that they should in a local inertial frame have the same form as in
Minkowski space; this is referred to as the {\em{strong equivalence
principle}}. It amounts to replacing all partial derivatives by
the corresponding {\em{covariant derivatives}}\index{covariant derivative} $\nabla\label{nabla}$ of the
Levi-Civita connection; we write symbolically
\begin{equation} \label{egauge}
\partial \;\longrightarrow\; \nabla \:.
\end{equation}
We thus obtain the equations
\begin{equation} \label{maxclass}
m \:\nabla_\tau u^j \;=\; -F^{jk} u_k \;,\spc
\nabla_k F^{kl} \;=\; 4 \pi e^2\: j^l
\end{equation}
with $F_{jk} = (dA)_{jk} = \nabla_j A_k - \nabla_k A_j$.

The gravitational field is described via the curvature of
space-time. More precisely, the Riemannian {\em{curvature
    tensor}}\index{curvature}
is defined by the relations \label{Rijkl}
\beq \label{RT}
R^i_{jkl} \:u^l \;=\; \nabla_j \nabla_k u^i - \nabla_k \nabla_j u^i \:.
\eeq
Contracting indices, one obtains the {\em{Ricci tensor}} $R_{jk} =
R^i_{jik}$ and {\em{scalar curvature}} $R = R^j_j$.
The relativistic generalization of Newton's gravitational law are
the Ein\-stein equations \index{Einstein's equations}
\begin{equation} \label{einstein}
R_{jk} - \frac{1}{2}\:R\: g_{jk} \;=\;
8  \pi \kappa\: T_{jk}\:,
\end{equation}
where~$\kappa$ is the gravitational constant. Here the {\em{energy-momentum
tensor}}\index{energy-momentum tensor} $T_{jk}\label{Tjk}$ gives the distribution of matter and energy in
space-time.

It is very convenient that the physical equations can all be derived
from a variational principle. To this end, one considers the
{\em{action}} (see e.g.~\cite{LL2}) \label{S}
\begin{equation} \label{action} \index{action!classical}
S \;=\; \int \left( m\:g_{jk} \, u^j u^k + A_j u^j \right) d\tau
\:+\: \int_M \left( -\frac{1}{16 \pi e^2}\:
F_{jk} F^{jk} - \frac{1}{16 \pi \kappa}\:R \right) d\mu\:,
\end{equation}
where $u=c'(\tau)$ is the tangent vector of a timelike curve,
and $d\mu := \sqrt{-\det g} \:d^4x$ is the integration measure on $M$.
This action is not bounded below, but one can nevertheless look for
stationary points and derive the corresponding Euler-Lagrange equations.
Varying the space-time curve, the electromagnetic potential and the
metric yield the equations of motion, the Maxwell equations and
the Einstein equations, respectively.

\section{Relativistic Quantum Mechanics} \label{isec2}
\setcounter{equation}{0}
We now give an elementary introduction to relativistic quantum
mechanics in Min\-kow\-ski space (for more details see~\cite{BD, Tha}).
According to the Heisenberg Uncertainty Principle, the position and
momentum of a quantum mechanical particle cannot be determined
simultaneously, making it impossible to describe the particle by
a trajectory in space-time. Instead, one uses a {\em{wave
function}}\index{wave function}
$\Psi(t, \vec{x})$, whose absolute square has the interpretation as the
probability density that the particle is at position $\vec{x}$. The
simplest relativistic wave equation is the {\em{Klein-Gordon equation}}
\index{Klein-Gordon equation}
\begin{equation} \label{KG1}
(-\Box - m^2)\: \Psi \;=\; 0 \:,
\end{equation}
where $\Box\label{Box} \equiv \partial_j \partial^j$ is the wave operator\index{wave operator}.
This equation describes a scalar particle (=particle without spin)
of mass $m$. If the particle has electric charge, one needs to suitably
insert the electromagnetic potential $A$ into the Klein-Gordon equation.
More precisely, one finds empirically that the equation
\begin{equation} \label{KG2}
-(\partial_k - i A_k) (\partial^k - i A^k)\:\Psi \;=\; m^2 \Psi
\end{equation}
describes a scalar particle of mass $m$ and charge $e$ in the presence
of an electromagnetic field.

In order to describe a particle with spin, it was Dirac's idea to work
with a first order differential operator whose square is the wave
operator. One introduces the {\em{Dirac matrices}}\index{Dirac matrix} $\gamma^j\label{gammaj}$ as
$4 \times 4$-matrices which satisfy the {\em{anti-commutation
relations}}\index{anti-commutation relations}
\begin{equation}
2\: g^{jk}\:\1 \;=\; \{ \gamma^j,\: \gamma^k\} \;\equiv\; \gamma^j \gamma^k +
\gamma^k \gamma^j \:. \label{f:0b}
\end{equation}\label{{.,.}}
Then the square of the operator $\gamma^j \partial_j$ is
\begin{equation}
(\gamma^j \partial_j)^2 \;=\;
\gamma^j \gamma^k\: \partial_j \partial_k
\;=\; \frac{1}{2}\: \{\gamma^j, \gamma^k \}\: \partial_{jk}
\;=\; \Box \:. \label{ac}
\end{equation}
For convenience, we shall always work in the Dirac representation
\beq \label{Dirrep}
\gamma^0 \;=\; \left( \begin{array}{cc} \1 & 0 \\ 0 & -\1 \end{array} \right) ,\qquad \vec{\gamma} \;=\; \left( \begin{array}{cc}
0 & \vec{\sigma} \\ -\vec{\sigma} & 0 \end{array} \right) \:,
\eeq
where $\vec{\sigma}\label{vecsigma}$ are the three Pauli matrices\index{Pauli matrix}
\[ \sigma^1 \;=\; \left( \begin{array}{cc} 0 & 1 \\ 1 & 0 \end{array} \right), \qquad
\sigma^2 \;=\; \left( \begin{array}{cc} 0 & -i \\ i & 0 \end{array} \right), \qquad
\sigma^3 \;=\; \left( \begin{array}{cc} 1 & 0 \\ 0 & -1 \end{array} \right). \]
The Dirac equation in the vacuum\index{Dirac equation!in the vacuum} reads
\begin{equation}
    \left( i \gamma^k \frac{\partial}{\partial x^k} - m \right)
    \Psi(x) \;=\; 0 \;, \label{Dirac1}
\end{equation}
where $\Psi(x)$, the {\em{Dirac spinor}}\index{Dirac spinor}, has four complex
components. The leptons and quarks in the standard model are
Dirac particles, and thus one can say that all matter is on the fundamental
level described by the Dirac equation\index{Dirac equation}. Multiplying~(\ref{Dirac1}) by the
operator $(i \gamma^j \partial_j + m)$ and using~(\ref{ac}), one finds
that each component of $\Psi$ satisfies the Klein-Gordon equation~(\ref{KG1}).
In the presence of an electromagnetic field, the Dirac equation must be modified to
\begin{equation} \label{Dirac2} \index{Dirac equation!in the
electromagnetic field}
i \gamma^k (\partial_k -i A_k) \Psi \;=\; m \Psi \:.
\end{equation}
Multiplying by the operator
$(i \gamma^j (\partial_j -i A_j) + m)$ and using the anti-commutation
relations, we obtain the equation
\[ \left[ -(\partial_k - i A_k) (\partial^k - i A^k)
+ \frac{i}{2}\: F_{jk} \gamma^j \gamma^k - m^2 \right] \Psi \;=\; 0\:. \]
This differs from the Klein-Gordon equation~(\ref{KG2}) by the extra term
$\frac{i}{2} F_{jk} \gamma^j \gamma^k$, which describes
the coupling of the spin to the electromagnetic field.
We also denote the contraction with
Dirac matrices by a slash, i.e.\ $u \!\:\!\slsh = \gamma^j u_j$ for $u$ a
vector of Minkowski space and $\Pdd\label{Pdd} = \gamma^j \partial_j$.

The wave functions at every space-time point are
endowed with an indefinite scalar product of signature $(2,2)$, which we call
{\em{spin scalar product}}\index{spin scalar product} and denote by
\label{Sl.|.Sr}
\begin{equation}
\Sl \Psi \:|\: \Phi \Sr(x) \;=\; \sum_{\alpha=1}^4 s_\alpha\: \Psi^\alpha(x)^*\:
\Phi^\alpha(x) \:,\qquad s_1=s_2=1,\; s_3=s_4=-1\:,
    \label{f:0c}
\end{equation}
where $\Psi^*$ is the complex conjugate wave function (this scalar product is
often written as $\overline{\Psi} \Phi$ with the so-called adjoint spinor
$\overline{\Psi} \equiv \Psi^* \gamma^0$). By the {\em{adjoint}}\index{adjoint}~$A^*\label{Aast}$
of a matrix~$A$ we always mean the adjoint with respect to the spin
scalar product as defined via the relations
\[ \Sl A^* \Psi \:|\: \Phi \Sr \;=\;
\Sl \Psi \:|\: A \Phi \Sr \qquad {\mbox{for all $\Psi, \Phi$}}. \]
In an obvious way, this definition of the adjoint gives rise to the notions
``{\em{selfadjoint}},''\index{selfadjoint} ``anti-selfadjoint'' and ``{\em{unitary}}.''\index{unitary}
With these notions, the Dirac matrices are selfadjoint, meaning that
\[ \Sl \gamma^l \Psi \:|\: \Phi \Sr \;=\;
\Sl \Psi \:|\: \gamma^l \Phi \Sr \qquad {\mbox{for all $\Psi, \Phi$}}. \]
To every solution $\Psi$ of the Dirac equation
we can associate a time-like vector field $J$ by \label{Jk}
\begin{equation} \label{dc}
J^k \;=\; \Sl \Psi \:|\: \gamma^k\: \Psi \Sr \;,
\end{equation}
which is called the {\em{Dirac current}}\index{Dirac current}. The Dirac current is divergence-free,
\begin{eqnarray*}
\partial_k J^k &=& \partial_k \:\Sl \Psi \:|\: \gamma^k\: \Psi \Sr
\;=\; \Sl \partial_k \Psi \:|\: \gamma^k\: \Psi \Sr +
\Sl \Psi \:|\: \gamma^k \partial_k \: \Psi \Sr \\
&=& i \left( \Sl i \Pdd \Psi \:|\: \Psi \Sr - \Sl \Psi \:|\: i \Pdd \Psi \Sr \right) \\
&=& i \left( \Sl (i \Pdd+\Aslsh-m) \Psi \:|\: \Psi \Sr
- \Sl \Psi \:|\: (i \Pdd+\Aslsh-m) \Psi \Sr \right) \;=\; 0\:,
\end{eqnarray*}
this is referred to as {\em{current conservation}}\index{current conservation}.

So far Dirac spinors were introduced in a given reference frame. In order to
verify that our definitions are coordinate independent, we consider two
reference frames $(x^j)$ and $(\tilde{x}^l)$ with the same orientation of time
and space. Then the reference frames are related to each other by
an orthochronous proper Lorentz transformation\index{Lorentz transformation}~$\Lambda$,
i.e. in components
\[ \tilde{x}^l \;=\; \Lambda^l_j\: x^j \;,\spc
\Lambda^l_j\: \frac{\partial}{\partial \tilde{x}^l} \;=\; \frac{\partial}
{\partial x^j} \:, \]
and~$\Lambda$ leaves the Minkowski metric invariant,
\begin{equation} \label{Lt}
\Lambda^l_j \: \Lambda^m_k\: g_{lm} \;=\; g_{jk}\:.
\end{equation}
Under this change of space-time coordinates, the Dirac
operator $i \gamma^j (\partial_{x^j} - i A_j)$ transforms to
\begin{equation} \label{transdir}
i \tilde{\gamma}^l \left( \frac{\partial}{\partial \tilde{x}^l}
- i \tilde{A}_l \right) \spc {\mbox{with}} \spc
\tilde{\gamma}^l \;=\; \Lambda^l_j \gamma^j\:.
\end{equation}
This transformed Dirac operator does not coincide with the Dirac operator
$i \gamma^l (\partial_{\tilde{x}^l}-i \tilde{A}_l)$
as defined in the reference frame $(\tilde{x}^l)$ because the Dirac matrices have
a different form. However, the next lemma shows that the two Dirac operators
do coincide up to a suitable unitary transformation of the spinors.
\begin{Lemma} \label{Dirtrans}
For any orthochronous proper Lorentz transformation $\Lambda$
there is a unitary matrix $U(\Lambda)$ such that
\[ U(\Lambda)\: \Lambda^l_j \gamma^j\: U(\Lambda)^{-1} \;=\; \gamma^l\:. \]
\end{Lemma}
{\Proof} Since~$\Lambda$ is orthochronous and proper, we can write it
in the form~$\Lambda=\exp(\lambda)$, where~$\lambda$ is a suitable
generator of a rotation and/or a Lorentz boost.
Then $\Lambda(t):=\exp(t \lambda)$, $t \in \R$, is a family of Lorentz
transformations, and differentiating~(\ref{Lt}) with respect to~$t$
at~$t=0$, we find that
\[ \lambda^l_j\:g_{lk} \;=\; -g_{jm}\: \lambda^m_k \:. \]
Using this identity together with the fact that the Dirac matrices are
selfadjoint, it is straightforward to verify that the matrix
\[ u \;:=\; \frac{1}{4}\: \lambda^l_k\: \gamma_l \gamma^k \]
is anti-selfadjoint. As a consequence, the family of matrices
\[ U(t) \;:=\; \exp \left( t u \right) \]
is unitary. We now consider for a fixed index~$l$ the family of matrices
\[ A(t) \;:=\; U(t)\: \Lambda(t)^l_j \gamma^j\: U(t)^{-1}\:. \]
Clearly, $A(0)=\gamma^l$. Furthermore, differentiating with respect to~$t$
gives
\[ \frac{d}{dt}\:A(t) \;=\; U\: \Lambda^l_j \left\{ u \:\gamma^j
- \gamma^j \:u + \lambda^j_k \gamma^k \right\} U^{-1}\:, \]
and a short calculation using the commutation relations
\[ \left[ \gamma_l \gamma_k, \gamma^j \right]
\;=\; 2 \left( \gamma_l\: g^{kj} - \delta^j_l\: \gamma^k \right) \]
shows that the curly brackets vanish.
We conclude that $A(1)=A(0)$, proving the lemma.
\hspace*{2cm} \QED
Applying this lemma to the Dirac operator in~(\ref{transdir}), one
sees that the Dirac operator is invariant under the joint transformation
of the space-time coordinates and the spinors
\begin{equation} \label{Lorentz}
x^j \;\longrightarrow\; \Lambda^j_k x^k\;,\spc
\Psi \;\longrightarrow\; U(\Lambda)\: \Psi \:.
\end{equation}
Moreover, since the matrix~$U(\Lambda)$ is unitary,
the representation of the spin scalar product~(\ref{f:0c}) is valid in any
reference frame. We conclude that our definition of spinors is indeed
coordinate invariant.

Out of the Dirac matrices one can form the {\em{pseudoscalar
    matrix}}\index{pseudoscalar matrix}
$\rho\label{rho}$ by
\beq \label{rhodef}
\rho \;=\; \frac{i}{4!}\: \epsilon_{jklm} \gamma^j \gamma^k \gamma^l
\gamma^m
\eeq
(this matrix in the physics literature is usually denoted by $\gamma^5$).
Here $\epsilon_{jklm}\label{epsilonjklm}$ is the totally antisymmetric symbol
(i.e.\ $\epsilon_{jklm}$ is equal to $\pm 1$ if $(j,k,l,m)$ is an even and
odd permutation of $(0,1,2,3)$, respectively, and vanishes otherwise).
A short calculation shows that the pseudoscalar matrix is anti-selfadjoint
and $\rho^2=\1$. As a consequence, the matrices \label{chiL}\label{chiR}
\beq \label{chidef}
\chi_L \;=\; \frac{1}{2} \left(\1 - \rho \right) \;,\spc
\chi_R \;=\; \frac{1}{2} \left(\1 + \rho \right)
\eeq
satisfy the relations
\[ \chi_{L\!/\!R}^2 = \chi_{L\!/\!R} \;,\qquad
\rho \:\chi_{L} = -\chi_{L} \;,\qquad
\rho \:\chi_{R} = \chi_{R} \;,\qquad
\chi_L^* \;=\; \chi_R \;,\qquad \chi_L+\chi_R=\1\:. \]
They can be regarded as the spectral projectors of the
matrix~$\rho$ and are called the {\em{chiral projectors}}
\index{chiral projector}.
The projections $\chi_L \Psi$ and $\chi_R \Psi$ are referred to
as the {\em{left-}}\index{spinor matrix!left-handed} and
{\em{right-handed}}\index{spinor matrix!right-handed} components of the
spinor. A matrix is said to be {\em{even}}\index{even matrix} and {\em{odd}}\index{odd matrix} if it commutes and anti-commutes with $\rho$, respectively. It is straightforward to verify that the Dirac matrices are
odd, and therefore
\[ \gamma^j \: \chi_{L\!/\!R} \;=\; \chi_{R\!/\!L}\: \gamma^j \:. \]
Using this relation, one can rewrite the Dirac equation~(\ref{Dirac2})
as a system of equations for the left- and right-handed components of
$\Psi$,
\[ i \gamma^k (\partial_k -i A_k) \:\chi_L \Psi \;=\; m \:\chi_R \Psi \;,\qquad
i \gamma^k (\partial_k -i A_k) \:\chi_R \Psi \;=\; m \:\chi_L \Psi \:. \]
If $m=0$, these two equations decouple, and we get separate equations
for the left- and right-handed components of $\Psi$.
This observation is the starting point of the 2-component Weyl spinor
formalism. We shall not use this formalism here, but will instead
describe chiral massless particles (like neutrinos)
by the left- or right-handed component of a Dirac spinor.

For the probabilistic interpretation of the Dirac wave function,
we need to distinguish a direction of time and work in a particular
reference frame. Then the zero component of the Dirac current
$J^0(t, \vec{x})$ has for a given time $t$ the interpretation as the
{\em{probability density}}\index{probability density} of the particle to be at position $\vec{x}$
(and is thus the relativistic analogue of the absolute square $|\Psi|^2$
of the Schr{\"o}dinger or Pauli wave functions). Clearly, for this
probabilistic interpretation the wave function must be properly
normalized. More precisely, physical states must satisfy the
{\em{normalization condition}}\index{normalization condition}
\begin{equation}
\int_{\sR^3} \Sl \Psi \:|\: \gamma^0\: \Psi \Sr(t, \vec{x}) \:d\vec{x}
\;=\; 1 \;. \label{f:0d}
\end{equation}
The integral in~(\ref{f:0d}) is also called the {\em{probability
integral}}\index{probability integral}.
Using Gauss' (divergence) theorem and the current conservation, one sees that
the normalization integral is time independent,
\begin{eqnarray}
\lefteqn{ \int_{\sR^3} \Sl \Psi \:|\: \gamma^0 \: \Psi
\Sr(t_2, \vec{x}) \:d\vec{x}
- \int_{\sR^3} \Sl \Psi \:|\: \gamma^0 \: \Psi \Sr(t_1, \vec{x})
\:d\vec{x} } \nonumber \\
&=& \int_{t_1}^{t_2} dt \int_{\sR^3} d\vec{x}\;
\partial_k \Sl \Psi \:|\: \gamma^k\: \Psi \Sr(t, \vec{x}) \;=\; 0 \:,
\label{tind}
\end{eqnarray}
and thus it suffices to satisfy~(\ref{f:0d}) for example at $t=0$.

In a given reference frame, it is convenient to introduce a positive
scalar product by polarizing the normalization integral, \label{(.|.)}
\begin{equation}
(\Psi \:|\: \Phi) \;:=\; \int_{\sR^3} \Sl \Psi \:|\: \gamma^0 \: \Phi \Sr(t,
\vec{x})\: d\vec{x} \:. \label{f:0e}
\end{equation}
We denote the Hilbert space corresponding to this scalar product
by $\mathcal{H} = L^2(\R^3)^4$.
Multiplying the Dirac equation~(\ref{Dirac2})
by $\gamma^0$ and bringing the $t$-derivative on a separate side of the
equation, we can write the Dirac equation as
\begin{equation} \label{Hamilton}
i \partial_t \Psi \;=\; h \Psi
\end{equation}
with a purely spatial operator $h$. Clearly, this equation is not
manifestly covariant. In analogy to nonrelativistic
quantum mechanics, it is referred to as the Dirac
equation in {\em{Hamiltonian form}}\index{Dirac equation!in
  Hamiltonian form}, and $h\label{h}$ is the {\em{Hamiltonian}}\index{Hamiltonian}.
If $\Psi$ and $\Phi$ are solutions of the Dirac equation, one sees
similar to~(\ref{tind}) that the scalar product~(\ref{f:0e}) is
independent of time. Hence
\[ 0 \;=\; \partial_t ( \Psi \:|\: \Phi) \;=\; i \left(
( h \Psi \:|\: \Phi) - ( \Psi \:|\: h \Phi) \right) \:. \]
This shows that the Hamiltonian is a symmetric operator on $\mathcal{H}$.

We conclude this section by a brief discussion of the solutions of the
free Dirac equation (=the Dirac equation without electromagnetic
field) in the case $m \neq 0$. Taking the Fourier transform of the wave function,
\[ \Psi(x) \;=\; \int \frac{d^4k}{(2 \pi)^4}\: \hat{\Psi}(k)\: e^{-ikx} \;,\qquad
\hat{\Psi}(k) \;=\; \int d^4x \: \Psi(x)\: e^{ikx} \:, \]
the Dirac equation $(i \Pdd - m) \Psi=0$ reduces to the algebraic equation
in momentum space
\begin{equation} \label{kDirac}
(k\slsh - m)\: \hat{\Psi}(k) \;=\; 0 \:.
\end{equation}
Multiplying by $k\slsh+m$ and using the identity
$(k\slsh-m)(k\slsh+m)=k^2-m^2$, one sees that if $k^2 \neq m^2$,
the matrix $k\slsh-m$ is invertible and thus~(\ref{kDirac}) has no solutions.
If conversely~$k^2=m^2$, we have the relation $(k\slsh-m)^2 = -2 m (k \slsh-m)$,
showing that the matrix $k\slsh-m$ is diagonalizable and that its
eigenvalues are either $-2m$ or zero. Taking the trace, $\Tr(k\slsh-m) =-4m$,
it follows that the matrix~$k\slsh-m$ has a two-dimensional kernel.
A short calculation shows that the projector onto this kernel is given by
\begin{equation} \label{projkern}
\Pi(k) \;:=\; \frac{k\slsh+m}{2m}\:.
\end{equation}
We conclude that~(\ref{kDirac}) has a solution only if $k$ is on the
{\em{mass shell}} $\{ k \:|\: k^2=m^2 \}$.
For each $k$ on the mass shell,
(\ref{kDirac}) has exactly two linearly independent solutions.
In order to give these solutions more explicitly, we
choose a reference frame $(t, \vec{x})$ and denote the corresponding
momentum variables by $k=(\omega, \vec{p})$. The momenta on the mass
shell are then given by
\[ \omega \;=\; \omega(\vec{p}, \epsilon) :=
\epsilon \: \sqrt{|\vec{p}|^2+m^2} \]
with parameters~$\vec{p} \in \R^3$ and~$\epsilon \in \{\pm 1 \}$.
The momenta with~$\epsilon=1$ and~$\epsilon=-1$ are said to be on the
{\em{upper}} and {\em{lower mass shell}}, respectively.
For any given~$(\vec{p}, \epsilon)$, we label the two linearly independent
solutions of~(\ref{kDirac}) by a parameter
$s \in \{1,2 \}$ and denote them by~$\chi_{\vec{p} s \epsilon}$. It is most
convenient to choose them pseudo-orthonormal with respect to the spin
scalar product,
\begin{equation} \label{spinnorm}
\Sl \chi_{\vec{p} s \epsilon} | \chi_{\vec{p} s' \epsilon} \Sr
\;=\; \epsilon\:\delta_{s,s'} \spc {\mbox{for all~$\vec{p} \in \R^3,
\epsilon \in \{\pm 1\}$~and~$s,s' \in \{1,2\}$  }}.
\end{equation}
Here the factor~$\epsilon$ reflects that the solution spaces on the
upper and lower mass shell are positive and negative definite,
respectively. Using a bra/ket notation in the spin scalar product,
we get a simple representation of the projector~(\ref{projkern}),
\beq \label{exppr}
\sum_{s=1,2} \epsilon\: | \chi_{\vec{p} s \epsilon} \Sr \Sl
\chi_{\vec{p} s \epsilon} | \;=\; \Pi(\omega(\vec{p},\epsilon), \vec{p})\:.
\eeq
The spinors~$\chi_{\vec{p} s \epsilon}$ form a complete set of solutions
of~(\ref{kDirac}). Taking their suitably normalized Fourier transform,
we obtain the~{\em{plane wave solutions}}\index{plane wave solution}
\begin{equation} \label{pws}
\Psi_{\vec{p} s \epsilon}(t,\vec{x})
\;=\; \frac{1}{(2 \pi)^{\frac{3}{2}}}\; e^{-i \omega(\vec{p}, \epsilon) t + i \vec{p} \vec{x}}\:
\chi_{\vec{p} s \epsilon} \:.
\end{equation}\label{Psivecpsepsilon}
Each solution of the free Dirac equation is a superposition
of plane wave solutions.

In the Hamiltonian framework~(\ref{Hamilton}), the plane wave solutions
are eigenfunctions of the Hamiltonian with eigenvalue~$\omega(\vec{p}, \epsilon)$,
\[ h\: \Psi_{\vec{p} s \epsilon} \;=\; \omega(\vec{p}, \epsilon)\:
\Psi_{\vec{p} s \epsilon}\:. \]
Since the eigenvalue of the Hamiltonian has the interpretation as the
physical energy of the state, we conclude that the plane wave solutions
on the upper and lower mass shell have positive and negative energy,
respectively. Expressed more mathematically, the plane wave solutions
correspond to points in the essential spectrum of the Hamiltonian, and
thus
\[ \sigma_{\mbox{\scriptsize{ess}}}(h) \;=\; \{ \pm \sqrt{\vec{p}^2+m^2},\:
\vec{p} \in \R^3 \} \;=\; (-\infty, -m] \cup [m, \infty) \;. \]
In particular, we conclude that the Dirac equation has solutions of
negative energy and that the Hamiltonian is not bounded below.
This was originally considered a serious problem
of Dirac theory, mainly because a system with unbounded Hamiltonian has no stable
ground state.  Dirac resolved these problems by
introducing the so-called Dirac sea\index{Dirac sea}~\cite{Di2}. The concept of the Dirac sea
plays a crucial role in the present work. At this point, we merely explain
Dirac's idea in words (in the next section we shall explain how it is
implemented mathematically in the framework of quantum field theory, and in
Chapter~{\S}\ref{sec3} we will come back to it in greater detail).
Thinking of many-particle quantum
mechanics (and assuming that the particles do not interact with each other), the
solutions of the Dirac equation can be regarded as one-particle
states, which can be occupied by the particles of the system.  According to the
Pauli Exclusion Principle, each state may be occupied by at most one particle.
If one assumes that no states are occupied in the vacuum, a system of $n$
particles is unstable because the energy of the system can be made negative and
arbitrarily small by occupying $n$ negative-energy states.  However, this
problem disappears if one assumes that in the vacuum all states of negative
energy are already occupied.  Then the $n$ additional particles must occupy
states of positive energy, and the system becomes stable.  This consideration
led Dirac to no longer think of the vacuum as ``empty space,'' but instead
to conjecture that the vacuum should be formed of a ``sea'' of quantum
mechanical particles of
negative energy.  Dirac's conception was that the effects of all the particles
of the sea should compensate each other in such a way that the sea cannot be
observed.  Likewise, in this picture an interacting system of $n$ particles
corresponds to the Dirac sea and $n$ additional particles of positive energy
which all interact with each
other.  This intuitive concept of the Dirac sea as a ``sea of interacting
particles'' was not only useful for resolving the problem of the negative-energy
solutions, but furthermore led to the prediction of anti-particles and pair
creation/annihilation.  To this end, Dirac considered an
interacting system which at initial time $t=0$ is the vacuum.  Then at a later
time, one of the particles of the sea may no longer occupy a state of
negative energy, but be instead in a positive-energy state.
In this case, the system consists of one particle and one ``hole'' in the Dirac
sea.  Since the completely filled Dirac sea should be invisible, the hole appears
as a virtual particle of energy and electric charge opposite to that of the
unoccupied negative-energy state.  Thus the virtual particle has positive energy,
but its charge is opposite to that of an ordinary particle.  This virtual
particle is referred to as anti-particle.  In the above process, particles and
anti-particles are always generated in pairs, this explains the physical effect
of pair creation.  Conversely, a particle and a hole can recombine in a process
called pair annihilation.

\section{Fock Space Quantization of the Free Dirac Field} \label{isec4}
\setcounter{equation}{0}
In this section we outline the canonical quantization\index{canonical quantization} of the free
Dirac field (for details see~\cite{BD2, IZ}). For clarity, we first quantize
without taking into account the Dirac sea and explain afterwards how the
construction is to be modified in order to cure the problem of the negative-energy
states. We begin with the one-particle Hilbert space~$({\mathcal{H}}, (.,.))$ in
the Hamiltonian framework~(\ref{Hamilton}). Clearly, the plane-wave
solutions~(\ref{pws}) are not square integrable, but we can normalize them
in the distributional sense. More precisely,
\begin{equation} \label{normalize}
(\Psi_{\vec{p} s \epsilon} \:|\: \Psi_{\vec{p}' s' \epsilon'}) \;=\;
\delta^3(\vec{p}-\vec{p'}) \: \Sl \chi_{\vec{p} s \epsilon}
\:|\: \gamma^0 \chi_{\vec{p} s' \epsilon'} \Sr \:.
\end{equation}
In order to compute the inner product~$\Sl \chi_{\vec{p} s \epsilon}
\:|\: \gamma^0 \chi_{\vec{p} s' \epsilon'} \Sr$, we first plug in the
spectral projectors~(\ref{projkern}), which for convenience we now denote
by~$\Pi_{\vec{p} \epsilon} := \Pi(\omega(\vec{p}, \epsilon), \vec{p})$,
\[ \Sl \chi_{\vec{p} s \epsilon}
\:|\: \gamma^0 \chi_{\vec{p} s' \epsilon'} \Sr \;=\;
\Sl \Pi_{\vec{p} \epsilon} \chi_{\vec{p} s \epsilon}
\:|\: \gamma^0 \:\Pi_{\vec{p} \epsilon'} \chi_{\vec{p} s' \epsilon'} \Sr
\;=\; \Sl \chi_{\vec{p} s \epsilon} \:|\: \Pi_{\vec{p} \epsilon} \:\gamma^0
\:\Pi_{\vec{p} \epsilon'} \chi_{\vec{p} s' \epsilon'} \Sr \:. \]
The matrix product~$\Pi_{\vec{p} \epsilon} \gamma^0 \Pi_{\vec{p} \epsilon'}$
is computed in the cases~$\epsilon=\epsilon'$ and~$\epsilon \neq \epsilon'$
as follows,
\begin{eqnarray*}
\Pi_{\vec{p} \epsilon} \:\gamma^0\: \Pi_{\vec{p}\:-\epsilon} &=&
\frac{\omega \gamma^0 - \vec{p} \vec{\gamma} + m}{2m}
\:\gamma^0\: \frac{-\omega \gamma^0 - \vec{p} \vec{\gamma} + m}{2m} \\
&=& \frac{(k\slsh+m)(-k\slsh+m)}{4m^2}\:\gamma^0 \;=\; 0 \\
\Pi_{\vec{p} \epsilon} \:\gamma^0\: \Pi_{\vec{p} \epsilon}
&=& \frac{\omega \gamma^0 - \vec{p} \vec{\gamma} + m}{2m}
\:\gamma^0\: \frac{\omega \gamma^0 - \vec{p} \vec{\gamma} + m}{2m} \\
&=& \frac{\omega \gamma^0 - \vec{p} \vec{\gamma} + m}{4m^2}\: 2 \omega\:
+ \frac{(k\slsh+m)(-k\slsh+m)}{4m^2}\:\gamma^0 \;=\;
\frac{\omega}{m}\: \Pi_{\vec{p} \epsilon} \:,
\end{eqnarray*}
where we set $\omega=|\omega(\vec{p}, \epsilon)|$ and $k=(\omega, \vec{p})$.
Hence the matrix products reduce to a multiple of the identity, and
we can use the normalization~(\ref{spinnorm}) to obtain
\begin{equation}
    (\Psi_{\vec{p} s \epsilon} \:|\: \Psi_{\vec{p}' s' \epsilon'}) \;=\;
    \frac{\omega(\vec{p})}{m}\:
    \delta^3(\vec{p}-\vec{p}')\: \delta_{\epsilon \epsilon'}\: \delta_{s s'}\;,
    \label{f:0hh}
\end{equation}
with~$\omega(\vec{p}) := |\omega(\vec{p}, \epsilon)| = \sqrt{|\vec{p}|^2+m^2}$.
Readers who dislike this $\delta$-normalization can also state~(\ref{f:0hh})
by saying that, similar to a Fourier transformation, the mapping
\[ L^2\!\left(\R^3, \frac{d\vec{p}}{2 \omega(\vec{p})} \right)^4
\:\longrightarrow\: {\mathcal{H}} \;:\;
f_{s \epsilon}(\vec{p}) \:\longmapsto\: \sqrt{2m} \:\sum_{s, \epsilon} \int_{\sR^3}
\frac{d\vec{p}}{2 \omega(\vec{p})}\: f_{s \epsilon}(\vec{p})\:
\Psi_{\vec{p} s \epsilon}(t, \vec{x}) \]
is an isometry of Hilbert spaces. The factor
$d\vec{p}/(2 \omega(\vec{p}))$ appearing here can be interpreted
as the Lorentz invariant measure on the mass shell
(i.e.~formally~$d\vec{p}/(2 \omega(\vec{p})) = \delta(k^2-m^2)\:d^4k$),
and we abbreviate it in what follows by~$d\mu_{\vec{p}}\label{dmuvecp}$.

In many-particle quantum mechanics, the system where the~$n$ one-particle
states $\Psi_{\vec{p}_1 s_1 \epsilon_1},\ldots, \Psi_{\vec{p}_n s_n \epsilon_n}$
are occupied is described by the {\em{Hartree-Fock
    state}}\index{Hartree-Fock state}
\begin{equation}
\Psi \;=\; \Psi_{\vec{p}_1 s_1 \epsilon_1} \:\land\: \cdots \:\land\:
\Psi_{\vec{p}_n s_n \epsilon_n} \;.    \label{f:0h1}
\end{equation}
Here the wedge product~$\land\label{land}$ is the anti-symmetrized tensor product.
Due to the anti-symmetry, the wedge product vanishes if two of the
one-particle wave functions~$\Psi_{\vec{p}_i, s_i, \epsilon_i}$ coincide. This corresponds to the
\begin{quote}
{\em{Pauli Exclusion Principle}}\index{Pauli Exclusion Principle}: \quad\!\!\!
\parbox[t]{5.4cm}{each quantum
mechanical state can be occupied by at most one particle.} \\[0em]
\end{quote}
Particles which obey the Pauli Exclusion Principle are called {\em{fermions}}
(whereas for bosons one uses instead of~(\ref{f:0h1}) the symmetric tensor
product). Working with the $n$-particle state~(\ref{f:0h1}) also implies
that the~$n$ particles are {\em{indistinguishable}} in the sense that if we
exchange two particles, the wave function~$\Psi$ changes only by a physically
irrelevant phase.

A general $n$-particle state corresponds to a linear combination of
Hartree-Fock states and is thus a vector of the
Hilbert space ${\mathcal{F}}^n \label{mathcalFn}= \land^n {\mathcal{H}}$.  In quantum field
theory, the number of particles is not fixed, and therefore the Dirac particles
are described more generally by a vector of the {\em{fermionic Fock
space}} \index{Fock space}
${\mathcal{F}}\label{mathcalF}=\oplus_{n=0}^\infty {\mathcal{F}}^n$.  Notice that the scalar
product on ${\mathcal{F}}$ is induced by that on ${\mathcal{H}}$; we denote
it for clarity by $(.|.)_{\mathcal{F}}$.  On the Fock space, we introduce the
field operators
$\hat{\Psi}^\dagger_{\vec{p} s \epsilon}$ by
\[ \hat{\Psi}^\dagger_{\vec{p} s \epsilon}\label{hatPsidagger} \;:\;
{\mathcal{F}}^n \:\longrightarrow\: {\mathcal{F}}^{n+1} \;:\;
\Psi \:\longmapsto\: \Psi_{\vec{p} s \epsilon} \land \Psi \]
and denote their adjoint with respect to the scalar product $(.|.)_{\mathcal{F}}$
by $\hat{\Psi}_{\vec{p} s \epsilon}\label{hatPsivec}$, $\hat{\Psi}_{\vec{p} s \epsilon} =
(\hat{\Psi}^\dagger_{\vec{p} s \epsilon})^*$.
The operators $\hat{\Psi}^\dagger_{\vec{p} s \epsilon}$ and
$\hat{\Psi}_{\vec{p} s \epsilon}$ are referred to as the {\em{creation}}\index{creation operator} and
{\em{annihilation operators}}\index{annihilation operator}, respectively. A straightforward calculation
using our definitions and the normalization condition~(\ref{f:0hh}) yields
that the field operators satisfy the {\em{canonical anticommutation
relations}}\index{anti-commutation relations!canonical}
\begin{equation} \begin{split}
\{ \hat{\Psi}_{\vec{p} s \epsilon},\hat{\Psi}_{\vec{p}' s'
\epsilon'} \} &\;=\;0 \;=\; \{ \hat{\Psi}^\dagger_{\vec{p} s \epsilon},
\hat{\Psi}^\dagger_{\vec{p}' s' \epsilon'} \} \\
\{ \hat{\Psi}_{\vec{p} s \epsilon}, \hat{\Psi}^\dagger_{\vec{p}'
s' \epsilon'} \} &\;=\; 2 \omega(\vec{p})\: \delta^3(\vec{p}-\vec{p}')
\: \delta_{\epsilon \epsilon'}\: \delta_{s s'} \;. \end{split} \label{f:0n}
\end{equation}
The vacuum corresponding to these field operators, denoted by $|0\ket$, is a
unit vector of $\mathcal{F}$ on which all annihilation operators vanish,
\begin{equation}
\hat{\Psi}_{\vec{p} s \epsilon} \:|0\ket \;=\; 0 \spc
{\mbox{for all $\vec{p}, s, \epsilon$}}.
    \label{f:0j}
\end{equation}
The Hartree-Fock states can be obtained from it by applying the creation
operators,
\[ (\ref{f:0h1}) \;=\; \hat{\Psi}^\dagger_{\vec{p}_1 s_1 \epsilon_1} \:
\cdots \: \hat{\Psi}^\dagger_{\vec{p}_n s_n \epsilon_n} \: |0\ket \;, \]
and taking linear combinations, we can build up the whole Fock space from
the vacuum. The energy of the Hartree-Fock state~(\ref{f:0h1}) is the sum of
the energies $\omega(\vec{p}, \epsilon)$ of all particles, and
a short calculation shows that this coincides with the
eigenvalue of the following operator on the Fock space,
\begin{equation}
H_0 \;=\; \sum_{\epsilon, s} \int_{\sR^3} \omega(\vec{p}, \epsilon)\;
\hat{\Psi}^\dagger_{\vec{p} s \epsilon} \hat{\Psi}_{\vec{p} s \epsilon}\:
d\mu_{\vec{p}} \;.    \label{f:0i}
\end{equation}
Thus $H_0$ is the Hamiltonian of the free many-particle theory.

Since the factor $\omega(\vec{p}, \epsilon)$ in the integrand can be negative,
the Hamiltonian~(\ref{f:0i}) is not bounded from below. This is precisely the
problem of the negative-energy solutions of the Dirac equation which we described at the end of the previous section.  This problem disappears in quantum field theory as follows.  According to the concept of the Dirac
sea, all negative-energy states should be occupied in the vacuum.  This is
implemented here by redefining the vacuum; namely we replace~(\ref{f:0j}) by
the conditions
\begin{equation}
\hat{\Psi}_{\vec{p}s +} \:|0\ket \;=\; 0 \;=\; \hat{\Psi}^\dagger_{\vec{p}s -} \:|0\ket
\spc {\mbox{for all $\vec{p}, s$}}.    \label{f:0k}
\end{equation}
Since the anti-particles correspond to ``holes'' in the Dirac sea,
we reinterpret the creation operators for the negative-energy states as
annihilation operators and vice versa, i.e.\ we perform the formal replacements
\begin{equation}
\hat{\Psi}_{\vec{p}s -} \;\longleftrightarrow\; \hat{\Psi}^\dagger_{\vec{p}s -} \;.
    \label{f:0l2}
\end{equation}
This is convenient because after the reinterpretation, the new
vacuum~(\ref{f:0k}) again satisfies the usual conditions~(\ref{f:0j}). The
Hamiltonian~(\ref{f:0i}) transforms under the replacements~(\ref{f:0l2}) into
\begin{equation}
    H_0 \;=\; \sum_{\epsilon, s} \int_{\sR^3} \omega(\vec{p})\;
    \hat{\Psi}^\dagger_{\vec{p} s \epsilon} \hat{\Psi}_{\vec{p} s \epsilon}\:
    d\mu_{\vec{p}}
    \:-\: \sum_{s} \int_{\sR^3} \omega(\vec{p})\ \{
    \hat{\Psi}^\dagger_{\vec{p} s -},
    \hat{\Psi}_{\vec{p} s -} \} \: d\mu_{\vec{p}} \:. \label{f:0m}
\end{equation}
The first part of this Hamiltonian is positive. Using the anti-commutation
relations~(\ref{f:0n}), one sees
that the second term in~(\ref{f:0m}) is an infinite negative constant. Using the
argument that adding a constant to the total energy of a system is nothing more
than introducing a new convention for the zero point of energy measurements,
one drops this second term and redefines the Hamiltonian by
\begin{equation}
    H_0 \;=\; \sum_{\epsilon, s} \int_{\sR^3} |\omega(\vec{p}, s)|\;
    \hat{\Psi}^\dagger_{\vec{p} s \epsilon} \hat{\Psi}_{\vec{p} s \epsilon}
    \: d\mu_{\vec{p}} \:.
    \label{f:0nn}
\end{equation}
This Hamiltonian is positive and vanishes on the vacuum, giving rise to
a satisfying physical theory. However, dropping the second summand
in~(\ref{f:0m}) was a problematic step in the construction. We postpone
the discussion of this point to~{\S}\ref{jsec2}.

\section{Classical Gauge Theories} \label{isec3}
\setcounter{equation}{0}
We now briefly introduce the framework of local gauge theories (for a
more detailed introduction see for example~\cite{gaugeth}). In order
to avoid confusion between covariant derivatives~$\nabla$ and
gauge-covariant derivatives~$D$ we restrict attention to Minkowski space.
The generalization to curved space-time will be described in connection with
the Dirac equation in~{\S}\ref{isec5}.
The starting point for gauge theories is the observation that changing
the electromagnetic potential by the gradient of a real-valued function
$\Lambda$,
\begin{equation} \label{gauge}
A \;\longrightarrow\; A + \partial \Lambda \:,
\end{equation}
leaves the field tensor unchanged,
\[ F_{jk} \;\longrightarrow\; F_{jk} + \partial_j \partial_k \Lambda
- \partial_k \partial_j \Lambda \;=\; F_{jk} \:. \]
The equations of classical electrodynamics~(\ref{maxwell}, \ref{eom}) do
not involve the electromagnetic potential, only its field tensor. Therefore,
these equations are obviously invariant under the transformation~(\ref{gauge}).
In the quantum mechanical wave equations~(\ref{KG2}, \ref{Dirac2}) the
electromagnetic potential does appear, but only in combination with a partial
derivative in the operators $\partial_k -i A_k$. These operators transform
under~(\ref{gauge}) as follows,
\[ \partial_k -i A_k \;\longrightarrow\;
\partial_k -i A_k - i \partial_k \Lambda \;=\;
e^{i \Lambda}\:(\partial_k -i A_k)\: e^{-i \Lambda}\:. \]
Writing the transformation law with the multiplication operators
$e^{\pm i \Lambda}$ reveals that the equations of quantum mechanics
are invariant under~(\ref{gauge}) if at the same time the local phase of
the wave functions is transformed according to
\begin{equation} \label{phase}
\Psi \;\longrightarrow\; e^{i \Lambda}\: \Psi\:.
\end{equation}
Finally, these local phase transformations leave the Dirac current~(\ref{dc})
unchanged. We conclude that classical field theory and relativistic quantum
mechanics are invariant under the transformation~(\ref{gauge}, \ref{phase}), which
is referred to as a {\em{local gauge transformation}}\index{local
  gauge transformation} of electrodynamics.
The invariance of the physical equations under local gauge transformations can
be interpreted as a physical symmetry, the {\em{local gauge
symmetry}}\index{gauge symmetry!local}.

Extending the above concept leads to the mathematical framework of
gauge theories. We first note that the phase factor~$e^{i \Lambda}$ in~(\ref{phase})
can be interpreted as the operation of an element of the Lie group~$U(1)$ on~$\Psi$.
Likewise, the factors~$\partial_j \Lambda = i e^{i \Lambda} \partial_j e^{-i \Lambda}$
can be regarded as elements of the corresponding Lie algebra~$u(1)$.
Since in~(\ref{gauge}) this factor is added to the components~$A_j$ of the
electromagnetic potential, it is natural to also consider the~$A_j$ as
$u(1)$-valued functions. In generalization, we let the
{\em{gauge group}}\index{gauge group}~${\mathcal{G}}\label{mathcalG}$
be an arbitrary Lie group in a given matrix representation on the wave functions
(the wave functions may have more than four components; a typical example
is~${\mathcal{G}}=U(p)$ and~$\Psi(x) \in \C^4 \otimes \C^p$). The corresponding Lie algebra
in its representation on the wave functions is denoted
by~$\mathfrak{g}$. We introduce the {\em{gauge potentials}}~$A_j$ as
$\mathfrak{g}$-valued functions on $M$. For any smooth function
$U \::\: M \to {\mathcal{G}}$, the transformation of the wave functions
\begin{equation} \label{lgt}
\Psi(x) \;\longrightarrow\; U(x)\: \Psi(x)
\end{equation}
is referred to as a {\em{local gauge transformation}}.
Clearly, partial derivatives of~$\Psi$
do not behave well under gauge transformations because we pick up derivatives
of $U$. This problem disappears if instead of partial derivatives we consider
{\em{gauge-covariant derivatives}}\index{gauge-covariant derivative}
\beq \label{gcd}
D_j \;=\; \partial_j - i A_j \:,
\eeq
provided that the gauge potentials transform according to
\beq \label{Atrans}
A_j \;\longrightarrow\; U A_j U^{-1} + i U\:(\partial_j U^{-1})\:.
\eeq
Namely, a short calculation shows that the gauge-covariant derivative behaves
under gauge transformations according to
\beq \label{Dlgt}
D_j \;\longrightarrow\; U\:D_j \:U^{-1} \:,
\eeq
and thus the gauge-covariant derivatives of $\Psi$ obey the simple transformation
rule
\[ D_j \Psi \;\longrightarrow\; U\: D_j \Psi\:. \]

Next we need to introduce the gauge potentials into the physical equations
and formulate the equations that describe the dynamics of the gauge fields.
We just saw that in order to ensure gauge invariance, one should work with gauge-invariant derivatives instead of partial derivatives. The simplest method
for making the physical theory gauge invariant is to replace all partial
derivatives by the corresponding gauge-invariant derivatives,
\begin{equation} \label{minicoupling}
\partial \;\longrightarrow\; D\:.
\end{equation}
This ad-hoc method is in physics called the {\em{minimal
    coupling}}\index{minimal coupling} procedure.
For the equations of quantum mechanics it can be motivated
if one keeps in mind that with a local gauge transformation of the form
$U(x) = \1 - i A_j\:(x-p)^j + o(x-p)$ we can always arrange that~$A(p)=0$.
In this gauge, the gauge-covariant derivatives coincide at~$p$ with the
partial derivatives, and thus we can state minimal coupling as follows,
\beq \label{mc}
\begin{tabular}{l}
{\mbox{Around each space-time point~$p$ there is a gauge such that the}} \\
{\mbox{quantum mechanical equations coincide at~$p$ with the equations}} \\
{\mbox{without gauge fields.}}
\end{tabular}
\eeq
In this formulation, minimal coupling can be understood similar to the strong
equivalence principle; we only need to replace ``coordinate system'' by ``gauge'' and ``gravitational field'' by ``gauge field.''
In the example of the free Dirac equation~$(i \Pdd -m) \Psi=0$,
minimal coupling yields the equation
\[ i \gamma^j (\partial_j - i A_j)\:\Psi \;=\; m \:\Psi\:, \]
which describes a behavior of a Dirac particle in the presence of the
gauge field. This equation can also be derived by varying~$\Psi$ in the
corresponding Dirac action
\[ S_{\mbox{\scriptsize{D}}} \;=\; \int_M \Sl \Psi \:| \left(i \gamma^j (\partial_j - i A_j) - m \right)
\Psi\Sr \: d\mu . \index{action!Dirac} \]
In order to get the equations for the gauge field, we construct out of
the gauge-covariant derivative the {\em{field tensor}} by
\[ F_{jk} \;=\; i \:[D_j, D_k] \;=\; \partial_j A_k - \partial_k A_j -i
[A_j, A_k]\:. \]
Since its behavior under gauge transformation is simply
\[ F_{jk} \;\longrightarrow\; U\: F_{jk}\:U^{-1} \:, \]
we can generalize the action of the electromagnetic field in~(\ref{action})
by the {\em{Yang-Mills action}}
\[ S_{\mbox{\scriptsize{YM}}} \;=\; -\frac{1}{16 \pi e^2}
\int_M \Tr(F_{jk} F^{jk})\: d\mu \:,
\index{action!Yang-Mills} \]
where ``$\Tr$'' is a suitably normalized matrix trace.
The total action is simply the sum of the Dirac and Yang-Mills actions,
\[ S \;=\; S_{\mbox{\scriptsize{D}}} + S_{\mbox{\scriptsize{YM}}}\:. \]
Varying~$(\Psi, A)$ we obtain the coupled Dirac-Yang/Mills equations which
describe the classical dynamics.

\section{Dirac Spinors in Curved Space-Time} \label{isec5}
\setcounter{equation}{0}
Dirac spinors are often formulated on a manifold using frame bundles, either
an orthonormal frame~\cite{Baum, Friedrich} or a Newman-Penrose null
frame~\cite{PR, Ch}. We here outline an equivalent formulation of
spinors in curved space-time in the framework of a~$U(2,2)$ gauge
theory (for details see~\cite{F1}). We restrict attention to the Dirac operator
in local coordinates; for global issues like topological obstructions
for the existence of spin structures see e.g.~\cite{LawsonMichelson}. We
let~$M$ be a 4-dimensional manifold (without Lorentz metric) and
define the {\em{spinor bundle}}\index{spinor bundle}~$SM$ as a vector bundle over $M$ with
fibre~$\C^4$. The fibres are endowed with a scalar product $\Sl .|. \Sr$
of signature~$(2,2)$, which is again referred to as the~{\em{spin
scalar product}}\index{spin scalar product}.
Sections in the spinor bundle are called {\em{spinors}}\index{spinor} or wave functions.
In local coordinates, a spinor is represented by a
4-component complex function on space-time, usually denoted by $\Psi(x)$.
Choosing at every space-time point a pseudo-orthonormal basis
$(e_\alpha)_{\alpha=1,\ldots,4}$ in the fibres,
\begin{equation} \label{spinbasis}
\Sl e_\alpha| e_\beta \Sr \;=\; s_\alpha \:\delta_{\alpha \beta}
\:,\qquad s_1=s_2=1,\; s_3=s_4=-1
\end{equation}
and representing the spinors in this basis, $\Psi = \Psi^\alpha e_\alpha$,
the spin scalar product takes again the form~(\ref{f:0c}).
Clearly, the basis~$(e_\alpha)$ is not unique, but at every space-point
can be transformed according to
\[ e_\alpha \;\longrightarrow\; (U^{-1})^\beta_\alpha\:e_\beta \:, \]
where~$U$ is an isometry of the spin scalar product, $U \in U(2,2)$.
Under this basis transformation the spinors behave as follows,
\beq \label{psigauge}
\Psi^\alpha(x) \;\longrightarrow\; U^\alpha_\beta(x)\: \Psi^\beta(x)\:.
\eeq
Due to the analogy to~(\ref{lgt}) we interpret this transformation of the
wave functions as a local gauge transformation with gauge group
${\mathcal{G}}=U(2,2)$. We refer to a choice of the
spinor basis~$(e_\alpha)$ as a {\em{gauge}}\index{gauge}.

Our goal is to formulate classical Dirac theory in
such a way that the above $U(2,2)$ gauge transformations correspond to
a physical symmetry, the {\em{$U(2,2)$ gauge symmetry}}.
\index{gauge symmetry!$U(2,2)$}%
To this end, we shall introduce the Dirac operator as the basic
object on~$M$, from which we will later deduce the Lorentz metric and the
gauge potentials. We define a {\em{differential operator}}~${\mathcal{D}}$
{\em{of first order}} on the wave functions by the requirement that in a chart and gauge it should be of the form
\begin{equation}
\label{4_6}
{\mathcal{D}} \;=\; i G^j(x) \frac{\partial}{\partial x^j} + B(x)
\end{equation}
with suitable~$(4 \times 4)$-matrices $G^j$ and~$B$.
This definition does not depend on coordinates and gauge, although
the form of the matrices~$G^j$ and~$B$ clearly does. More precisely,
under a change of coordinates $x^i \to \tilde{x}^i$
the operator~(\ref{4_6}) transforms into
\beq \label{ctrans}
i \left( G^k(\tilde{x})\: \frac{\partial \tilde{x}^j}{\partial x^k} \right)
\frac{\partial}{\partial \tilde{x}^j} + B(\tilde{x})\:,
\eeq
whereas a gauge transformation $\Psi \to U \Psi$ yields the operator
\beq \label{gtrans}
U {\mathcal{D}} U^{-1} \;=\; i \left( U G^j U^{-1} \right)
\frac{\partial}{\partial x^j} + \left(
U B U^{-1} + i U G^j (\partial_j U^{-1}) \right) .
\eeq
We define the Dirac operator by the requirement that by choosing suitable
coordinates and gauge, one can arrange that the matrices~$G^j$ in front of the
partial derivatives ``coincide locally'' with the Dirac matrices of
Minkowski space.
\begin{Def}
\label{def1n}
A differential operator ${\mathcal{D}}\label{mathcalD}$ of first order is called {\bf{Dirac
operator}}\index{Dirac operator} if for every $p \in M$ there is a chart $(x^i, U)$ around $p$
and a gauge $(e_\alpha)_{\alpha=1,\ldots, 4}$ such that ${\mathcal{D}}$ is
of the form~(\ref{4_6}) with
\begin{equation}
\label{4_21}
G^j(p) \;=\; \gamma^j \:,
\end{equation}
where the~$\gamma^j$ are the Dirac matrices of Minkowski space in the
Dirac representation~(\ref{Dirrep}).
\end{Def}

It may seem unconventional that in this definition the
zero order term~$B$ of the Dirac operator is not at all specified.
Furthermore, our formulation as a gauge theory seems incomplete
because we introduced local gauge transformations~(\ref{psigauge},
\ref{lgt}), but not a corresponding gauge-covariant derivative~(\ref{gcd}).
In order to clarify the situation, we shall now
construct from the Dirac operator a gauge-covariant
derivative~$D$, also referred to as {\em{spin derivative}}.
To this end, we must find matrices~$A_j$ which transform under local
gauge transformations according to~(\ref{Atrans}).
This construction will also reveal the structure of the matrix~$B$,
and this will finally lead us to the definition of the so-called
{\em{physical Dirac operator}}, which involves precisely the gravitational and electromagnetic fields.

In the chart and gauge where~(\ref{4_21}) holds, it is obvious
from~(\ref{f:0b}) that the anti-commutator of the matrices~$G^j(p)$ gives
the Minkowski metric. Using the transformation rules~(\ref{ctrans}, \ref{gtrans}), one sees that in a general coordinate system and gauge, their
anti-commutator defines a Lorentz metric,
\begin{equation}
\label{4_4}
g^{jk}(x) \:\1 \;=\; \frac{1}{2} \: \{ G^j(x),\: G^k(x) \} \: .
\end{equation}
In this way, the Dirac operator induces on the manifold a
Lorentzian structure. We refer to the matrices~$G^j$ as the
Dirac matrices in curved space-time.
Since we can arrange that these matrices coincide locally with
the Dirac matrices of Minkowski space, all manipulations of Dirac
matrices can be performed at any given space-time point in an obvious
way. In particular, the pseudoscalar matrix~(\ref{rhodef}) now takes the
more general form
\[ \rho(x) \;=\; \frac{i}{4!}\: \varepsilon_{jklm} \:G^j
G^k G^l G^m\:, \]
where the anti-symmetric tensor~$\varepsilon_{jklm}$ differs from
the anti-symmetric symbol~$\epsilon_{jklm}$ by the volume density,
$\varepsilon_{jklm} = \sqrt{|\det g|}\: \epsilon_{jklm}$.
The pseudoscalar matrix gives us again the notion of even and odd
matrices and of chirality~(\ref{chidef}). Furthermore, we introduce the
{\em{bilinear matrices}}\index{bilinear matrix} $\sigma^{jk}\label{sigmajk}$ by
\[ \sigma^{jk}(x) \;=\; \frac{i}{2} \: [G^j,\: G^k] \: . \]
As in Minkowski space, the matrices
\begin{equation}
    G^j\;,\;\;\;\; \rho G^j \;,\;\;\;\; \1 \;,\;\;\;\; i \rho
    \;,\;\;\;\; \sigma^{jk}
    \label{5_10}
\end{equation}
form a basis of the 16-dimensional (real) vector space
of selfadjoint matrices (with respect to $\Sl .|. \Sr$).
The matrices $G^j$ and $\rho G^j$ are odd, whereas
$\1$, $i\rho$ and $\sigma^{jk}$ are even.

For the construction of the spin connection we must clearly consider
derivatives.
The Lorentzian metric (\ref{4_4}) induces the Levi-Civita connection
$\nabla$, which defines the covariant derivative of tensor
fields. Taking covariant derivatives of the Dirac
matrices, $\nabla_k G^j = \partial_k G^j + \Gamma^j_{kl} \:G^l$,
we obtain an expression which behaves under coordinate transformations
like a tensor. However, it is not gauge covariant, because
a gauge transformation~(\ref{lgt}) yields contributions
involving first derivatives of $U$. More precisely, according to~(\ref{gtrans}),
\begin{eqnarray}
    \nabla_k G^j \;\longrightarrow\; \nabla_k (U G^j U^{-1}) & = &
    U (\nabla_k G^j) U^{-1} \:+\: (\partial_k U) G^j U^{-1} \:+\:
    U G^j (\partial_k U^{-1}) \nonumber \\
& = & U (\nabla_k G^j) U^{-1} \:-\: \left[ U (\partial_k U^{-1}),\:
     U G^j U^{-1} \right] \: .
    \label{5_5a}
\end{eqnarray}
We can use the second summand in (\ref{5_5a}) to partially fix the gauge.
\begin{Lemma}
\label{lemma1}
For every space-time point $p \in M$ there is a gauge such that
\begin{equation}
    \nabla_k G^j(p) \;=\; 0
    \label{5_5}
\end{equation}
(for all indices~$j,k$).
\end{Lemma}
{\Proof}
We start with an arbitrary gauge and construct the desired gauge with
two subsequent gauge transformations:
\begin{enumerate}
\item
The matrix $\partial_j \rho$ is odd, because
\[ 0 \;=\; \partial_j \1 \;=\; \partial_j (\rho \rho) \;=\;
(\partial_j \rho) \rho + \rho (\partial_j \rho) \: . \]
As a consequence, the matrix $i \rho (\partial_j \rho)$ is
selfadjoint. We can thus perform a gauge
transformation $U$ with $U(p)=\1$, $\partial_j U(p)=\frac{1}{2} \rho
(\partial_j \rho)$. In the new gauge the matrix $\partial_j \rho(p)$ vanishes,
\[ \partial_j \rho_{|p} \;\longrightarrow\; \partial_j (U \rho U^{-1})_{|p}
\;=\; \partial_j \rho_{|p} + \frac{1}{2} \left[ \rho (\partial_j \rho), \:
\rho  \right]_{|p} \;=\; \partial_j \rho_{|p} - \rho^2 (\partial_j
\rho)_{|p} \;=\; 0 \: . \]
Differentiating the relation $\{ \rho, G^j \}=0$, one sees that
the matrix $\nabla_k G^j_{|p}$ is odd. We can thus represent it in the form
\begin{equation}
    \nabla_k G^j_{|p} \;=\; \Lambda^j_{km} \:G^m_{|p} \:+\:
\Theta^j_{km} \: \rho G^m
    \label{5_7}
\end{equation}
    with suitable coefficients~$\Lambda^j_{km}$ and~$\Theta^j_{km}$.

    This representation can be further simplified: According to Ricci's
    Lemma, $\nabla_n g^{jk}=0$. Expressing the metric via the
    anti-commutation relations and differentiating through with the
    Leibniz rule, we obtain
\begin{eqnarray}
    0 & = &  \{ \nabla_n G^j,\:G^k\} \:+\: \{ G^j,\:
    \nabla_n G^k\} \nonumber \\
     & = & 2 \Lambda^j_{nm} \: g^{mk} \:-\: \Theta^j_{nm} \: 2i\rho
\sigma^{mk}
     \:+\: 2 \Lambda^k_{nm} \: g^{mj} \:-\: \Theta^k_{nm} \: 2i\rho
\sigma^{mj} \label{5_9}
\end{eqnarray}
    and thus
\begin{equation}
    \Lambda^j_{nm} \:g^{mk}_{|p} \;=\; -\Lambda^k_{nm} \: g^{mj}_{|p} \:.
    \label{5_8}
\end{equation}
    In the case $j=k \neq m$, (\ref{5_9}) yields that $\Theta^j_{nm}=0$.
    For $j \neq k$, we obtain $\Theta^j_{nj} \:\sigma^{jk} +
    \Theta^k_{nk} \:\sigma^{kj} = 0$ and thus $\Theta^j_{nj} =
\Theta^k_{nk}$
    ($j$ and $k$ denote fixed indices, no summation is performed).
    We conclude that there are coefficients $\Theta_k$ with
\begin{equation}
    \Theta^j_{km} \;=\; \Theta_k \: \delta^j_m \: .
    \label{5_9a}
\end{equation}
    \item  We perform a gauge transformation $U$ with $U(p)=\1$ and
\[ \partial_k U \;=\; -\frac{1}{2} \:\Theta_k \: \rho \:-\: \frac{i}{4} \:
\Lambda^m_{kn} \: g^{nl} \: \sigma_{ml} \: . \]
    Using the representation (\ref{5_7}) together with
    (\ref{5_8}, \ref{5_9a}),
    the matrix $\nabla_k G^j$ transforms into
\begin{eqnarray*}
\nabla_k G^j & \longrightarrow & \nabla_k G^j \:+\: [\partial_k U,\:
G^j]  \\
     &  & =\; \Lambda^j_{km} \:G^m \:+\: \Theta_k \: \rho G^j
     \:-\: \Theta_k \: \rho G^j \:-\:
     \frac{i}{4}\: \Lambda^m_{kn} \:g^{nl} \: [\sigma_{ml},\: G^j] \\
     &  & =\; \Lambda^j_{km} \:G^m \:-\: \frac{i}{4}\: \Lambda^m_{kn}
     \:g^{nl} \:2i \: (G_m \: \delta^j_l - G_l \:\delta^j_m) \\
     &  & =\; \Lambda^j_{km} \:G^m \:+\: \frac{1}{2}\: \Lambda^m_{kn} \:
     g^{nj} \: G_m \:-\: \frac{1}{2}\: \Lambda^j_{km} \: G^m \;=\; 0 \: .
\end{eqnarray*}
\end{enumerate}
\vspace*{-1.85em}
\QED

We call a gauge satisfying condition~(\ref{5_5}) a {\em{normal
gauge}}\index{gauge!normal} around $p$. In order to analyze the remaining gauge freedom, we let~$U$ be a
transformation between two normal gauges. Then according to~(\ref{5_5a})
and~(\ref{5_5}), the commutator
$[U (\partial_k U^{-1}),\: UG^jU^{-1}]$ vanishes at~$p$ or, equivalently,
\[ [i (\partial_k U^{-1})\:U,\: G^j]_{|p} \;=\; 0 \:. \]
As is easily verified in the basis~(\ref{5_10}) using the
commutation relations between the Dirac matrices, a matrix which
commutes with all Dirac matrices is a multiple of the identity matrix.
Moreover, the matrix~$i (\partial_j U^{-1}) \:U$ is selfadjoint because
$(i (\partial_j U^{-1}) \:U)^* = -i U^{-1}\:(\partial_j U)
= -i \partial_j (U^{-1} U) + i (\partial_j U^{-1}) \:U
= i (\partial_j U^{-1}) \:U$. We conclude that the matrix
$i (\partial_j U^{-1}) \:U$ is a real multiple of the identity matrix,
and transforming it unitarily with~$U$ we see that it also
coincides with the matrix~$i U\:(\partial_j U^{-1})$.
Under this strong constraint for the gauge transformation it is easy to find
expressions with the required behavior~(\ref{Atrans}) under
gauge transformations. Namely, setting \label{Tr}
\beq \label{adef}
a_j \;=\; \frac{1}{4}\: {\mbox{Re }} \Tr (G_j \:B) \: \1 \:,
\eeq
where ``Tr'' denotes the trace of a $4\times 4$-matrix,
one sees from~(\ref{gtrans}) that
\[ a_j  \;\longrightarrow \;
a_j \:+\: \frac{1}{4}\: {\mbox{Re }} \Tr \left( G_j G^k \: i(\partial_k
U^{-1})\:U \right) \1 \;=\; a_j + i U (\partial_j U^{-1}) \: . \]
We can identify the~$a_j$ with the gauge potentials~$A_j$ and
use~(\ref{gcd}) as the definition of the spin connection.
\begin{Def} \label{def_sd}
The {\bf{spin derivative}}\index{spin derivative} $D\label{D}$ is defined by the condition that
it behaves under gauge transformations~(\ref{lgt}) according to~(\ref{Dlgt})
and in normal gauges around $p$ has the form
\begin{equation}
    D_j(p) \;=\; \frac{\partial}{\partial x^j} \:-\: i a_j
    \label{5_11}
\end{equation}
with the potentials~$a_j$ according to~(\ref{adef}).
\end{Def}
In general gauges, the spin derivative can be written as
\begin{equation}
    D_j \;=\; \frac{\partial}{\partial x^j} \:-\: i E_j \:-\: i a_j
    \label{5_12}
\end{equation}
with additional matrices $E_j(x)$, which involve the Dirac matrices and
their first derivatives.
A short calculation shows that the trace of the matrix~$E_j$ does
not change under gauge transformations, and since it vanishes in
normal gauges, we conclude that the matrices~$E_j$ are trace-free.
A straightforward calculation yields that they are explicitly given by
\[ E_j \;=\; \frac{i}{2}\: \rho\: (\partial_j \rho) \:-\: \frac{i}{16}\: \Tr
    (G^m \:\nabla_j G^n) \: G_m G_n \:+\: \frac{i}{8}\: \Tr (\rho G_j \:
    \nabla_m G^m) \:\rho \:. \]

In the next two theorems we collect the basic properties of the spin connection.
\begin{Thm}
\label{thm_2}
The spin derivative satisfies for all wave
functions~$\Psi, \Phi$ the equations
\begin{eqnarray}
[D_k, G^j] \:+\: \Gamma^j_{kl} \: G^l &=& 0 \label{5_13} \\
\partial_j \:\Sl \Psi \:|\: \Phi \Sr &=& \Sl D_j \Psi \:|\: \Phi \Sr
    \:+\: \Sl \Psi \:|\: D_j \Phi \Sr \:.    \label{5_17}
\end{eqnarray}
\end{Thm}
{\Proof}
The left side of (\ref{5_13}) behaves under gauge transformations
according to the adjoint representation $. \to U \:.\: U^{-1}$ of the gauge group.
Thus it suffices to check (\ref{5_13}) in a normal gauge, where
\[ [D_k,G^j] + \Gamma^j_{kl} \: G^l \;=\; \nabla_k G^j \:-\:
\frac{i}{4}\: {\mbox{Re }} \Tr (G_j B) \: [\1,G^j] \;=\; 0 \: . \]

Since both sides of~(\ref{5_17}) are gauge invariant, it again suffices to
consider a normal gauge. The statement is then an immediate consequence
of the Leibniz rule for partial derivatives and the fact that the spin derivative differs from the partial derivative by
an imaginary multiple of the identity matrix~(\ref{5_11}).
\QED
The identity~(\ref{5_13}) means that the coordinate and gauge invariant
derivative of the Dirac matrices vanishes.
The relation~(\ref{5_17}) shows that the spin connection is
compatible with the spin scalar product.

We define {\em{torsion}}\index{torsion}~${\mathcal{T}}$ and {\em{curvature}}\index{curvature}~${\mathcal{R}}$ of the spin connection
as the following 2-forms,
\[ {\mathcal{T}}_{jk} \;=\; \frac{i}{2} \left( [D_j, G_k] - [D_k, G_j] \right) \;,\spc
    {\mathcal{R}}_{jk} \;=\; \frac{i}{2} \: [D_j, D_k] \:. \]
\begin{Thm}
\label{thm2}
The spin connection is torsion-free. Curvature has the form
\begin{equation}
    {\mathcal{R}}_{jk} \;=\; \frac{1}{8} \: R_{mnjk} \:\sigma^{mn} \:+\: \frac{1}{2} \:
    (\partial_j a_k - \partial_k a_j)
    \label{5_21}
\end{equation}
where~$R_{mnjk}$ is the the Riemannian curvature tensor and
the~$a_j$ are given by~(\ref{adef}).
\end{Thm}
{\Proof} The identity~(\ref{5_13}) yields that
\[ [D_j,\: G_k] \;=\; [D_j, \:g_{kl} \: G^l] \;=\; (\partial_j g_{kl})
\: G^l
   \:-\: g_{kl} \: \Gamma^l_{jm} \: G^m \;=\; \Gamma^m_{jk} \: G_m \]
and thus, using that the Levi-Civita connection is torsion-free,
\[ {\mathcal{T}}_{jk} \;=\; \frac{i}{2} \: (\Gamma^m_{jk} - \Gamma^m_{kj}) \:
G_m \;=\; 0 \: . \]

Again using~(\ref{5_13}), we can rewrite the covariant derivative as a
spin derivative,
\[ G_l\: \nabla_k u^l \;=\; [D_k,\: G_l u^l]\:. \]
Iterating this relation, we can express the Riemann tensor~(\ref{RT}) by
\begin{eqnarray*}
G_i \: R^i_{jkl} \:u^l &=&
[D_j,\: [D_k,\: G_l u^l ]] - [D_k,\: [D_j,\: G_l u^l ]] \\
& = & [[D_j, \: D_k],\: G_l u^l ] \;=\; -2i \: [{\mathcal{R}}_{jk},\: G_l u^l] \: .
\end{eqnarray*}
This equation determines curvature up to a multiple of the identity
matrix,
\[ {\mathcal{R}}_{jk}(x) \;=\; \frac{1}{8} \: R_{mnjk} \:\sigma^{mn} \:+\:
   \lambda_{jk} \1 \: . \]
Thus it remains to compute the trace of curvature,
\[ \frac{1}{4}\: \Tr ({\mathcal{R}}_{jk})\:\1 \;=\; \frac{1}{8}\:
     \Tr (\partial_j A_k - \partial_k A_j)\:\1  \;=\; \frac{1}{2} \:
     (\partial_j a_k - \partial_k a_j) \: , \]
where we used~(\ref{5_12}) and the fact that the matrices $E_j$ are
trace-free.
\QED

We come to the physical interpretation of the above construction.
According to Lemma~\ref{lemma1} we can choose a gauge around~$p$ such that the
covariant derivatives of the Dirac matrices vanish at~$p$.
Moreover, choosing normal coordinates and making a global (=constant) gauge
transformation, we can arrange that~$G(p) = \gamma^j$
and~$\partial_j g_{kl}(p)=0$. Then the covariant derivatives at~$p$
reduce to partial derivatives, and we conclude that
\beq \label{nrf}
G^j(p) \;=\; \gamma^j \;,\spc \partial_k G^j(p) \;=\; 0 \:.
\eeq
These equations have a large similarity with the conditions for normal
coordinates (\ref{gnc}), only the role of the metric
is now played by the Dirac matrices. Indeed, differentiating~(\ref{4_4}) one sees
that~(\ref{nrf}) implies~(\ref{gnc}). Therefore, (\ref{nrf}) is a
stronger condition which not only gives a constraint for the coordinates,
but also for the gauge. We call a coordinate system and gauge where~(\ref{nrf})
is satisfied a {\em{normal reference frame}}\index{reference frame!normal} around~$p$.

In a normal reference frame, the Dirac matrices, and via~(\ref{4_4}) also
the metric, are the same as in Minkowski space up to the order
$o(x-p)$. According to the strong equivalence principle, the Dirac equation
at~$p$ should coincide with that in Minkowski space. Now we use
minimal coupling in the formulation~(\ref{mc}) to conclude that there should
be a normal gauge such that all gauge potentials vanish at~$p$, and thus the
Dirac operator at~$p$ should coincide with the free Dirac operator~$i \Pdd$.
This physical argument allows us to specify the zero order term
in~(\ref{4_6}).
\begin{Def}
\label{def_pdo}
A Dirac operator ${\mathcal{D}}\label{mathcalD2}$ is called {\bf{physical Dirac
    operator}}\index{Dirac operator!physical}
if for any $p \in M$ there is a normal reference frame around $p$ such that $B(p)=0$.
\end{Def}
Equivalently, the physical Dirac operator could be defined as a
differential operator of first order~(\ref{4_6}) with the additional
structure that for any~$p \in M$ there is a coordinate chart and gauge
such that the following three conditions are satisfied,
\[ G^j(p) = \gamma^j \;,\spc \partial_k G^j(p) = 0 \;,\spc B(p) = 0\:. \]
This alternative definition has the disadvantage that it is a-priori not
clear whether the second condition~$\partial_k G^j(p) = 0$
can be satisfied for a general metric. This is the reason why we
preferred to begin with only the first condition
(Def.~\ref{def1n}), then showed that the second condition can be
arranged by choosing suitable coordinates and gauge, and satisfied the
third condition at the end (Def.~\ref{def_pdo}).

In general coordinates and gauge, the physical Dirac operator can be
written as
\[ {\mathcal{D}} \;=\; i G^j D_j \;=\; i G^j\:(\partial_j - i E_j - i a_j)
\:, \]
where~$D$ is the spin connection of Def.~\ref{def_sd}.
The matrices~$E_j$ take into account the gravitational field and are
called {\em{spin coefficients}}, whereas the~$a_j$ can be identified with
the {\em{electromagnetic potential}} (compare~(\ref{Dirac2})).
We point out that the gravitational field cannot be introduced into
the Dirac equation by the simple replacement rule~(\ref{minicoupling})
because gravity has an effect on both the Dirac matrices and the
spin coefficients. But factorizing the gauge group as
$U(2,2) = U(1) \times SU(2,2)$, the~$SU(2,2)$-gauge transformations
are linked to the gravitational field because they
influence~$G^j$ and~$E_j$, whereas the~$U(1)$ can be identified with
the gauge group of electrodynamics. In this sense, we obtain a
unified description of electrodynamics and general relativity as a $U(2,2)$
gauge theory. The Dirac equation\index{Dirac equation}
\[ ({\mathcal{D}}-m)\: \Psi \;=\; 0 \index{Dirac equation!in the
gravitational field} \]
describes a Dirac particle in the gravitational and electromagnetic
field. According to Theorem~\ref{thm2}, the curvature of the spin
connection involves both the Riemann tensor and the electromagnetic field
tensor. We can write down the classical action in terms of these tensor
fields, and variation yields the classical Einstein-Dirac-Maxwell
equations.

For the probabilistic interpretation of the Dirac equation in
curved space-time, we choose a space-like hypersurface~${\mathcal{H}}$
(corresponding to ``space'' for some observer) and consider
in generalization of~(\ref{f:0e}) on solutions of the Dirac
equation the scalar product
\beq \label{nicsp}
(\Psi \:|\: \Phi) \;=\; \int_{\mathcal{H}} \Sl \Psi \:|\: G^j \nu_j\:
\Phi \Sr \: d\mu_{\mathcal{H}} \:,
\eeq
where~$\nu$ is the future-directed normal on~${\mathcal{H}}$
and~$d\mu_{\mathcal{H}}$ is the invariant measure on the Riemannian
manifold~${\mathcal{H}}$. \label{(.|.)b}
Then~$(\Psi | \Psi)$ is the {\em{normalization
    integral}}\index{normalization integral}, which
we again normalize to one.
Its integrand has the interpretation as the {\em{probability
    density}}\index{probability density}.
In analogy to~(\ref{dc}) the {\em{Dirac current}} is introduced
by~$J^k = \Sl \Psi \:|\: G^k \Psi \Sr$. Using Theorem~\ref{thm_2} one
sees similar as in Minkowski space that the Dirac current is
divergence-free, $\nabla_k J^k = 0$. From Gauss'
theorem one obtains that the scalar product~(\ref{nicsp}) does not
depend on the choice of the hypersurface~${\mathcal{H}}$.

We finally remark that using Theorem~\ref{thm_2} together with
Gauss' theorem, one easily verifies that the
physical Dirac operator is Hermitian with respect to the inner product
\label{langle.|.rangle}
\beq \label{defsp}
\bra \Psi \:|\: \Phi \ket \;:=\; \int_M \Sl \Psi \:|\: \Phi \Sr\: d\mu\:,
\eeq
in which the wave functions (which need not satisfy the Dirac
equation but must have a suitable decay at infinity)
are integrated over the whole space-time.
This inner product is not positive, but it will nevertheless
play an important conceptual role in the next chapters.

\chapter{The Fermionic Projector in the Continuum} \label{sec3}
\setcounter{equation}{0}
In the previous chapter we introduced the concept of the Dirac sea
in order to give the negative-energy solutions of the free Dirac
equation a physical meaning as anti-particle states (see~{\S}\ref{isec2},
{\S}\ref{isec4}). Now we shall extend this concept to the
case with interaction. We will see that the Dirac sea can still be
introduced as a universal object in space-time, described mathematically
by the so-called {\em{fermionic projector}} ({\S}\ref{jsec3}).
We develop the mathematical methods for an explicit analysis of the
fermionic projector in position space ({\S}\ref{jsec5}) and finally
consider the normalization of the fermionic states ({\S}\ref{jsec6}).

\section{The External Field Problem}\index{external field problem}
\label{jsec1} \setcounter{equation}{0}
We begin with the simplest interaction: a classical external field in
Minkowski space. In this situation the Dirac wave function is a solution
of the Dirac equation
\begin{equation}
(i \Pdd + {\mathcal{B}} - m) \:\Psi \;=\; 0 \: ,
\label{1}
\end{equation}
where the operator ${\mathcal{B}}\label{mathcalB}$ is composed of
the external potentials (as an example one may choose
${\mathcal{B}}=\Aslsh$ with $A$ the electromagnetic
potential~(\ref{Dirac2})). In order to have current conservation
(i.e.\ the identity $\partial_k J^k =0$ with~$J$ according
to~(\ref{dc})), we always assume that~${\mathcal{B}}$ is Hermitian
(with respect to the spin scalar product). If~${\mathcal{B}}$ is
{\em{static}} (=time independent), we can separate the time
dependence of the wave function with a plane wave ansatz, \beq
\label{pwa} \index{potential!static} \Psi(t,\vec{x}) \;=\; e^{-i
\omega t}\: \psi(\vec{x})\:. \eeq The separation constant~$\omega$
has the interpretation as the energy of the solution. Thus the
energy is a conserved quantity, and its sign distinguishes between
solutions of positive and negative energy. In more mathematical
terms, for a static potential the Hamiltonian in~(\ref{Hamilton}) is
time independent, and the sign of the spectrum of $h$ gives a
splitting of the solution space of the Dirac equation into the
subspaces of positive and negative energy, respectively. As a
consequence, our previous construction of the Dirac sea can be
adapted: When building up the Fock space from the one-particle
states, we cure the problem of the negative-energy solutions similar
to~(\ref{f:0k}--\ref{f:0m}) by redefining the vacuum and by formally
exchanging the creation and annihilation operators corresponding to
the negative-energy solutions.

The situation becomes much more difficult when~${\mathcal{B}}$ is
{\em{time-dependent}}. In this case, the separation ansatz~(\ref{pwa}) no longer
works. The energy is not conserved, and it is even possible that a
solution which has positive energy at initial time will have negative
energy at a later time. Expressed more mathematically, the Hamiltonian
in~(\ref{Hamilton}) now depends explicitly on time, and therefore the
sign of the spectrum of~$h$ no longer gives a canonical splitting of the solution
space of the Dirac equation (this splitting would also depend on time).
As a consequence, it is not clear which solutions have the interpretation
as ``negative-energy solutions'' and thus correspond to anti-particle
states~(\ref{f:0l2}).  For this reason, it is no longer obvious how to
quantize the Dirac field in a canonical way.
This difficulty is usually referred to as the {\em{external field problem}}.
It becomes most evident in the setting of Klein's
paradox\index{Klein's paradox}, where
one considers a step potential whose amplitude is larger than the mass
gap (see~\cite{BD, Tha}).
However, we point out that the external field problem appears already for
arbitrarily weak external fields, simply because the time dependence
of~${\mathcal{B}}$ leads to a complicated mixing of the solutions of positive
and negative energy. We could speak of a
``solution of negative energy'' only if it were a superposition of
states which {\em{all}} had negative energy, and there seems no
reason why such solutions should exist.

It is instructive to discuss the external field problem in the setting
of pertubation theory. Consider a first order perturbation of the plane-wave
solution~$\Psi_{\vec{p} s \epsilon}$,
\beq \label{1ord}
\Psi \;=\; \Psi_{\vec{p} s \epsilon} + \Delta \Psi
+ {\mathcal{O}}({\mathcal{B}}^2)\:.
\eeq
Substituting this ansatz into the Dirac equation~(\ref{1}), we
obtain to first order in~${\mathcal{B}}$ the inhomogeneous Dirac equation
\beq \label{iDe}
(i \Pdd - m)\: \Delta \Psi \;=\; - {\mathcal{B}}\: \Psi_{\vec{p} s
\epsilon} \:.
\eeq
If $s_m\label{sm}(x,y)$ is a {\em{Green's function}}\index{Green's function} of the free Dirac equation,
characterized by the distributional equation
\begin{equation}
(i \Pdd_x - m) \: s_m(x,y) \;=\; \delta^4(x-y) \: , \label{4}
\end{equation}
we can construct a solution of~(\ref{iDe}) by
\beq \label{DPsi}
\Delta \Psi \;=\; - \int d^4y\; s(x,y)\: {\mathcal{B}}(y)\:
\Psi_{\vec{p} s \epsilon}(y)
\eeq
(in order not to distract from the main ideas, we here
calculate on a formal level; the analytic justification will be given
at the end of~{\S}\ref{jsec2}).
If the Green's function were unique, (\ref{1ord}, \ref{DPsi}) would
give a unique procedure for perturbing the negative-energy
solutions of the vacuum, making it possible to extend the notion of
``negative-energy state'' to the interacting theory (at least in
first order perturbation theory).

The problem is that the Green's function is {\em{not unique}}, as we now
briefly recall (for details see~\cite{BD}). Taking the Fourier transform of~(\ref{4}),
\beq \label{Ft}
s_m(x,y) \;=\; \int \frac{d^4k}{(2 \pi)^4} \:s_m(k)\: e^{-ik (x-y)} \:,
\eeq
we obtain the algebraic equation
\[ (k\slsh - m)\: s_m(k) \;=\; \1\:. \]
Since the matrix~$k \slsh - m$ is singular on the mass shell
(see the argument after~(\ref{kDirac})), this equation can be solved
for~$s_m(k)$ only after using a $\pm i \varepsilon$-regularization on
the mass shell. The most popular choices are the
{\em{advanced}}\index{Green's function!advanced} and the {\em{retarded}} Green's
functions\index{Green's function!retarded}
defined by \label{slorm}
\begin{equation}
        s^{\lor}_{m}(k) \;=\; \lim_{\varepsilon \searrow 0}
            \frac{k \slsh + m}{k^{2}-m^{2}-i \varepsilon k^{0}}
            \qquad {\mbox{and}} \qquad
           s^{\land}_{m}(k) \;=\; \lim_{\varepsilon \searrow 0}
            \frac{k \slsh + m}{k^{2}-m^{2}+i \varepsilon k^{0}} \:,
        \label{8b}
\end{equation}
respectively  \label{slandm} (with the limit~$\varepsilon \searrow 0$ taken
in the distributional sense). Computing their Fourier transform~(\ref{Ft}) with
residues, one sees that they are {\em{causal}} in the sense that
their supports lie in the upper and lower light cone, respectively, \label{supps}
\beq \label{GF}
{\mbox{supp}}\, s^\lor_m(x,.) \subset J_x^\lor \;,\spc
{\mbox{supp}}\, s^\land_m(x,.) \subset J_x^\land\:.
\eeq
Another commmon choice is the {\em{Feynman propagator}}\index{Feynman propagator}
\beq \label{Fp}
s^F_m(k) \;:=\; \lim_{\varepsilon \searrow 0}
\frac{k \slsh + m}{k^{2}-m^{2}+i \varepsilon} \:.
\eeq
\label{sFm}
Taking the Fourier transform~(\ref{Ft}) with residues, one finds
\beq \label{frc}
s^F_m(x,y) \;=\; \frac{i}{(2 \pi)^3} \int_{\sR^3} \left. (k\slsh+m)\:
e^{-ik(x-y)} \right|_{k = (\epsilon(t) \,\omega(\vec{p}),\: \vec{p})}
d\mu_{\vec{p}}
\eeq
with~$t \equiv (y-x)^0$ and~$\omega(\vec{p})$,~$d\mu_{\vec{p}}$ as
introduced after~(\ref{f:0hh}). Here~$\epsilon$ denotes the step function $\epsilon(x)=1$
for $x \geq 0$ and $\epsilon(x)=-1$ otherwise. Thus for positive~$t$
we get the integral over the upper mass shell, whereas for negative~$t$
we integrate over the lower mass shell.
As a consequence, the Feynman propagator is not causal, but it is instead
characterized by the {\em{frequency conditions}}\index{frequency condition}  that it is for
positive and negative time~$t$ composed only of positive and negative
frequencies, respectively. More systematically,
the defining equation for the Green's function~(\ref{4}) determines~$s_m$
only up to a solution of the homogeneous Dirac equation. Thus we can
write a general Green's function $s_m$ in the form
\beq \label{arbi}
s_m(x,y) \;=\; s^{\lor}_{m}(x,y) \:+\: a(x,y) \: ,
\eeq
where $a(x,y)$ is  a linear combination of
plane-wave solutions in the variable~$x$, i.e.
\[ a(x,y) \;=\; \sum_{s,\epsilon} \int_{\sR^3} \Psi_{\vec{p} s \epsilon}(x) \:
c_{\vec{p} s \epsilon}(y) \; d\mu_{\vec{p}} \]
with a suitable complex-valued function $c_{\vec{p} s \epsilon}(y)$.

Due to the non-uniqueness of the Green's function~(\ref{arbi}),
the relations~(\ref{1ord}, \ref{DPsi}) do {\em{not}} give a unique
procedure for perturbing the plane-wave solutions~$\Psi_{\vec{k} s \epsilon}$.
\index{perturbation expansion!non-uniqueness}
In particular, it is not clear how to extend the notion of
``negative energy state'' to the interacting theory. This corresponds
precisely to the external field problem. We conclude that in a perturbative approach, the external field problem becomes manifest in the non-uniqueness of the perturbation expansion.

Feynman~\cite{Fey} gave the frequency conditions in the Feynman
propagator the physical interpretation as ``positive-energy states
moving to the future'' and ``ne\-ga\-tive-energy states moving to the past''.
Identifying ``negative-energy states moving to the past'' with
``antiparticle states'' he concluded that the Feynman propagator
is distinguished from all other Green's functions in that it
takes into account the particle/anti-particle interpretation of the
Dirac equation in the physically correct way.
Feynman proposed to perform the perturbation expansion
exclusively with the Feynman propagator, thereby making the perturbation
expansion unique. The flaw is that the frequency conditions are not
invariant under general coordinate and gauge transformations
(simply because such transformations ``mix'' positive and negative
frequencies), and therefore Feynman's method is not compatible
with the equivalence principle and the local gauge principle.
This is not a problem for most calculations in physics,
but it is not satisfying conceptually.
Another approach is to work with the so-called~{\em{Hadamard
states}}\index{Hadamard state}~\cite{Haag, Wald2}. The disadvantage of this approach is
that the states of the quantum field no longer have a particle
interpretation. In other words, the notions of ``particle'' and ``anti-particle'' depend on the local observer,
and therefore also the notion of the Dirac sea loses its universal meaning.
We proceed in~{\S}\ref{jsec2} by showing that the Dirac sea can
indeed be introduced as a global object of space-time, even in the
presence of a general interaction.

We finally remark that the above arguments apply in the same way for
second quantized fields: we only need to replace~${\mathcal{B}}$ by
an operator on a suitable bosonic Fock space. Also, our assumption
that~${\mathcal{B}}$ is an {\em{external}} field\index{external field} is merely a technical
simplification (more precisely, we disregard the dynamical equations for the bosonic
fields, thereby also avoiding the divergences of QFT and the
renormalization procedure), but it is not essential for our arguments.
Namely, in a time-dependent interacting system, the Dirac wave
functions satisfy~(\ref{1}), where~${\mathcal{B}}(t,\vec{x})$ is
determined by the dynamical equations of the whole system.
Solving these equations we can (at least in principle) compute~${\mathcal{B}}$,
and applying our above arguments with this~${\mathcal{B}}$ as an
external field, we conclude that the notion of ``negative-energy state''
ceases to exist. In what follows we will for simplicity again consider
an external field, but we shall come back to coupled systems in~{\S}\ref{jsec4}.

\section{The Causal Perturbation Expansion}\index{perturbation expansion!causal}
\label{jsec2} \setcounter{equation}{0}
We saw in the previous section that the external field problem for the
Dirac equation~(\ref{1}) is equivalent to the non-uniqueness of the
perturbation expansion for the individual states of the Dirac sea.
We shall now solve this problem by considering the collection of all states
of the Dirac sea. This will reveal an underlying causal structure,
which will enable us to make the perturbation expansion
unique\footnote{We remark for clarity that our ``causal perturbation
expansion'' does not seem to be related to Scharf's
``causal approach'' to QED~\cite{Scharf}.
Scharf uses causality to avoid the ultraviolet divergences of
perturbative QED, whereas in our setting of an external field all Feynman
diagrams are finite anyway. On the other hand, Scharf is
interested only in the scattering states, whereas our goal is to
describe the dynamics also for intermediate times.}.
We closely follow the constructions given in~\cite{F3}.

In the vacuum, out of all plane-wave solutions of
negative energy we form the object\label{Psea}
\beq \label{Pvac}
P^{\mbox{\scriptsize{sea}}}(x,y) \;:=\; -\frac{m}{\pi} \sum_s \int_{\sR^3}
|\Psi_{\vec{p}s-} \Sr \Sl \Psi_{\vec{p}s-} | \; d\mu_{\vec{p}}\:.
\eeq
Using the explicit form of the plane-wave solutions~(\ref{pws},
\ref{exppr}, \ref{projkern}), we obtain the covariant formula
\begin{equation}
P^{\mbox{\scriptsize{sea}}}(x,y) \;=\; \int \frac{d^4k}{(2 \pi)^4}\:
(k\slsh+m)\: \delta(k^2-m^2)\: \Theta(-k^0)\: e^{-ik(x-y)}\:,  \label{f:2b}
\end{equation}
which also shows that $P^{\mbox{\scriptsize{sea}}}(x,y)$ is a well-defined
distribution. In order to get a connection to causality, we
decompose~$P^{\mbox{\scriptsize{sea}}}(x,y)$ in the form
\begin{equation}
P^{\mbox{\scriptsize{sea}}}(x,y) \;=\; \frac{1}{2} \left( p_{m}(x,y)
\:-\: k_{m}(x,y) \right) ,      \label{12b}
\end{equation}
where~$p_m$ and~$k_m$ are the distributions \label{pm}
\begin{eqnarray}
p_{m}(x,y) &=& \int \frac{d^{4}k}{(2 \pi)^{4}} \:(k \slsh + m) \:
  \delta(k^{2}-m^{2}) \: e^{-ik(x-y)} \label{x20} \\
k_{m}(x,y) &=& \int \frac{d^{4}k}{(2 \pi)^{4}} \:(k \slsh + m) \:
  \delta(k^{2}-m^{2}) \:\epsilon(k^{0}) \: e^{-ik(x-y)} \:. \label{x21}
\end{eqnarray}
In order to relate the \label{km}
distribution~$k_m$ to the advanced and retarded Green's
functions, we substitute the distributional equation
\[ \lim_{\varepsilon \searrow 0} \left( \frac{1}{x-i\varepsilon}
\:-\: \frac{1}{x+i \varepsilon} \right) \;=\; 2 \pi i \: \delta(x) \]
into the formula for $k_{m}$ in momentum space,
\begin{eqnarray}
k_{m}(p) &=& (p \slsh + m) \: \delta(p^{2}-m^{2}) \: \epsilon(p^{0})
\nonumber \\
&=& \frac{1}{2 \pi i} \:(p \slsh + m) \; \lim_{\varepsilon
\searrow 0} \left[ \frac{1}{p^{2}-m^{2}-i\varepsilon} \:-\:
\frac{1}{p^{2}-m^{2}+i\varepsilon} \right] \epsilon(p^{0}) \nonumber \\
&=& \frac{1}{2 \pi i} \:(p \slsh + m) \; \lim_{\varepsilon
\searrow 0} \left[ \frac{1}{p^{2}-m^{2}-i\varepsilon p^{0}} \:-\:
\frac{1}{p^{2}-m^{2}+i\varepsilon p^{0}} \right] .
\label{2pole}
\end{eqnarray}
Using~(\ref{8b}) we get the simple formula
\begin{equation}
        k_{m} \;=\; \frac{1}{2\pi i} \: \left( s^{\lor}_{m} \:-\: s^{\land}_{m}
        \right) . \label{32}
\end{equation}
The support property~(\ref{GF}) yields that~$k_m$ is causal in the sense that
\[ {\mbox{supp}}\, k_m(x,.) \;\subset\; J_x\:. \]
The distribution~$p_m$ is {\em{not}} causal, but it can be deduced from~$k_m$
as follows. For a diagonalizable matrix $A$ with real eigenvalues
we can uniquely define its {\em{absolute value}} $|A|$ as the
diagonalizable matrix with non-negative eigenvalues and~$|A|^2=A^2$.
The matrix~$k\slsh+m$ in~(\ref{x20}, \ref{x21}) is diagonalizable with
non-negative eigenvalues and thus
\beq \label{absv}
\left| \epsilon(k^0)\: (k\slsh+m) \right| \;=\; k\slsh+m \:.
\eeq
Before applying this relation to~(\ref{x20}, \ref{x21}),
it is useful to consider the above distributions~$P^{\mbox{\scriptsize{sea}}}$,
$p_m, \ldots$ as integral kernels of corresponding operators on the wave functions in
space-time, for example
\begin{equation}
(P^{\mbox{\scriptsize{sea}}} \:\Psi)(x) \;:=\; \int
P^{\mbox{\scriptsize{sea}}}(x,y)\: \Psi(y)\: d^4y \:.
    \label{f:0x}
\end{equation}
Then the operators~$k_m$ and~$p_m$ are diagonal in momentum space, and
so~(\ref{absv}) gives rise to the formal identity
\begin{equation}
p_m \;=\; |k_m| \: , \label{51}
\end{equation}
where~$|\,.\,|$ now is the absolute value of an operator on
wave functions in Minkowski space.
With~(\ref{12b}) and~(\ref{32}, \ref{51}) we have related the fermionic
projector to the causal Green's functions in a way which can be generalized
to the interacting theory, as we shall now make precise.

We begin with the perturbation expansion for the causal Green's functions.
The retarded Green's function in the presence of the external field~${\mathcal{B}}$,
denoted by~$\tilde{s}^\land_m$, is characterized by the conditions
\beq \label{rGf}
(i \Pdd + {\mathcal{B}} - m) \: \tilde{s}^\land_m(x,y) \;=\; \delta^4(x-y)
\;,\spc {\mbox{supp}}\, \tilde{s}^\land_m(x,.) \;\subset\;
J_x^\land \:.
\eeq
The existence and uniqueness of the advanced Green's functions
follows from the general theory of linear hyperbolic PDEs~\cite{John,
Taylor}. In short, for the existence proof one considers
the solution of the Cauchy problem
\[ (i \Pdd + {\mathcal{B}} - m) \: \Psi \;=\; f \in C^\infty_0((t_0,\infty) \times \R^3)^4 \;,
\spc \Psi(t_0, \vec{x}) \;=\; 0 \:; \]
by linearity it can be expressed as an integral over the inhomogeneity,
\[ \Psi(x) \;=\; \int_{\sR^4} \tilde{s}^\land_m(x,y)\: f(y)\: d^4y\:. \]
To prove uniqueness, one considers the difference of two retarded
Green's functions~$\tilde{s}^\land_{m,1}$ and~$\tilde{s}^\land_{m,2}$,
\[ \Psi(x) \;=\; \tilde{s}^\land_{m,1}(x,y) -
\tilde{s}^\land_{m,2}(x,y) \:. \]
Then~$\Psi(x)$ is for fixed~$y$ a solution of the homogeneous Dirac equation which
vanishes identically on the half space $x^0<y^0$. The uniqueness of the
solution of the Cauchy problem yields that $\Psi \equiv 0$.

Expanding~(\ref{rGf}) in powers of~${\mathcal{B}}$, one obtains the
perturbation series $\tilde{s}^\land_m = \sum_{n=0}^\infty s^\land_{(n)}$,
where~$s^\land_{(0)}=s^\land_m$ is the advanced Green's function of the
vacuum, and the other summands are determined by the conditions
that they are causal, ${\mbox{supp}}\, s^\land_{(n)}(x,.) \in J_x^\land$, and satisfy the
inductive relations
\[ (i \Pdd - m)\: s^\land_{(n)} \;=\; - {\mathcal{B}}\: s^\land_{(n-1)}
\spc (n \geq 1). \]
Here we again used the operator notation~(\ref{f:0x}) and
considered~${\mathcal{B}}$ as a multiplication operator.
The operator product
\begin{equation}
(-s^\land_m \:{\mathcal{B}}\: s^\land_m)(x,y) \;=\;
-\int d^4z \; s^\land_m(x,z) \: {\mathcal{B}}(z) \: s^\land_m(z,y)
        \label{2t1}
\end{equation}
is causal in the sense that~${\mathcal{B}}(z)$ enters
only for $z \in L^{\land}_{x} \cap L^{\lor}_{y}$
(the analytic justification of this and all other operator products
in this section will be given in Lemma~\ref{l:lemma0} below).
In particular, the support of (\ref{2t1}) is again in the
past light cone. Furthermore, it satisfies the relation
\[ (i \Pdd - m)\: (-s^\land_m \:{\mathcal{B}}\: s^\land_m)
\;=\; -{\mathcal{B}}\: s^\land_m \:, \]
and can thus be identified with the operator~$s^\land_{(1)}$.
By iteration, we obtain for the other terms of the perturbation
series the explicit formulas
\[ s^\land_{(n)} \;=\; (-s^\land_m \:{\mathcal{B}})^n\: s^\land_m\:. \]
We conclude that the retarded Green's function can be represented as
\begin{equation}
\tilde{s}_m^\land \;=\; \sum_{k=0}^\infty \left(- s_m^\land \: {\mathcal{B}}
\right)^k s_m^\land \: . \label{2t2}
\end{equation}
Similarly, we introduce the advanced Green's function~$\tilde{s}^\lor_m$
by the conditions
\beq \label{aGf}
(i \Pdd - m + {\mathcal{B}}) \: \tilde{s}^\lor_m(x,y) \;=\; \delta^4(x-y)
\;,\spc {\mbox{supp}}\, \tilde{s}^\lor_m(x,.) \;\subset\; J_x^\lor \:.
\eeq
It has the perturbation expansion
\begin{equation}
\tilde{s}_m^\lor \;=\; \sum_{k=0}^\infty \left(- s_m^\lor \: {\mathcal{B}}                \right)^k s_m^\lor \: . \label{2t3}
\end{equation}

Having uniquely introduced the causal Green's functions,
we can now extend~(\ref{32}) to the case with interaction.
Namely, we define the operator $\tilde{k}_{m}$ by \label{tildekm}
\begin{eqnarray}
\label{2tm}
\tilde{k}_m &=& \frac{1}{2 \pi i} \: \left(\tilde{s}_m^\lor -
        \tilde{s}_m^\land \right)
\end{eqnarray}
with the causal Green's functions as given by~(\ref{2t2}, \ref{2t3}).
Finally, we also extend~(\ref{51}) to the case with interaction
by setting \label{tildepm}
\begin{equation}
        \tilde{p}_m \;\stackrel{\mbox{\scriptsize{formally}}}{:=}\;
        \sqrt{\tilde{k}_m^2} \: .
        \label{51x}
\end{equation}
In the next theorem we will give this last relation a precise mathematical
meaning and show that it gives rise to a unique perturbation expansion.
It is most convenient to work with the Green's function \label{sm2}
\begin{equation}
        s_{m} \;:=\; \frac{1}{2} (s^{\lor}_{m} + s^{\land}_{m}) \: .
        \label{51a}
\end{equation}
Furthermore, we introduce the series of operator products
\[ b_m^< \;=\; \sum_{k=0}^\infty (-s_m \:{\mathcal{B}})^k \;\;\;,\;\;\;\;\;\;
b_m \;=\; \sum_{k=0}^\infty (-{\mathcal{B}} \: s_m)^k \:{\mathcal{B}}
\;\;\;,\;\;\;\;\;\;
b_m^> \;=\; \sum_{k=0}^\infty (-{\mathcal{B}} \:s_m)^k \]
and set for $Q \subset \N$
\[ F_m(Q,n) \;=\; \left\{ \begin{array}{ll}
        p_m & {\mbox{if $n \in Q$}} \\
        k_m & {\mbox{if $n \not \in Q$}} \end{array} \right. . \]

\begin{Thm}
\label{thm1}
The relations (\ref{2tm}, \ref{51x}) uniquely determine the
perturbation expansions for $k_m$ and $p_m$. We have the explicit
formulas
\begin{eqnarray}
\tilde{k}_m &=& \sum_{\beta=0}^\infty (-i \pi)^{2 \beta} \; b_m^< \:k_m\:
(b_m \: k_m)^{2\beta} \: b_m^>
\label{52} \\
\tilde{p}_m &=& \sum_{\beta=0}^\infty \sum_{\alpha=0}^{\left[
\frac{\beta}{2}
\right]} c(\alpha,\beta) \: G_m(\alpha,\beta)
\label{53}
\end{eqnarray}
with the coefficients
\begin{eqnarray}
c(0,0) &=& 1 \\
c(\alpha,\beta) &=& \sum_{n=\alpha+1}^\beta (-1)^{n+1} \:\frac{(2n-3)!!}
{n! \: 2^n} \: \left( \!\! \begin{array}{c} \beta-\alpha-1 \\ n-\alpha-1
\end{array} \!\! \right)
        \quad {\mbox{for $\beta \geq 1$}}
        \label{54}
\end{eqnarray}
and the operator products
\begin{eqnarray}
\,\,\lefteqn{ G_m(\alpha,\beta) \;=\; \sum_{Q \in {\mathcal{P}}(\beta+1) , \;\;
\# Q=2\alpha+1}
    (-i \pi)^{2\beta} }  \qquad     \nonumber \\
  &\!\!\!\!\!\!\!\!\!\!\!\!\!\!\!\!\!\times \: b_m^< \:F_m(Q,1) \:b_m k_m b_m\: F_m(Q,2) \:b_m k_m b_m
        \:\cdots\: b_m k_m b_m \:F_m(Q,\beta+1)\: b_m^> \: , \spc
        \label{55}
\end{eqnarray}
where ${\mathcal{P}}(n)$ denotes the set of subsets of $\{1,\ldots,n\}$
(we use the convention $l!!=1$ for $l \leq 0$).
\end{Thm}
{\Proof}
An explicit calculation using~(\ref{4}) shows
that $(i \Pdd + {\mathcal{B}} - m) \: b_m^< = 0$. Since all
operator products in (\ref{52}) and~(\ref{55}) have a factor $b_m^<$ at
the left, the operators $\tilde{p}_m$, $\tilde{k}_m$ are solutions of the
Dirac equation,
\[ (i \Pdd + {\mathcal{B}} - m) \: \tilde{p}_m \;=\; 0 \;=\;
   (i \Pdd + {\mathcal{B}} - m) \: \tilde{k}_m \:. \]
It remains to
verify that the conditions (\ref{2tm}, \ref{51x}) are satisfied and
to show uniqueness.

According to (\ref{32}, \ref{51a}), the advanced and retarded Green's function
can be written as
\begin{equation}
        s^{\lor}_{m} \;=\; s_{m} + i \pi \:k_{m} \;\;\;,\spc
        s^{\land}_{m} \;=\; s_{m} - i \pi \:k_{m} \: .
        \label{32a}
\end{equation}
We substitute the series (\ref{2t2}, \ref{2t3}) into (\ref{2tm}),
\begin{equation}
        \tilde{k}_m \;=\; \frac{1}{2 \pi i} \sum_{k=0}^\infty \left(
                (- s^\lor_m \:{\mathcal{B}})^k \: s^\lor_m \:-\:
                (- s^\land_m \:{\mathcal{B}})^k \: s^\land_m \right) \: ,
        \label{2t7}
\end{equation}
insert (\ref{32a}) and expand. This gives a sum of operator products of the
form
\[ C_{1} \:{\mathcal{B}}\: C_{2} \:{\mathcal{B}}\: \cdots \:{\mathcal{B}}\: C_{l+1} \qquad {\mbox{with}} \qquad C_j=k_m {\mbox{ or }} C_j=s_m \: . \]
The contributions with an even number of factors $k_{m}$ have the same
sign for the advanced and retarded Green's functions and cancel in
(\ref{2t7}). The contributions with an odd number of $k_{m}$'s occur
in each Green's function exactly once and have the opposite sign. Using
the notation
\[ C_m(Q, n) \;=\; \left\{ \begin{array}{ll}
        k_m & {\mbox{if $n \in Q$}} \\
        s_m & {\mbox{if $n \not \in Q$}} \end{array} \right.
        , \spc Q \subset \N \: , \]
we can thus rewrite~(\ref{2t7}) in the form
\begin{eqnarray*}
\tilde{k}_m &=& \sum_{l=0}^\infty \: (-1)^l \!\!\! \sum_{ Q \in
{\mathcal{P}}(l+1) , \;\; \# Q \; {\mbox{\scriptsize odd}} }  (i \pi)^{\#Q-1} \\
&& \hspace*{1cm} \times \;
        C_m(Q,1) \: {\mathcal{B}} \: C_m(Q,2) \: {\mathcal{B}} \cdots
        {\mathcal{B}} \: C_m(Q,l) \: {\mathcal{B}} \: C_m(Q,l+1) \:.\spc
\end{eqnarray*}
After reordering the sums, this coincides with (\ref{52}).

Next we want to give the relation (\ref{51}) a mathematical meaning.
To this end, we consider $m \geq 0$ as a variable mass parameter.
Then we can form products of the operators $p_m, k_m$ by manipulating
the arguments of the distributions in momentum space. For example,
using~(\ref{x20}) we obtain
\begin{eqnarray}
p_m(k) \: p_{m^\prime}(k) &=& (k \slsh + m) \: \delta(k^2 - m^2) \;
   (k \slsh + m^\prime) \: \delta(k^2 - (m^\prime)^2) \nonumber \\
&=& (k^2 + (m+m^\prime) k \slsh + m m^\prime) \: \delta(m^2 -
   (m^\prime)^2) \; \delta(k^2 - m^2) \nonumber \\
&=& (k^2 + (m+m^\prime) k \slsh + m m^\prime) \: \frac{1}{2m} \:\delta(m -
   m^\prime) \; \delta(k^2 - m^2) \nonumber \\
&=& \delta(m-m^\prime) \: p_m(k) \: ,
\label{56}
\end{eqnarray}
and similarly from~(\ref{x21}),
\begin{eqnarray}
p_m \: k_{m^\prime} &=& k_{m^\prime} \: p_m \;=\; \delta(m-m^\prime) \: k_m \\
\label{57} \\
k_m \: k_{m^\prime} &=& \delta(m-m^\prime) \: p_m \: .
\label{58}
\end{eqnarray}
We remark that this formalism has some similarity with the bra/ket notation
in  quantum mechanics, if the position variable~$\vec{x}$ is replaced by
the mass parameter $m$. Equation~(\ref{56}) can be interpreted that
the~$p_m$ are the spectral projectors of the free Dirac operator; the
relations~(\ref{57}, \ref{58}) reflect the relative minus sign in
$k_m$ for the states on the upper and lower mass shell. In particular,
one sees that $k_m \:k_{m^\prime} = p_m \:p_{m^\prime}$. This relation can be
extended to the case with interaction,
\begin{equation}
        \tilde{p}_m \: \tilde{p}_{m^\prime} \;=\; \tilde{k}_m \:
        \tilde{k}_{m^\prime} \: ,
        \label{59}
\end{equation}
and gives a meaningful square of (\ref{51}) (we will see in a moment that
$\tilde{k}_m \: \tilde{k}_{m^\prime}$ vanishes for
$m \neq m^\prime$). If our construction
ensures that $\tilde{p}_m$ is a positive operator, (\ref{59}) is even
equivalent to (\ref{51}).

Let us compute the product $\tilde{k}_m \:\tilde{k}_{m^\prime}$
explicitly. The definitions~(\ref{x20}, \ref{x21}) and (\ref{51a}, \ref{8b})
yield in analogy to~(\ref{56}) the formulas\footnote{{\textsf{Online version}:}
As noticed by A.\ Grotz, in~\eqref{62} the summand~$\pi^2 \delta(m-m^\prime)$
is missing. This error is corrected in the paper~\cite{grotz}
(listed in the references in the preface to the second online edition).}
\begin{eqnarray}
p_m \: s_{m^\prime} &=& s_{m^\prime} \: p_m \;=\; {\mbox{PP}} \left(
  \frac{1}{m-m^\prime} \right) \: p_m
\label{60} \\
k_m \: s_{m^\prime} &=& s_{m^\prime} \: k_m \;=\; {\mbox{PP}} \left(
  \frac{1}{m-m^\prime} \right) \: k_m
\label{61} \\
s_m \: s_{m^\prime} &=& {\mbox{PP}} \left(
  \frac{1}{m-m^\prime} \right) \: (s_m - s_{m^\prime}) \: ,
\label{62}
\end{eqnarray}
where ${\mbox{PP}}(x^{-1}) = \frac{1}{2} \: \lim_{\varepsilon
\searrow 0} [ (x+i \varepsilon)^{-1} + (x-i \varepsilon)^{-1}]$
denotes the principal value.
As a consequence, the operator products involving the
factor~$s_m \!\cdot\! s_{m^\prime}$ are telescopic,
\begin{equation}
        \sum_{p=0}^n k_m \:({\mathcal{B}} \:s_m)^p \: (s_{m^\prime}
        \:{\mathcal{B}})^{n-p} \: k_{m^\prime} \;=\; 0 \spc {\mbox{for $n \geq 1$}} .
        \label{63a}
\end{equation}
This allows us to evaluate the following product,
\begin{equation}
        k_m \:b_m^> \:b_{m^\prime}^< \: k_{m^\prime} \;=\;
        \delta(m-m^\prime) \: p_m \:.
        \label{63}
\end{equation}
With this formula, we can compute the square of (\ref{52}),
\begin{equation}
        \tilde{k}_m \: \tilde{k}_{m^\prime} \;=\; \delta(m-m^\prime) \:
           \sum_{\beta_1, \beta_2 = 0}^\infty (-i \pi)^{2\beta_1 + 2\beta_2} \:
           b_m^< \:(k_m \:b_m)^{2\beta_1} \: p_m\: (b_m \: k_m)^{2\beta_2} \:
           b_m^> \:.
        \label{64}
\end{equation}

We could continue the proof by verifying explicitly that the product
$\tilde{p}_m \: \tilde{p}_{m^\prime}$ with $\tilde{p}_m$ according to
(\ref{53}) coincides with (\ref{64}). This is a
straightforward computation, but it is rather lengthy and not very
instructive. We prefer to describe how the operator
products (\ref{55}) and the coefficients (\ref{54}) can be derived.
In order to make the proof more readable, we make the following simplifications.
Since the factors $b_m^<$, $b_m^>$ cancel similar to (\ref{63})
in telescopic sums, we can omit them in all formulas without changing
the multiplication rules for the operator products. Then all operator
products have $k_m$ or $p_m$ as their first and last factor, and we
can multiply them with the rules (\ref{56}--\ref{58}).
Since all these rules give a factor $\delta(m-m^\prime)$, we will in
any case get the prefactor $\delta(m-m^\prime)$ as in (\ref{64}).
Therefore, we can just forget about all factors $\delta(m-m^\prime)$
and consider all expressions at the same value of $m$.
Furthermore, we will omit the subscript `$_m$' and write the intermediate
factors $b$ as a dot `.'.
After these simplifications, we end up with formal products of the form
\begin{equation}
        F_1 \:.\: F_2 \:.\: F_3 \:.\: \cdots \:.\:F_n \spc {\mbox{with}} \spc
        F_j=k {\mbox{ or }} F_j=p
        \label{41c}
\end{equation}
and have the multiplication rules
\begin{equation}
        p^2 \;=\; k^2 \;=\; 1 \;\;\;,\spc p \:k \;=\; k \:p \;=\; k \: .
        \label{71}
\end{equation}
We must find a positive operator $\tilde{p}$ being a formal sum of
operator products (\ref{41c}) such that
\begin{equation}
        \tilde{p}^2 \;=\; \sum_{\beta_1, \beta_2 = 0}^\infty
        (-i \pi)^{2\beta_1 + 2\beta_2} \: (k \:.)^{2 \beta_1} \:p\:
        (. \:k)^{2 \beta_2} \: .
        \label{65}
\end{equation}
In this way, we have reduced our problem to the combinatorics of
the operator products. As soon as we have found a solution $\tilde{p}$ of
(\ref{65}), the expression for $\tilde{p}_m$ is
obtained by adding the subscripts `$_m$' and by inserting the factors
$b_m^<$, $b_m$, $b_m^>$.
Relation (\ref{59}) follows as an immediate consequence of (\ref{65}).

The basic step for the calculation of $\tilde{p}$ is to rewrite
(\ref{65}) in the form
\begin{equation}
\tilde{p}^2 \;=\; p + A \spc{\mbox{with}}\spc A \;=\;
\sum_{(\beta_1,\beta_2) \neq (0,0)} (-i \pi)^{2 \beta_1 + 2 \beta_2} \:
(k \:.)^{2\beta_1} \:p\: (. \:k)^{2\beta_2} \: .
\label{44d}
\end{equation}
The operator $p$ is idempotent and acts as the identity on $A$,
$Ap=pA=A$. Therefore, we can take the square root of $p+A$ with a
formal Taylor expansion,
\begin{equation}
        \tilde{p} \;=\; \sqrt{p+A} \;=\; p \:+\: \sum_{n=1}^\infty
        (-1)^{n+1} \: \frac{(2n-3)!!}{n! \: 2^n} \: A^n \: ,
        \label{74}
\end{equation}
which uniquely defines $\tilde{p}$ as a positive operator.

It remains to calculate $A^n$. If we take the $n$th power of the sum
in (\ref{44d}) and expand, we end up with one sum over more complicated
operator products. We first consider how these operator products look like:
The operator products in (\ref{44d}) all contain an even number of factors
$k$ and exactly one factor $p$. The factor $p$ can be the 1st, 3rd,\ldots\
factor of the product. Each combination of this type occurs in $A$
exactly once. If we multiply $n$ such terms,
the resulting operator product consists of a total odd number of factors
$p, k$. It may contain several factors $p$, which all occur at odd
positions in the product. Furthermore, the total number of factors $p$
is odd, as one sees inductively. We conclude that $A^n$ consists of
a sum of operator products of the form
\begin{equation}
        (k \:.\: k \:.)^{q_1} \;p\:.\:k\:.\; (k\:.\:k\:.)^{q_2}
        \;p\:.\:k\:.\; (k\:.\:k\:.)^{q_3} \:\cdots\:
        (k\:.\:k\:.)^{q_{2\alpha+1}} \;p\; (.\:k\:.\:k)^{q_{2\alpha+2}}
        \label{72}
\end{equation}
with $\alpha, q_j \geq 0$. We set $\beta=2\alpha+\sum_j q_j$. Notice that the
number of factors $p$ in (\ref{72}) is $2\alpha+1$; the total number of
factors $p, k$ is $2\beta+1$. The form of the operator product gives the
only restriction $0 \leq 2\alpha \leq \beta$ for the choice of the parameters
$\alpha, \beta$.

Next we count how often each operator product (\ref{72}) occurs in the sum:
The easiest way to realize (\ref{72}) is to form the
product of the $\alpha+1$ factors
\begin{eqnarray}
&&\left[ (k.k.)^{q_1} \:p\: (.k.k)^{q_2+1} \right] \:
\left[ (k.k.)^{q_3+1} \:p\: (.k.k)^{q_4+1} \right] \nonumber \\
&& \hspace*{4cm} \:\cdots\:
\left[ (k.k.)^{q_{2\alpha+1}+1} \:p\: (.k.k)^{q_{2\alpha+2}} \right] \: .
\label{73}
\end{eqnarray}
However, this is not the only way to factor~(\ref{72}).
More precisely, to each factor in~(\ref{73}) we can apply the
identities
\begin{eqnarray*}
(k\:.\:k\:.)^q \:p\: (.\:k\:.\:k)^r &=& \left[ (k\:.\:k\:.)^q \:p \right] \:
   \left[ p\: (.\:k\:.\:k)^r \right] \\
(k\:.\:k\:.)^q \:p\: (.\:k\:.\:k)^r &=& \left[ (k\:.\:k\:.)^s \:p \right] \:
   \left[ (k\:.\:k\:.)^{q-s} \:p\: (.\:k\:.\:k)^r \right] \\
(k\:.\:k\:.)^q \:p\: (.\:k\:.\:k)^r &=& \left[ (k\:.\:k\:.)^q \:p\:
   (.\:k\:.\:k)^{r-s}  \right] \: \left[ p\:(.\:k\:.\:k)^s \right] \: .
\end{eqnarray*}
By iteratively substituting these identities into (\ref{73}), we can
realize every factorization of (\ref{72}). Each substitution step increases
the number of factors by one. The steps are independent in the sense that we
can fix at the beginning at which positions in (\ref{73}) the product shall
be split up, and can  then apply the steps in arbitrary order. There are
$(\alpha+1)+(q_1-1)+\sum_{j=2}^{2\alpha+1} q_j + (q_{2\alpha+2}-1) =
\beta-(\alpha+1)$ positions
in (\ref{73}) where we could split up the product (in the case $q_1=0$ or
$q_{2\alpha+2}=0$, the counting of the positions is slightly different, but
yields the same result). Since we want to
have $n$ factors at the end, we must choose $n-(\alpha+1)$ of these
positions, which is only possible for $\alpha+1 \leq n \leq \beta$ and then
gives $(\beta-\alpha-1)!/((n-\alpha-1)! \: (\beta-n)!)$ possibilities.

Combining these combinatorial factors with the constraints $0\leq
2\alpha \leq \beta$ and $\alpha+1 \leq n \leq \beta$, we obtain for $n \geq 1$ the identity
\begin{eqnarray}
A^n &=& \sum_{\beta=n}^\infty \sum_{\alpha=0}^{\min \left( n-1,
\left[ \frac{\beta}{2} \right] \right)}
   \left( \!\! \begin{array}{c} \beta-\alpha-1 \\ n-\alpha-1 \end{array} \!\!
      \right) \: \sum_{Q \in {\mathcal{P}}(\beta+1), \;\; \# Q = 2\alpha+1}
          \nonumber \\
\hspace*{1cm} &&\times  \:(-i \pi)^{2\beta} \: F(Q,1)
\:.\:k\:.\: F(Q,2) \:.\:k\:.\: \cdots \:.\:k\:.\: F(Q, \beta+1) \spc
\label{47a}
\end{eqnarray}
with $F(Q,n) = p$ for $n \in Q$ and $F(Q,n) = k$ otherwise.
Notice that the last sum in (\ref{47a}) runs over all possible
configurations of the factors $p, k$ in the operator product (\ref{72})
for fixed $\alpha, \beta$.
We finally substitute this formula into (\ref{74}) and pull the sums
over $\alpha, \beta$ outside. This gives the desired formula for $\tilde{p}$.
\QED

We call the perturbation expansion of the above theorem the {\em{causal
perturbation expansion}}. It allows us to define the Dirac
sea\index{Dirac sea} in the
presence of an external field canonically by
\[ P^{\mbox{\scriptsize{sea}}}(x,y) \;=\; \frac{1}{2} \:(\tilde{p}_m -
\tilde{k}_m)(x,y) \: . \]
In the next section the causal perturbation expansion will be extended
to systems of Dirac seas, and in~{\S}\ref{jsec4} we will discuss it in detail.

We conclude this section by showing that, under suitable regularity and decay
assumptions on the external potentials, all operator products
which appeared in this section are well-defined and finite.

\begin{Lemma}
\label{l:lemma0}
Let $(C_j)$, $0 \leq j \leq n$, be a choice of operators $C_j \in \{ k_m, p_m,
s_m\}$. If the external potential~${\mathcal{B}}$
is smooth and decays so fast at infinity that the functions
${\mathcal{B}}(x)$, $x^i {\mathcal{B}}(x)$, and $x^i x^j
{\mathcal{B}}(x)$ are integrable, then the operator product
\begin{equation}
        (C_n \:{\mathcal{B}} \:C_{n-1}\: {\mathcal{B}} \cdots {\mathcal{B}} \:C_0)(x,y)
        \label{l:2}
\end{equation}
is a well-defined tempered distribution on $\R^4 \times \R^4$.
\end{Lemma}
{\Proof}
Calculating the Fourier transform of (\ref{l:2}) gives the formal
expression
\begin{eqnarray}
\lefteqn{M(q_2,q_1) \;:=\; \int \frac{d^4 p_1}{(2 \pi)^4}
\cdots \int \frac{d^4 p_{n-1}}{(2 \pi)^4} C_n(q_2) \;\hat{\mathcal{B}}(q_2-p_{n-1}) } \nonumber \\
&\times&\!\!\!\!\!
\:C_{n-1}(p_{n-1})\: \hat{\mathcal{B}}(p_{n-1}-p_{n-2})
\:\cdots\: C_1(p_1) \:\hat{\mathcal{B}}(p_1-q_1) \:C_0(q_1) \; , \label{l:3}
\end{eqnarray}
where we consider the $C_j$ as multiplication operators in momentum
space and where $\hat{\mathcal{B}}$ denotes the Fourier transform of
the function~${\mathcal{B}}$ (it is more convenient to work in momentum
space because the operators $C_j$ are then diagonal).
We will show that $M(q_2,q_1)$ is a well-defined tempered distribution;
the Lemma then immediately follows by transforming back to position space.

The assumptions on ${\mathcal{B}}$ yield
that $\hat{\mathcal{B}}$ is $C^2$ and has rapid decay at infinity, i.e.
\[ \sup_{q \in \sR^4, \;|\kappa| \leq 2} |q^{i_1} \cdots q^{i_n} \:
\partial_\kappa {\hat{\mathcal{B}}}(q)| \;<\; \infty \]
for all~$n$, all tensor indices $i_1,\ldots, i_n$ and multi-indices
$\kappa$ (with $\kappa=(\kappa^1,\ldots,\kappa^q)$, $|\kappa|:=q$).
As is verified explicitly in momentum space, the
distributions $k_m$, $p_m$ or~$s_m$ are bounded in the Schwartz norms
of the test functions involving derivatives of only first order, more precisely
\[ |C(f)| \;\leq\; {\mbox{const}}\: \|f\|_{4,1} \qquad
{\mbox{with $C=k_m$, $p_m$ or $s_m$ and $f \in {\mathcal{S}}$,}} \]
where the Schwartz norms are as usual defined by
\[ \|f\|_{p,q} \;=\; \max_{|I| \leq p,\; |J| \leq q} \;\;\sup_{x \in \sR^4}
|x^I \:\partial_J f(x)| \: . \]
As a consequence, we can apply the corresponding operators even to
functions with rapid decay which are only $C^1$. Furthermore, we can form the convolution of such
functions with~$C$; this gives continuous functions (which will no
longer have rapid decay, however). Since~$C$ involves first derivatives,
a convolution decreases the order of
differentiability of the function by one.

We consider the combination of multiplication and convolution
%\mpar{die Notation ist verwirrend!}
\begin{equation}
        F(p_2) \;:=\; \int \frac{d^4p_1}{(2 \pi)^4} \;
        f(p_2-p_1) \:C(p_1) \:g(p_1) \: ,
        \label{l:3a}
\end{equation}
where we assume that $f \in C^2$ has rapid decay and $g \in C^1$ is
bounded together with its first derivatives, $\|g\|_{0,1}<\infty$.
For any fixed $p_2$,
the integral in (\ref{l:3a}) is well-defined and finite because $f(p_2-.) \:g(.)$
is $C^1$ and has rapid decay. The resulting function $F$ is $C^1$ and
bounded together with its first derivatives, more precisely
\begin{eqnarray}
\|F\|_{0,1} &\leq& {\mbox{const}} \;\|f\|_{4,2} \:\|g\|_{0,1} \: .
        \label{l:3c}
\end{eqnarray}

After these preparations, we can estimate the integrals in (\ref{l:3}) from
the right to the left: We choose two test functions $f, g \in
{\mathcal{S}}(\R^4, \C^{4})$ and introduce the functions
\begin{eqnarray}
\hspace*{-1cm}F_1(p_1) &\!\!\!=\!\!\!& \int \frac{d^4q_2}{(2 \pi)^4} \; \hat{\mathcal{B}}(p_1-q_1)
        \:C_0(q_1)\:g(q_1)\label{l:4z} \\
\hspace*{-1cm}F_j(p_j) &\!\!\!=\!\!\!& \int \frac{d^4p_{j-1}}{(2 \pi)^4}
        \:\hat{\mathcal{B}}(p_j-p_{j-1})
        \:C_{j-1}(p_{j-1}) \:F_{j-1}(p_{j-1}) \;\;, \;\; 1 < j \leq n
        \:. \label{l:4a}
\end{eqnarray}
The integral~(\ref{l:4z}) is of the form~(\ref{l:3a}) and satisfies the
above considered assumptions on the integrand. Using the bound
(\ref{l:3c}), we can proceed inductively in (\ref{l:4a}).
Finally, we perform the $q_2$-integration,
\[ M(f,g) \;=\; \int \frac{d^4q_2}{(2 \pi)^4} \:f(q_2) \:C_n(q_2)
\:F_n(q_2) \: . \]
We conclude that $M$ is a linear functional on
${\mathcal{S}}(\R^4,\C^{4}) \times {\mathcal{S}}(\R^4, \C^{4})$, which is
bounded in the Schwartz norm $\|.\|_{4,1}$ of the test functions.
\QED

Using the language of quantum field theory, we also refer to the summands
of the perturbation expansions as {\em{Feynman
    diagrams}}\index{Feynman diagram}. Then the
result of the last lemma can be understood from the fact that in an
external field one only encounters tree diagrams, which are all finite.
Clearly, the existence of the perturbation expansion to every order
does not imply the convergence of the perturbation series, and we will
come back to this problem in~{\S}\ref{jsec5}.

\section{Definition of the Fermionic Projector}
\label{jsec3} \setcounter{equation}{0}
In this section we introduce the mathematical framework for describing
a many-fermion system in the presence of an external field.
To this end, we first extend the construction of~{\S}\ref{jsec2}
to a system of Dirac seas of in general different masses,
which may involve chiral massless Dirac seas.
Then we introduce particles and anti-particles
by occupying additional states and creating ``holes'' in the Dirac seas,
respectively.
Our construction is intended to be so general that it allows us to model
the fermion configuration of the standard model (see~{\S}\ref{esec1}).
For clarity, we postpone the question of how the fermionic states are
to be normalized to~{\S}\ref{jsec6}.

First, we need to introduce a distribution~$P^{\mbox{\scriptsize{sea}}}(x,y)$ which
describes the system in the vacuum. The most general ansatz is to
take a direct sum of sums of Dirac seas, \label{Psea2}
\beq \label{vs1}
P^{\mbox{\scriptsize{sea}}} \;=\; \bigoplus_{a=1}^N \sum_{\alpha=1}^{g(a)}
P^{\mbox{\scriptsize{sea}}}_{a \alpha} \:,
\eeq
where~$g(a)$ are positive integers and the summands~$P^{\mbox{\scriptsize{sea}}}_{a \alpha}$
are Dirac seas of a form similar to~(\ref{f:2b}).
The direct sum increases the total number of components of
the wave functions, the so-called {\em{spin dimension}}\index{spin dimension}, to~$4N$.
The direct summands are called sectors, and we refer to the
indices~$a$ and~$\alpha$ as the {\em{sector}}\index{sector} and
{\em{generation index}}\index{generation index}, respectively.
For each Dirac sea we introduce a mass parameter~$m_{n \alpha} \geq 0$.
In order to allow for chiral massless Dirac seas, we introduce
$(4 \times 4)$-matrices~$X_{a \alpha}$ with
\[ X_{a \alpha} \;=\; \left\{ \begin{array}{cl} \1 & {\mbox{if $m_{a \alpha} > 0$}} \\
\1, \chi_L {\mbox{ or }} \chi_R & {\mbox{if $m_{a \alpha} = 0$}}
\end{array} \right. \]
and set
\beq \label{vs2}
P^{\mbox{\scriptsize{sea}}}_{a \alpha} \;=\; \frac{1}{2}\: X_{a \alpha} \left(
p_{m_{a \alpha}} - k_{m_{a \alpha}} \right) .
\eeq
We refer to~$P^{\mbox{\scriptsize{sea}}}$ as defined by~(\ref{vs1}, \ref{vs2}) as
the {\em{fermionic projector of the vacuum}}.
It is sometimes useful to consider~$P^{\mbox{\scriptsize{sea}}}$ as a matrix in the sectors indices,
\[ (P^{\mbox{\scriptsize{sea}}})^a_b \;=\; \delta^a_b\: \frac{1}{2}\: \sum_{\alpha=1}^{g(a)} X_{a \alpha} \left(
p_{m_{a \alpha}} - k_{m_{a \alpha}} \right)  \]
with~$a,b=1,\ldots,N$.

Since each sector may involve several Dirac seas of
different masses, it seems impossible to write the fermionic projector of the
vacuum as a solution of a suitable Dirac equation, and thus we have no
starting point for a perturbation expansion.
In order to bypass this problem, we replace the sum in~(\ref{vs1}) by a
direct sum and introduce the so-called {\em{auxiliary fermionic
    projector}}\index{fermionic projector!auxiliary} by
\beq \label{vafp1}
P^{\mbox{\scriptsize{sea}}} \;=\; \bigoplus_{a=1}^N \bigoplus_{\alpha=1}^{g(a)}
P^{\mbox{\scriptsize{sea}}}_{a \alpha} \:.
\eeq
Using the same notation as for the fermionic projector is usually
no problem because it will be clear from the context whether
the fermionic projector or the auxiliary fermionic projector is meant.
In case of potential confusion we write the auxiliary fermionic projector as
a matrix in the sector and generation indices,
\[ (P^{\mbox{\scriptsize{sea}}})^{(a \alpha)}_{(b \beta)} \;=\;
\delta^a_b\:\delta^\alpha_\beta \frac{1}{2}\: X_{a \alpha} \left(
p_{m_{a \alpha}} - k_{m_{a \alpha}} \right) \]
with $a,b=1,\ldots,N$, $\alpha=1,\ldots,g(a)$, $\beta=1,\ldots,g(b))$.
In this notation, one also sees that the fermionic projector can be
obtained from the auxiliary fermionic projector by taking the
so-called {\em{partial trace}}\index{partial trace}\footnote{{\textsf{Online version}:}
In more recent works on the fermionic projector,
the partial trace is referred to as the {\em{sectorial projection}}.}\
over the generations,
\beq \label{pt}
(P^{\mbox{\scriptsize{sea}}})^a_b \;=\; \sum_{\alpha=1}^{g(a)} \sum_{\beta=1}^{g(b)}
(P^{\mbox{\scriptsize{sea}}})^{(a \alpha)}_{(b \beta)}\:.
\eeq

We introduce the operators \label{p} \label{k} \label{X} \label{Y}
\[ p \;=\; \bigoplus_{a=1}^N \bigoplus_{\alpha=1}^{g(a)} \:p_{m_{a \alpha}} \:,\spc
   k \;=\; \bigoplus_{a=1}^N \bigoplus_{\alpha=1}^{g(a)} \:k_{m_{a \alpha}}  \]
and define the matrices
\[ X \;=\; \bigoplus_{a=1}^N \bigoplus_{\alpha=1}^{g(a)} \:X_{a \alpha}
\:,\spc Y \;=\; \frac{1}{m}
\:\bigoplus_{a=1}^N \bigoplus_{\alpha=1}^{g(a)} m_{a \alpha}\: , \]
which are called {\em{chiral asymmetry matrix}}\index{chiral asymmetry
  matrix} and {\em{mass matrix}}\index{mass matrix},
respectively. Here~$m$ is an arbitrary mass parameter; a convenient choice is
$m=\max_{a, \alpha} m_{a \alpha}$.
These operators act on direct sums of Dirac wave functions, i.e.\
on functions of the form~$\Psi=(\Psi^{(a \alpha)}(x))$
with~$\Psi^{(a \alpha)}$ a~$4$-component Dirac spinor. On these wave funtions,
we introduce the {\em{spin scalar product}} by
\beq \label{spsps}
\Sl \Psi \:|\: \Phi \Sr(x) \;=\;
\sum_{a=1}^n \sum_{\alpha=1}^{g(a)} \Sl \Psi^{(a \alpha)}
\:|\: \Phi^{(a \alpha)} \Sr_{\mbox{\tiny{Dirac}}} \;,
\eeq
where~$\Sl .|. \Sr_{\mbox{\tiny{Dirac}}}$ is the usual spin scalar
product on Dirac spinors~(\ref{f:0c}). In generalization of~(\ref{defsp})
we also introduce the indefinite inner product
\beq \label{defsps}
\bra \Psi \:|\: \Phi \ket \;=\;
\int_M \Sl \Psi \:|\: \Phi \Sr\: d\mu\:.
\eeq
Then the operators~$p$ and~$k$ are Hermitian with respect to~$\bra .|. \ket$,
and the mass matrix~$Y$ is Hermitian with respect to the spin scalar product.
Using the above notation, we can write the auxiliary fermionic projector as
\begin{equation}
        P^{\mbox{\scriptsize{sea}}}(x,y) \;=\; X \: \frac{1}{2} \:(p(x,y) - k(x,y)) \: .
        \label{81n}
\end{equation}
Since $m_l=0$ for $X_l \neq \1$ and since the operators~$p_{m=0}$, $k_{m=0}$
are odd, we have alternatively
\begin{equation}
        P^{\mbox{\scriptsize{sea}}}(x,y) \;=\; \frac{1}{2} \:(p(x,y) - k(x,y)) \: X^* \: ,
        \label{49d}
\end{equation}
where $X^*$ is the adjoint with respect to the spin scalar product.
The auxiliary fermionic projector is a solution of
the free Dirac equation
\beq \label{fDe}
(i \Pdd_{x} - m Y) \:P^{\mbox{\scriptsize{sea}}}(x,y) \;=\; 0 \: .
\eeq
Our strategy is to extend the definition of the auxiliary fermionic projector
to the interacting system and then to get back to the fermionic projector
by taking the partial trace~(\ref{pt}).

In order to describe the system of Dirac seas in the presence of an
external field, we insert a differential
operator~${\mathcal{B}}=({\mathcal{B}}^{(a \alpha)}_{(b \beta)})$ into
the Dirac equation~(\ref{fDe}),
\begin{equation}
        (i \Pdd_x + {\mathcal{B}} - mY) \: P^{\mbox{\scriptsize{sea}}}(x,y)
\;=\; 0 \: . \label{b}
\end{equation}
We always assume that~${\mathcal{B}}$ is Hermitian with respect to
the inner product~$\bra .|. \ket$.
The causal perturbation expansion for the operators~$k$ and~$p$ can be carried
out exactly as in~{\S}\ref{jsec2}: We define the advanced and retarded Green's
functions by
\[ s^\lor \;=\;
\bigoplus_{a=1}^N \bigoplus_{\alpha=1}^{g(a)} \:s^\lor_{m_{a \alpha}} \:,\spc
s^\land \;=\;
\bigoplus_{a=1}^N \bigoplus_{\alpha=1}^{g(a)} \:s^\land_{m_{a \alpha}}\: . \]
Their perturbation expansion is, in analogy to (\ref{2t2}, \ref{2t3}),
uniquely given by
\begin{equation}
\tilde{s}^\lor \;=\; \sum_{k=0}^\infty (-s^\lor \:{\mathcal{B}})^k \: s^\lor
        \:,\spc
\tilde{s}^\land \;=\; \sum_{k=0}^\infty (-s^\land \:{\mathcal{B}})^k
\: s^\land \:.
\label{79}
\end{equation}
The method of Theorem~\ref{thm1} now yields the following result.
\begin{Thm} \label{Thm2}
The perturbation expansion for $p$ and $k$ is uniquely determined by the
conditions
\begin{equation}
\tilde{k} \;=\; \frac{1}{2 \pi i} \: (\tilde{s}^\lor -
\tilde{s}^\land) \:,\spc
\tilde{p} \;\stackrel{\mbox{\scriptsize{formally}}}{=}
\sqrt{\tilde{k}^2} \:.
\label{80}
\end{equation}
We have the explicit formulas
\begin{eqnarray*}
\tilde{k} &=& \sum_{\beta=0}^\infty (-i \pi)^{2\beta} \; b^< \:k\: (b \:
k)^{2\beta} \: b^>
\:,\spc
\tilde{p} \;=\; \sum_{\beta=0}^\infty \sum_{\alpha=0}^{\left[
\frac{\beta}{2}
\right]} c(\alpha,\beta) \: G(\alpha,\beta)
\end{eqnarray*}
with
\begin{eqnarray*}
c(0,0) &=& 1 \: , \\
c(\alpha,\beta) &=& \sum_{n=\alpha+1}^\beta (-1)^{n+1} \:\frac{(2n-3)!!}
{n! \: 2^n} \:
        \left( \!\! \begin{array}{c} \beta-\alpha-1 \\ n-\alpha-1 \end{array} \!\!
        \right) \spc {\mbox{for $\beta \geq 1$}}
\end{eqnarray*}
and
\begin{eqnarray*}
G(f,g) &=&
\sum_{Q \in {\mathcal{P}}(\beta+1) , \;\; \# Q=2\alpha+1} (-i \pi)^{2\beta} \\
&& \quad \times \: b^< \:F(Q,1) \:b k b\: F(Q,2) \:b k b
        \:\cdots\: b k b \:F(Q,\beta+1)\: b^> \: ,
\end{eqnarray*}
where ${\mathcal{P}}(n)$ is the set of subsets of $\{1,\ldots, n\}$
and where we use the notation
\begin{eqnarray*}
s &=& \frac{1}{2} \:(s^\lor + s^\land) \:,\spc
F(Q,n) \;=\; \left\{ \begin{array}{ll}
        p & {\mbox{if $n \in Q$}} \\
        k & {\mbox{if $n \not \in Q$}} \end{array} \right. \\
b^< &=& \sum_{k=0}^\infty (-s \:{\mathcal{B}})^k \:,\spc
b \;=\; \sum_{k=0}^\infty (-{\mathcal{B}} \:s)^k \:{\mathcal{B}} \:,\spc
b^> \;=\; \sum_{k=0}^\infty (-{\mathcal{B}} \:s)^k \: .
\end{eqnarray*}
\end{Thm}
The contributions to this perturbation expansion are all well-defined
according to Lemma~\ref{l:lemma0}.

After this straightforward generalization, we come to the more subtle
question of how to define~$P^{\mbox{\scriptsize{sea}}}$ when a chiral
asymmetry is present. The obvious idea is to set
in generalization of (\ref{81n})
\begin{equation}
P^{\mbox{\scriptsize{sea}}}(x,y) \;=\; X \: \frac{1}{2} \: (\tilde{p}
-\tilde{k})(x,y) \: .
\label{84}
\end{equation}
This is not convincing, however, because we could just as well have defined
$P^{\mbox{\scriptsize{sea}}}(x,y)$ in analogy to (\ref{49d}) by
$P^{\mbox{\scriptsize{sea}}} =
\frac{1}{2} (\tilde{p}-\tilde{k}) \: X^*$, which does not coincide with
(\ref{84}) as soon as $X, X^*$ do not commute with ${\mathcal{B}}$.
Actually, this arbitrariness in defining the Dirac sea reflects a
basic problem of the causal perturbation expansion for systems with chiral
asymmetry. In order to describe the problem in more detail, we consider
the perturbation calculation for $k$ to first order. According to (\ref{79}, \ref{80}),
\begin{eqnarray}
\tilde{k} &=& k \:-\: \frac{1}{2 \pi i} \:(s^\lor \:{\mathcal{B}}\: s^\lor \:-\:
        s^\land \:{\mathcal{B}}\: s^\land) \:+\: {\mathcal{O}}({\mathcal{B}}^2)
        \label{81} \\
&=& k \:-\: s \:{\mathcal{B}}\: k \:-\: k \:{\mathcal{B}}\: s \:+\:
{\mathcal{O}}({\mathcal{B}}^2) \: . \nonumber
\end{eqnarray}
This expansion is causal in the sense that $\tilde{k}(x,y)$ only
depends on ${\mathcal{B}}$ in the ``diamond'' $(L^\lor_x \cap L^\land_y) \cup
(L^\lor_y \cap L^\land_x)$, as is obvious from~(\ref{81}).
It is not clear, however, how to insert the chiral asymmetry matrix into this
formula. It seems most natural to replace all factors $k$ by $Xk$,
\begin{equation}
\tilde{(Xk)} \;=\; Xk \:-\: s \:{\mathcal{B}}\: Xk \:-\: Xk \:{\mathcal{B}}\: s \:+\:
        {\mathcal{O}}({\mathcal{B}}^2) \: .
\label{55m}
\end{equation}
Unfortunately, this expression cannot be written similar to (\ref{81})
with the advanced and retarded Green's functions, which means that the causality
of the expansion is in general lost. In order to avoid this problem, one might
want to insert $X$ at every factor $s, k$,
\begin{eqnarray}
\tilde{(Xk)} &=& Xk \:-\: Xs \:{\mathcal{B}}\: Xk \:-\: Xk \:{\mathcal{B}}\: Xs
        \:+\: {\mathcal{O}}({\mathcal{B}}^2) \nonumber \\
&=& Xk \:-\: \frac{1}{2 \pi i} \:(Xs^\lor \:{\mathcal{B}}\: Xs^\lor \:-\:
        Xs^\land \:{\mathcal{B}}\: Xs^\land) \:+\: {\mathcal{O}}({\mathcal{B}}^2)\: .
        \label{83}
\end{eqnarray}
This expansion is causal similar to (\ref{81}). In general, however, it does
not satisfy the Dirac equation
$(i \Pdd + {\mathcal{B}} - m) \:\tilde{(X k)}=0$,
which does not seem to be what we want.

The only way to resolve this problem is to impose that the perturbation
expansions~(\ref{55m}) and~(\ref{83}) should coincide. This yields a
condition for the operator ${\mathcal{B}}$, which can be characterized as follows. We demand that
\begin{equation}
X s^\lor \:{\mathcal{B}}\: X s^\lor \;=\; s^\lor \:{\mathcal{B}}\: X s^\lor
\;=\; X s^\lor \:{\mathcal{B}}\: s^\lor \: .
\label{90}
\end{equation}
Since the operator $s^\lor_{m=0}$ is odd, we have $X s^\lor = s^\lor X^*$.
Substituting into the second equation of (\ref{90}) yields
the condition $X^* \:{\mathcal{B}} \;=\; {\mathcal{B}} \:X$.
Since $X$ is idempotent, this condition automatically implies the first equation
of (\ref{90}).
We formulate the derived condition for the whole Dirac operator $i \Pdd +
{\mathcal{B}}-m Y$ and thus combine it with the fact that chiral fermions
are massless (i.e. $X^* Y = Y X = Y$) and that $X$ is composed of
chiral projectors (which implies that $X^* \Pdd = \Pdd X$).
\begin{Def} \label{ccc}
The Dirac operator is called {\bf{causality
compatible}}\index{causality compatible}
with $X$ if
\begin{equation}
X^* \: (i \Pdd + {\mathcal{B}} - m Y) \;=\; (i \Pdd + {\mathcal{B}} - m Y) \: X \: .
\label{89}
\end{equation}
\end{Def}
In the perturbation expansion to higher order, the condition (\ref{89})
allows us to commute $X$ through all operator products. Using idempotence
$X^2=X$, we can moreover add factors $X$ to the product; in particular,
\[ X \;C_1 \:{\mathcal{B}}\: C_1 \:{\mathcal{B}}\:  \cdots \:{\mathcal{B}} \:C_n \;=\;
  XC_1 \:{\mathcal{B}}\: XC_1 \:{\mathcal{B}}\:  \cdots \:{\mathcal{B}} \:XC_n
\;\;\;\;\; {\mbox{with}} \;\;\;\;\; C_j=p,\: C_j=k {\mbox{ or }} C_j=s \: . \]
This ensures that the perturbation expansion is also well-defined to
higher order. For a Dirac operator which is causality compatible with $X$,
the~{\em{auxiliary fermionic projector}} is defined canonically by (\ref{84}).

So far the auxiliary fermionic projector describes a system of Dirac seas
in the presence of an external field. In order to insert particles and
anti-particles into the system, we add the projectors on particle states
and substract the projectors on anti-particle states,
\begin{eqnarray}
\lefteqn{ P(x,y) \;=\; P^{\mbox{\scriptsize{sea}}}(x,y) } \nonumber \\
&&+ c_{\mbox{\scriptsize{norm}}}
\sum_{k=1}^{\np} |\Psi_k(x) \Sr \Sl \Psi_k(y)| \:-\:
c_{\mbox{\scriptsize{norm}}}
\sum_{l=1}^{\na} |\Phi_l(x) \Sr \Sl \Phi_l(y)| \;, \label{1g}
\end{eqnarray}
where~$\Psi_k$ and~$\Phi_l$ are an orthogonal set of solutions of the Dirac equation,
and the~$\Phi_l$ must lie in the image of~$P^{\mbox{\scriptsize{sea}}}$
(for the normalization constant~$c_{\mbox{\scriptsize{norm}}}$ see~{\S}\ref{jsec6}).
The parameters~$\np$ and~$\na$ denote the total number of particles and
anti-particles, respectively. We usually avoid the issue of convergence
of the sums in~(\ref{1g}) by assuming that $\np, \na < \infty$, but one could
clearly also consider an infinite number of particles and/or anti-particles.
Finally, the {\em{fermionic projector}}\index{fermionic projector} is obtained from this expression
by taking similar to~(\ref{pt}) the partial trace\index{partial trace},
\beq \label{part}
(P)^a_b \;=\; \sum_{\alpha=1}^{g(a)} \sum_{\beta=1}^{g(b)} (P)^{(a \alpha)}_{(b \beta)}\:.
\eeq

Theorem~\ref{Thm2} together with~(\ref{84}, \ref{1g}, \ref{part}) yields a
mathematical framework for describing a general many-fermion system in the
presence of an external field. Our construction makes Dirac's concept
of a ``sea of interacting particles'' mathematically precise.
Apart from the causality compatibility condition~(\ref{89}) and the
regularity conditions in Lemma~\ref{l:lemma0},
the operator~${\mathcal{B}}$ is completely arbitrary.
We point out that we do not use the fermionic Fock space formalism of
canonical quantum field theory; the connection to this formalism will
be explained in~{\S}\ref{psec12} and Appendix~\ref{pappA}.

\section{Interpretation and Consequences}
\label{jsec4} \setcounter{equation}{0}
With the definition of the fermionic projector we radically departed from
the usual concept of the Dirac sea as ``all negative-energy solutions''
of the Dirac equation. Instead, we used causality in a particular way.
In order to clarify the connection between our definition and
the usual concept of the Dirac sea, we now describe how the
above constructions simplify in the special
situation that~${\mathcal{B}}$ is {\em{static}}.
If considered as multiplication operators, static
potentials map functions of positive and negative frequency into functions
of positive and negative frequency, respectively.
Since the operators~$p$, $k$ and~$s$
are diagonal in momentum space, they clearly preserve the sign
of the frequency too. Thus
\begin{equation}
[\Pi^\pm,p] \;=\; [\Pi^\pm,k] \;=\; [\Pi^\pm,s] \;=\; [\Pi^\pm,{\mathcal{B}}]
\;=\; 0 \: , \label{1n}
\end{equation}
where the operators~$\Pi^\pm$ are the projectors onto the states of positive
and negative frequency, respectively
(i.e.\ in momentum space, $\Pi^{\pm}$ are the operators of
multiplication by the functions~$\Theta(\pm k^0)$).
The operators~$p$ and~$k$ differ only by a relative
minus sign for the states of positive and negative frequency,
\[ \Pi^\pm \:p \;=\; \pm \:\Pi^\pm \:k \: . \]
Using this relation together with~(\ref{1n}), we can replace pairs of
factors $p$ by pairs of factors $k$. For example,
\begin{eqnarray}
\cdots p \:{\mathcal{B}} \cdots p \:{\mathcal{B}} \cdots &=&
        \cdots p \:{\mathcal{B}} \cdots p \:{\mathcal{B}} \cdots \:(\Pi^+ + \Pi^-) \nonumber \\
&=& \Pi^+ (\cdots k \:{\mathcal{B}} \cdots k \:{\mathcal{B}} \cdots)
        \:+\: \Pi^- (\cdots (-k) \:{\mathcal{B}} \cdots (-k) \:{\mathcal{B}} \cdots) \nonumber \\
&=& \cdots k \:{\mathcal{B}} \cdots k \:{\mathcal{B}} \cdots \: ,
\label{prrep}
\end{eqnarray}
where the dots `$\cdots$' denote any combination of the operators $s$,
$k$, $p$ and ${\mathcal{B}}$. This allows us to simplify the formula for
$\tilde{p}$ by using only one factor $p$ in every operator product.
After going through the details of the combinatorics, one obtains the formula
\[ \tilde{p} \;=\; \sum_{b=0}^\infty (-i \pi)^{2b} \:b^< \:p\:(b
\:k)^{2b} \:b^> \: . \]
Thus the fermionic projector~(\ref{84}) can be written as
\[ P^{\mbox{\scriptsize{sea}}}(x,y) \;=\;\sum_{b=0}^\infty (-i \pi)^{2b} \:b^<
\:\left[\frac{1}{2} \: X \:(p-k) \right]\:(b
\:k)^{2b} \:b^> \: . \]
This equation shows that $P^{\mbox{\scriptsize{sea}}}$ is composed of all
negative-frequency eigenstates of the Dirac operator (notice that the
expression in the brackets $[ \cdots ]$ is the fermionic projector of
the vacuum and that all other factors preserve the sign of the
frequency). We conclude that for static potentials our definition
reduces to the usual concept of the Dirac sea as
``all negative-energy states.''

In order to get a better understanding of the time-dependent
situation, we next consider a {\em{scattering
    process}}\index{scattering process}. For simplicity,
we consider a system of one Dirac sea and
assume that the scattering takes place in finite time $t_0<t<t_1$.
This means that the wave functions~$\Psi$ satisfy the Dirac equation~(\ref{1})
with ${\mathcal{B}}$ supported in a finite time interval,
\begin{equation}
{\mathcal{B}}(t, \vec{x}) \;=\; 0 \qquad {\mbox{if $t \not \in [t_0,
t_1]$}} \: .        \label{s1}
\end{equation}
As a consequence, $\Psi(t, \vec{x})$ is for~$t<t_0$ a solution of the
free Dirac equation.  We uniquely extend this free solution to the
whole Minkowski space and denote  it by $\Psi_{\mbox{\scriptsize{in}}}$,
\[ (i \Pdd - m) \:\Psi_{\mbox{\scriptsize{in}}}\label{Psiin} \;=\; 0
\spc {\mbox{and}} \spc \Psi_{\mbox{\scriptsize{in}}}(t,\vec{x}) \;=\;
\Psi(t, \vec{x}) \;\;\;{\mbox{for $t<t_0$}}. \]
Similarly, $\Psi(t, \vec{x})$ is also for $t>t_1$ a solution of the free
Dirac equation; we denote its extension by
$\Psi_{\mbox{\scriptsize{out}}}$,
\[ (i \Pdd - m) \:\Psi_{\mbox{\scriptsize{out}}}\label{Psiout} \;=\; 0
\spc {\mbox{and}} \spc \Psi_{\mbox{\scriptsize{out}}}(t,\vec{x}) \;=\;
\Psi(t, \vec{x}) \;\;\;{\mbox{for $t>t_1$}}. \]
The wave functions $\Psi_{\mbox{\scriptsize{in}}}$ and
$\Psi_{\mbox{\scriptsize{out}}}$ are called the incoming and outgoing
{\em{scattering states}}, respectively. Recall that
the dynamics of the wave functions is described infinitesimally by the Dirac
equation in the Hamiltonian form~(\ref{Hamilton}),
where~$h$ is a symmetric operator on the Hilbert space
$({\mathcal{H}}, (.,.))$ with scalar product~(\ref{f:0e}).
Integrating this equation from~$t_0$ to~$t_1$, we obtain a unitary operator~$S$ which maps
the incoming scattering states to the corresponding
outgoing states,
\beq \label{Smatrix}
\Psi_{\mbox{\scriptsize{out}}} \;=\;
S \:\Psi_{\mbox{\scriptsize{in}}} \: .
\eeq
The operator~$S$ is called scattering operator or {\em{$S$-matrix}}\index{$S$-matrix}.

Using the scattering states, we can introduce fermionic projectors
which describe the vacua in the asymptotic past and future:
For an observer in the past $t<t_0$, the external potential is zero.
Thus it is natural for him
to describe the vacuum with the free Dirac sea~(\ref{Pvac}). If this
Dirac sea is extended to the whole Minkowski space with external
potential, one gets the object
\beq \label{Pret}
P^\land(x,y) \;=\; -\frac{m}{\pi} \sum_s \int_{\sR^3}
|\Psi^\land_{\vec{p}s-} \Sr \Sl \Psi^\land_{\vec{p}s-} | \; d\mu_{\vec{p}}\:,
\eeq
where the wave functions $\Psi^\land_{\vec{p}s \epsilon}$ are the
solutions of the Dirac equation~(\ref{1}) whose incoming scattering
states are the plane wave solutions $\Psi_{\vec{p}s \epsilon}$,
\[ (i \Pdd + {\mathcal{B}} - m) \:\Psi^\land_{\vec{p}s \epsilon} \;=\; 0
\spc {\mbox{and}} \spc (\Psi^\land_{\vec{p}s \epsilon})_{\mbox{\scriptsize{in}}} \;=\; \Psi_{\vec{p}s \epsilon} \: . \]
Using the support conditions~(\ref{s1}, \ref{GF}),
we can express the state $\Psi^\land_{\vec{p}s \epsilon}$ in a perturbation
series,
\[ \Psi^\land_{\vec{p}s \epsilon} \;=\; \sum_{n=0}^\infty (-s^\land\:{\mathcal{B}})^n \:\Psi_{\vec{p}s \epsilon} \: . \]
Substituting this formula into~(\ref{Pret}) we obtain
for~$P^\land$ a perturbation expansion which involves only the retarded
Green's functions,
\beq \label{Prete}
P^\land \;=\; \sum_{n_1, n_2=0}^\infty  (-s^\land\:{\mathcal{B}})^{n_1}\: P^{\mbox{\scriptsize{vac}}}\:
(-{\mathcal{B}}\: s^\land)^{n_2} \:,
\eeq
where~$P^{\mbox{\scriptsize{vac}}}$ stands for the free Dirac sea~(\ref{Pvac}).
Accordingly, an observer in the future $t>t_0$ describes the vacuum by
the fermionic projector
\beq \label{Padv}
P^\lor(x,y) \;=\; -\frac{m}{\pi} \sum_s \int_{\sR^3}
|\Psi^\lor_{\vec{p}s-} \Sr \Sl \Psi^\lor_{\vec{p}s-} | \; d\mu_{\vec{p}}\:,
\eeq
where
\[ (i \Pdd + {\mathcal{B}} - m) \:\Psi^\lor_{\vec{p}s \epsilon} \;=\; 0
\spc {\mbox{and}} \spc (\Psi^\lor_{\vec{p}s \epsilon})_{\mbox{\scriptsize{out}}}
\;=\; \Psi_{\vec{p}s \epsilon} \: . \]
Its perturbation expansion involves only the advanced Green's function,
\beq \label{Padve}
P^\lor \;=\; \sum_{n_1, n_2=0}^\infty  (-s^\lor\:{\mathcal{B}})^{n_1}\: P^{\mbox{\scriptsize{vac}}}\:
(-{\mathcal{B}}\: s^\lor)^{n_2} \:.
\eeq
Using~(\ref{Smatrix}) in~(\ref{Pret}, \ref{Padv}), we can describe
the fermionic projectors in the asymptotic past and future
with the S-matrix by
\begin{equation}
        P^\land_{\mbox{\scriptsize{in}}} \;=\; P^{\mbox{\scriptsize{vac}}} \;=\;
        P^\lor_{\mbox{\scriptsize{out}}} \:,\qquad
        P^\land_{\mbox{\scriptsize{out}}} \;=\; S
        P^{\mbox{\scriptsize{vac}}} S^{-1} \:,\qquad
        P^\lor_{\mbox{\scriptsize{in}}} \;=\; S^{-1}
        P^{\mbox{\scriptsize{vac}}} S \:.
        \label{s3}
\end{equation}

What makes the scattering process interesting is the fact that
the vacua in the asymp\-to\-tic past and future in general do not coincide.
Consider for example the physical system described by
the fermionic projector~$P:=P^\land$.
For the observer in the past, the system is in the vacuum.
However, if $P^\land \neq P^\lor$, the system will {\em{not}}
be in the vacuum for the observer in the future.
This means that for him, positive frequency states
are occupied and negative frequency states are unoccupied
and thus the system contains particles and anti-particles.
More precisely, if we write the fermionic projector in analogy
to~(\ref{1g}) as
\begin{eqnarray}
\lefteqn{ P(x,y) \;=\; P^\lor(x,y) } \nonumber \\
&&+\:
c_{\mbox{\scriptsize{norm}}} \sum_{k=1}^{\np} |\Psi_k(x) \Sr \Sl \Psi_k(y)|
\:-\: c_{\mbox{\scriptsize{norm}}} \sum_{l=1}^{\na} |\Phi_l(x) \Sr \Sl \Phi_l(y)| \;, \label{Pfut}
\end{eqnarray}
then the~$\Psi_k$ and $\Phi_l$ are the wave functions of the particles
and anti-particles, respectively.
These particles and anti-particles are physical reality; the observer
in the future can detect them by making suitable experiments.
This is the physical effect of {\em{pair creation}}.
Using~(\ref{s3}) one can express the pair creation completely
in terms of the $S$-matrix.
Other scattering processes are described similarly.

We point out that describing the scattering process with
the two observers in the past and future is merely a matter
of convenience. The physical process can be described
equivalently (although maybe less conveniently) in the reference frame
of any other observer. To give a concrete example, we consider an
observer in the future who is in a reference frame moving with
constant acceleration. This leads to the so-called
{\em{Unruh effect}}\index{Unruh effect} , which we now briefly outline (for details see e.g.~\cite{Wald2}).
For the accelerated observer, space-time is
stationary (i.e.\ his time direction is a Killing field, but it is
not a unit normal to the hypersurfaces $t={\mbox{const}}$), and this
allows him to use the separation ansatz~(\ref{pwa}) with~$t$ his
proper time. The sign of~$\omega$ gives him a splitting of the solution
space into solutions of positive and negative energy.
Using Dirac's hole interpretation corresponding to this splitting,
he finds for the many-fermion system described by $P$
an infinite number of particles
and anti-particles in a thermal equilibrium.
This bizarre effect shows that the
interpretation of the physical system in terms of particles and
anti-particles does depend on the observer.
Nevertheless, the Unruh effect does not contradict
the pair creation experiments made by the future observer at rest.
Namely, if the accelerated observer wants to explain the experiments
by the future observer at rest, he must take into account that he
himself is feeling a gravitational field, and that for him the
experimental apparatus used by the observer at rest is in accelerated
motion. It turns out that these additional effects just compensate
the Unruh effect, so that the predictions by the accelerated
observer are in complete agreement with the observations of
the particles and anti-particles in~(\ref{Pfut}) by the future observer
at rest. More generally, all quantities which can be measured
in experiments can be expressed in terms of the S-matrix.
Since the S-matrix does not depend on the particle/anti-particle
interpretation, it is clear that all experiments can be explained
equivalently in any reference frame.

We just saw that the particle/anti-particle interpretation of a
fermionic system may depend on the observer. Actually, the
situation is even worse for an observer in the time period
$t_0 < t < t_1$ when the interaction takes place.
For him, the system is neither static nor stationary.
Therefore, he has no notion of ``negative-energy state'', and thus for him
the particle/anti-particle interpretation completely breaks down.
Taking into account that a scattering process is an idealized
process and that in a real physical situation there will be no
region of space-time where no interaction takes place, we come to the disillusioning conclusion that for a local observer under generic
conditions, a many-fermion system has no interpretation in terms
of particles and anti-particles.

The causal perturbation expansion yields a canonical object $P^{\mbox{\scriptsize{sea}}}$ which describes the Dirac sea in the
scattering process, even in the region with interaction $t_0<t<t_1$.
Its construction is explicitly covariant and
independent of a local observer.
Decomposing the fermionic projector in the form~(\ref{1g}), we obtain
a canonical interpretation of the many-fermion system
in terms of particles and anti-particles. One should keep in mind
that~$P^{\mbox{\scriptsize{sea}}}$ does not correspond to the
vacuum of any local observer, but is a global object of space-time.
As a consequence, also the particle/anti-particle interpretation
in~(\ref{1g}) can be associated only to an abstract ``global observer''
in space-time.
More specifically, comparing Theorem~\ref{Thm2} and~(\ref{84})
with~(\ref{Prete}, \ref{Padve}), one sees that~$P^{\mbox{\scriptsize{sea}}}$
coincides neither with~$P^\land$ nor with~$P^\lor$. Since its perturbation
expansion involves both retarded and advanced Green's functions, it can
be considered as being some kind of ``interpolation'' between~$P^\land$
and~$P^\lor$.

Let us now discuss our assumption on the potential~${\mathcal{B}}$
as being {\em{external}}\index{external field}. As explained at the end of~{\S}\ref{jsec1}, this is no
restriction in principle because one can first solve the physical equations
of the coupled system and then can define~$P^{\mbox{\scriptsize{sea}}}$
for the external potential~${\mathcal{B}}$ as given by the solution
of the coupled system.
Clearly, this procedure cannot be carried out in practice,
but this is of no relevance for the theoretical considerations here.
The important point is that~$P^{\mbox{\scriptsize{sea}}}$ is not defined
locally; for its definition we need to know~${\mathcal{B}}$ in the
whole space-time. This is puzzling because the conventional physical
equations are local and causal, and this is the first time that an object
appears which is defined in a non-local way.
One might conclude that~$P^{\mbox{\scriptsize{sea}}}$ is an object which
is not compatible with causality and should therefore have no physical
significance. Our concept is the opposite: We regard the appearance
of a non-local object as a first hint that locality and causality
should be given up in the strict sense.
In order to formulate physical equations which could replace the
conventional local and causal equations, we shall consider the fermionic projector as the fundamental object in space-time.

Before we can make these ideas precise in Chapter~\ref{secpfp}, we
need to analyze the fermionic projector in the continuum in more detail.
One task is to understand what ``causality'' of the
causal perturbation expansion means precisely. At the moment,
we know that causality was used for the definition of~$P^{\mbox{\scriptsize{sea}}}$,
but that nevertheless the fermionic projector is a nonlocal
and non-causal object. We need to find out how these
seemingly contradicting facts fit together.
Also, we must understand better how~$P^{\mbox{\scriptsize{sea}}}$
depends on~${\mathcal{B}}$. More specifically, we need to analyze
what information on the external potentials is encoded
in~$P^{\mbox{\scriptsize{sea}}}$, and how this information can be
extracted. Finally, we must specify how the fermionic states
in~(\ref{1g}) are to be normalized.
The next sections provide the mathematical tools for answering these
questions.

\section{The Light-Cone Expansion}\index{light-cone expansion}
\label{jsec5} \setcounter{equation}{0}
The light-cone expansion is a very useful technique for analyzing the
fermionic projector near the light cone. In order to give a brief
but self-contained introduction, we will explain the methods and results of~\cite{F5} leaving out many proofs and technical details.
Our setting is that of~{\S}\ref{jsec3} with several sectors
and generations~(\ref{vs1}). It suffices to consider the {\em{auxiliary
fermionic projector}} because the fermionic projector is obtained from it
simply by taking the partial trace~(\ref{part}).
We again assume that the Dirac operator in~(\ref{b}) is causality
compatible~(\ref{89}) and that the operator~${\mathcal{B}}$
is Hermitian with respect to the inner product~(\ref{defsps}).
Furthermore, we assume as in~\cite{F5} that~${\mathcal{B}}$ is a
multiplication operator composed of {\em{chiral}}\index{chiral} and {\em{scalar/pseudoscalar
potentials}}\index{potential!scalar/pseudoscalar}, \label{AslshR} \label{AslshL}
\begin{equation}
{\mathcal{B}}(x) \;=\; \chi_L \:\Aslsh_R(x) \:+\: \chi_R \:\Aslsh_L(x) \:+\:
\Phi(x) \:+\: i \rho \:\Xi(x) \:.
        \label{l:1}
\end{equation} \label{Phi} \label{Xi}
We note for clarity that these potentials may act non-trivially on the
sectors and generations (e.g.\ writing the right-handed potential
as a matrix, $A_R = (A_R)^{(a \alpha)}_{(b \beta)}$, the
matrix elements are independent vector fields with the only constraint
that the matrix must be Hermitian). In particular, the potentials
in~(\ref{l:1}) do in general not commute, and thus we must in products
be careful with the order of multiplication. For an
operator~${\mathcal{B}}$ which involves bilinear and gravitational
potentials see~\cite{F4}.

\begin{Def}
\label{l:def1}
A distribution~$A(x,y)$ on~$\R^4 \times \R^4$ is
of the order~${\mathcal{O}}((y-x)^{2p})$, $p \in \Z$, if the product
\[ (y-x)^{-2p} \: A(x,y) \]
is a regular distribution (=a locally integrable function).
It has the {\bf{light-cone expansion}}
\begin{equation}
        A(x,y) \;=\; \sum_{j=g}^{\infty} A^{[j]}(x,y)
        \label{l:6a}
\end{equation}
with $g \in \Z$ if the distributions $A^{[j]}(x,y)$ are of the order
${\mathcal{O}}((y-x)^{2j})$ and if $A$ is approximated by the partial sums
in the sense that for all $p \geq g$,
\begin{equation}
        A(x,y) \;-\; \sum_{j=g}^p A^{[j]}(x,y) \qquad
        {\mbox{is of the order~${\mathcal{O}}((y-x)^{2p+2})$}}\:.
        \label{l:6b}
\end{equation}
\end{Def}

The light-cone expansion describes the
behavior of a distribution near the light cone. More precisely,
the expansion parameter~$(y-x)^2$ vanishes if~$y$ lies on
the light cone centered at~$x$, and thus the distributions~$A^{[j]}(x,y)$ approximate~$A(x,y)$ for $y$ in a neighborhood of~$L_x$.
The first summand~$A^{[g]}(x,y)$ gives the leading order of~$A(x,y)$
on the light cone, $A^{[g+1]}$ gives the next order on the light cone,
etc. If the distribution~$A$ is singular on the light cone,
the parameter~$g$ will be negative.
Note that the distributions~$A^{[j]}$ are determined only up to contributions of higher order~${\mathcal{O}}((y-x)^{2j+2})$, but this ambiguity
will not lead to any problems in what follows.
We point out that we do not demand that the
infinite series in~(\ref{l:6a}) converges. This series is defined
only via the approximation by the partial sums~(\ref{l:6b}).
Despite this formal character of the series, the light-cone expansion
completely describes the behavior of~$A(x,y)$ near the light cone.
This situation can be seen in analogy to the
Taylor expansion of a smooth, nonanalytic function.
Although the Taylor series does in general not converge, the
Taylor polynomials give local approximations of the function.
An important difference to a Taylor expansion
is that the $A^{[j]}(x,y)$ approximate~$A(x,y)$ even for points~$x$ and~$y$ which are far apart. We only need that~$y$ is close to the light cone~$L_x$, which is an unbounded hypersurface in~$\R^4$. In this sense, the light-cone
expansion is a {\em{non-local expansion}}.

For clarity, we begin with the light-cone expansion for the
causal Green's functions, and we will later extend the results to
the fermionic projector.
In order to get a first idea of how the light-cone expansion can be
carried out, we consider the free advanced Green's function $s^\lor_m$
as defined by~(\ref{Ft}, \ref{8b}). We can pull the Dirac matrices out
of the Fourier integral by setting
\begin{equation}
        s^\lor_m(x,y) \;=\; (i \Pdd_x + m) \: S^\lor_{m^2}(x,y) \: ,
        \label{l:10}
\end{equation}
where $S^\lor_{m^2}$ is the advanced Green's function of the
Klein-Gordon operator, \label{Slorm2}
\begin{equation}
S^\lor_{m^2}(x,y) \;=\; \lim_{\varepsilon \searrow 0} \int
\frac{d^4k}{(2 \pi)^4} \:\frac{1}{k^2-m^2-i \varepsilon k^0} \:
e^{-ik(x-y)} \: .        \label{l:11}
\end{equation}
This Fourier integral can be computed explicitly; we expand the
resulting Bessel function in a power series,
\begin{eqnarray}
\lefteqn{ S^\lor_{m^2}(x,y) \;=\; -\frac{1}{2 \pi} \:\delta(\xi^2) \:
\Theta(\xi^0) \nonumber \: +\:\frac{m^2}{4 \pi} \:\frac{J_1(\sqrt{m^2
\xi^2})}{\sqrt{m^2 \xi^2}} \:\Theta(\xi^2) \:\Theta(\xi^0) } \nonumber \\
&=&-\frac{1}{2 \pi} \:\delta(\xi^2) \: \Theta(\xi^0)
\:+\:\frac{m^2}{8 \pi} \sum_{j=0}^\infty \frac{(-1)^j}{j! \:(j+1)!}
\: \frac{(m^2 \xi^2)^j}{4^j} \: \Theta(\xi^2) \:\Theta(\xi^0)\:,
\label{l:12}
\end{eqnarray}
where we used the abbreviation~$\xi \equiv y-x$.
This calculation shows that $S^\lor_{m^2}(x,y)$ has a $\delta((y-x)^2)$-like
singularity on the light cone. Furthermore, one sees that $S^\lor_{m^2}$
is a power series in $m^2$. The important point for us is
that the contributions of higher order in $m^2$ contain more
factors $(y-x)^2$ and are thus of higher order on the light cone.
More precisely,
\begin{equation}
\left( \frac{d}{dm^2} \right)^n S^\lor_{m^2 \:|m^2=0}(x,y) \spc
{\mbox{is of the order $\;\;\;{\mathcal{O}}((y-x)^{2n-2})$}}\:.
        \label{l:24b}
\end{equation}
According to (\ref{l:10}), the Dirac Green's function is obtained by
taking the first partial derivatives of (\ref{l:12}). Thus
$s^\lor_m(x,y)$ has a singularity on the light cone which is even
$\sim \delta^\prime((y-x)^2)$.
The higher order contributions in $m$ are again of increasing order on
the light cone. This means that we can view the Taylor expansion of
(\ref{l:10}) in $m$,
\beq \label{lcsf}
s^\lor_m(x,y) \;=\; \sum_{n=0}^\infty \frac{m^{2n}}{n!}\; (i \Pdd + m)
\left( \frac{d}{dm^2} \right)^n S^\lor_{m^2 \:|\: m^2=0}(x,y) \:,
\eeq
as a light-cone expansion of the Green's function.

Writing the light-cone expansion of~$s^\lor_m$
in the form~(\ref{lcsf}) clearly is more convenient than working with
the explicit formula~(\ref{l:12}). This is our motivation for using
an expansion with respect to the mass parameter also in the presence of the external field.
Expanding the perturbation expansion~(\ref{79}) in~$m$
gives a double series in powers of~$m$ and~${\mathcal{B}}$.
In order to combine these two expansions in a single
perturbation series, we write the mass matrix and the
scalar/pseudoscalar potentials together by setting \label{YL} \label{YR}
\begin{equation}
        Y_L(x) \;=\; Y \:-\: \frac{1}{m} \:(\Phi(x) + i \Xi(x)) \:,\spc
        Y_R(x) \;=\; Y \:-\: \frac{1}{m} \:(\Phi(x) - i \Xi(x)) \: .
        \label{l:n20}
\end{equation}
The matrices $Y_{L\!/\!R}(x)$ are called {\em{dynamical mass
    matrices}}\index{mass matrix!dynamical};
notice that $Y_L^*=Y_R$. With this notation, we can rewrite the Dirac
operator as
\begin{eqnarray}
i \Pdd + {\mathcal{B}} - m Y  &=& i \Pdd + B \spc {\mbox{with}} \\
B &=& \chi_L \:(\Aslsh_R -m\:Y_R) \:+\: \chi_R \:(\Aslsh_L - m\:Y_L) \: .
\label{l:18a}
\end{eqnarray}
For the light-cone expansion of the Green's functions, we shall always
consider~$B$ as the perturbation of the Dirac operator. This has the advantage
that the free theory consists of zero-mass fermions, and thus the Green's
functions of the free Dirac operator have the simple form
\begin{equation}
        s^\lor(x,y) \;=\; i \Pdd_x \:S^\lor_{m^2=0}(x,y) \:,\spc
        s^\land(x,y) \;=\; i \Pdd_x \:S^\land_{m^2=0}(x,y)\:.
        \label{l:11a}
\end{equation}
The Green's functions with interaction are given in analogy to~(\ref{79})
by the  perturbation series
\begin{equation}
        \tilde{s}^\lor \;=\; \sum_{k=0}^\infty (-s^\lor \:B)^k
        s^\lor \:,\spc \tilde{s}^\land \;=\; \sum_{k=0}^\infty
        (-s^\land \:B)^k s^\land \:.
        \label{l:11b}
\end{equation}
We remark that this perturbation expansion around zero mass is most convenient,
but not essential for the light-cone expansion; see~\cite{F4} for
a light cone expansion of a massive Dirac sea.

Our first goal is to perform the light-cone expansion
of each Feynman diagram in~(\ref{l:11b}).
Using an inductive construction based on Lemma~\ref{l:lemma1} below,
this will give us the result of Theorem~\ref{l:thm1}.
Since the construction is exactly the same for
the advanced and retarded Green's functions, we will omit all superscripts~`$^\lor$' and~`$^\land$'.
The formulas for the advanced and retarded Green's functions are obtained by adding either superscripts~`$^\lor$' or~`$^\land$' to all operators~$s$
and~$S$. For the mass expansion of the operator~$S_{m^2}$, we
set~$a=m^2$ and introduce the notation \label{Sl}
\begin{equation}
        S^{(l)} \;=\; \left( \frac{d}{da} \right)^l S_{a | a=0}
        \spc (l \geq 0) .
        \label{l:23b}
\end{equation}
Let us derive some computation rules for the $S^{(l)}$.
$S_a$ satisfies the defining equation of a Klein-Gordon Green's function
\[ (-\OBox_x - a) \:S_a(x,y) \;=\; \delta^4(x-y) \: . \]
Differentiating with respect to $a$ yields
\begin{equation}
        -\OBox_x S^{(l)}(x,y) \;=\; \delta_{l,0} \:\delta^4(x-y)
        \:+\: l \:S^{(l-1)}(x,y) \spc (l \geq 0) . \label{l:5}
\end{equation}
For $l=0$, this formula does not seem to make sense because $S^{(-1)}$
is undefined. However, the expression is meaningful if one keeps
in mind that in this case the factor $l$ is zero, and thus the whole
second summand vanishes. We will also use this convention in the following
calculations. Next, we differentiate the formula for~$S_a$ in momentum space,
\begin{equation}
        S_a^\lor(p)\;=\; \frac{1}{p^2-a-i \varepsilon p^0} \:,\spc
        S_a^\land(p)\;=\; \frac{1}{p^2-a+i \varepsilon p^0} \:,
        \label{l:21x}
\end{equation}
with respect to both~$p$ and~$a$. Comparing the results gives the
relation
\[ \frac{\partial}{\partial p^k} S_{a}(p) \;=\; -2p_k \:\frac{d}{da}
S_a(p) \:, \]
or, after expanding in the parameter $a$,
\begin{equation}
        \frac{\partial}{\partial p^k} S^{(l)}(p) \;=\; -2 p_k
        \:S^{(l+1)}(p) \spc (l \geq 0) .
        \label{l:21a}
\end{equation}
This formula also determines the derivatives of $S^{(l)}$ in position space.
Namely,
\begin{eqnarray}
\lefteqn{ \frac{\partial}{\partial x^k} S^{(l)}(x,y) \;=\;
\int \frac{d^4p}{(2 \pi)^4} \:S^{(l)}(p) \:(-i p_k) \:e^{-ip(x-y)} }
\nonumber \\
& \stackrel{(\ref{l:21a})}{=} & \frac{i}{2} \int \frac{d^4p}{(2 \pi)^4} \:
\frac{\partial}{\partial p^k} S^{(l-1)}(p) \; e^{-ip(x-y)} \nonumber \\
&=& -\frac{i}{2} \int \frac{d^4p}{(2 \pi)^4} \:
S^{(l-1)}(p) \;\frac{\partial}{\partial p^k} e^{-ip(x-y)} \nonumber \\
&=&\frac{1}{2} \: (y-x)_k \: S^{(l-1)}(x,y) \spc (l \geq 1) .
\label{l:7z}
   \end{eqnarray}
We iterate this relation to compute the Laplacian,
\begin{eqnarray*}
        -\OBox_x S^{(l)}(x,y) & = & -\frac{1}{2} \:\frac{\partial}{\partial
        x^k} \left( (y-x)^k \:S^{(l-1)}(x,y) \right) \\
        &=& 2 \:S^{(l-1)}(x,y) \:-\: \frac{1}{4} \:(y-x)^2 \:S^{(l-2)}(x,y)
        \spc (l \geq 2) .
\end{eqnarray*}
After comparing with (\ref{l:5}), we conclude that
\begin{equation}
        (y-x)^2 \:S^{(l)}(x,y) \;=\; -4l\: S^{(l+1)}(x,y) \spc (l \geq 0) .
        \label{l:22a}
\end{equation}
Furthermore, $S^{(l)}(x,y)$ is only a function of $(y-x)$, and thus
\begin{equation}
        \frac{\partial}{\partial x^k} S^{(l)}(x,y) \;=\;
        -\frac{\partial}{\partial y^k} S^{(l)}(x,y) \spc (l \geq 0) .
        \label{l:20a}
\end{equation}
Finally, it is convenient to use the identity~(\ref{l:7z}) also in the
case~$l=0$ and to use it as the definition of~$S^{-1}$,
\beq \label{l:7}
\frac{\partial}{\partial x^k} S^{(l)}(x,y) \;=\; \frac{1}{2} \: (y-x)_k \: S^{(l-1)}(x,y) \spc (l \geq 0) .
\eeq
Notice that~$S^{(-1)}$ itself remains undefined, only the combination~$(y-x)_k S^{(-1)}$ is given a mathematical meaning as the partial derivative of the distribution~$2 S^{(0)}$.

The next lemma gives the light-cone expansion of an operator product
where a potential~$V$ is sandwiched between two mass-derivatives of the
Green's function. This expansion is the key for the subsequent iterative light-cone
expansion of all Feynman diagrams. We always assume without saying that
the potentials satisfy the regularity conditions of Lemma~\ref{l:lemma0}.

\begin{Lemma} \label{l:lemma1}
The operator product $S^{(l)} \:V\: S^{(r)}$ with $l, r \geq
0$ has the light-cone expansion
\begin{eqnarray}
(S^{(l)} \:V\: S^{(r)})(x,y) &=&
\sum_{n=0}^\infty \frac{1}{n!} \int_0^1 \alpha^{l} \:(1-\alpha)^{r} \:
(\alpha - \alpha^2)^n \nonumber \\
&&\qquad \times\: (\OBox^n V)_{|\alpha y + (1-\alpha) x} \:d\alpha \;
S^{(n+l+r+1)}(x,y) \:.\spc \label{l:4}
\end{eqnarray}
\end{Lemma}
The fact that line integrals appear in this lemma can be understood
in analogy to the method of {\em{integration along
    characteristics}}\index{integration along characteristics} (see
e.g.~\cite{Taylor, Friedlander}) for a solution of an
inhomogeneous wave equation (for a more detailed discussion of this
point see~\cite{F4}). The advantage of the above lemma is
that it gives a whole series of line integrals. A further difference
is that the left side is an operator product, making it unnecessary
to specify initial or boundary values.\\[-.8em]

{\em{Proof of Lemma~\ref{l:lemma1}. }}
Our method is to first compute the Laplacian of both
sides of (\ref{l:4}). Comparing the structure of the
resulting formulas, it will be possible to proceed by induction in~$l$.

On the left side of (\ref{l:4}), we calculate the Laplacian with the
help of (\ref{l:5}) to
\begin{equation}
        -\OBox_x (S^{(l)} \:V\: S^{(r)})(x,y) \;=\; \delta_{l,0}
        \:V(x) \: S^{(r)}(x,y) \:+\:
        l \: (S^{(l-1)} \:V\: S^{(r)})(x,y) \: .
        \label{l:6}
\end{equation}

The Laplacian of the integral on the right side of (\ref{l:4}) can be
computed with (\ref{l:7}) and (\ref{l:5}),
\begin{eqnarray}
\lefteqn{-\OBox_x \int_0^1 \alpha^{l} \:(1-\alpha)^{r} \:(\alpha-\alpha^2)^n
\: (\OBox^n V)_{|\alpha y + (1-\alpha) x} \:d\alpha \;
S^{(n+l+r+1)}(x,y) } \label{l:8} \\
&=& -\int_0^1 \alpha^{l} \:(1-\alpha)^{r+2} \:(\alpha-\alpha^2)^n \: (\OBox^{n+1} V)_{|\alpha y + (1-\alpha) x} \:d\alpha \;
S^{(n+l+r+1)}(x,y) \nonumber \\
&&-\int_0^1 \alpha^{l} \:(1-\alpha)^{r+1} \:(\alpha-\alpha^2)^n \nonumber \\
&&\spc\qquad\quad\quad \times \;(\partial_k \OBox^{n} V)_{|\alpha y + (1-\alpha) x}
\:d\alpha \; (y-x)^k \: S^{(n+l+r)}(x,y) \nonumber \\
&&+(n+l+r+1) \nonumber \\
&&\quad\times \int_0^1 \alpha^{l} \:(1-\alpha)^{r} \:(\alpha-\alpha^2)^n
(\OBox^{n} V)_{|\alpha y + (1-\alpha) x}
\:d\alpha \;S^{(n+l+r)}(x,y) \: . \nonumber
\end{eqnarray}
In the second summand, we rewrite the partial derivative as a
derivative with respect to $\alpha$ and integrate by parts,
\begin{eqnarray*}
\lefteqn{\int_0^1 \alpha^{l} \:(1-\alpha)^{r+1} \:(\alpha-\alpha^2)^n \:
(\partial_k \OBox^{n} V)_{|\alpha y + (1-\alpha) x}
\:d\alpha \; (y-x)^k } \\
&=& \int_0^1 \alpha^{l} \:(1-\alpha)^{r+1} \:(\alpha-\alpha^2)^n \:
\frac{d}{d\alpha}(\OBox^{n} V)_{|\alpha y + (1-\alpha) x}
\:d\alpha \\
&=& -\delta_{n,0} \:\delta_{l,0}\:V(x) \\
&&-(n+l) \int_0^1 \alpha^{l} \:(1-\alpha)^{r+2} \:(\alpha-\alpha^2)^{n-1} \:
(\OBox^{n} V)_{|\alpha y + (1-\alpha) x} \:d\alpha \\
&&+(n+r+1) \int_0^1 \alpha^{l} \:(1-\alpha)^{r} \:(\alpha-\alpha^2)^n \:
(\OBox^{n} V)_{|\alpha y + (1-\alpha) x} \:d\alpha \\
&=& -\delta_{n,0} \:\delta_{l,0}\:V(x) \\
&&-n \int_0^1 \alpha^{l} \:(1-\alpha)^{r+2} \:(\alpha-\alpha^2)^{n-1} \:
(\OBox^{n} V)_{|\alpha y + (1-\alpha) x} \:d\alpha \\
&&+(n+l+r+1) \int_0^1 \alpha^{l} \:(1-\alpha)^{r} \:(\alpha-\alpha^2)^n \:
(\OBox^{n} V)_{|\alpha y + (1-\alpha) x} \:d\alpha \\
&&-l \int_0^1 \alpha^{l-1} \:(1-\alpha)^{r} \:(\alpha-\alpha^2)^{n} \:
(\OBox^{n} V)_{|\alpha y + (1-\alpha) x} \:d\alpha \: .
\end{eqnarray*}
We substitute back into the original equation to obtain
\begin{eqnarray*}
\lefteqn{ (\ref{l:8}) \;=\; \delta_{n,0} \:\delta_{l,0} \: V(x) \:
S^{(r)}(x,y) } \\
&&+l \int_0^1 \alpha^{l-1} \:(1-\alpha)^{r} \:(\alpha-\alpha^2)^{n} \:
(\OBox^{n} V)_{|\alpha y + (1-\alpha) x}
\:d\alpha \; S^{(n+l+r)}(x,y) \\
&&-\int_0^1 \alpha^{l} \:(1-\alpha)^{r+2} \:(\alpha-\alpha^2)^{n} \:
(\OBox^{n+1} V)_{|\alpha y + (1-\alpha) x}
\:d\alpha \; S^{(n+l+r+1)}(x,y) \\
&&+n\int_0^1 \alpha^{l} \:(1-\alpha)^{r+2} \:(\alpha-\alpha^2)^{n-1} \:
(\OBox^{n} V)_{|\alpha y + (1-\alpha) x}
\:d\alpha \; S^{(n+l+r)}(x,y) \: .
\end{eqnarray*}
After dividing by~$n!$ and summing over~$n$, the last two summands
are telescopic and cancel each other. Thus we get
\begin{eqnarray}
\lefteqn{-\OBox \sum_{n=0}^\infty \frac{1}{n!} \int_0^1
\alpha^{l} \:(1-\alpha)^{r} \:(\alpha-\alpha^2)^n \: (\OBox^n
V)_{|\alpha y + (1-\alpha) x} \:d\alpha \; S^{(n+l+r+1)}(x,y) }
\nonumber \\
&=& \delta_{l, 0} \:V(x)\:S^{(r)}(x,y) \;+\; l \sum_{n=0}^\infty \frac{1}{n!} \int_0^1
\alpha^{l-1} \:(1-\alpha)^{r} \:(\alpha-\alpha^2)^n \spc \nonumber \\
&&\quad \times\; (\OBox^n
V)_{|\alpha y + (1-\alpha) x} \:d\alpha \; S^{(n+l+r)}(x,y) \:.\spc
\label{l:9a}
\end{eqnarray}

We now compare the formulas (\ref{l:6}) and (\ref{l:9a}) for the Laplacian of
both sides of (\ref{l:4}). In the special case $l=0$, these formulas
coincide, and we can use a uniqueness argument for the solutions of the
wave equation to prove (\ref{l:4}): We assume that
we consider the advanced Green's function (for the
retarded Green's function, the argument is analogous). For given $y$,
we denote the difference of both sides of (\ref{l:4}) by $F(x)$.
Since the support of $F(x)$ is in the past light cone $x \in
L^\land_y$, $F$ vanishes in a neighborhood of the
hypersurface ${\mathcal{H}}=\{z \in \R^4 \:|\: z^0 = y^0 + 1\}$.
Moreover, the Laplacian of $F$ is identically equal to zero according to
(\ref{l:6}) and (\ref{l:9a}). We conclude that
\[      \OBox F \;=\; 0 \spc {\mbox{and}} \spc F_{|{\mathcal{H}}}\;=\; \partial_k
        F_{|{\mathcal{H}}} \;=\; 0 \:. \]
Since the wave equation has a unique solution for given initial data
on the Cauchy surface ${\mathcal{H}}$, $F$ vanishes identically.

The general case follows by induction in $l$: Suppose that
(\ref{l:4}) holds for given $\hat{l}$ (and arbitrary $r$).
Then, according to (\ref{l:6}, \ref{l:9a}) and the induction hypothesis,
the Laplacian of both sides of (\ref{l:4}) coincides for $l = \hat{l} + 1$.
The above uniqueness argument for the solutions of the wave equation again
gives (\ref{l:4}).
\QED

The above lemma can be used iteratively for the light-cone expansion of
more complicated operator products. To explain the method, we look at
the example of three factors $S^{(0)}$ and two potentials $V$, $W$,
\begin{equation}
(S^{(0)} \:V\: S^{(0)} \:W\: S^{(0)})(x,y) \;=\; \int d^4z \:
S^{(0)}(x,z) \: V(z)\; (S^{(0)}\:W\:S^{(0)})(z,y) \:.
\label{l:28a}
\end{equation}
Having split up the operator product in this form, we can apply
Lemma \ref{l:lemma1} to the factor $S^{(0)} W S^{(0)}$,
\[ \;=\; \sum_{n=0}^\infty \frac{1}{n!} \int d^4z \;
S^{(0)}(x,z) \left\{ V(z) \int_0^1 (\alpha-\alpha^2)^n \:
(\OBox^n W)_{|\alpha y + (1-\alpha) z} \:d\alpha \right\} S^{(n+1)}(z,y) \:. \]
Now we rewrite the $z$-integral as the operator product
$(S^{(0)} g_y S^{(0)})(x,y)$, where $g_y(z)$ is the function in the
curly brackets. The $y$-dependence of $g_y$ causes no problems because we can
view $y$ as a fixed parameter throughout. Thus we can simply
apply Lemma \ref{l:lemma1} once again to obtain
\begin{eqnarray*}
&=& \sum_{m,n=0}^\infty \frac{1}{m!\:n!} \int_0^1 d\beta
\;(1-\beta)^{n+1} \:(\beta-\beta^2)^m \:\int_0^1 d\alpha
\;(\alpha-\alpha^2)^n \\
&&\hspace*{1.5cm}
\times\; \OBox^m_z \left( V(z) \:(\OBox^n W)_{|\alpha y + (1-\alpha)
z} \right)_{|z=\beta y + (1-\beta) x} \; S^{(m+n+2)}(x,y) \:.
\end{eqnarray*}
Now the Laplacian $\OBox^m_z$ can be carried out with the Leibniz
rule. Notice that the manipulations of the infinite series are not
problematic because the number of terms is finite to every order on the
light cone.

The Feynman diagrams in~(\ref{l:11b}) can be expanded with this iterative
method, but they are a bit more difficult to handle.
One complication is that pulling the Dirac matrices out of the Green's
functions according to~(\ref{l:11a}) gives one additional partial derivative
per Green's function. However, this causes no major problems, because these partial derivatives
can be carried out after each induction step by differentiating through the
light-cone expansion of Lemma~\ref{l:lemma1}.
Another issue is to keep track of the {\em{chirality}} of the potentials.
To this end, one must use that the zero mass Green's function~$s$
and the factors $\Aslsh_{L\!/\!R}$ are odd, whereas the dynamical mass matrices
are even. Thus the chirality of the potentials changes each time
a dynamical mass matrix appears, as in the example of the operator product
\beq \label{chiprod}
\chi_L \:s\: \Aslsh_L \:s \cdots s\: \Aslsh_L \:s\: Y_L \:s\: \Aslsh_R \:s \cdots s\:
\Aslsh_R \:s\: Y_R \:s\: \Aslsh_L \:s \cdots .
\eeq
The last difficulty is that partial derivatives inside the line integrals
may be contracted with a factor~$(y-x)$, like for example in the expression
\[ \int_0^1 (y-x)^j \: \partial_j V_{|\alpha y + (1-\alpha) x}\:d\alpha \:. \]
Such derivatives act in direction of the line integral and are thus called
{\em{tangential}}. Writing them as derivatives with
respect to the integration variable, we can integrate by parts, e.g.
\[ \int_0^1 (y-x)^j \: \partial_j V_{|\alpha y + (1-\alpha) x}\:d\alpha \;=\; \int_0^1 \frac{d}{d\alpha}
V(\alpha y + (1-\alpha) x)\:d\alpha \;=\; V(y) - V(x) \:. \]
Going through the calculations and the combinatorics in detail,
one finds that with such integrations by parts we can indeed get rid of
all tangential derivatives, and one ends up with terms of th following
structure (for the proof see~\cite{F5}).
\begin{Thm} {\bf{(light-cone expansion of the $k^{\mbox{\scriptsize{th}}}$ order
Feynman diagram)}}
\label{l:thm1} \index{light-cone expansion!of the Feynman diagram}
Using a multi-index notation and the abbreviation
\begin{equation}
        \int_x^y [l,r\:|\: n] \:f(z) \:dz  \;:=\; \int_0^1 d\alpha \;
        \alpha^{l}\:(1-\alpha)^{r} \:(\alpha-\alpha^2)^n \; f(\alpha
        y + (1-\alpha) x) \:,
        \label{l:29x}
\end{equation}
the light-cone expansion of the~$k^{\mbox{\scriptsize{th}}}$
order  contribution to the perturbation series
(\ref{l:11b}) can be written as an infinite sum of expressions
of the form
\begin{eqnarray}
&& \chi_c\:C \:(y-x)^K \:W^{(0)}(x) \int_x^y [l_1, r_1 \:|\: n_1] \:dz_1 \;
W^{(1)}(z_1) \int_{z_1}^y [l_2, r_2 \:|\: n_2] \:dz_2 \; W^{(2)}(z_2)
\nonumber \\
&&\hspace*{1cm} \cdots \int_{z_{\alpha-1}}^y [l_\alpha, r_\alpha \:|\:
n_\alpha] \:dz_\alpha \; W^{(\alpha)}(z_\alpha) \;
\gamma^J \; S^{(h)}(x,y) \;\;\;\;\;, \;
\alpha \leq k \:. \spc
\label{l:70}
\end{eqnarray}
Here the factors $W^{(\beta)}$ are composed of the potentials and
their partial derivatives,
\begin{equation}
        W^{(\beta)} \;=\; (\partial^{K_{a_\beta}} \OBox^{p_{a_\beta}}
        V^{(a_\beta)}_{J_{a_\beta}, c_{a_\beta}})
        \cdots (\partial^{K_{b_\beta}} \OBox^{p_{b_\beta}}
        V^{(b_\beta)}_{J_{b_\beta}, c_{b_\beta}})
        \label{l:71}
\end{equation}
with $a_1=1$, $a_{\beta+1}=b_\beta+1$, $b_\beta \geq a_\beta-1$ (in
the case $b_\beta=a_\beta-1$, $W^{(\beta)}$ is identically equal to one)
and $b_\alpha=k$. Furthermore, $c, c_a \in \{L, R\}$ are chiral indices,
$C$ is a complex number, and the parameters
$l_a$, $r_a$, $\na$, and $p_a$ are non-negative integers.
The functions $V^{(a)}_{J_a, c_a}$
coincide with any of the individual potentials in (\ref{l:18a}) with
chirality $c_a$, i.e.
\begin{eqnarray}
V^{(a)}_{c_a} &=& A_{c_a} \;\;\;\: \;\;\:\;\;\;\;\;
{\mbox{(in which case $|J_a|=1$)}} \quad {\mbox{or}} \nonumber \\
V^{(a)}_{c_a} &=& m Y_{c_a} \;\;\;\;\;\;\;\;\; {\mbox{(in which case
$|J_a|=0$)}} \: .
\label{l:42e}
\end{eqnarray}
The chirality $c_a$ of the potentials is determined by the following rules:
\begin{enumerate}
\item[\rm{\em{(i)}}] The chirality $c_1$ of the first potential
coincides with the chirality of the factor $\chi_c$.
\item[\rm{\em{(ii)}}] The chirality of the potentials is reversed at
every mass matrix, i.e.
\[ {\mbox{$c_a$ and $c_{a+1}$}}\;\;\;\left\{ \begin{array}{cl}
\mbox{coincide} & \mbox{if $V^{(a)}_{c_a}=A_{c_a}$} \\
\mbox{are opposite} & \mbox{if $V^{(a)}_{c_a}=m Y_{c_a} \:.$}
\end{array} \right. \]
\end{enumerate}
The tensor indices of the multi-indices are all
contracted with each other, according to the following rules:
\index{contraction rules}
\begin{enumerate}
\item[\rm{\em{(a)}}] No two tensor indices of the same multi-index are
contracted with each other.
\item[\rm{\em{(b)}}] The tensor indices of the factor $\gamma^{J}$ are
all contracted with different multi-indices.
\item[\rm{\em{(c)}}] The tensor indices of the factor~$(y-x)^K$ are all
contracted with the tensor indices of the factors~$V^{(a)}_{J_a}$ or~$\gamma^J$,
but not with the partial derivatives~$\partial^{K_a}$.
\end{enumerate}
To every order~$h$ on the light cone, the number of
terms of the form~(\ref{l:70}) is finite. Furthermore,
\beq \label{71x}
2h \;=\; k-1 - |K| + \sum_{a=1}^k (|K_a| + 2 p_a) \:.
\eeq
\end{Thm}
The rules~{\em{(i)}} and~{\em{(ii)}} correspond precisely to our observation
that the chirality changes at each dynamical mass matrix~(\ref{chiprod}).
The restrictions~{\em{(a)}} and~{\em{(b)}} on the possible contractions
of tensor indices prevent an abuse of our multi-index notation.
More precisely, {\em{(a)}} avoids factors
$(y-x)^2$ in $(y-x)^I$, an unnecessary
large number of $\gamma$-matrices in $\gamma^J$ and ``hidden'' Laplacians
inside the partial derivatives $\partial_{z_a}^{I_a}$.
The rule {\em{(b)}}, on the other hand, prevents factors $(y-x)^2$ and
hidden Laplacians in combinations of the form
$(y-x)_i \:(y-x)_j \: \gamma^i \:\gamma^j$ and
$\partial_{ij} V^{(a)}_{J_a} \: \gamma^i \:\gamma^j$, respectively.
The rule {\em{(c)}} means that no tangential derivatives appear.
The rules~{\em{(a)}}--{\em{(c)}} imply that the tensor indices
of the multi-index~$K$ are all contracted with the chiral potentials,
except for one index which may be contracted with the factor~$\gamma^J$.
Since at most~$k$ chiral potentials appear, we get the inequality
$|K| \leq k+1$. Using this inequality in~(\ref{71x}) we get
the bound
\beq \label{hbd}
h \;\geq\; -1 + \frac{1}{2} \sum_{a=1}^k (|K_a| + 2 p_a) \:.
\eeq
This shows that~$h$ never becomes smaller than~$-1$ and that
derivatives of the potentials increase the order on the light cone.
In the case~$h=-1$, it follows from~(\ref{71x}) that~$|K| \geq 1$,
meaning that at least one factor~$(y-x)$ appears in~(\ref{l:70}).
We conclude that the factor $S^{(h)}$ in~(\ref{l:70}) is always
well-defined by either~(\ref{l:23b}) or~(\ref{l:7}).

So far the Green's function was described perturbatively by
a sum of Feynman diagrams~(\ref{l:11b}). In order to get from this
perturbative description to {\em{non-perturbative}} formulas of the
light-cone expansion, we shall now carry out the sum over all
Feynman diagrams to any fixed order on the light cone. This procedure
is called {\em{resummation}} of the light-cone
expansion\index{light-cone expansion!resummation of the}.
In order to give a first idea of how the resummation works,
we consider the leading singularity on the light cone
by neglecting all terms of the order~${\mathcal{O}}((y-x)^{-2})$.
According to~(\ref{l:24b}), we need to take into account only
the contributions~(\ref{l:70}) with~$h=-1$. The
inequality~(\ref{hbd}) implies that no
derivatives of the potentials appear. Moreover, we obtain from~(\ref{71x})
that $|K|=k+1$. Again using the contraction rules~{\em{(a)}}--{\em{(c)}},
we conclude that one tensor index of the multi-index~$K$ is
contracted with a Dirac matrix, whereas the remaining $k$ indices of~$K$
are all contracted with chiral potentials.
Therefore, all~$k$ potentials are chiral, and
no dynamical mass matrices appear.
A detailed calculation yields for the~$k^{\mbox{\scriptsize{th}}}$ order
Feynman diagram a term of precisely this structure,
\begin{eqnarray*}
\lefteqn{ \chi_c \:((-s \:B)^k s)(x,y) \;=\; \chi_c \:(-i)^k \int_x^y dz_1 \;
(y-x)_{j_1} \:A^{j_1}_c(z_1) } \nonumber \\
&& \times \int_{z_1}^y dz_2 \:(y-z_1)_{j_2}\: A^{j_2}_c(z_2)
\cdots \int_{z_{k-1}}^y dz_k \:(y-z_k)_{j_2}\: A^{j_k}_c(z_k) \:
s(x,y) \\
&&\:+\: {\mathcal{O}}((y-x)^{-2})\:,
\end{eqnarray*}
where we used for the line integrals the short notation
\beq \label{snot}
\int_x^y f(z)\: dz \;:=\; \int_x^y [0,0 \:|\: 0]\: f(z)\:dz
\;=\; \int_0^1 f(\alpha y + (1-\alpha) x)\: d\alpha\:.
\eeq
The obtained nested line integrals can be identified with the
summands of the familiar Dyson series\index{Dyson series}.
This allows us to carry out the sum over all Feynman diagrams,
\begin{equation} \label{sho}
\chi_c \:\tilde{s}(x,y) \;=\; \chi_c \:\Pexp \left( -i \int_x^y
(y-x)_j \:A^j_c(z)\:dz \right) s(x,y) \:+\:
{\mathcal{O}}((y-x)^{-2}) \:,
\end{equation}
where we again used the notation~(\ref{snot}) and the following definition
of~$\Pexp$.
\begin{Def}\label{2.5.4}
For a smooth one-parameter family of matrices $F(\alpha)$, $\alpha
\in \R$, the {\bf{ordered exponential}}\index{ordered exponential}~$\Pexp\label{Pexp} (\int F(\alpha) \:d\alpha)$
is given by the Dyson series
\begin{eqnarray}
\Pexp \left( \int_a^b F(\alpha) \:d\alpha \right)
&=& \1 \:+\: \int_a^b dt_0 \:F(t_0) \:dt_0 \:+\: \int_a^b dt_0
        \:F(t_0) \int_{t_0}^b dt_1 \: F(t_1) \nonumber \\
&&\hspace*{-2.5cm} +\int_a^b dt_0
        \:F(t_0) \int_{t_0}^b dt_1 \: F(t_1) \int_{t_1}^b dt_2 \:F(t_2)
        \:+\: \cdots \: . \label{defPexp}
\end{eqnarray}
\end{Def}

The appearance of the ordered exponential in~(\ref{sho}) can be understood
from the local gauge invariance. We explain this relation for simplicity
in the example of dynamical mass matrices and chiral potentials of the form
\[ Y_L \;=\; Y_R \;=\; 0 \;,\qquad
A_L(x) \;=\; A_R(x) \;=\; i U(x) \:(\partial U^{-1}(x)) \:, \]
where~$U=U^{(a \alpha)}_{(b \beta)}$ is a unitary matrix on the
sectors and generations. In this case, the Dirac operator is related to the
free Dirac operator simply by a local unitary transformation,
\[ i \Pdd + B \;=\; U \:i \Pdd\: U^{-1}\:. \]
Interpreting this local transformation as in~{\S}\ref{isec3} as
a gauge transformation~(\ref{lgt}),
we can say that the external potential can be transformed
away globally by choosing a suitable gauge. Using the well-known
behavior of the Green's function under gauge transformations, we obtain
the simple formula
\beq \label{gt}
\tilde{s}(x,y) \;=\; U(x)\: s(x,y)\: U^{-1}(y)\:.
\eeq
Let us verify that this is consistent with~(\ref{sho}).
Setting~$V(\alpha) = U(\alpha y + (1-\alpha) x)$ and using the
relation~$V (V^{-1})' =
(V V^{-1})' - V' V^{-1} = - V' V^{-1}$, we can write the
integrand of the ordered exponential as
\beq \label{spot}
-i (y-x)_j \:A^j_c(z) \;=\;
- V'(\alpha)\: V^{-1}(\alpha)\:.
\eeq
Differentiating~(\ref{defPexp}) with respect to~$a$ as well as evaluating
it for~$a=b$, one sees that the ordered exponential can be characterized
as the solution of the initial value problem
\[ \frac{d}{da} \Pexp \left( \int_a^b F \right)
\;=\; -F(a)\: \Pexp \left( \int_a^b F \right) \:,\qquad
\Pexp \left( \int_b^b F \right) \;=\; \1\:. \]
In the case~$F=-V'V^{-1}$, it is easily verified that the solution to
this initial value problem is
\[ \Pexp \left( -\int_a^b V'(\alpha)\:V^{-1}(\alpha)\:d\alpha
\right) \;=\; V(a)\: V^{-1}(b)\:. \]
Using~(\ref{spot}), we conclude that
\[ \Pexp \left( -i \int_x^y (y-x)_j \:A^j_c(z)\:dz \right)
\;=\; U(x)\: U^{-1}(y)\:. \]
Thus the ordered exponential in~(\ref{sho}) gives precisely the
factor~$U(x)\; U(y)^{-1}$ needed for the correct behavior under gauge transformations~(\ref{gt}).

To higher order on the light cone, the situation clearly is more
complicated. Nevertheless, it is very helpful to imagine
that after rearranging the summands of the light-cone expansion in
a suitable way, certain subseries can be summed up explicitly giving rise to
ordered exponentials of the chiral potentials. As in~(\ref{chiprod}),
the chirality of the potentials should change each time a dynamical
mass matrix appears. This conception is made precise by the following
definition and theorem, proving that the
light-cone expansion of the Green's function can be obtained to any given
order on the light cone by taking a finite number of terms of
the form~(\ref{l:29x}) and inserting suitable ordered exponentials.
\begin{Def}
\label{l:def_pf}
A contribution (\ref{l:29x}) to the light-cone expansion of Theorem
\ref{l:thm1} is called {\bf{phase-free}}\index{phase-free} if all the tangential potentials
$V^{(a)}_{J_a}$ are differentiated, i.e.
\[ |K_a|+2 p_a > 0 \spc{\mbox{whenever}}\spc {\mbox{$J_a$ is contracted with
$(y-x)^K$.}} \]
From every phase-free contribution the corresponding
{\bf{phase-inserted}}\index{phase-inserted} contribution is obtained as follows:
We insert ordered exponentials according to the replacement rule
\[ W^{(\beta)}(z_\beta) \;\longrightarrow\;
W^{(\beta)}(z_\beta) \:\Pexp \left(-i \int_{z_\beta}^{z_{\beta+1}} A^{j_\beta}_{c_\beta}
\: (z_{\beta+1} - z_\beta) \right),\quad \beta=0,\ldots,\alpha\:,  \]
where we set~$z_0=x$ and~$z_{\alpha+1}=y$. The chiralities~$c_\beta$
are determined by the relations~$c_0 = c$ and
\[ {\mbox{$c_{\beta-1}$ and $c_\beta$}} \left\{ \!\!\!\begin{array}{c}
{\mbox{coincide}} \\ {\mbox{are opposite}} \end{array} \!\!\right\}
{\mbox{if $W^{(\beta-1)}$ contains an}} \left\{ \!\!\!\begin{array}{c}
{\mbox{even}} \\ {\mbox{odd}} \end{array} \!\!\right\}
{\mbox{number of factors $Y_.$.}} \]
\end{Def}

\begin{Thm}
\label{l:thm3}
To every order on the light cone, the number of phase-free contributions
is finite.
The light-cone expansion of the Green's function $\tilde{s}(x,y)$
is given by the sum of the corresponding phase-inserted contributions.
\end{Thm}
For the proof we refer to~\cite{F5}. In short, the first statement
follows directly from the contraction rules~{\em{(a)}}--{\em{(c)}}
and~(\ref{71x}), whereas for the second part one uses for
fixed~$x$ and~$y$ the behavior of the Green's function under
non-unitary local transformations of the spinors.

Our constructions have led to a convenient procedure for performing
the light-cone expansion of the Green's function. One only needs to
compute to any order on the light cone
the finite number of phase-free contributions. Then one inserts
ordered exponentials according to Def.~\ref{l:def_pf}.
For the computation of the phase-free contributions we use a
computer program. Appendix~\ref{pappLC} lists those phase-free
contributions which will be of relevance in
Chapters~\ref{esec3}--\ref{secegg}.

In the remainder of this section we describe how the above methods
can be adapted to the {\em{fermionic projector}}.
We begin for simplicity with the fermionic projector corresponding to
one Dirac sea in the vacuum~(\ref{Pvac}). Similar to~(\ref{l:10}) we pull out
the Dirac matrices,
\beq \label{Ta}
P^{\mbox{\scriptsize{sea}}}(x,y) \;=\; (i \Pdd_x + m)\:
T_{m^2}(x,y) \:,
\eeq
where~$T_{m^2}$ is the Fourier transform of the lower mass shell, \label{Tm2}
\beq \label{Taf}
T_{m^2}(x,y) \;=\; \int \frac{d^4k}{(2 \pi)^4} \:
\delta(k^2-m^2)\: \Theta(-k^0) \:e^{-ik(x-y)} \: .
\eeq
Computing this Fourier integral and expanding the resulting
Bessel functions gives
\begin{eqnarray}
\lefteqn{ T_{m^2}(x,y) \;=\; -\frac{1}{8 \pi^3} \:
\left( \frac{\mbox{PP}}{\xi^2} \:+\: i \pi \delta (\xi^2) \:
\varepsilon (\xi^0) \right) } \nonumber \\
&&+\: \frac{m^2}{32 \pi^3} \left( \log |m^2 \xi^2| + C_E
+ i \pi \:\Theta(\xi^2) \:\epsilon(\xi^0) \right)
\sum_{j=0}^\infty \frac{(-1)^j}{j! \: (j+1)!} \: \frac{(m^2
\xi^2)^j}{4^j} \nonumber \\
&& -\frac{m^2}{32 \pi^3} \sum_{j=0}^\infty \frac{(-1)^j}{j! \:
(j+1)!} \: \frac{(m^2 \xi^2)^j}{4^j} \: (\Phi(j+1) + \Phi(j)) \:.
\label{l:3.1}
\end{eqnarray}
Here~$\xi \equiv y-x$, $C_E =2C-2\log 2$ with Euler's constant~$C$,
and~$\Phi$ is the function
\[ \Phi(0)=0 \:,\spc \Phi(n)=\sum_{k=1}^n \frac{1}{k} \quad
{\mbox{ for $n \geq 1$}}\: . \]
Similar to~(\ref{l:12}), this expansion involves distributions
which are singular on the light cone. But in addition to
singularities~$\sim \delta((y-x)^2$ and~$\sim \Theta((y-x)^2$, now also
singularities of the form form~${\mbox{PP}}/(y-x)^2$ and~$\log|(y-x)^2|$
appear. In particular, $T_{m^2}$ is {\em{not causal}} in the sense that~${\mbox{supp}}\, T_{m^2}(x,.) \not \subset L_x$.
Another similarity to~(\ref{l:12}) is that power series
in~$m^2 (y-x)^2$ appear. This suggests that in analogy to~(\ref{l:24b}),
the higher orders in~$m^2$ should be of higher order on the light cone.
However, due to the term~$\log |m^2 \xi^2|$, the distribution~$T_{m^2}$
is {\em{not}} a power series in~$m^2$. This means that the higher mass
derivatives of~$T_{m^2}$ do not exist, and the analog of~(\ref{l:24b})
breaks down.
This so-called {\em{logarithmic mass problem}}\index{logarithmic mass problem} reflects a basic
infrared problem in the light-cone expansion of the fermionic
projector. In the vacuum, it can be resolved with the following
simple construction. We subtract the problematic~$\log |m^2|$-terms by
setting \label{Tregm2}
\beq \label{Tadef}
T_{m^2}^{\mbox{\scriptsize{reg}}}(x,y) \;=\; T_{m^2}(x,y) \:-\:
\frac{m^2}{32 \pi^3} \:\log|m^2| \: \sum_{j=0}^\infty
\frac{(-1)^j}{j! \: (j+1)!} \: \frac{(m^2 \xi^2)^j}{4^j} \:.
\eeq
This distribution is a power series in~$m^2$, and in analogy to~(\ref{l:23b})
we can set \label{T(l)}
\beq \label{Tldef}
T^{(l)} \;=\;  \left( \frac{d}{da} \right)^l \left.
T^{\mbox{\scriptsize{reg}}}_a \right|_{a=0} \spc (l \geq 0).
\eeq
Furthermore, we introduce~$T^{(-1)}$ similar to~(\ref{l:7}) as the
distributional derivative of~$T^{(0)}$. Similar to~(\ref{l:24b}),
\[ T^{(n)}(x,y) \spc
{\mbox{is of the order $\;\;\;{\mathcal{O}}((y-x)^{2n-2})$}}\:, \]
and thus the mass expansion of~$T_{m^2}^{\mbox{\scriptsize{reg}}}$ gives
us its light-cone expansion. The point is that the difference
of~$T_{m^2}$ and~$T^{\mbox{\scriptsize{reg}}}_{m^2}$,
\[ T_{m^2} - T_{m^2}^{\mbox{\scriptsize{reg}}} \;=\;
\frac{m^2}{32 \pi^3} \:\log|m^2| \: \sum_{j=0}^\infty
\frac{(-1)^j}{j! \: (j+1)!} \: \frac{(m^2 \xi^2)^j}{4^j} \:, \]
is an absolutely convergent power series in~$\xi^2$ and is thus
a {\em{smooth}} function in position space. This smooth contribution is
of no relevance as long as the singularities on the light cone are
concerned. This leads us to write the fermionic projector in the form
\beq \label{Plcev}
P^{\mbox{\scriptsize{sea}}}(x,y) \;=\; \sum_{n=0}^\infty
\frac{m^{2n}}{n!} (i \Pdd_x + m)\: T^{(n)}(x,y) \:+\:
P^{\mbox{\scriptsize{lc}}}(x,y) \:,
\eeq
with
\[ P^{\mbox{\scriptsize{lc}}} \;:=\; (i \Pdd_x + m)
\left( T_{m^2} - T_{m^2}^{\mbox{\scriptsize{reg}}} \right). \]
The series in~(\ref{Plcev}) is a light-cone expansion which completely
describes the singular behavior of the fermionic projector on the light
cone. The so-called {\em{low-energy contribution}}~$P^{\mbox{\scriptsize{lc}}}$,
on the other hand, is a smooth function.

This method of performing a light-cone expansion modulo smooth functions
on the light cone also works for the general fermionic projector with
interaction. But the situation is more complicated and at the same time
more interesting, in particular because the space-time dependence
of the involved external potentials reveals the causal structure of
the fermionic projector. The first construction step is to use
the identity on the left side of~(\ref{80}) to carry over the
light-cone expansion from the Green's function to the
distribution~$\tilde{k}$. Next, comparing~(\ref{2pole}) with the formula
\begin{eqnarray*}
p_m(k) &=& (k \slsh + m) \: \delta(k^{2}-m^{2}) \\
&=& \frac{1}{2 \pi i} \:(k \slsh + m) \; \lim_{\varepsilon
\searrow 0} \left[ \frac{1}{k^{2}-m^{2}-i\varepsilon} \:-\:
\frac{1}{k^{2}-m^{2}+i\varepsilon} \right] ,
\end{eqnarray*}
one sees that the distributions~$p$ and~$k$ differ from each only by
the $i \varepsilon$-regularization in momentum space.
The key step of the construction is the so-called
{\em{residual argument}}\index{residual argument}, which relates the light-cone expansion of~$\tilde{k}$
to an expansion in momentum space. Using that the latter
expansion remains unchanged if the poles of the distributions are
suitably shifted in momentum space, one obtains the light-cone expansion
for an operator~$\tilde{p}^{\mbox{\scriptsize{res}}}$, which
can be regarded as a perturbation of~$p$, but with a
different combinatorics than in the expansion of Theorem~\ref{Thm2}.
More precisely,  $\tilde{p}$ can be obtained from~$\tilde{p}^{\mbox{\scriptsize{res}}}$
by replacing pairs of factors~$k$ in the perturbation expansion by
corresponding factors~$p$. The argument~(\ref{prrep}) shows that
the difference $\tilde{p}-\tilde{p}^{\mbox{\scriptsize{res}}}$ vanishes
in the static case, and more generally one can say that
$\tilde{p}-\tilde{p}^{\mbox{\scriptsize{res}}}$ will be of significance
only if the frequency of the external potentials is of the order of the
mass gap. Therefore, the operator
\[ P^{\mbox{\scriptsize{he}}}\label{Phe} \;:=\; \frac{1}{2}\:X\:(\tilde{p}-\tilde{p}^{\mbox{\scriptsize{res}}}) \]
is called the {\em{high-energy contribution}}\index{high-energy contribution} to the fermionic projector.
Moreover, resolving the logarithmic mass problem we obtain again a
{\em{low-energy contribution}}\index{low-energy contribution} $P^{\mbox{\scriptsize{le}}}$. We thus obain
a representation of the fermionic projector of the form \label{Ple}
\begin{eqnarray}
P^{\mbox{\scriptsize{sea}}}(x,y) &=& \sum_{n=-1}^\infty
{\mbox{(phase-inserted line integrals)}} \times  T^{(n)}(x,y) \nonumber \\
&&+ P^{\mbox{\scriptsize{le}}}(x,y) + P^{\mbox{\scriptsize{he}}}(x,y) \:.
\label{fprep}
\end{eqnarray}
Here the series is a light-cone expansion which describes
the singular behavior of the fermionic projector on the light cone
non-perturbatively. It is obtained from the light-cone expansion of the
Green's functions by the simple replacement rule
\[ S^{(n)} \;\longrightarrow\; T^{(n)}\:. \]
In particular, the phase-inserted line integrals are exactly
the same as those for the Green's functions (see Def.~\ref{l:def_pf}).
The contributions $P^{\mbox{\scriptsize{le}}}$
and~$P^{\mbox{\scriptsize{he}}}$, on the other hand,
are given perturbatively by a
series of terms which are all smooth on the light cone
(we expect that the perturbation series for~$P^{\mbox{\scriptsize{le}}}$
and~$P^{\mbox{\scriptsize{le}}}$ should converge, but this has not
yet been proven).
The ``causality'' of the causal perturbation expansion
can be understood from the fact that the phase-inserted line integrals
in~(\ref{fprep}) are all bounded integrals along the line segment joining
the points~$x$ and~$y$ (whereas the light-cone expansion of general operator
products involves unbounded line integrals~\cite{Fprep}).
In particular, when~$y$ lies in the causal future or past of~$x$,
the light-cone expansion in~(\ref{fprep}) depends
on the external potential only inside the ``diamond''
$(J^\lor_x \cap J^\land_y) \cup (J^\land_x \cap J^\lor_y)$.
Nevertheless, the light-cone expansion is not causal in the strict
sense because there are contributions for~$y \not \in J_x$.
Furthermore, the low- and high-energy contributions cannot be described with
line integrals and violate locality as well as causality.
This non-locality can be understood from the fact that
the fermionic projector is a global object in space-time (see the
discussion in~{\S}\ref{jsec4}).
Mathematically, it is a consequence of the
non-local operation of taking the absolute value of an
operator~(\ref{51x}) in the definition of the fermionic projector.
We conclude that the singular behavior of the
fermionic projector on the light-cone can be described explicitly by
causal line integrals, whereas the smooth contributions to the fermionic projector are governed by non-local effects.

Inspecting the explicit formulas of Appendix~\ref{pappLC}, one sees
immediately that from the line integrals of the light-cone expansion
one can reconstruct the chiral and scalar/pseu\-do\-sca\-lar potentials.
In this sense, $P^{\mbox{\scriptsize{sea}}}$ encodes all information
on the external potential. Furthermore, the fermionic projector gives
via its representation~(\ref{1g}) all information on the fermions and
anti-fermions of the system. We thus come to the important conclusion
that the fermionic projector describes the physical system completely.

\section{Normalization of the Fermionic States}
\label{jsec6} \setcounter{equation}{0}
In~{\S}\ref{jsec2} we disregarded that the fermionic states are in general
not normalizable in infinite volume. We avoided this problem using
a $\delta$-normalization in the mass parameter (see e.g.~(\ref{56})).
In this section we will analyze the normalization in detail
by considering the system in finite volume
and taking the infinite volume limit.
Apart from justifying the formalism introduced in~{\S}\ref{jsec2}, this
will clarify in which sense the fermionic projector is idempotent
(Theorem~\ref{thm261}). Furthermore, we will see that the probability
integral has a well-defined infinite volume limit (Theorem~\ref{thm262}),
and this will also determine the normalization
constant~$c_{\mbox{\scriptsize{norm}}}$ in~(\ref{1g}) (see~(\ref{cnorm})).
We postpone the complications related to the chiral fermions to
Appendix~\ref{appcf} and thus assume here that the chiral asymmetry
matrix $X=\1$. We work again in the setting of~{\S}\ref{jsec3}
and assume that the external potential~${\mathcal{B}}$ satisfies
the regularity assumptions of Lemma~\ref{l:lemma0}.
Furthermore, we make the physically reasonable assumption that the
{\em{masses}} \index{masses} are {\em{non-degenerate in the
    generations}}\index{masses!non-degenerate!in the generations}, meaning that
\begin{equation} \label{eq:2c}
m_{a \alpha} \neq m_{a \beta} \spc
{\mbox{for all $a$ and $\alpha \neq \beta$}}\;.
\end{equation}

In order to ensure that all normalization integrals are finite, we need to
introduce an {\em{infrared regularization}}\index{infrared regularization}. For clarity, we explain the
construction for a single Dirac sea of mass~$m$ in the vacuum.
First, we replace space by the three-dimensional box
\begin{equation}\label{7a}
T^3 \;=\; [-l_1, l_1] \times [-l_2, l_2] \times [-l_3, l_3] \spc {\mbox{with $0<l_i<\infty$}}
\end{equation}
and set $V = |T^3| = 8 \:l_1 l_2 l_3$. We impose periodic boundary conditions; this
means that we restrict the momenta $\vec{k}$ to the lattice $L^3$
given by
\[ L^3 \;=\; \frac{\pi \:\Z}{l_1} \times \frac{\pi \:\Z}{l_2} \times \frac{\pi \:\Z}{l_3}
\;\subset\; \R^3 \;. \]
In order to carry over the operators~$p_m$, $k_m$ and~$s_m$
(see~{\S}\ref{jsec2}) to finite volume, we leave the distributions in momentum space unchanged. In the transformation to position space, we replace the Fourier integral over $3$-momentum by a Fourier series according to
\begin{equation} \label{eq:A}
\int \frac{d\vec{k}}{(2 \pi)^3} \;\longrightarrow\; \frac{1}{V}\: \sum_{\vec{k} \in L^3} \;.
\end{equation}
When taking products of the resulting operators, we must take into account
that the spatial integral is now finite. For example, we obtain that
\begin{eqnarray}
(p_m \: p_{m'})(x,y) &=& \int_{\sR \times T^3} p_m(x,z) \:p_{m'}(z,y)\: d^4z \nonumber \\
&=& \int_{\sR \times T^3} d^4z \int_{-\infty}^\infty \frac{dk^0}{2\pi} \; \frac{1}{V}\:
\sum_{\vec{k} \in L^3} p_m(k)\: e^{-ik(x-z)} \nonumber \\
&&\quad \times
\int_{-\infty}^\infty \frac{dl^0}{2\pi} \; \frac{1}{V}\:
\sum_{\vec{l} \in L^3} p_{m'}(l)\: e^{-il(z-y)} \nonumber \\
&=& \int_{-\infty}^\infty \frac{dk^0}{2\pi}\: \frac{1}{V} \sum_{\vec{k} \in L^3} p_m(k)\: e^{-ikx} \nonumber \\
&& \quad \times
\int_{-\infty}^\infty \frac{dl^0}{2\pi}\: \frac{1}{V} \sum_{\vec{l} \in L^3} p_{m'}(l)\: e^{ily}
\;2 \pi\: \delta(k^0 - l^0)\; V\: \delta_{\vec{k}, \vec{l}} \nonumber \\
&=& \int_{-\infty}^\infty \frac{dk^0}{2\pi}\: \frac{1}{V} \sum_{\vec{k} \in L^3} p_m(k)\: p_{m'}(k)\; e^{-ik(x-y)} \nonumber \\
&=& \delta(m-m')\: p_m(x,y) \;,\quad \label{eq:a}
\end{eqnarray}
where $p_m(k) = (k \slsh + m)\: \delta(k^2-m^2)$. More generally, the
calculation rules~(\ref{56}--\ref{62}) for products of the operators
$k_m$, $p_m$ and $s_m$ remain valid in finite $3$-volume.

In~(\ref{eq:a}) we are still using a $\delta$-normalization in the mass parameter.
In order to go beyond this formalism and to get into the position where we can
multiply operators whose mass parameters coincide, we ``average'' the mass
over a small interval $[m, m+\delta]$.
More precisely, we set
\begin{equation} \label{eq:b}
\bar{p}_m \;=\; \frac{1}{\delta}\: \int_m^{m+\delta} p_\mu \: d\mu \spc{\mbox{and}}\spc
\bar{k}_m \;=\; \frac{1}{\delta}\: \int_m^{m+\delta} k_\mu \: d\mu \;.
\end{equation}
Then
\begin{eqnarray*}
\bar{p}_m \: \bar{p}_m &=& \frac{1}{\delta^2} \: \int_m^{m+\delta} d\mu \int_m^{m+\delta} d\mu' \;
p_\mu \: p_{\mu'} \\
&=& \frac{1}{\delta^2} \: \int_m^{m+\delta} d\mu \int_m^{m+\delta} d\mu' \;
\delta(\mu-\mu') \: p_\mu
\;=\; \frac{1}{\delta^2} \: \int_m^{m+\delta} p_\mu\: d\mu \;=\; \frac{1}{\delta}\: \bar{p}_m \;,
\end{eqnarray*}
and thus, apart from the additional factor $\delta^{-1}$, $\bar{p}_m$ is idempotent.
Similarly, we have the relations
\[ \bar{k}_m \: \bar{k}_m \;=\; \frac{1}{\delta}\: \bar{p}_m \spc {\mbox{and}} \spc
\bar{k}_m \: \bar{p}_m \;=\; \bar{p}_m \: \bar{k}_m \;=\; \frac{1}{\delta}\: \bar{k}_m
\;. \]
Thus, introducing the {\em{infrared regularized fermionic
    projector}}\index{fermionic projector!infrared regularized}
corresponding to a Dirac sea of mass $m$ by
\begin{equation} \label{eq:c}
P^{\mbox{\scriptsize{sea}}} \;=\; \frac{\delta}{2} \: (\bar{p}_m - \bar{k}_m)\;,
\end{equation}
this operator is indeed a projector,
\begin{equation} \label{eq:d}
(P^{\mbox{\scriptsize{sea}}})^* \;=\; P^{\mbox{\scriptsize{sea}}} \spc{\mbox{and}}\spc (P^{\mbox{\scriptsize{sea}}})^2 \;=\; P^{\mbox{\scriptsize{sea}}} \;.
\end{equation}
The {\em{infinite volume limit}}\index{infinite volume limit} corresponds to taking the limits
$l_1, l_2, l_3 \to \infty$ and $\delta \searrow 0$.

Let us discuss the above construction.
Clearly, our regularization method relies on special assumptions
($3$-dimensional box with periodic boundary conditions, averaging
of the mass parameter). This is partly a matter of
convenience, but partly also a necessity, because much more
general regularizations would lead to unsurmountable technical
difficulties. Generally speaking, infrared regularizations change
the system only on the macroscopic scale near spatial infinity and
possibly for large times. Due to the decay assumptions on the external potentials in Lemma~\ref{l:lemma0}, in this
region the system is only weakly interacting. This should make infrared
regularizations insensitive to the details of the regularization
procedure, and it is reasonable to expect (although it is certainly
not proven) that if the infinite volume limit exists, it
should be independent of which regularization method is used. Here
we simply take this assumption for granted and thus restrict
attention to a special regularization scheme. But at least we
will see that the infinite volume limit is independent of how the
limits $l_i \to \infty$ and $\delta \searrow 0$ are taken.

It is worth noting that not {\em{every}} infrared regularization has
a well-defined infinite volume limit. To see this, we consider the
example of a regularization in a $4$-dimensional box. Restricting the time
integral to the finite interval $t \in [-T, T]$, we obtain
\begin{eqnarray*}
\lefteqn{ (p_m \: p_m)(x,y) } \\
&=& \int_{-T}^T dt \int_{-\infty}^\infty \frac{dk^0}{2\pi} \int_{-\infty}^\infty \frac{dl^0}{2\pi} \:
\frac{1}{V} \sum_{\vec{k}, \vec{l} \in L^3,\; \vec{k}=\vec{l}} e^{-i (k^0 - l^0) t}
\: p_m(k) \:p_m(l)\; e^{ikx - ily} %\\ &=&
\end{eqnarray*}
\begin{eqnarray*}
&=&\int_{-T}^T dt \int_{-\infty}^\infty \frac{dk^0}{2\pi} \int_{-\infty}^\infty \frac{dl^0}{2\pi} \:
\frac{1}{V} \sum_{\vec{k}, \vec{l} \in L^3,\; \vec{k}=\vec{l}} e^{-i (k^0 - l^0) t} \\
&& \hspace*{1.5cm} \times \: (k \slsh+m)\: (l \:\! \slsh+m)\: \delta((k^0)^2 - (l^0)^2)\:
\delta(k^2-m^2)\: e^{ikx - ily} \\
&=& \int_{-\infty}^\infty \frac{dk^0}{2\pi} \frac{1}{V} \sum_{\vec{k} \in L^3} \frac{mT}{|k^0|}\:
p_m(k)\: e^{-ik(x-y)} \:+\: {\mathcal{O}}(T^0) \;,
\end{eqnarray*}
and due to the factor $|k^0|^{-1}$ in the last line, this is not a
multiple of $p_m(x,y)$. This problematic factor $|k^0|^{-1}$ also
appears under more general circumstances (e.g.\ when we introduce
boundary conditions at $t=\pm T$ and/or take averages of the mass
parameter), and thus it seems impossible to arrange that the
fermionic projector is idempotent. We conclude that a
$4$-dimensional box does not give a suitable regularization scheme.

The mass averaging in~(\ref{eq:b}) leads to the bizarre effect that for
fixed $\vec{k}$, a whole continuum of states of the fermionic projector,
namely all states with
\begin{equation} \label{eq:E1}
k^0 \in \left[-\sqrt{|\vec{k}|^2 + (m+\delta)^2},\:
-\sqrt{|\vec{k}|^2 + m^2} \right] ,
\end{equation}
are occupied. If one prefers to occupy only a
finite number of states for every $\vec{k}$,
one can achieve this by taking the mass averages for the bra- and ket-states
separately. For example, instead
of~(\ref{eq:c}) we could define the fermionic projector by
\begin{equation} \label{eq:e}
P(x,y) \;=\; \delta \: \int_{-\infty}^\infty \frac{dk^0}{2\pi} \int_{-\infty}^\infty \frac{dl^0}{2\pi} \:\frac{1}{V}
\sum_{\vec{k}, \vec{l} \in L^3,\; \vec{k}=\vec{l}} \bar{t}_m(k)\: \bar{t}_m(l)\:
e^{-ikx + ily}
\end{equation}
with $\bar{t}_m = \frac{1}{2}(\bar{p}_m - \bar{k}_m)$. This fermionic projector
is for every $\vec{k}$ composed of a finite number of states. Furthermore,
it is a projector in the sense of~(\ref{eq:d}).
In contrast to~(\ref{eq:c}), (\ref{eq:e}) is
not homogeneous in time, but decays on the scale $t \sim \delta^{-1}$. However, if
we restrict attention to a fixed region of space-time for which $t \ll \delta^{-1}$,
then~(\ref{eq:c}) and~(\ref{eq:e}) differ only by terms of higher order in
$\delta$, and therefore we can expect that~(\ref{eq:b})
and~(\ref{eq:d}) should have
the same infinite volume limit. The definition~(\ref{eq:c}) has the advantage
that it will be easier to introduce the interaction.

After these preparations, we come to the general construction of the
fermionic projector in the three-dimensional box~$T^3$.
Since we want to ``smear out'' the mass similar to~(\ref{eq:b}), the mass parameter needs to be variable. To this end, we introduce a parameter $\mu>0$
which shifts all masses by the same amount. Thus we describe the
system in the vacuum by the Dirac operator
\begin{equation} \label{eq:dir1}
i \Pdd - mY - \mu \1 \;.
\end{equation}
The external field is described by an operator~${\mathcal{B}}$
in the space-time~$\R \times T^3$, which we again insert into
the Dirac operator,
\[ i \Pdd + {\mathcal{B}} - mY - \mu \1 \;. \]
Now we can introduce the operators $p$, $k$ and their
perturbation expansions exactly as in~{\S}\ref{jsec3}. For clarity,
we denote the dependence on the parameter~$\mu$ by an additional
index $+\mu$. In particular, we denote the operator products in
Theorem~\ref{Thm2} by~$\tilde{p}_{+\mu}$ and~$\tilde{k}_{+\mu}$.
Since the multiplication rules~(\ref{56}--\ref{62}) also hold
in finite~$3$-volume, all the computations of~{\S}\ref{jsec3}
are still true. In particular, the operators~$\tilde{p}_{+\mu}$ and~$\tilde{k}_{+\mu}$ satisfy the~$\delta$-normalization conditions\footnote{{\textsf{Online version}:}
As noticed by A.\ Grotz, these relations are in general violated
to higher order in perturbation theory. In order to cure the problem,
one needs to rescale the states of the fermionic projector, as
is worked out in the paper~\cite{grotz} 
(listed in the references in the preface to the second online edition).}
\begin{eqnarray}
\tilde{p}_{+\mu}\: \tilde{p}_{+\mu'} &=& \tilde{k}_{+\mu}\: \tilde{k}_{+\mu'} \;=\;
\delta(\mu-\mu')\: \tilde{p}_{+\mu} \label{eq:2a1} \\
\tilde{p}_{+\mu}\: \tilde{k}_{+\mu'} &=& \tilde{k}_{+\mu}\: \tilde{p}_{+\mu'} \;=\;
\delta(\mu-\mu')\: \tilde{k}_{+\mu} \label{eq:2a2} \;.
\end{eqnarray}
In analogy to~(\ref{eq:b}) and~(\ref{eq:c}) we define the auxiliary fermionic
projector by
\begin{equation} \label{eq:2b}
P^{\mbox{\scriptsize{sea}}} \;=\; \frac{1}{2} \int_0^\delta
(\tilde{p}_{+\mu} - \tilde{k}_{+\mu})\: d\mu \;,
\end{equation}
and the fermionic projector is again obtained by taking the partial
trace~(\ref{part}),
\beq \label{eq:2bb}
(P^{\mbox{\scriptsize{sea}}})^a_b \;=\; \sum_{\alpha=1}^{g(a)} \sum_{\beta=1}^{g(b)} (P^{\mbox{\scriptsize{sea}}})^{(a \alpha)}_{(b \beta)}\:.
\eeq
The next theorem shows that the fermionic projector is idempotent in the
infinite volume limit, independent of how the limits $l_i \to \infty$ and $\delta \searrow 0$ are taken.
\begin{Thm} \label{thm261} {\bf{(idempotence of the fermionic
      projector)}}\index{fermionic projector!idempotence of the}
Consider a system composed of massive fermions with non-degenerate masses~(\ref{eq:2c}). Then the fermionic projector defined by~(\ref{eq:2b})
and~(\ref{eq:2bb}) satisfies the relations
\[ \int_{\sR \times T^3} d^4z \sum_{b=1}^N (P^{\mbox{\scriptsize{sea}}})^a_b(x,z)\: (P^{\mbox{\scriptsize{sea}}})^b_c(z,y) \;=\; (P^{\mbox{\scriptsize{sea}}})^a_c(x,y) \:+\:
\delta^2\: Q^a_c(x,y) \;, \]
where $Q$ has an expansion as a sum of operators which all have a
well-defined infinite volume limit.
\end{Thm}
{\Proof} For simplicity we omit the superscript `sea'. It follows
immediately from~(\ref{eq:2a1}--\ref{eq:2b}) that the auxiliary fermionic projector is idempotent,
\beq \label{2did}
\sum_{b, \beta} P^{(a \alpha)}_{(b \beta)}\: P^{(b \beta)}_{(c \gamma)} \;=\; P^{(a \alpha)}_{(c \gamma)} \;.
\eeq
Thus it remains to show that
\begin{equation} \label{eq:2d}
\sum_b \sum_{\alpha, \gamma} \;\sum_{\beta, \beta' {\mbox{\scriptsize{ with }}} \beta \neq \beta'} P^{(a \alpha)}_{(b \beta)}\: P^{(b \beta)}_{(c \gamma)} \;=\;
\delta^2\: Q^a_c(x,y) \;.
\end{equation}
According to the non-degeneracy assumption~(\ref{eq:2c}), there are
constants $c,\delta>0$ such that for all sufficiently small $\delta$,
\begin{equation} \label{eq:2e}
|(m_{b \beta} + \mu) - (m_{b \beta'} + \mu')| \;\geq\; c \spc {\mbox{for all $b$, $\beta \neq \beta'$, and
$0<\mu, \mu'<\delta$}}\;.
\end{equation}
On the left side of~(\ref{eq:2d}) we substitute~(\ref{eq:2b}) and the
operator product expansions of Theorem~\ref{Thm2}. Using~(\ref{eq:2e}), the resulting operator products are all finite and can be estimated using the relations
\beq \label{2dif}
\int_0^\delta d\mu \int_0^\delta d\mu' \left(\cdots A_{+\mu} \right)^{(a \alpha)}_{(b \beta)} \: \left(A_{+\mu'} \cdots
\right)^{(b \beta')}_{(c \gamma)} \;=\; c^{-1}\: {\mathcal{O}}(\delta^2) \;,
\eeq
where each factor~$A$ stands for $p$, $k$ or $s$.
This gives~(\ref{eq:2d}).
\QED

At this point we can make a remark on the name ``partial
trace.''\index{partial trace}  The notion of
a trace suggests that two matrix indices should be set equal and summed over;
thus one may want to define the fermionic projector instead of~(\ref{part}) by\footnote{{\textsf{Online version}:}
This potential source of confusion is the reason why in more recent works on the fermionic projector,
the partial trace is referred to as the {\em{sectorial projection}}.}
\begin{equation}
P^a_b(x,y) \;=\; \sum_{\alpha=1}^3 P^{(a \alpha)}_{(b \alpha)}(x,y)\:.
    \label{eq:Ap}
\end{equation}
This alternative definition suffers from the following problem.
The off-diagonal elements of~$P^{(a \alpha)}_{(b \beta)}$, $\alpha \neq \beta$,
are important to make the auxiliary fermionic projector idempotent, because
\[ P^{(a \alpha)}_{(c \gamma)} \;\stackrel{(\ref{2did})}{=}\;
\sum_{b=1}^8 \:\sum_{\beta=1}^3 P^{(a \alpha)}_{(b \beta)} \:P^{(b \beta)}_{(c \gamma)}
\;\stackrel{\mbox{\scriptsize{in general}}}{\neq}\;
\sum_{b=1}^8 P^{(a \alpha)}_{(b \alpha)} \:P^{(b \gamma)}_{(c \gamma)} \:. \]
But these off-diagonal elements do not enter the definition~(\ref{eq:Ap}), and
this makes it difficult to arrange that $P^a_b$ is idempotent. In more
technical terms, defining the fermionic projector by~(\ref{eq:Ap}) would in the
proof of the above theorem lead instead of~(\ref{eq:2d}) to the conditions
\[ \sum_{b=1}^8 \:\sum_{\alpha, \beta=1}^3 \left(
P^{(a \alpha)}_{(b \alpha)} \:P^{(b \beta)}_{(c \beta)} \:-\:
P^{(a \alpha)}_{(b \beta)} \:P^{(b \beta)}_{(c \alpha)} \right) \;=\;
\delta^2\: Q^a_c(x,y)\:. \]
As a consequence, we would in~(\ref{2dif}) get contributions with
$\beta=\beta'$, which are singular. The only way to avoid these singular
contributions would be to consider perturbations which are diagonal on the
generations. But in this special case, also the auxiliary fermionic
projector is diagonal on the generations, and so the
definitions~(\ref{eq:Ap}) and~(\ref{part}) coincide. We conclude
that~(\ref{part}) is the more general and thus preferable definition of
the partial trace.

We next consider the normalization of the individual states of
the fermionic projector.
In finite $3$-volume in the vacuum, a Dirac sea of mass~$m$
is composed of a discrete number of fermionic states. More precisely,
\begin{eqnarray}
P^{\mbox{\scriptsize{sea}}}(x,y)
&=& \int \frac{dk^0}{2 \pi} \:\frac{1}{V} \sum_{\vec{k}
\in L^3} (k \slsh + m)\:
\delta(k^2-m^2)\:\Theta(-k^0)\: e^{-ik(x-y)} \nonumber \\
&=& \frac{1}{2 \pi V} \sum_{\vec{k} \in L^3} \frac{1}{2\: |k^0|}\: (k \slsh + m)
\left. e^{-ik(x-y)} \right|_{k^0=-\sqrt{|\vec{k}|^2+m^2}} \;. \label{eq:D}
\end{eqnarray}
Here the image of $(k \slsh+m)$ is two-dimensional; it is spanned
by the two plane-wave solutions of the Dirac equation of momentum
$k$ with spin up and down, respectively. Thus we can write the
fermionic projector in analogy to~(\ref{Pvac}) as
\begin{equation} \label{eq:B}
P^{\mbox{\scriptsize{sea}}}(x,y) \;=\; \sum_{\vec{k} \in L^3} \sum_{s=\pm1}
-|\Psi_{\vec{k}s}(x) \Sr \Sl \Psi_{\vec{k}s}(y)| \;,
\end{equation}
where $\Psi_{\vec{k}s}$ are the suitably normalized
negative-energy plane-wave solutions of the Dirac equation, and
$s$ denotes the two spin orientations.
If an external field is present, it is
still possible to decompose the fermionic projector similar
to~(\ref{eq:B}) into individual states. But clearly, each of these
states is perturbed by~${\mathcal{B}}$; we denote these perturbed
states by a tilde. The next theorem shows that the probability
integral for these states is independent of the interaction and of
the size of $T^3$.

\begin{Thm} \label{thm262}
{\bf{(probability integral)}}\index{probability integral}
Under the assumptions of Theorem~\ref{thm261}, every state
$\tilde{\Psi}$ of the fermionic projector is normalized according to
\begin{equation}
\int_{T^3} \Sl \tilde{\Psi} | \gamma^0 \:\tilde{\Psi}
\Sr(t,\vec{x})\: d\vec{x} \;=\; \frac{1}{2 \pi} \;. \label{eq:C}
\end{equation}
\end{Thm}
{\Proof} Since $\tilde{\Psi}$ is a solution of the Dirac equation
$(i \Pdd + {\mathcal{B}} - mY - \mu \1) \tilde{\Psi}=0$, it follows
from current conservation (see~\ref{tind}) that the probability
integral~(\ref{eq:C}) is time independent. Thus it suffices to
compute it in the limits $t \to \pm\infty$, in which according to our
decay assumptions on ${\mathcal{B}}$ the system is non-interacting.
Since in the vacuum, the fermionic projector splits into a direct sum of Dirac
seas, we may restrict attention to a single Dirac
sea~(\ref{eq:B}). Using that the probability integral is the same
for both spin orientations,
\[ \int_{T^3} \Sl \Psi_{\vec{k}s} | \gamma^0\: \Psi_{\vec{k}s} \Sr(t,\vec{x})\:d\vec{x}
\;=\; \int_{T^3} \:\frac{1}{2} \sum_{s=\pm} \Tr \left( \gamma^0\: |\Psi_{\vec{k}s} \Sr \Sl
\Psi_{\vec{k}s} | \right) d\vec{x} \;, \]
and comparing with~(\ref{eq:D}) gives
\begin{eqnarray*}
\int_{T^3} \Sl \Psi_{\vec{k}s} | \gamma^0 \Psi_{\vec{k}s}
\Sr(t,\vec{x})\:d\vec{x} &=& \frac{1}{4 \pi\:V} \int_{T^3}
\frac{1}{2 k^0}\left. \Tr(\gamma^0 \:(k \slsh + m))
\right|_{k^0 = -\sqrt{|\vec{k}|^2+m^2}} \:d\vec{x} \\
&=& \frac{1}{4 \pi\:V} \int_{T^3} \frac{4k^0}{2k^0}\:d\vec{x} \;=\; \frac{1}{2 \pi}\;.
\end{eqnarray*}

\vspace*{-.55cm}
\QED

Let us discuss what this result means for the states of the fermionic
projector~(\ref{eq:2b}, \ref{eq:2bb}). As pointed out in the
paragraph of~(\ref{eq:E1}), the fermionic projector of the vacuum
is composed for each $\vec{k} \in L^3$ of a continuum of states~(\ref{eq:E1}).
However, if we choose the space-time points in the fixed time interval $-T<t<T$
and let $\delta \searrow 0$, we need not
distinguish between the frequencies in~(\ref{eq:E1}) and obtain that only the
discrete states with $\vec{k} \in L^3$, $k^0=-\sqrt{|\vec{k}|^2+m^2}$ are occupied (see
the discussion after~(\ref{eq:E1})).
In the causal perturbation expansion, each of these states is perturbed,
and thus also the interacting fermionic projector for small $\delta$ can be
regarded as being composed of discrete states. We write in analogy
to~(\ref{eq:B}),
\[ P(x,y) \;=\; \sum_a - |\tilde{\Psi}_a \Sr \Sl \tilde{\Psi}_a | \;, \]
where $a$ runs over all the quantum number of the fermions. According
to~(\ref{eq:2b}) and Theorem~\ref{thm2}, the probability integral is
\begin{equation}
\int_{T^3} \Sl \tilde{\Psi}_a \:|\: \gamma^0 \: \tilde{\Psi}_a \Sr(t,\vec{x})\: d\vec{x}
\;=\; \frac{\delta}{2 \pi}\;.
\end{equation}
By substituting the formulas of the light-cone expansion of~{\S}\ref{jsec5}
into~(\ref{eq:2b}), one sees that the contributions of the light-cone
expansion to the fermionic projector all involve at least one factor~$\delta$.
Thus after rescaling $P$ by a factor $\delta^{-1}$, the probability
integral~(\ref{eq:D}) as well as the formulas of the
light-cone expansion have a well-defined and non-trivial continuum limit.
In particular, using that the particle and anti-particle states are to be
normalized in accordance with the states of the sea, we can specify
the normalization constant~$c_{\mbox{\scriptsize{norm}}}$ in~(\ref{1g}).
If the wave functions~$\Psi_k$ and~$\Phi_l$ in~(\ref{1g}) are
normalized according to~(\ref{f:0d}), we must choose
\beq c_{\mbox{\scriptsize{norm}}} \;=\; - \frac{1}{2 \pi}\:.
\label{cnorm}
\eeq

We finally remark that Theorem~\ref{thm2} can be generalized in a
straightforward way to include a gravitational field, if~(\ref{eq:C}) is
replaced by~(\ref{nicsp}),
with ${\mathcal{H}}$ a space-like hypersurface with future-directed
normal~$\nu$. However, we need to assume that the gravitational field decays at
infinity in the sense that space-time is asymptotically flat and for
$t \to \pm\infty$ goes over asymptotically to Minkowski space. In particular,
realistic cosmological models like the Friedman-Robertson-Walker space-times
are excluded. We do not expect that the large-scale structure of space-time
should have an influence on the normalization\footnote{{\textsf{Online version}:}
This picture has been confirmed by the paper~arXiv:0901.0602 [math-ph].}, but
this is an open problem which remains to be investigated.

% The Principle of the Fermionic Projector
\chapter{The Principle of the Fermionic Projector}
\label{secpfp} \setcounter{equation}{0}
In this chapter we introduce a new mathematical framework intended
for the formulation of physical theories.
We first generalize the notions of relativistic quantum
mechanics and classical field theory in several construction
steps. This will be done in a very intuitive way. The aim is to work out the
essence of the underlying physical principles by dropping
all additional and less important structures. This will lead us to a quite
abstract mathematical framework. The ``principle of the fermionic projector''
states that the fundamental physical equations should be formulated within
this framework. We conclude this chapter with a brief physical overview and discussion.

\section{Connection between Local Gauge Freedom and the\\
Measurability of Position and Time} \label{psec11}
\setcounter{equation}{0}
\markboth{3. THE PRINCIPLE OF THE FERMIONIC PROJECTOR}
{3.1. LOCAL GAUGE FREEDOM $\longleftrightarrow$ MEASURABILITY OF POSITION
AND TIME}
In this section we give a possible explanation as to why local gauge
freedom occurs in nature. This physical consideration will provide a
formalism which will be the starting
point for the constructions leading to the principle of the fermionic projector.
We begin for simplicity with the example of the~$U(1)$ gauge
transformations of the magnetic field for a Schr{\"o}dinger wave function
$\Psi$ in nonrelativistic quantum mechanics. Since it will be
sufficient to consider the situation for fixed time, we only write out
the spatial dependence of the wave function,
 $\Psi = \Psi(\vec{x})$ with $\vec{x} \in \R^3$.
In the nonrelativistic and static limit, the gauge freedom of
electrodynamics~(\ref{gauge}, \ref{phase}) reduces to the transformations
\begin{eqnarray}
\vec{A}(\vec{x}) &\longrightarrow& \vec{A}(\vec{x})
+ \vec{\nabla} \Lambda(\vec{x}) \label{p:5} \\
\Psi(\vec{x}) &\longrightarrow& e^{i \Lambda(\vec{x})}\: \Psi(\vec{x})\:,
\label{p:1}
\end{eqnarray}
where the so-called vector potential~$\vec{A}$ consists of the
three spatial components of the electromagnetic potential~$A$.
Similar to~(\ref{gcd}), we introduce the gauge-covariant
derivative\index{gauge-covariant derivative} by
\beq \label{nrgcd}
\vec{D} \;=\; \vec{\nabla} - i \vec{A}\:.
\eeq
With the transformation~(\ref{p:1}) we can arbitrarily change the phase of the wave
function~$\Psi$ at any point~$\vec{x}$. This is consistent with the quantum
mechanical interpretation of the wave function, according to which the phase
of a wave function is not an observable quantity, only its absolute square
$|\Psi(\vec{x})|^2$ has a physical meaning as the probability density of the
particle. One can even go one step further and take the point of view
that the inability to determine the local phase of a quantum mechanical
wave function is the physical reason for the local gauge freedom
(\ref{p:5}, \ref{p:1}). Then the $U(1)$ gauge transformations of the
magnetic field become a consequence of the principles of quantum mechanics.
This argument becomes clearer when stated in more mathematical terms:
We consider on the Schr{\"o}dinger wave functions the usual scalar product
\[ \bra \Psi \:|\: \Phi \ket \;=\; \int_{\sR^3}
\overline{\Psi(\vec{x})}\: \Phi(\vec{x}) \: d\vec{x} \]
and denote the corresponding Hilbert space by~$H$. On~$H$, the
position operators $\vec{X}$ are given as the multiplication operators
with the coordinate functions,
\[ \vec{X} \:\Psi(\vec{x}) \;=\; \vec{x} \:\Psi(\vec{x}) \:. \]
As it is common in quantum mechanics, we consider $H$ as an abstract
Hilbert space (i.e.\ we forget about the fact that $H$ was introduced
as a space of functions). Then the wave function $\Psi(\vec{x})$
corresponding to a vector $\Psi \in H$ is obtained by constructing a
{\em{position representation}} of the Hilbert space. In bra/ket
notation, this is done by choosing an ``eigenvector
basis'' $|\vec{x} \ket$ of the position operators,
\begin{equation}
\vec{X} \:|\vec{x} \ket \;=\; \vec{x} \:|\vec{x} \ket \:,\spc
\bra \vec{x} \:|\: \vec{y} \ket \;=\; \delta^3(\vec{x}-\vec{y}) \:,
        \label{p:6}
\end{equation}
and the wave function is then introduced by
\begin{equation}
\Psi(\vec{x}) \;=\; \bra \vec{x} \:|\: \Psi \ket
        \label{p:7}
\end{equation}
(we remark that the formal bra/ket notation can be made
mathematically precise using
spectral measures \cite{F2}).
The important point for us is that the
``eigenvectors'' $|\vec{x} \ket$ of the position operators are
determined only up to a phase. Namely, the transformation
\begin{equation}
|\vec{x} \ket \;\longrightarrow\; \exp \left(-i \Lambda(\vec{x}) \right)
|\vec{x} \ket \label{p:8}
\end{equation}
leaves invariant the conditions (\ref{p:6}) for the ``eigenvector basis.''
If we substitute (\ref{p:8}) into (\ref{p:7}), we obtain precisely the
transformation (\ref{p:1}) of the wave function.
The transformation properties of the gauge-covariant derivative~(\ref{nrgcd})
and of the gauge potentials in~(\ref{p:5}) follow from
(\ref{p:1}) if one assumes that the gauge-covariant derivatives~$\vec{D}$
are operators on~$H$ (and thus do not depend on the representation of
$H$ as functions in position space). In physics, the operators
$\vec{\pi}=-i \vec{D}$ are called the ``canonical momentum operators.''

The relation just described between the position representation of
quantum mechanical states and the~$U(1)$ gauge transformations of the
magnetic field was noticed long ago. However, the idea of
explaining local gauge freedom from quantum mechanical principles was
not regarded as being of general significance. In particular, it was
never extended to the relativistic setting or to more general gauge
groups. The probable reason for this is that these generalizations are not
quite straightforward; they make it necessary to formulate relativistic
quantum mechanics in a particular way as follows.
We again consider on the four-component
Dirac spinors $(\Psi^\alpha(x))_{\alpha=1,\ldots,4}$ in Minkowski
space the spin scalar product~(\ref{f:0c}) and denote the vector
space of all Dirac wave functions by $H$. Integrating the spin
scalar product over space-time, we obtain an indefinite scalar product
on $H$,
\begin{equation}
    \bra \Psi \:|\: \Phi \ket \;=\; \int_{\sR^4} \Sl \Psi \:|\: \Phi
    \Sr(x) \: d^4x \:.
    \label{p:18a}
\end{equation}
Furthermore, we introduce on $H$ time/position operators
$(X^i)_{i=0,\ldots,3}$ by multiplication with the coordinate
functions, \label{X^i}
\[ X^i \:\Psi(x) \;=\; x^i \:\Psi(x) \:. \]
We now consider $(H,\: \bra . | . \ket)$ as an abstract scalar product
space. In order to construct a {\em{time/position representation}} of $H$, we must
choose an ``eigenvector basis'' of the time/position operators. Since the
wave functions have four components, an ``eigenvector basis'' has in
bra/ket notation the form $|x \alpha \ket$, $x \in \R^4$, $\alpha=1,\ldots,4$;
it is characterized by the conditions \label{vertxalpha}
\begin{equation}
X^i \:|x \alpha \ket \;=\; x^i \:|x \alpha \ket \:,\spc
\bra x \alpha \:|\: y \beta \ket \;=\; s_\alpha \:\delta_{\alpha
\beta}\:\delta^4(x-y) \label{p:10}
\end{equation}
with $s_\alpha$ as in (\ref{f:0c}).
The wave function corresponding to a vector $\Psi \in H$ is defined by
\begin{equation}
\Psi^\alpha(x) \;=\; \bra x \alpha \:|\: \Psi \ket \:.
        \label{p:11}
\end{equation}
The conditions~(\ref{p:10}) determine the basis~$|x \alpha \ket$
only up to local isometries of a scalar product of signature $(2,2)$, i.e.\
up to transformations of the form
\begin{equation}
|x \alpha \ket \;\longrightarrow\; \sum_{\beta=1}^4 (U(x)^{-1})^\alpha_\beta
\:|x \beta \ket \qquad {\mbox{with}} \qquad U(x) \in
U(2,2) \:.
        \label{p:12}
\end{equation}
If we identify these transformations with gauge transformations
and substitute into (\ref{p:11}), we obtain local gauge freedom of
the form
\beq \label{gtr}
\Psi(x) \;\longrightarrow\; U(x)\: \Psi(x)\:.
\eeq
Since gauge transformations correspond to changes of the
``eigenvector basis'' $|x \alpha \ket$, we also refer to~$|x \alpha \ket$ as a {\em{gauge}}\index{gauge}.

From the mathematical point of view, (\ref{p:10}--\ref{p:12}) is
a straightforward generalization of (\ref{p:6}--\ref{p:8}) to the
four-dimensional setting and four-component wave functions, taking
into account that the spin scalar product has signature $(2,2)$.
However, our construction departs from the
usual description of physics, because the time operator~$X^0$ is not commonly
used in relativistic quantum mechanics and because
the scalar product~(\ref{p:18a}) is unconventional.
In particular, one might object that the scalar product~(\ref{p:18a})
may be infinite for physical states,
because the time integral diverges. However, this is not a serious
problem, which could be removed for example by considering the system in
finite 4-volume and taking a suitable infinite volume limit.
Furthermore, one should keep in mind that the scalar product~(\ref{p:18a})
gives us the spin scalar product, and using the spin scalar product
one can introduce the usual positive scalar product~$(.|.)$ by
integrating over a space-like hypersurface (see~(\ref{f:0e}) or
more generally~(\ref{nicsp})). Therefore, it causes no principal
problems to consider~$\bra .|. \ket$ instead
of~$(.|.)$ as the fundamental scalar product.
We conclude that~(\ref{p:18a}--\ref{p:12}) is certainly an
unconventional point of view, but it is nevertheless consistent and
indeed mathematically equivalent to the usual description of
relativistic quantum mechanics as outlined in~{\S}\ref{isec2}.

The above explanation of local gauge freedom
fits together nicely with our description of Dirac spinors
in the gravitational field in~{\S}\ref{isec5}: We
let~$(H, \bra .|. \ket)$ be the vector space of wave functions
on a manifold~$M$, endowed with the indefinite scalar
product~(\ref{defsp}). For every coordinate system~$x^i$ we
introduce the corresponding multiplication operators
$X^i$. Considering~$H$ as an abstract vector space, the
arbitrariness of the time/position representation of~$H$
again yields the local~$U(2,2)$ gauge freedom~(\ref{gtr}).
We thus obtain precisely the gauge transformations~(\ref{psigauge}).
In this way, (\ref{p:10}--\ref{p:12}) is not only consistent with
all the constructions in~{\S}\ref{isec5}, but it also gives a simple
explanation for the gauge group~$U(2,2)$.

The~$U(2,2)$ gauge symmetry describes gravitation and
electrodynamics, but it does not include the weak and strong interactions.
In order to obtain additional gauge freedom, we must extend the
gauge group. Since our gauge group
is the isometry group of the spin scalar product, this
can be accomplished only by increasing the number of components of the
wave functions. In general, one can take wave functions with $p+q$
components and a spin scalar product\index{spin scalar product} of
signature $(p,q)$, \label{SlPsi|PhiSr}
\begin{eqnarray}
\Sl \Psi \:|\: \Phi \Sr(x) &=& \sum_{\alpha=1}^{p+q}
    s_\alpha \;\Psi^\alpha(x)^* \:\Phi^\alpha(x) \spc {\mbox{with}}
    \nonumber \\
&&\hspace*{-3cm} s_1=\cdots=s_p=1 \:,\qquad
s_{p+1}=\cdots=s_{p+q}=-1\:. \label{p:14}
\end{eqnarray}
We call $(p,q)$ the {\em{spin dimension}}\index{spin dimension}. Repeating the above
construction (\ref{p:18a}--\ref{p:11}) for this spin scalar
product yields local gauge freedom with gauge group~$U(p,q)$.
Unfortunately, it is not possible to introduce
the Dirac operator in this generality. Therefore, we will always
assume that the spin dimension is $(2N, 2N)$ with
$N \geq 1$. In this case, one can regard the $4N$ component wave
functions as the direct sum of $N$ Dirac spinors, exactly as we
did in the general definition of the fermionic projector~(\ref{vs1}).
Then our above argument yields the gauge group~$U(2N, 2N)$. The interaction
can be described for example as in~{\S}\ref{jsec3} by inserting a
multiplication operator~${\mathcal{B}}$ into the Dirac
operator~(\ref{b}) and taking the partial trace~(\ref{part}).
More generally, one can modify the first order terms of the Dirac
operator in analogy to Def.~\ref{def1n}. Our concept is that the~$U(2n,2n)$
gauge symmetry should be related to corresponding gauge potentials
in the Dirac operator, and that this should, in the correct model,
give rise to the gravitational, strong and electroweak gauge fields.

For clarity, we finally point out the differences of our approach to standard
gauge theories. Usually, the gauge groups (e.g.\ the $SU(2)_{\mbox{w}}$
or $SU(3)_{\mbox{s}}$ in the standard model) act on separate indices of the
wave functions (called the isospin and color indices, respectively).
In contrast to this, our $U(2,2)$ gauge transformations simply act
on the spinor index. In our generalization to
higher spin dimension (\ref{p:14}), we make no distinction between
the spinor index and the index of the gauge fields; they are both
combined in one index $\alpha=1,\ldots,4N$.
In our setting, the gauge group and the coupling of
the gauge fields to the Dirac particles are
completely determined by the spin dimension. Compared to standard gauge
theories, where the gauge groups and their couplings can be chosen arbitrarily,
this is a strong restriction for the formulation of physical models.

\section{Projection on Fermionic States} \label{psec12}
\setcounter{equation}{0}
The fermionic projector was introduced in Chapter~\ref{sec3}
in order to resolve the external field problem, and we used it
to describe a general many-fermion system~(\ref{1g}).
We now discuss the concept of working with a ``projector'' in
a more general context. A single Dirac particle is clearly
described by its wave function $\Psi^\alpha(x) =
\bra x \alpha | \Psi \ket$, or, in a gauge-independent way, by a
vector $\Psi \in H$.
Since the phase and normalization of $\Psi$ have no physical significance, we
prefer to describe the Dirac particle by the one-dimensional
subspace $\bra \Psi \ket \equiv \{ \lambda \Psi,\: \lambda \in \C\} \subset H$.
Now consider the system of $n$ Dirac particles, which occupy the
one-particle states $\Psi_1,\ldots,\Psi_n \in H$. Generalizing the
subspace $\bra \Psi \ket$ of the one-particle system, it seems natural
to describe the many-particle system by the
subspace $\bra \Psi_1,\ldots,\Psi_n \ket \subset H$ spanned by $\Psi_1,
\ldots, \Psi_n$. We consider for simplicity only the generic case
that this subspace is non-degenerate (i.e.\ there should
be no vectors $0 \neq \Psi \in Y$ with $\bra \Psi | \Phi \ket = 0$
for all $\Phi \in Y$). Just as in positive
definite scalar product spaces, every non-degenerate
subspace $Y \subset H$ uniquely determines a projector
$P_Y$ on this subspace, characterized by the conditions
$P_Y^* = P_Y = P_Y^2$ and ${\mbox{Im}}(P_Y)= Y$, where the star denotes the
adjoint with respect to the scalar product $\bra .|. \ket$. Instead of working with the subspace $\bra \Psi_1,\ldots,\Psi_n \ket \subset H$,
it is more convenient for us to consider the corresponding projector $P$,
\[ P \;=\; P_{\bra \Psi_1,\ldots,\Psi_n \ket} \:. \]
We call $P$ the {\em{fermionic projector}}\index{fermionic projector}. In this work we will always
describe the Dirac particles of the system by a fermionic projector.

The concept of the fermionic projector departs from the usual description
of a many-particle state by a vector of the
fermionic Fock space (as introduced in~{\S}\ref{isec4}).
Let us discuss this difference in detail.
In many-particle quantum mechanics, the system of Dirac particles
$\Psi_1, \ldots, \Psi_n$ is described by the anti-symmetric product
wave function
\beq
\Psi \;=\; \Psi_1 \land \cdots \land \Psi_n \:. \label{p:1_16}
\eeq
The wave functions of the form
(\ref{p:1_16}) are called $n$-particle Hartree-Fock
states. They span the $n$-particle Fock space $F^n = \bigwedge^n H$.
In the fermionic Fock space formalism, a quantum state is a
linear combination of Hartree-Fock states\index{Hartree-Fock state}, i.e.\ a vector of
the Fock space $F=\bigoplus_{n=0}^\infty F^n$ (see~{\S}\ref{isec4} for
details). In order to connect the fermionic projector with the
Fock space formalism, we can associate to a projector $P_Y$ on a subspace
$Y=\bra \Psi_1, \ldots, \Psi_n \ket \subset H$ the
wave function (\ref{p:1_16}). This mapping clearly depends on the choice
of the basis of~$Y$. More precisely, choosing another basis
$\Phi_i = \sum_{j=1}^n \kappa^j_i \Psi_j$, we have
\[ \Phi_1 \land \cdots \land \Phi_n \;=\;
\det (\kappa) \; \Psi_1 \land \cdots \land \Psi_n \:. \]
This shows that, due to the anti-symmetrization, the mapping is unique
up to a complex factor. Therefore, with the mapping
\[ P_{\bra \Psi_1,\ldots,\Psi_n \ket} \;\to\;
\bra \Psi_1 \land \cdots \land \Psi_n \ket \subset F^n \]
we can associate to every projector a unique one-dimensional subspace
of the Fock space. Since the image of this mapping is always a
Hartree-Fock state, we obtain a one-to-one correspondence between the
projectors $P_Y$ on $n$-dimensional subspaces $Y \subset H$ and
$n$-particle Hartree-Fock states. In this way, one sees that
the description of a many-particle state with the fermionic projector is
equivalent to using a Hartree-Fock state. With this correspondence, the
formalism of the fermionic projector becomes
a special case of the Fock space formalism, obtained by restricting to
Hartree-Fock states. In particular, we conclude that the physical concepts
behind fermionic Fock spaces, namely the Pauli Exclusion Principle and
the fact that quantum mechanical particles are indistinguishable
(see page~\pageref{f:0h1}),
are also respected by the fermionic projector. However, we point
out that the the fermionic projector is not mathematically
equivalent to a state of the Fock space, because a vector of the
Fock space can in general be represented only by a linear
combination of Hartree-Fock states.

Let us analyze what this mathematical difference means physically.
If nature is described by a fermionic projector, the joint wave
function of all fermions of the Universe must be a Hartree-Fock
state. However, this condition cannot be immediately verified in
experiments, because measurements can never take into account all
existing fermions. In all realistic situations, one must restrict
the observations to a small subsystem of the Universe. As is
worked out in Appendix~\ref{pappA}, the effective wave function of
a subsystem need {\em{not}} be a Hartree-Fock state; it
corresponds to an arbitrary vector of the Fock space of the
subsystem, assuming that the number of particles of the whole
system is sufficiently large. From this we conclude that the
description of the many-particle system with the fermionic
projector is indeed physically equivalent to the Fock space
formalism. For theoretical considerations, it must be taken into
account that the fermionic projector merely corresponds to a
Hartree-Fock state; for all practical purposes, however, one can
just as well work with the whole Fock space.

We saw after (\ref{p:1_16}) that the description of a many-particle state
with the fermionic projector implies the Pauli Exclusion Principle.
This can also be understood directly in a non-technical
way as follows. For a given state $\Psi \in H$, we can
form the projector $P_{\bra \Psi \ket}$ describing the
one-particle state, but there is no projector which would correspond
to a two-particle state (notice that the naive generalization $2 P_{\bra \Psi
\ket}$ is not a projector). More generally, every vector $\Psi \in H$
either lies in the image of $P$, $\Psi \in P(H)$, or it does not.
Via these two conditions, the fermionic projector encodes for every state
$\Psi \in H$ the occupation numbers $1$ and $0$, respectively, but it is
impossible to describe higher occupation numbers. In this way, the
fermionic projector naturally incorporates the Pauli Exclusion Principle
in its formulation that each quantum mechanical state may be occupied by at
most one fermion.

As explained at the end of~{\S}\ref{jsec5}, the fermionic projector
contains all the information about the physical system in the sense
that from a given fermionic projector one can uniquely reconstruct
the fermionic states as well as the Dirac operator with interaction.
Therefore, it is consistent to consider
the fermionic projector as the basic object in space-time and to
regard the Dirac operator merely as an auxiliary object which is
useful in describing the interaction of the fermions via classical
fields.

\section{Discretization of Space-Time}
\label{psec13}  \setcounter{equation}{0}
The ultraviolet divergences of perturbative QFT
indicate that the current description of physics should break down
at very small distances. It is generally believed that the length
scale where yet unknown physical effects should become important
is given by the Planck length. Here we will assume that space-time
consists on the Planck scale of discrete space-time points. The
simplest way to discretize space-time would be to replace the
space-time continuum by a four-dimensional lattice (as it is e.g.\
done in lattice gauge theories). In the following construction, we
go much further and discretize space-time in a way where
notions like ``lattice spacing'' and ``neighboring lattice
points'' are given up. On the other hand, we will retain the
principles of general relativity and our local gauge freedom.

We first consider the situation in a given coordinate system $x^i$ in
space-time\footnote{We assume for simplicity that the chart $x^i$ describes
all space-time. The generalization to a non-trivial space-time topology
is done in a straightforward way by gluing together different charts;
for details see~\cite{F2}.}. For the
discretization we replace the time/position operators $X^i$\label{X^i2} by mutually
commuting operators with a {\em{purely discrete spectrum}}.
We take the joint spectrum of these operators, i.e.\ the set \label{M}
\[ M \;=\; \{ x \in \R^4 \;|\; {\mbox{there is $u \in H$ with
$X^i u = x^i u$ for all $i=0,\ldots,3$}} \} \:, \]
as our discrete space-time points. We assume that the joint eigenspaces
$e_x$ of the $X^i$,
\[  e_x \;=\; \{ u \;|\; X^i u = x^i u {\mbox{ for all
$i=0,\ldots,3$}} \} \:,\spc x \in M \:, \]
are $4N$-dimensional subspaces of $H$, on which the scalar product
$\bra . | . \ket$ has the signature $(2N,2N)$. Then we can choose a
basis $|x \alpha \ket$, $x \in M$, $\alpha=1,\ldots, 4N$ satisfying
\begin{eqnarray}
X^i \: |x \alpha \ket &=& x^i \: |x \alpha \ket \:,\spc
        \bra x \alpha \:|\: y \beta \ket = s_\alpha\: \delta_{\alpha \beta}
        \: \delta_{xy} \spc {\mbox{with}} \nonumber \\
&&\hspace*{-2.2cm}
s_1=\cdots=s_{2N}=1 \:,\;\;\;\;\; s_{2N+1}=\cdots=s_{4N}=-1 \: .
\label{p:19}
\end{eqnarray}
These relations differ from (\ref{p:10}) only by the replacement
$\delta^4(x-y) \to \delta_{xy}$. It is useful to
introduce the projectors $E_x$ on the eigenspaces $e_x$ by \label{E_x}
\begin{equation}
E_x \;=\; \sum_{\alpha=1}^{p+q} s_\alpha \: |x \alpha \ket \bra x \alpha |
\: ; \label{p:1_8a}
\end{equation}
they satisfy the relations
\begin{eqnarray}
X^i \:E_x &=& x^i \:E_x \spc {\mbox{and}} \label{p:18c} \\
E_x^* &=& E_x \:,\spc
E_x \:E_y \;=\; \delta_{xy} \:E_x \:,\spc
\sum_{x \in M} E_x \;=\; \1 \:, \label{p:18d}
\end{eqnarray}
where the star denotes the adjoint with respect to the scalar product
$\bra . | . \ket$ (these relations immediately follow from (\ref{p:19}) and
the fact that $|x \alpha \ket$ is a basis).
The operators $E_x$ are independent of the
choice of the basis $|x \alpha \ket$, they are uniquely characterized
by (\ref{p:18c}, \ref{p:18d}) as the spectral projectors of the
operators $X^i$.

If we change the coordinate system to $\tilde{x}^i = \tilde{x}^i(x)$, the
discrete space-time points $M \subset \R^4$ are mapped to different
points in $\R^4$, more precisely
\begin{equation}
    \tilde{M}=\tilde{x}(M) \:,\spc \tilde{E}_{\tilde{x}(x)} = E_x \:.
    \label{p:19a}
\end{equation}
With such coordinate transformations, the relative position of the discrete
space-time points in~$\R^4$ can be arbitrarily changed. Taking general
coordinate invariance seriously on the Planck scale, this is
consistent only if we forget about the fact
that $M$ and $\tilde{M}$ are subsets of $\R^4$ and consider them merely
as index sets for the spectral projectors. In other words, we give up
the ordering of the discrete space-time points, which is
inherited from the ambient vector space $\R^4$, and consider $M$ and $\tilde{M}$
only as point sets.
After this generalization, we can identify $M$ with $\tilde{M}$ (via
the equivalence relation $\tilde{x}(x) \simeq x$). According to
(\ref{p:19a}), the spectral projectors
$(E_p)_{p \in M}$ are then independent of the choice of coordinates.

We regard the projectors $(E_p)_{p \in M}$ as the basic objects describing
space-time. The time/position operators can be deduced from them.
Namely, every coordinate system yields an injection of the
discrete space-time points
\begin{equation}
x \;:\; M \:\hookrightarrow\: \R^4 \:,
\label{p:19c}
\end{equation}
and the corresponding time/position operators $X^i$ can be written as
\begin{equation}
X^i \;=\; \sum_{p \in M} x^i(p) \: E_p \:.
\label{p:19d}
\end{equation}
Since every injection of the discrete space-time points into $\R^4$
can be realized by a suitable choice of coordinates (i.e.\ for every
injection $\iota : M \hookrightarrow \R^4$ there is a chart $x^i$
such that $x(M) = \iota(M)$), we can drop the
condition that $x$ is induced by a coordinate system. We can thus take for
$x$ in (\ref{p:19c}, \ref{p:19d}) any embedding of $M$ into $\R^4$.

Let us summarize the result of our construction. We shall describe
space-time by an indefinite scalar product space
$(H, \:\bra . | . \ket)$ and projectors $(E_p)_{p \in M}$ on
$H$, where $M$ is a (finite or countable) index set.
The projectors $E_p$ are characterized by the conditions (\ref{p:18d}).
Furthermore, we assume that the {\em{spin dimension}} is $(2N,2N)$, i.e.\
$E_p(H) \subset H$ is for all $p \in M$ a subspace of signature $(2N,2N)$.
We call $(H, \:\bra . | . \ket,\: (E_p)_{p \in M})$
{\em{discrete space-time}}\index{discrete space-time}. The equivalence principle is taken into
account via the freedom in choosing the embeddings
(\ref{p:19c}, \ref{p:19d})
of the discrete space-time points. Moreover, one can choose a basis
$|p \alpha \ket$, $p \in M$, $\alpha=1,\ldots, 4N$, of $H$ satisfying
the conditions
\[ E_p \:|q \alpha \ket \;=\; \delta_{pq} \:|p \alpha \ket \:,\spc
\bra p \alpha \:|\: q \beta \ket \;=\; s_\alpha \:\delta_{\alpha
\beta} \:\delta_{pq} \]
with $s_\alpha$ as in (\ref{p:19});
such a basis is called a {\em{gauge}}\index{gauge}. It is determined only
up to transformations of the form
\begin{equation}
|p \alpha \ket \;\to\; \sum_{\beta=1}^{2N} (U(p)^{-1})^\alpha_\beta
\;|p \beta \ket \;\;\;\;{\mbox{with}}\;\;\;\; U(p) \in
U(2N,2N) \: . \label{p:36}
\end{equation}
These are the local gauge transformations of discrete space-time.

\section{The Principle of the Fermionic Projector}
\setcounter{equation}{0}
For the complete description of a physical system we must
introduce additional objects in discrete space-time $(H,\: \bra
.|. \ket,\: (E_p)_{p \in M})$. As mentioned at the end of {\S}\ref{psec12},
one can in the space-time continuum regard the
fermionic projector as the basic physical object.
Therefore, it seems promising to carry over the fermionic
projector to discrete space-time. We introduce the {\em{fermionic
projector of discrete space-time}}\index{fermionic projector!of
discrete space-time} $P$ as a projector acting on
the vector space $H$ of discrete space-time.

In analogy to the situation for the continuum, we expect that a physical
system can be completely characterized by a fermionic projector in discrete
space-time. At this stage, however, it is not at all clear whether this
description makes any physical sense. In particular, it seems problematic
that neither the Dirac equation nor the classical field equations can
be formulated in or extended to discrete space-time; thus it becomes
necessary to replace them by equations of completely different type.
We take it as an ad-hoc postulate that this can actually be done;
namely we assert
\begin{quote}
\centerline{\em{The Principle of the Fermionic Projector:}}\index{principle
of the fermionic projector}

A physical system is completely
described by the fermionic projector in discrete space-time.
The physical equations should be formulated exclusively
with the fermionic projector in discrete space-time, i.e.\ they must be
stated in terms of the operators $P$ and $(E_p)_{p \in M}$ on $H$.
\end{quote}
Clearly, the validity and consequences of this postulate still need to
be investigated; this is precisely the aim of the present work.
The physical equations formulated with $P$ and $(E_p)_{p \in M}$ are
called the {\em{equations of discrete space-time}}\index{equations of
  discrete space-time}.

\section{A Variational Principle}
\label{psec15}  \setcounter{equation}{0}
Before coming to the general discussion of the
principle of the fermionic projector, we want to give
an example of a variational principle in discrete space-time. This
is done to give the reader an idea of how one can formulate
equations in discrete space-time. This example will serve as our
model variational principle, and we will often come back to it.
A more detailed motivation of our Lagrangian is given in
Chapter~\ref{esec2}.

Let us first discuss the general mathematical form of possible equations
in discrete space-time. The operators $P$ and $(E_p)_{p \in M}$ all have
a very simple structure in that they are projectors acting on $H$.
Therefore, it is not worth studying these operators separately; for
physically promising equations, we must combine the projectors $P$ and $(E_p)_{p
\in M}$ in a mathematically interesting way. Composite expressions in these
operators can be manipulated using the idempotence of
$P$ and the relations (\ref{p:18d}) between the projectors $(E_p)_{p
\in M}$: First of all, the identities $\sum_{p \in M} E_p = \1$ and
$E_p^2=E_p$ allow us to insert factors $E_p$ into the formulas; e.g.
\[ E_x \:P\: \Psi \;=\; E_x \:P \left(\sum_{y \in M} E_y \right) \Psi
\;=\; \sum_{y \in M} (E_x \:P\: E_y) \:E_y\:\Psi \:. \]
Writing \label{P(x,y)}
\[ P(x,y) \;\equiv\; E_x \:P\: E_y \:, \]
we obtain the identity
\[ E_x \:(P \:\Psi) \;=\; \sum_{y \in M} P(x,y) \;E_y \:\Psi \:. \]
This representation of $P$ by a sum over the discrete space-time
points resembles the integral representation of an operator in the
continuum with an integral kernel. Therefore, we call $P(x,y)$ the
{\em{discrete kernel}}\index{discrete kernel} of the fermionic projector. The discrete kernel can
be regarded as a canonical representation of the fermionic projector of
discrete space-time, induced by the projectors $(E_p)_{p \in M}$.
Now consider a general product of the operators $P$ and
$(E_p)_{p \in M}$. Using the relations $P^2=P$ and $E_x \:E_y =
\delta_{xy} \:E_x$, every operator product can be simplified to one
with alternating factors $P$ and $E_p$, i.e.\ to an operator product
of the form
\begin{equation}
E_{x_1} \:P\: E_{x_2} \:P\: E_{x_3} \cdots E_{x_{n-1}} \:P\: E_{x_n} \;\;\;\;\;
{\mbox{with $x_j \in M$.}}
    \label{p:33}
\end{equation}
Again using that~$E_p^2 = E_p$, we can rewrite this product
with the discrete kernel as
\begin{equation}
P(x_1, x_2) \:P(x_2, x_3) \:\cdots\: P(x_{n-1}, x_n) \:. \label{p:34}
\end{equation}
We conclude that the equations of discrete space-time should be formed
of products of the discrete kernel, where the second argument of each
factor must coincide with the first argument of the following factor.
We refer to (\ref{p:34}) as a {\em{chain}}\index{chain}.

In analogy to the Lagrangian formulation of classical field theory,
we want to set up a variational principle. Our ``action'' should
be a scalar functional depending on the operators $P$ and $E_p$. Most
scalar functionals on operators (like the trace or the determinant)
can be applied only to endomorphisms (i.e.\ to operators which map
a vector space into itself). The chain (\ref{p:34}) is a mapping from
the subspace $E_{x_n}(H) \subset H$ to $E_{x_1}(H)$. This makes it
difficult to form a scalar, unless $x_1=x_n$. Therefore, we will
only consider {\em{closed chains}}\index{chain!closed}
\[ P(x, y_1)\: P(y_1, y_2) \cdots P(y_k, x) \;:\; E_x(H) \rightarrow
E_x(H) \:. \]
In the simplest case $k=0$, the closed chain
degenerates to a single factor $P(x,x)$. This turns out to be too
simple for the formulation of a physically interesting action,
mainly because the light-cone structure of the fermionic projector
(see~{\S}\ref{jsec5}) would then not enter the variational principle.
Thus we are led to considering closed chains of two factors, i.e.\ to the
operator product $P(x,y)\: P(y,x)$. Suppose that we are given a
real-valued functional ${\mathcal{L}}$ on the endomorphisms of
$E_x(H) \subset H$ (this will be discussed and specified below).
Then ${\mathcal{L}}[P(x,y)\:P(y,x)]$ is a real function depending
on two space-time arguments, and we get a scalar by summing over
$x$ and $y$. Therefore, we take for our action $S$ the ansatz \label{S2}
\begin{equation}
S \;=\; \sum_{x, y \in M} {\mathcal{L}}[P(x,y) \:P(y,x)] \:.
    \label{p:35}
\end{equation}
This ansatz is called a {\em{two-point action}}\index{action!two-point}, and in analogy to
classical field theory we call ${\mathcal{L}}$\label{mathcalL} the corresponding
{\em{Lagrangian}}\index{Lagrangian}.

We shall now introduce a particular Lagrangian ${\mathcal{L}}$.
The requirement which will lead us quite naturally to this
Lagrangian is that ${\mathcal{L}}$ should be {\em{positive}}.
Positivity of the action is desirable because it is a more
convincing concept to look for a local minimum of the action than
merely for a critical point of an action which is unbounded below.

Let us first consider how one can form a positive functional on
$P(x,y)\:P(y,x)$. The closed chain $P(x,y) \:P(y,x)$ is an
endomorphism of $E_x(H)$; we abbreviate it in what follows by $A$\label{A}.
In a given gauge, $A$ is represented by a $4N \times 4N$ matrix.
Under gauge transformations (\ref{p:36}), this matrix transforms
according to the adjoint representation,
\[ A \;\to\; U(x) \:A\: U(x)^{-1} \:. \]
Furthermore, $A$ is Hermitian on $E_x(H)$, i.e.
\begin{equation}
    \bra A\:\Psi \:|\: \Phi \ket \;=\; \bra \Psi \:|\: A \:\Phi \ket
    \spc {\mbox{for $\Psi, \Phi \in E_x(H)$}}\:, \label{p:37}
\end{equation}
or simply $A^*=A$.
In positive definite scalar product spaces, the natural positive
functional on operators is an operator norm, e.g.\ the Hilbert-Schmidt
norm $\|B\|_2=\tr (B^* B)^{\frac{1}{2}}$. In our setting, the situation
is more difficult because our scalar product $\bra .|. \ket$
is indefinite on $E_x(H)$ (of signature $(2N,2N)$).
As a consequence, Hermitian matrices do not have the same nice properties as
in positive definite scalar product spaces; in particular, the
matrix $A$ might have complex eigenvalues, and it is in general not even
diagonalizable. Also, the operator product $A^* A$ need not be positive,
so that we cannot introduce a Hilbert-Schmidt norm.
In order to analyze the situation more systematically, we
decompose the characteristic polynomial of $A$ into linear factors
\begin{equation}
    \det (\lambda-A) \;=\; \prod_{k=1}^K (\lambda - \lambda_k)^{n_k}
    \:. \label{p:55}
\end{equation}
This decomposition is useful because every functional on $A$ can be
expressed in terms of the roots and multiplicities of the characteristic
polynomial; thus it is sufficient to consider the $\lambda_k$'s and
$n_k$'s in what follows. Each root $\lambda_k$ corresponds to an
$n_k$-dimensional $A$-invariant subspace of $E_x(H)$, as one sees
immediately from a Jordan representation of $A$. The roots $\lambda_k$ may be
complex. But since~$A$ is Hermitian~(\ref{p:37}), we know at least
that the characteristic polynomial of $A$ is real,
\[ \overline{ \det (\lambda-A)} \;=\;
\det (\lambda-A) \spc {\mbox{for $\lambda \in \R$.}} \]
This means that the complex conjugate of every root
is again a root with the same multiplicity (i.e.\ for every
$\lambda_k$ there is a $\lambda_l$ with
$\overline{\lambda_k}=\lambda_l$ and $n_k=n_l$).
The reality of the characteristic polynomial is verified in detail as
follows. In a given gauge, we can form the transposed, complex conjugated
matrix of $A$, denoted by $A^\dagger$. For clarity, we point out that
$A^\dagger$ is {\em{not}} an endomorphism of $E_x(H)$, because it has the wrong
behavior under gauge transformations (in particular, the trace $\tr
(A^\dagger A)$ depends on the gauge and is thus ill-defined).
Nevertheless, the matrix $A^\dagger$ is useful because
we can write the adjoint of $A$ in the form $A^* = S A^\dagger S$, where $S$
is the spin signature matrix, $S={\mbox{diag}}((s_\alpha)_{\alpha=1,\ldots,
4N})$. Since $S^2=\1$, and since the determinant is multiplicative,
we conclude that for any real $\lambda$,
\begin{eqnarray*}
\overline{ \det (\lambda-A)} &=& \det (\lambda - A^\dagger)
\;=\; \det (\lambda - S^2 \:A^\dagger) \\
&=& \det (\lambda - S A^\dagger S)
\;=\; \det(\lambda - A^*) \;=\; \det (\lambda-A) \:.
\end{eqnarray*}

It is worth noting that every Lagrangian is symmetric in
the two arguments~$x$ and~$y$, as the following consideration shows.
For any two quadratic matrices~$B$ and~$C$, we choose~$\varepsilon$ not in the spectrum of~$C$ and
set~$C^\varepsilon = C-\varepsilon\1$. Taking the determinant of the
relation $C^\varepsilon (B C^\varepsilon - \lambda) = (C^\varepsilon B - \lambda) C^\varepsilon$,
we can use that the determinant is multiplicative and that~$\det C^\varepsilon \neq 0$
to obtain the equation~$\det(B C^\varepsilon -\lambda) = \det(C^\varepsilon B -\lambda)$. Since
both determinants are continuous in~$\varepsilon$, this equation holds even for
all~$\varepsilon \in \R$, proving the elementary identity
\beq \label{element}
\det(B C -\lambda \1) \;=\; \det(C B -\lambda \1) \:.
\eeq
Applying this identity to the closed chain,
\[ \det(P(x,y)\, P(y,x) - \lambda \1)
\;=\; \det(P(y,x)\, P(x,y) - \lambda \1) \:, \]
we find that the characteristic polynomial of the matrix~$A$
remains the same if the two arguments~$x$ and~$y$ are interchanged.
Hence
\beq \label{symmetry}
{\mathcal{L}}[P(x,y)\, P(y,x)] \;=\; {\mathcal{L}}[P(y,x)\, P(x,y)]\:.
\eeq

An obvious way to form a positive functional is to add up the absolute
values of the roots, taking into account their multiplicities.  We
thus define the {\em{spectral weight}}\index{spectral weight} $|A|$ of
$A$ by \label{|.|}
\begin{equation}
|A| \;=\; \sum_{k=1}^K n_k \:|\lambda_k| \:.
\label{p:38}
\end{equation}
This functional depends continuously on the $\lambda_k$, and also it
behaves continuously when the roots of the characteristic polynomial
degenerate and the multiplicities $n_k$ change. Thus the spectral weight
$|\:.\:|$ is a continuous
functional. Furthermore, the spectral weight is zero if and only if the
characteristic polynomial is trivial, $\det (\lambda-A)=\lambda^{4N}$.
This is equivalent to $A$ being nilpotent (i.e.\ $A^k=0$ for some $k$).
Thus, in contrast to an operator norm, the vanishing of the spectral weight
does not imply that the operator is zero.
On the other hand, it does not seem possible to
define an operator norm in indefinite scalar product spaces, and therefore
we must work instead with the spectral weight.

Using the spectral weight, one can write down many positive
Lagrangians. The simplest choice would be~${\mathcal{L}}[A]=|A|$.
Minimizing the corresponding action (\ref{p:35}) yields a
variational principle which attempts to make the absolute values
of the roots $|\lambda_k|$ as small as possible. This turns out to
be a too strong minimizing principle. It is preferable to
formulate a variational principle which aspires to equalize the
absolute values of all roots. This can be accomplished by
combining the expressions $|A^2|$ and $|A|^2$. Namely, using
that the sum of the multiplicities equals the dimension of the
vector space, $\sum_{k=1}^K n_k = 4 N$, the Schwarz inequality
yields that
\[ |A^2| \;=\; \sum_{k=1}^K n_k \:|\lambda_k|^2 \;\geq\; \frac{1}{4N}
\left( \sum_{k=1}^K n_k \:|\lambda_k| \right)^2 \;=\; \frac{1}{4N}
\:|A|^2 \:, \]
and equality holds only if the absolute values of all roots are equal.
Thus it is reasonable to minimize $|A^2|$, keeping $|A|^2$ fixed. This is our
motivation for considering the two-point action:
\begin{equation}
    {\mbox{minimize }} \;\;\;
    S \;=\; \sum_{x,y \in M} \left| (P(x,y) \:P(y,x))^2 \right|
    \label{p:44}
\end{equation}
under the constraint
\begin{equation}
T \;:=\; \sum_{x,y \in M} |P(x,y) \:P(y,x)|^2 \;=\; {\mbox{const}}\:.
    \label{p:45}
\end{equation}
This is our model variational principle.
\index{Lagrangian!model}

We next consider a stationary point of the action and
derive the corresponding ``Euler-Lagrange equations.'' For
simplicity, we only consider the case that the matrix $P(x,y)\:P(y,x)$ can
be diagonalized. This is the generic
situation; the case of a non-diagonalizable matrix can be obtained
from it by an approximation procedure. Having this in mind, we may
assume that the endomorphism
$A=P(x,y)\:P(y,x)$ has a spectral decomposition\index{spectral decomposition} of the form
\begin{equation}
A \;=\; \sum_{k=1}^K \lambda_k \:F_k \:,
    \label{p:46}
\end{equation}
where $\lambda_k$ are the roots in (\ref{p:55}), and the $F_k$ are
operators mapping onto the corresponding eigenspaces ($A$, $K$,
the $\lambda_k$, and the $F_k$ clearly depend on $x$ and $y$, but we will,
for ease in notation, usually not write out this dependence).
Since the underlying scalar product space is indefinite, the spectral
decomposition (\ref{p:46}) requires a brief explanation. Suppose that we
choose a basis where $A$ is diagonal. In this basis, the operators $F_k$ are
simply the diagonal matrices with diagonal entries $1$ if the corresponding
diagonal elements of $A$ are $\lambda_k$, and $0$ otherwise. Clearly,
these operators map onto the eigenspaces and are orthonormal and complete, i.e.
\[ A \:F_k \;=\; \lambda_k \:F_k \:,\quad
F_k \:F_l \;=\; \delta_{kl} \:F_k \quad {\mbox{and}}\quad
\sum_{k=1}^K F_k \;=\; \1_{E_x(H)} \:. \]
However, the $F_k$ are in general {\em{not}} Hermitian (with respect to
the spin scalar product). More precisely, taking the adjoint swaps the
operators corresponding to complex conjugated eigenvalues,
\begin{equation}
F_k^* \;=\; F_l \spc{\mbox{when}} \spc
\overline{\lambda_k}=\lambda_l \:.
    \label{p:47}
\end{equation}
These relations can be understood immediately because they ensure that the
spectral decomposition (\ref{p:46}) is Hermitian,
\[ \left(\sum_{k=1}^K \lambda_k \:F_k \right)^*
\;=\; \sum_{k=1}^K \overline{\lambda_k} \:F_k^*
\;\stackrel{(\ref{p:47})}{=}\; \sum_{k=1}^K \lambda_k \:F_k \:. \]
Since the eigenvalues are in general complex, we can introduce a new
matrix by taking the complex conjugate of the eigenvalues but leaving
the spectral projectors unchanged,
\beq \label{e:2Abar}
\overline{A} \;=\; \sum_{k=1}^K \overline{\lambda_k} \:F_k
\eeq
We refer to~$\overline{A}$ \label{overlineA} as the {\em{spectral
    adjoint}}\index{spectral adjoint}
of~$A$.

We now consider continuous variations $P(\tau)$ and
$(E_p(\tau))_{p \in M}$, $-\varepsilon < \tau < \varepsilon$,
of our operators.
The structure of the operators must be respected by the variations;
this means that~$P(\tau)$ should be a projector and that the
relations~(\ref{p:18d}) between the operators~$(E_p)_{p \in M}$
should hold for all~$\tau$.
Continuity of the variation implies that the rank of
$P$ and the signature of its image do not
change. This implies that the variation of $P$ can be realized by a
unitary transformation
\begin{equation}
P(\tau) \;=\; U(\tau) \:P\:U(\tau)^{-1} \:, \label{p:50a}
\end{equation}
where~$U(\tau)$ is a unitary operator on~$H$ with~$U(0)=\1$.
Similarly, the variations of the projectors~$(E_p)_{p \in M}$ are
also unitary. From (\ref{p:18d}) we can conclude the stronger statement
that the variations of all operators
$(E_p)_{p \in M}$ can be realized by one unitary transformation, i.e.
\[ E_p(\tau) \;=\; V(\tau) \:E_p\: V(\tau)^{-1} \]
with a unitary operator $V(\tau)$ and $V(0)=\1$.
Since our action is invariant under unitary transformations of the
vector space $H$, we
can, instead of unitarily transforming both $P$ and $(E_p)_{p \in M}$,
just as well keep the $(E_p)_{p \in M}$ fixed and {\em{vary only the
fermionic projector}} by (\ref{p:50a}).
To first order in $\tau$, this variation becomes
\begin{equation}
\delta P \;\equiv\; \frac{d}{d\tau} P(\tau)_{|\tau=0} \;=\; i \:[B,\:P]
\:, \label{p:50}
\end{equation}
where~$B = -i U^\prime(0)$ is a Hermitian operator on~$H$. We
will only consider variations where~$B$ has {\em{finite
support}} \index{variation of finite support},
i.e.\ where the kernel~$B(x,y) \equiv E_x \:B \:E_y$ of~$B$ satisfies the
condition
\[ B(x,y) \;=\; 0 \qquad {\mbox{except for
$x,y \in N \subset M$ with $\#N$ finite.}} \]
This condition can be regarded as the analogue of
the assumption in the classical calculus
of variations that the variation should have compact support.

Let us compute the variation of the action (\ref{p:44}) (the
constraint (\ref{p:45}) will be considered afterwards).
Writing out the action with the eigenvalues $\lambda_k$ and
multiplicities $n_k$, we obtain
\[ S \;=\; \sum_{x,y \in M} \:\sum_{k=1}^K n_k
\:|\lambda_k|^2 \: . \]
The variation can be computed in perturbation theory to first order, \label{tr}
\begin{eqnarray*}
\delta S &=& 2 \:{\mbox{Re}} \sum_{x,y \in M} \:\sum_{k=1}^K
\overline{\lambda_k} \:\tr (F_k \:\delta A) \\
&=& 2 \:{\mbox{Re}} \sum_{x,y \in M} \:\sum_{k=1}^K
\overline{\lambda_k} \:\tr \left( F_k \:(\delta
P(x,y) \:P(y,x) \:+\: P(x,y) \:\delta P(y,x)) \right) ,
\end{eqnarray*}
where ``tr'' denotes the trace in the vector space $H$.
Exchanging the names of $x$ and $y$ in the first summand in the trace
and using cyclicity of the trace, this expression can be written as
an operator product,
\begin{equation}
\delta S \;=\; 2\: {\mbox{Re }} \tr (Q_1 \:\delta P) \:,
    \label{p:48}
\end{equation}
where the kernel $Q_1(x,y) \equiv E_x \:Q_1\: E_y$ of $Q_1$ has the form
\beq
Q_1(x,y) \;=\; \left[ \sum_{k=1}^K \overline{\lambda_k}
\:F_k \right]_{xy} \!\! P(x,y) \:+\: P(x,y)
\left[ \sum_{k=1}^K \overline{\lambda_k} \:F_k \right]_{yx} ,
    \label{p:49prel}
\eeq
and the subscripts ``$_{xy}$'' and ``$_{yx}$'' indicate that the
corresponding brackets contain the spectral decomposition of the
operators~$P(x,y)\: P(y,x)$ and~$P(y,x)\: P(x,y)$, respectively.
Note that the trace in~(\ref{p:48}) is well-defined because the
trace is actually taken only over a finite-dimensional subspace of~$H$.
At this point the following lemma is useful.
\begin{Lemma} \label{lemmaFBBF}
Let~$B$ and~$C$ be two matrices and assume that their
products~$A := BC$ and~$\tilde{A} := CB$ are both diagonalizable.
Then they have the same eigenvalues $\lambda_1, \ldots, \lambda_K$
with the same multiplicities~$n_1, \ldots, n_K$. The corresponding
spectral projectors~$F_k$ and~$\tilde{F}_k$ satisfy the relations
\beq \label{FBBF}
F_k\:B \;=\; B\: \tilde{F}_k \:.
\eeq
\end{Lemma}
{\Proof} It immediately follows from~(\ref{element}) that the
matrices~$A$ and~$\tilde{A}$ have the same eigenvalues with
the same multiplicities. For any~$\lambda$ not in the spectrum of~$A$,
we multiply the identity $B(CB-\lambda) = (BC-\lambda)B$ from the left
and right by~$(A-\lambda)^{-1}$ and~$(\tilde{A}-\lambda)^{-1}$, respectively.
This gives
\[ (A-\lambda)^{-1}\, B \;=\; B \, (\tilde{A}-\lambda)^{-1}\:. \]
Integrating~$\lambda$ over a contour around
any of the eigenvalues~$\lambda_k$ and using the Cauchy integral formulas
\[ F_k \;=\; -\frac{1}{2 \pi i} \oint_{\partial B_\epsilon(\lambda_k)}
(A-\lambda)^{-1}\: d\lambda \:, \qquad
\tilde{F}_k \;=\; -\frac{1}{2 \pi i} \oint_{\partial B_\epsilon(\lambda_k)}
(\tilde{A}-\lambda)^{-1}\: d\lambda \:, \]
we obtain~(\ref{FBBF}).
\QED
This lemma allows us to simplify~(\ref{p:49prel}) to
\beq
Q_1(x,y) \;=\; 2 \left[ \sum_{k=1}^K \overline{\lambda_k}
\:F_k \right]_{xy} \!\! P(x,y) \:.    \label{p:49}
\eeq
A short straightforward computation using~(\ref{p:47}) and
Lemma~\ref{FBBF} shows that
the operator~$Q_1$ is Hermitian. Thus the trace in
(\ref{p:48}) is real, and we conclude that
\[ \delta S \;=\; 2\: \tr (Q_1\:\delta P) \;\;\; . \]

The variation of our constraint (\ref{p:45}) can be computed
similarly, and one gets
\begin{eqnarray*}
\delta T &=& 2 \: \tr (Q_2 \:\delta P) \spc {\mbox{with}} \\
Q_2(x,y) &=& 2 \left[ \Big(\sum_{l=1}^K n_l \:|\lambda_l| \Big)
\sum_{k=1}^K \frac{\overline{\lambda_k}}{|\lambda_k|}
\:F_k \right]_{xy} \!\!\!\!\!P(x,y) \:.
\end{eqnarray*}

Now consider a local minimum of the action. Handling the constraint with a
Lagrange multiplier $\mu$, we obtain the condition
\[ 0 \;=\; \delta S - \mu \:\delta T \;=\;
2 \:\tr \left((Q_1 - \mu Q_2) \:\delta P
\right) \;\stackrel{(\ref{p:50})}{=}\; 2i \:\tr \left( (Q_1 - \mu
Q_2) \:[B, P] \right) . \]
Assume that the products $(Q_1 - \mu Q_2)\:P$ and $P\:(Q_1 - \mu Q_2)$
are well-defined operators. Since $B$ has finite support, we can then
cyclically commute the operators in the
trace and obtain
\[ 0 \;=\; 2i \:\tr\left( B \:[P, \:Q_1 - \mu Q_2] \right) . \]
Since $B$ is arbitrary, we conclude that $[P, \:Q_1 - \mu Q_2]=0$, where
our notation with the commutator implicitly contains the condition that
the involved operator products must be well-defined. Thus our
{\em{Euler-Lagrange equations}}\index{Euler-Lagrange (EL) equations} are
the commutator equations \label{Q}
\begin{eqnarray}
[P,\:Q] &=& 0 \spc {\mbox{with}} \spc
Q(x,y) \;=\; 2\,C_{xy} \:P(x,y) \:, \label{p:52} \\
C_{xy} & = &\sum_{k=1}^{K} \left[ \overline{\lambda_k}
\:-\: \mu \:\frac{\overline{\lambda_k}}{|\lambda_k|} \:
\sum_{l=0}^K n_l \:|\lambda_l| \right]_{xy} F_k \:.
    \label{p:54}
\end{eqnarray}
In the formula (\ref{p:54}) for $C_{xy}$, we consider the spectral
decomposition~(\ref{p:55}, \ref{p:46}) of the closed
chain $P(x,y)\:P(y,x)$. The equations (\ref{p:52}, \ref{p:54})
are the equations of discrete space-time corresponding to the
variational principle (\ref{p:44}, \ref{p:45}).

\section{Discussion}
\label{psec16}  \setcounter{equation}{0}
In the previous sections the principle of the
fermionic projector was introduced in a rather abstract
mathematical way. Our constructions departed radically from the
conventional formulation of physics, so much so that the precise
relation between the principle of the fermionic projector and the
notions of classical and quantum physics is not obvious. In order
to clarify the situation, we now describe the general
physical concept behind the principle of the fermionic projector
and explain in words the connection to classical field theory,
relativistic quantum mechanics and quantum field theory. Since we
must anticipate results which will be worked out later,
the description in this section is
clearly not rigorous and is intended only to give a brief
qualitative overview.

The constructions in~{\S}\ref{psec11} and~{\S}\ref{psec12} are
merely a reformulation of classical field theory and relativistic
quantum mechanics. Although they are an important preparation for
the following construction steps, they do not by themselves have
new physical implications. Therefore, we need not consider them
here and begin by discussing the concept of discrete space-time
of~{\S}\ref{psec13}. With our definition of discrete space-time,
the usual space-time continuum is given up and resolved into
discrete space-time points. A-priori, the discrete space-time
points are merely a point set without any relations (like
for example the nearest-neighbor relation on a lattice) between them.
Thus one may think of discrete space-time as a ``disordered
accumulation of isolated points.'' There exists no time parameter,
nor does it make sense to speak of the ``spatial distance''
between the space-time points. Clearly, this concept of a pure
point set is too general for a reasonable description of
space-time. Namely, we introduced discrete space-time with the
intention of discretizing the space-time continuum on the Planck
scale. Thus, for systems which are large compared to the Planck
length, the discrete nature of space-time should not be apparent.
This means that discrete space-time should, in a certain
{\em{continuum limit}}, go over to a Lorentzian manifold. However,
since $M$ is merely a point set, discrete space-time $(H,\: \bra
.|. \ket,\: (E_p)_{p \in M})$ is symmetric under permutations of
the space-time points. Taking a naive continuum limit would imply
that the points of space-time could be arbitrarily exchanged, in
clear contradiction to the topological and causal structure of a
Lorentzian manifold.

In order to avoid this seeming inconsistency, one must keep in mind
that we introduced an additional object space-time: the fermionic
projector $P$. Via its discrete kernel $P(x,y)$, the fermionic
projector yields relations between the discrete space-time points.
Our idea is that the discrete kernel should provide all structures
needed for a reasonable continuum limit. In more detail, our concept
is as follows. In the space-time continuum (see Chapter~\ref{sec3}),
the fermionic projector is built up of all quantum mechanical states
of the fermionic particles of the system. Closely following Dirac's
original concept, we describe the vacuum by the ``sea'' of all
negative-energy states; systems with particles and anti-particles
are obtained by occupying positive-energy states and removing states
from the Dirac sea, respectively. The fermionic projector of the
continuum completely characterizes the physical system. In
particular, its integral kernel~$P(x,y)$ is singular if and only
if~$y$ lies on the light cone centered at~$x$. In this way, the
fermionic projector of the continuum encodes the causal, and thus
also topological, structure of the underlying space-time. We have in
mind that the fermionic projector of discrete space-time should,
similar to a regularization on the Planck scale, approximate the
fermionic projector of the continuum. This means that on a
macroscopic scale (i.e.\ for systems comprising a very large number
of space-time points), the fermionic projector of discrete
space-time can, to good approximation, be identified with a
fermionic projector of the continuum. Using the just-mentioned
properties of the continuum kernel, we conclude that the discrete
kernel induces on discrete space-time a structure which is
well-approximated by a Lorentzian manifold. However, on the Planck
scale (i.e.\ for systems involving only few space-time points), the
discrete nature of space-time becomes manifest, and the notions of
space, time and causality cease to exit.

The critical step for making this concept precise is the formulation of
the physical equations intrinsically in discrete space-time.
Let us describe in principle how this is supposed to work.
In the continuum description of Chapter~\ref{sec3}, the fermionic projector
satisfies the Dirac equation~(\ref{b}); furthermore the
potentials entering the Dirac equation obey classical field
equations. As a consequence of these equations, the fermionic
projector of the continuum is an object with very specific
properties. Our idea is that, using the special form of the fermionic
projector, it should be possible to restate the Dirac
equation and classical field equations directly in
terms of the fermionic projector. Thus we wish to formulate
equations into which the fermionic projector enters as the basic object,
and which are equivalent to, or a generalization of, both the Dirac
equation and the classical field equations. It turns out that it
is impossible to state equations of this type in the space-time continuum, because
composite expressions in the fermionic projector are mathematically
ill-defined. But one can formulate mathematically meaningful
equations in discrete space-time, removing at the same time the
ultraviolet problems of the continuum theory. The variational
principle (\ref{p:44}, \ref{p:45}) leading to the Euler-Lagrange
equations (\ref{p:52}, \ref{p:54}) is an example for such equations.
Note that this variational principle and the corresponding
Euler-Lagrange equations in discrete space-time
are clearly not causal, but, for consistency with
relativistic quantum mechanics and classical field theory, we demand
that they should, in the continuum limit, reduce to local and causal
equations (namely, to the Dirac and classical field equations).
Since the fermionic projector is not an object which is commonly
considered in physics, it is difficult to give an immediate physical
interpretation for the equations of discrete space-time;
only a detailed mathematical
analysis can provide an understanding of the variational principle.
If one wishes, one can regard the equations of discrete
space-time as describing a direct particle-particle interaction
between all the states of the fermionic projector. The
collective interaction of the fermions of the Dirac sea with the
additional particles and holes should, in the continuum limit, give rise
to an effective interaction of fermions and anti-fermions via classical
fields. Ultimately, the collective particle-particle
interaction should even give a microscopic justification for the
appearance of a continuous space-time structure.

Let us now describe the relation to quantum
field theory. Since the coupled Dirac and classical field equations,
combined with the pair creation/annihilation of Dirac's hole theory,
yield precisely the Feynman diagrams of QFT (see e.g.\ \cite{BD}), it is
clear that all results of perturbative quantum field theory,
in particular the high precision tests of QFT, are also respected by our
ansatz (provided that the equations of discrete space-time have the
correct continuum limit).
Thus the only question is if the particular effects of
quantized fields, like the Planck radiation and the photo electric
effect, can be explained in our framework.
The basic physical assumption behind Planck's radiation law is that
the energy levels of an electromagnetic radiation mode do not take
continuous values, but are quantized in steps of $E=\hbar \omega$. While the
quantitative value $\hbar \omega$ of the energy steps can be understood via
the quantum mechanical identification of energy and frequency (which
is already used in classical Dirac theory), the crucial point of
Planck's assumption lies in the occurrence of discrete energy levels.
The photo electric effect, on the other
hand, can be explained by a ``discreteness'' of the electromagnetic
interaction: the electromagnetic wave tends not to transmit its energy
continuously, but prefers to excite few
atoms of the photographic material. We have the conception (which
will, however, not be worked out in this book) that these different
manifestations of ``discreteness'' should follow from the equations
of discrete space-time if one goes beyond the
approximation of an interaction via classical fields.

If this concept of explaining the effects of quantized fields from the
equations of discrete space-time were correct, it would even have consequences
for the interpretation of quantum mechanics. Namely, according to the statistical
interpretation, quantum mechanical particles are point-like; the
absolute value $|\Psi(\vec{x})|^2$ of the wave function gives the
probability density for the particle to be at the position $\vec{x}$. Here, we
could regard the wave function itself as the physical object; the
particle character would come about merely as a consequence of the
``discreteness'' of the interaction of the wave function with e.g.\
the atoms of a photographic material. The loss of determinism could be
explained naturally by the non-causality of the equations of discrete
space-time.

We conclude that the principle of the fermionic projector raises quite
fundamental questions on the structure of space-time, the nature of field
quantization and the interpretation of quantum mechanics.
Before entering the study of these general
questions, however, it is most essential to establish a quantitative
connection between the equations of discrete space-time and the Dirac
and classical field equations. Namely,
the principle of the fermionic projector can make physical sense only
if it is consistent with classical field theory and relativistic
quantum mechanics; thus it is of importance to first check this consistency.
Even this comparatively simple limiting case is of highest physical
interest. Indeed, the principle of the fermionic projector provides
a very restrictive framework for the formulation of physical
models; for example there is no freedom in choosing the gauge groups, the
coupling of the gauge fields to the fermions, or the masses of the gauge
bosons. This means that, if a connection could be established to
relativistic quantum mechanics and classical field theory, the
principle of the fermionic projector would give an explanation for
the interactions observed in nature and would yield theoretical
predictions for particle masses and coupling constants. We begin with
this study in the next chapters.

\chapter{The Continuum Limit}
\setcounter{equation}{0} \label{psec2} According to the principle
of the fermionic projector, we want to formulate physics with the
fermionic projector~$P$ in discrete space-time~$(H,\: \bra .|.
\ket, \:(E_p)_{p \in M})$. In this chapter we will establish a
mathematically sound connection between this description and the
usual formulation of physics in a space-time continuum. More
precisely, we will develop a general technique with which equations in
discrete space-time, like for example the Euler-Lagrange equations
(\ref{p:52}, \ref{p:54}), can be analyzed within the framework of
relativistic quantum mechanics and classical field theory.
Our approach is based on the assumption that the fermionic projector
of discrete space-time can be obtained from the
fermionic projector of the continuum
by a suitable regularization process on the Planck scale. The basic
difficulty is that composite expressions in the fermionic
projector (like in (\ref{p:52})) depend essentially on how
the regularization is carried out; our task is to analyze this
dependence in detail. We will show that, if we study the behavior
close to the light cone, the dependence on the regularization
simplifies considerably and can be described by a finite number of
parameters. Taking these parameters as free parameters, we will
end up with a well-defined effective continuum theory.

We point out that, since we deduce the fermionic projector of
discrete space-time from the fermionic projector of the continuum, the
causal and topological structure of the space-time continuum, as well
as the Dirac equation and Dirac's hole theory, will enter our construction from the
very beginning. Thus the continuum limit cannot give a justification or
even derivation of these structures from the equations of discrete space-time
(for such a justification one must go beyond the continuum limit;
see~{\S}\ref{esec24} for a first attempt in this direction).
The reason why it is nevertheless interesting to analyze the
continuum limit is that we do not need to specify the classical
potentials which enter the Dirac equation; in
particular, we do not assume that they satisfy the classical field
equations. Thus we can hope that an analysis of the equations of
discrete space-time should give constraints for the classical
potentials; this means physically that the equations of discrete
space-time should in the continuum limit yield a quantitative
description of the interaction of the Dirac particles via classical
fields. This quantitative analysis of the continuum limit of interacting
systems will be explained in Chapters~\ref{esec3}--\ref{secegg}.

For clarity we will mainly restrict attention to a fermionic
projector consisting of one Dirac sea of mass $m$. The
generalizations to systems of fermions with different masses and
to chiral fermions (as introduced in {\S}\ref{jsec3}) are given
in~{\S}\ref{psec26}. Having gauge fields in mind, which are in
quantum field theory described by bosons, we often refer to
the external potentials contained in the operator~${\mathcal{B}}$
\label{mathcalB2}
in the Dirac equation~(\ref{b}) as {\em{bosonic
    potentials}}\index{potential!bosonic}
and the corresponding fields as {\em{bosonic fields}}\index{bosonic!fields}.

\section{The Method of Variable Regularization}
\label{psec22}  \setcounter{equation}{0}
Let us consider how one can get a relation between
the continuum fermionic projector and the description of physics
in discrete space-time. As discussed in~{\S}\ref{psec16},
discrete space-time should for macroscopic systems go over
to the usual space-time continuum. For consistency with
relativistic quantum mechanics, the fermionic projector of
discrete space-time should in this limit coincide with the continuum
fermionic projector. Using furthermore that the discretization
length should be of the order of the Planck length, we conclude
that the fermionic projector of discrete space-time should
correspond to a certain ``regularization'' of the continuum
fermionic projector on the Planck scale. Thus it seems a physically
reasonable method to construct the fermionic projector of discrete
space-time from the fermionic projector of the continuum by a
suitable regularization process on the Planck scale.

Regularizations of the continuum theory are also used in perturbative
QFT in order to make the divergent Feynman diagrams finite. However,
there is the following major difference between the regularizations used
in QFT and our regularization of the fermionic projector. In
contrast to QFT, where the regularization is merely a mathematical
technique within the renormalization procedure, we here consider the
regularized fermionic projector as the object describing the physical
reality. The regularized fermionic projector should be
a model for the fermionic projector of discrete space-time, which we
consider as the basic physical object. As an important consequence,
it is not inconsistent for us if the effective continuum theory depends
on how the regularization is carried out. In
this case, we must regularize in such a way that the regularized
fermionic projector is a good microscopic approximation to the
``physical'' fermionic projector of discrete space-time; only such a
regularization can yield the correct effective continuum theory.
This concept of giving the regularization a physical significance
clearly suffers from the shortcoming that
we have no detailed information about the microscopic structure of the
fermionic projector in discrete space-time, and thus we do not know how
the correct regularization should look like. In order to deal with
this problem, we shall consider a general class of regularizations.
We will analyze in detail how the effective continuum theory depends
on the regularization. Many quantities will depend sensitively on the
regularization, so much so that they are undetermined and
thus ill-defined in the continuum limit. However, certain quantities
will be independent of the regularization and have a simple
correspondence in the continuum theory; we call these quantities
{\em{macroscopic}}\label{pl1}\index{macroscopic quantity}. We will try to express the effective
continuum theory purely in terms of macroscopic quantities. We cannot
expect that the effective continuum theory will be completely
independent of the regularization. But for a meaningful continuum
limit, it must be possible to describe the dependence on the
regularization by a small number of parameters, which we
consider as empiric parameters modelling the unknown microscopic
structure of discrete space-time. We refer to this general procedure
for constructing the effective continuum theory as the {\em{method of
variable regularization}}\index{method of variable regularization}.

In order to illustrate the method of variable regularization, we
mention an analogy to solid state physics. On the microscopic scale,
a solid is composed of atoms, which interact with each other quantum
mechanically. On the macroscopic scale, however, a solid can be
regarded as a continuous material, described by macroscopic quantities
like the density, the pressure, the conductivity, etc. The macroscopic
quantities satisfy macroscopic physical equations like the equations
of continuum mechanics, Ohm's law, etc. Both the macroscopic
characteristics of the solid and the macroscopic physical laws can, at
least in principle, be derived microscopically from many-particle
quantum mechanics. However, since the details of the microscopic
system (e.g.\ the precise form of the electron wave functions) are
usually not known, this derivation often does not completely
determine the macroscopic physical equations. For example, it may
happen that a macroscopic equation can be derived only up to a
proportionality factor, which depends on unknown microscopic
properties of the solid and is thus treated in the macroscopic theory
as an empirical parameter. The physical picture behind the method of
variable regularization is very similar to the physics of a solid, if
one considers on the microscopic scale our description of physics in
discrete space-time and takes as the macroscopic theory both
relativistic quantum mechanics and classical field theory. Clearly,
the concept of discrete space-time is more hypothetical than atomic
physics because it cannot at the moment be verified directly in experiments.
But we can nevertheless get indirect physical evidence for the principle of
the fermionic projector by studying whether or not the method of variable
regularization leads to interesting results for the continuum theory.

In the remainder of this section we will specify for which class of
regularizations we shall apply the method of variable
regularization. Our choice of the regularization scheme is an
attempt to combine two different requirements. On one hand, we
must ensure that the class of regularizations is large enough to
clarify the dependence of the effective continuum theory on the
regularization in sufficient detail; on the other hand, we must
keep the technical effort on a reasonable level. Consider the
integral kernel of the continuum fermionic projector
(\ref{1g}, \ref{fprep}). Under the reasonable assumption that
the fermionic wave functions $\Psi_k$ and $\Phi_l$ are smooth, the
projectors on the particle/anti-particle states in (\ref{1g}) are
smooth in $x$ and $y$. The non-causal low- and high-energy
contributions~$P^{\mbox{\scriptsize{le}}}$ and~$P^{\mbox{\scriptsize{he}}}$
as well as the phase-inserted line integrals
in~(\ref{fprep}) also depend smoothly on~$x$ and~$y$.
The factors $T^{(n)}$, however, have
singularities and poles on the light cone (see~(\ref{Tadef})
and~(\ref{Tldef})). Let us consider what would happen if we
tried to formulate a variational principle similar to that in
{\S}\ref{psec15} with the continuum kernel (instead of the discrete
kernel). The just-mentioned smooth terms in the kernel would not
lead to any difficulties; we could just multiply them with each
other when forming the closed chain $P(x,y) \:P(y,x)$, and the
resulting smooth functions would influence the eigenvalues
$\lambda_k(x,y)$ in (\ref{p:55}) in a continuous way. However, the
singularities of $T^{(n)}$ would cause
severe mathematical problems because the multiplication of
$T^{(n)}(x,y)$ with
$T^{(n)}(y,x)$ leads to singularities
which are ill-defined even in the distributional sense. For
example, the naive product $P(x,y) \:P(y,x)$ would involve
singularities of the form $\sim \delta^\prime((y-x)^2)
\:\delta((y-x)^2)$ and $\sim \delta((y-x)^2)^2$. This simple
consideration shows why composite expressions in the fermionic
projector make mathematical sense only after regularization.
Furthermore, one sees that the regularization is merely needed to
remove the singularities of $T^{(n)}$.
Hence, it seems reasonable to regularize only the factors
$T^{(n)}$ in (\ref{fprep}), but to leave
the fermionic wave functions $\Psi_a$, $\Phi_a$ as well as the
bosonic potentials unchanged. This regularization method implies
that the fermionic wave functions and the bosonic potentials are
well-defined also for the regularized fermionic projector; using
the notation of page~\pageref{pl1}, they are macroscopic
quantities. Therefore, we call our method of only regularizing
$T^{(n)}$ the {\em{assumption of macroscopic potentials and wave
functions}}\index{macroscopic potentials and wave functions, assumption of}.

The assumption of macroscopic potentials and wave functions means
physically that energy and momentum of all bosonic fields and of
each particle/anti-particle of the physical system should be small
compared to the Planck energy. In other words, we exclude the case
that the physical potentials and wave functions have oscillations or
fluctuations on the Planck scale. Namely, such microscopic inhomogeneities
could not be described by smooth functions in the continuum limit
and are thus not taken into account by our regularization method. If,
conversely, the potentials and wave functions are nearly constant on
the Planck scale, the unregularized and the (no matter by which
method) regularized quantities almost coincide, and it is thus a good
approximation to work in the regularized fermionic projector with the
unregularized potentials and wave functions.

According to the assumption of macroscopic potentials and wave
functions, it remains to regularize the factors
$T^{(n)}$ in (\ref{fprep}). Recall that we
constructed the distributions $T^{(n)}$
from the continuum kernel of the fermionic projector of the vacuum
(\ref{Pvac}) via (\ref{Ta}) and the expansion in the mass
parameter (\ref{Tldef}). An essential step for getting a meaningful
regularization scheme is to extend this construction to the case
with regularization. Namely, this extension makes it sufficient to
specify the regularization of the fermionic projector of the
vacuum; we can then deduce the regularized
$T^{(n)}$ and obtain, by substitution
into (\ref{fprep}), the regularized fermionic projector with
interaction (if it were, on the contrary, impossible to derive the
regularized~$T^{(n)}$ from the regularized fermionic projector of
the vacuum, the independent regularizations of all
functions~$T^{(n)}$, $n=-1,0,1,\ldots$, would
involve so many free parameters that the effective continuum
theory would be under-determined). Having in mind the extension of
(\ref{Ta}) and (\ref{Tldef}) to the case with regularization
(which will be carried out in {\S}\ref{psec26} and
Appendix~\ref{pappB}), we now proceed to describe our
regularization method for the fermionic projector of the vacuum.
In the vacuum and for one Dirac sea, the kernel of the continuum
fermionic projector $P(x,y)$ is given by the Fourier integral
(\ref{f:2b}),
\beq
P(x,y) \;=\; \int \frac{d^4k}{(2 \pi)^4}\:
(k\slsh+m)\: \delta(k^2-m^2)\: \Theta(-k^0)\: e^{-ik(x-y)}\:. \label{p:2b}
\eeq
This distribution
is invariant under translations in space-time, i.e.\ it depends
only on the difference vector $y-x$. It seems natural and is most
convenient to preserve the translation symmetry in the
regularization. We thus assume that the kernel of the regularized
fermionic projector of the vacuum, which we denote for simplicity
again by $P(x,y)$, is translation invariant,
\begin{equation}
P(x,y) \;=\; P(y-x) \;\;\;\;{\mbox{for}}\;\;\;\; x,y \in M \subset
\R^4 \:.
    \label{p:2h}
\end{equation}
We refer to (\ref{p:2h}) as a {\em{homogeneous regularization of the
vacuum}}\index{homogeneous regularization of the vacuum}. Notice that the assumption (\ref{p:2h}) allows for both
discrete and continuum regularizations. In the first case, the set $M$
is taken to be a discrete subset of $\R^4$ (e.g.\ a lattice), whereas
in the latter case, $M=\R^4$. According to our concept of discrete
space-time, it seems preferable to work with discrete regularizations.
But since continuous regularizations give the same results and are a
bit easier to handle, it is worth considering them too.
The assumption of a homogeneous regularization of the vacuum means
physically that the inhomogeneities of the fermionic projector on the
Planck scale should be irrelevant for the effective continuum theory.
Since such microscopic inhomogeneities can, at least in special
cases, be described by microscopic gravitational or gauge
fields, this assumption is closely related to the assumption of
macroscopic potentials and wave functions discussed above.

\begin{figure}[tb]
\begin{center}
\scalebox{0.9}
{\includegraphics{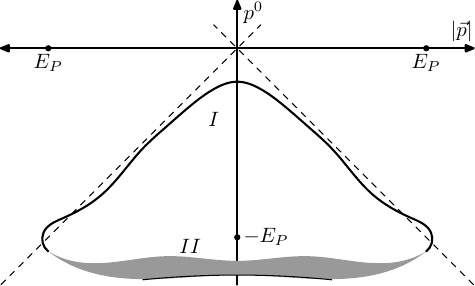}}
\caption{Example for $\hat{P}$, the regularized fermionic projector
of the vacuum in momentum space.} \label{fig1}
\end{center}
\end{figure}

Taking the Fourier transform in the variable $y-x$, we write
(\ref{p:2h}) as the Fourier integral \label{hatP}
\begin{equation}
P(x,y) \;=\; \int \frac{d^4p}{(2 \pi)^4} \:\hat{P}(p) \: e^{-i p (x-y)}
    \label{p:2k}
\end{equation}
with a distribution $\hat{P}$.
If one considers a discrete regularization, $\hat{P}$
may be defined only in a bounded region of $\R^4$ (for a lattice
regularization with lattice spacing $d$, for example, one can
restrict the momenta to the ``first Brillouin zone'' $p \in (-\frac{\pi}{d},
\frac{\pi}{d})^4$). In this case, we extend $\hat{P}$ to all~$\R^4$ by
setting it to zero outside this bounded region. Although it will be of no
relevance for what follows, one should keep in mind that for a
discrete regularization, $x$ and $y$ take values only in the discrete
set $M$. Let us briefly discuss the properties of the
distribution~$\hat{P}$. First of
all, $P(x,y)$ should be the kernel of a Hermitian operator; this
implies that $P(x,y)^* = P(y,x)$ and thus
\begin{equation}
    \hat{P}(p)^* \;=\; \hat{P}(p) \spc {\mbox{for all $p$}}
    \label{p:2ll}
\end{equation}
(where the star again denotes the adjoint with respect to the spin scalar
product). For consistency with the continuum theory, the regularized
kernel (\ref{p:2k}) should, for macroscopic systems, go over to the
continuum kernel (\ref{p:2b}). Thus we know that $\hat{P}(p)$
should, for small energy-momentum $p$ (i.e.\ when both the energy
$p^0$ and the momentum $|\vec{p}|$ are small compared to the Planck
energy), coincide with the distribution $(p \slsh + m) \: \delta(p^2 -
m^2) \:\Theta(-p^0)$. This is illustrated in the example of
Figure~\ref{fig1}.
In the region I close to the origin, $\hat{P}$
looks similar to a hyperbola on the lower mass shell. Furthermore, we
know that $\hat{P}$ is a regularization on the Planck scale. This
means that, in contrast to the integrand in (\ref{p:2b}), $\hat{P}$
should decay at infinity, at least so rapidly that the integral
(\ref{p:2k}) is finite for all $x$ and $y$. The length scale for
this decay in momentum space should be of the order of the Planck
energy $E_P= l^{-1}_P$ \label{E_P}. However, the precise form of $\hat{P}$ for large
energy or momentum is completely arbitrary\label{pl2}, as is indicated
in Figure~\ref{fig1} by the ``high energy region'' II. This
arbitrariness reflects our freedom in choosing the regularization.

We finally make an ansatz for $\hat{P}$ which seems general enough
to include all relevant regularization effects and which will
considerably simplify our analysis. According to
(\ref{p:2ll}), $\hat{P}(p)$ is a Hermitian $4 \times 4$ matrix
and can thus be written as a real linear combination of the basis of
the Dirac algebra $\1$, $i \rho$, $\rho \gamma^j$ and
$\sigma^{jk}$ (with the pseudoscalar matrix $\rho = i \gamma^0
\gamma^1 \gamma^2 \gamma^3$ and the bilinear covariants
$\sigma^{jk} = \frac{i}{2} [\gamma^j, \gamma^k]$). The
integrand of the continuum kernel (\ref{p:2b}) contains only vector
and scalar components. It is reasonable to assume that the regularized
kernel also contains no pseudoscalar and pseudovector components,
because the regularization would otherwise break the symmetry under
parity transformations. The inclusion of a bilinear component in
$\hat{P}$, on the other hand, would cause technical complications
but does not seem to give anything essentially new. Thus we make an ansatz
where $\hat{P}$ is composed only of a vector and a scalar component,
more precisely
\begin{equation}
\hat{P}(p) \;=\; (v_j(p) \:\gamma^j \:+\: \phi(p) \:\1) \:f(p)
    \label{p:2f2}
\end{equation}
with a vector field $v$ and a scalar field $\phi$; $f$ is a
distribution. We also need to assume that $\hat{P}$ is reasonably
regular and well-behaved; this will be specified in the following
sections. We refer to the ansatz (\ref{p:2f2}) as the assumption of a
{\em{vector-scalar structure}}\index{vector-scalar structure} for the fermionic projector of the
vacuum.

\section{The Regularized Product $P(x,y)\: P(y,x)$ in the Vacuum}
\label{psec23}  \setcounter{equation}{0}
According to the method of variable regularization,
we must analyze how the effective continuum theory depends on the
choice of the regularization. We shall now consider this problem
for the simplest composite expression in the fermionic projector,
the closed chain $P(x,y) \:P(y,x)$ in the vacuum. The discussion
of this example will explain why we need to analyze the fermionic
projector on the light cone. Working out this concept
mathematically will eventually lead us to the general formalism
described in {\S}\ref{psec26}.

Using the Fourier representation (\ref{p:2k}), we can calculate the
closed chain to be
\begin{eqnarray}
P(x,y) \:P(y,x) &=& \int \frac{d^4 k_1}{(2 \pi)^4} \int \frac{d^4
k_2}{(2 \pi)^4} \:\hat{P}(k_1) \:\hat{P}(k_2) \;e^{-i
(k_1-k_2)(x-y)} \nonumber \\
     & = & \int \frac{d^4p}{(2 \pi)^4} \left[ \int \frac{d^4q}{(2
     \pi)^4} \:\hat{P}(p+q) \:\hat{P}(q) \right] e^{-ip(x-y)} \:,
    \label{p:2n}
\end{eqnarray}
where we introduced new integration variables $p=k_1-k_2$ and
$q=k_2$. Thus the Fourier
transform of the closed chain is given by the convolution in the
square brackets. This reveals the following basic problem.
The convolution in the square bracket involves $\hat{P}$ for
small and for large energy-momentum. Even when $p$ is small, a
large $q$ leads to a contribution where both factors $\hat{P}(p+q)$
and $\hat{P}(q)$ are evaluated for large energy-momenta. If we look
at the example of Figure~\ref{fig1}, this means that (\ref{p:2n})
depends essentially on the behavior of $\hat{P}$ in the high-energy
region II and can thus have an arbitrary value. More generally, we
conclude that, since the form of $\hat{P}$ for large
energy or momentum is unknown, the value of (\ref{p:2n}) is undetermined.

At first sight, it might seem confusing that the pointwise product
$P(x,y) \:P(y,x)$ of the regularized fermionic projector should be
undetermined, although the unregularized kernel (\ref{p:2b}) is, for
$y-x$ away from the light cone, a smooth function, and so pointwise
multiplication causes no difficulties. In order to explain the
situation in a simple example, we briefly discuss the fermionic
projector $\tilde{P}$ obtained by adding to $P$ a plane wave,
\[ \tilde{P}(x,y) \;=\; P(x,y) \:+\: e^{-ik(x-y)}\:\1 \:. \]
If the energy or the momentum of the plane wave is of the order of
the Planck energy, the plane wave is highly oscillatory in
space-time. Such an oscillatory term is irrelevant on the macroscopic
scale. Namely, if $\tilde{P}$ acts on a macroscopic function $\eta$, the
oscillatory term is evaluated in the weak sense, and the resulting
integral $\int \exp(iky) \:\eta(y)\:d^4y$ gives almost zero because
the contributions with opposite signs compensate each other. This
``oscillation argument'' can be made mathematically precise using
integration by parts, e.g.\ in the case of high energy $k^0 \sim E_P$,
\[ \int e^{iky} \:f(y)\:d^4y \;=\; -\frac{1}{ik^0} \int e^{iky} \:
(\partial_t f) \:d^4y \;\sim\; \frac{1}{E_P} \:. \]
In the corresponding closed chain
\[ \tilde{P}(x,y) \:\tilde{P}(y,x) \;=\; P(x,y) \:P(y,x) \:+\: P(x,y)
\:e^{-ik(y-x)} \:+\: e^{-ik(x-y)} \:P(y,x) \:+\: \1 \:, \]
the second and third summands are also oscillatory. In the last
summand, however, the oscillations have dropped out, so that this
term affects the macroscopic behavior of the closed chain. This
elementary consideration illustrates why the unknown high-energy
contribution to the fermionic projector makes it impossible to
determine the closed chain pointwise.
We remark that for very special regularizations, for example the
regularization by convolution with a smooth ``mollifier'' function
having compact support, the pointwise product makes
sense away from the light cone and coincides approximately
with the product of the unregularized kernels. But such
regularizations seem too restrictive. We want to allow for
the possibility that the fermionic projector describes non-trivial
(yet unknown) high-energy effects. Therefore, the high-energy behavior of
the fermionic projector should not be constrained by a too simple
regularization method.

The fact that the product $P(x,y)\:P(y,x)$ is undetermined for fixed
$x$ and $y$ does not imply that a pointwise analysis of the closed
chain is mathematically or physically meaningless. But it means that a
pointwise analysis would essentially involve the unknown high-energy
behavior of $\hat{P}$; at present this is a problem out of reach.
Therefore, our strategy is to find a method for evaluating the closed
chain in a way where the high-energy behavior of $\hat{P}$ becomes so
unimportant that the dependence on the regularization can be described
in a simple way. We hope that this method will lead us to a certain
limiting case in which the equations of discrete space-time become
manageable.

The simplest method to avoid the pointwise analysis is to
evaluate the closed chain in the weak sense. The Fourier
representation (\ref{p:2n}) yields that
\begin{equation}
\int P(x,y) \:P(y,x) \:\eta(x) \:d^4x \;=\; \int \frac{d^4p}{(2
\pi)^4} \:\hat{\eta}(p) \left[ \int \frac{d^4q}{(2 \pi)^4}
\:\hat{P}(p+q) \:\hat{P}(q) \right] ,
    \label{p:2p}
\end{equation}
where $\hat{\eta}$ is the Fourier transform of a smooth function
$\eta$. For macroscopic $\eta$ (i.e.\ a function which is nearly
constant on the Planck scale), the function $\hat{\eta}(p)$ is
localized in a small neighborhood of $p=0$ and has rapid decay.
Thus, exactly as (\ref{p:2n}), the integral (\ref{p:2p}) depends on
the form of $\hat{P}$ for large energy-momentum. Hence this type
of weak analysis is not helpful. In order to find a better method,
we consider again the Fourier integral (\ref{p:2k}) in the example
of Figure~\ref{fig1}. We want to find a regime for $y-x$ where the
``low energy region'' I plays an important role, whereas the
region II is irrelevant. This can be accomplished only by
exploiting the special form of $\hat{P}$ in the low-energy region
as follows. The hyperbola of the lower mass shell in region I
comes asymptotically close to the cone $C=\{p^2=0\}$.
If we choose a vector $(y-x) \neq 0$ on the light cone
$L=\{(y-x)^2=0\}$, then the hypersurface ${\mathcal{H}}=\{p \:|\:
p (y-x)=0\}$ is null and thus tangential to the cone
$C$. This means that for all states on the hyperbola
which are close to the straight line $C \cap
{\mathcal{H}}$, the exponential in (\ref{p:2k}) is approximately
one. Hence all these states are ``in phase'' and thus yield a
large contribution to the Fourier integral (\ref{p:2k}). The
states in the high-energy region II, however, are not in phase;
they will give only a small contribution to (\ref{p:2k}), at least
when the vector $(y-x) \in L$ is large, so that the exponential in
(\ref{p:2k}) is highly oscillatory on the scale $p \sim E_P$. This
qualitative argument shows that by considering the fermionic
projector on the light cone, one can filter out information on the
behavior of $\hat{P}$ in the neighborhood of a straight line along
the cone $C$. This should enable us to analyze the
states on the lower mass shell without being affected too much by
the unknown high-energy behavior of $\hat{P}$. We point out that
if $P(x,y)$ depends only on the behavior of $\hat{P}$ close to the
cone $C$, then the same is immediately true for
composite expressions like the product $P(x,y) \:P(y,x)$. Thus
restricting our analysis to the light cone should simplify the
dependence on the regularization considerably, also for composite
expressions like the closed chain. Our program for the remainder
of this chapter is to make this qualitative argument mathematically
precise and to quantify it in increasing generality.

\section{The Regularized Vacuum on the Light Cone, Scalar Component}
\label{psec24}  \setcounter{equation}{0}
For simplicity we begin the analysis on the light
cone for the scalar component\index{scalar component} of (\ref{p:2f2}), i.e.\ we consider
the case
\begin{equation}
    \hat{P}(p) \;=\; \phi(p) \:f(p)
    \label{p:25s}
\end{equation}
(the vector component will be treated in the next
section). We can assume that the spatial component of the vector
$y-x$ in (\ref{p:2k}) points in the direction of the $x$-axis of our
Cartesian coordinate system, i.e.\ $y-x=(t,r,0,0)$ with $r>0$.
Choosing cylindrical coordinates $\omega$, $k$, $\rho$ and $\varphi$
in momentum space, defined by $p=(\omega, \vec{p})$ and $\vec{p}=(k,\:
\rho \:\cos \varphi,\: \rho \:\sin \varphi)$, the Fourier integral
(\ref{p:2k}) takes the form
\begin{equation}
P(x,y) \;=\; \frac{1}{(2 \pi)^4} \int_{-\infty}^\infty d\omega
\int_{-\infty}^\infty dk \int_0^\infty \rho\: d\rho \int_0^{2 \pi}
d\varphi \; \hat{P}(\omega, k, \rho, \varphi) \;e^{i \omega t - i k
r} \: .
    \label{p:2pq}
\end{equation}
Since the exponential factor in this formula is independent of $\rho$
and $\varphi$, we can write the fermionic projector as the
two-dimensional Fourier transform
\begin{equation}
P(x,y) \;=\; 2 \int_{-\infty}^\infty d\omega \int_{-\infty}^\infty dk \;
h(\omega, k) \:e^{i \omega t - i k r}
    \label{p:2pp}
\end{equation}
of a function $h$ defined by
\begin{equation}
h(\omega, k) \;=\; \frac{1}{2 \:(2 \pi)^4} \int_0^\infty \rho \:d\rho
\int_0^{2 \pi} d\varphi \;(\phi \:f)(\omega, k, \rho, \varphi) \:.
    \label{p:2q}
\end{equation}
We want to analyze $P(x,y)$ close to the light cone $(y-x)^2=0$ away from
the origin $y=x$. Without loss of generality, we can restrict
attention to the upper light cone $t=r$. Thus we are interested in
the region $t \approx r >0$. The ``light-cone
coordinates''\index{light-cone coordinates} \label{s} \label{l}
\label{u} \label{v}
\begin{equation}
s \;=\; \frac{1}{2} \:(t-r) \:,\spc l \;=\; \frac{1}{2} \:(t+r)
    \label{p:2q1}
\end{equation}
are well-suited to this region, because the ``small'' variable $s$
vanishes for $t=r$, whereas the ``large'' variable $l$ is positive and
non-zero. Introducing also the associated momenta
\begin{equation}
u \;=\; -k-\omega \:,\spc v \;=\; k-\omega \:,
    \label{p:2q2}
\end{equation}
we can write the fermionic projector as
\begin{equation}
P(s,l) \;=\; \int_{-\infty}^\infty du \int_{-\infty}^\infty dv
\;h(u,v) \:e^{-i(us + vl)} \:.
    \label{p:2r}
\end{equation}

\begin{figure}[tb]
\begin{center}
\scalebox{0.9}
{\includegraphics{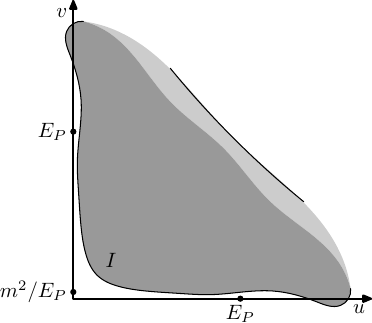}}
\caption{Example for $h(u,v)$, the reduced two-dimensional distribution.}
\label{fig2}
\end{center}
\end{figure}

Let us briefly discuss the qualitative form of the function $h$, (\ref{p:2q}).
According to the continuum kernel (\ref{p:2b}), the scalar component
(\ref{p:25s}) should, for energy and momentum small compared to the
Planck energy $E_P$,  go over to the
$\delta$-distribution on the lower mass shell $\hat{P} = m
\:\delta(p^2-m^2) \:\Theta(-p^0)$. In this limit, the integral
(\ref{p:2q}) can be evaluated to be
\begin{eqnarray}
h & = & \frac{m}{2 \:(2 \pi)^4} \int_0^\infty \rho\:d\rho
\int_0^{2 \pi} d\varphi \;\delta(\omega^2 - k^2 - \rho^2 - m^2)
\:\Theta(-\omega) \nonumber \\
     & = & \frac{m}{4 \:(2 \pi)^3} \:\Theta(\omega^2 - k^2 - m^2)
     \:\Theta(-\omega) \;=\; \frac{m}{32 \pi^3} \:\Theta(uv - m^2)
     \:\Theta(u) \:;
    \label{p:2rr}
\end{eqnarray}
thus integrating out $\rho$ and $\varphi$ yields a constant function
in the interior of the two-di\-men\-sio\-nal ``lower mass shell''
$\omega^2 - k^2 = m^2$, $\omega<0$.
From this we conclude that for $u,v \ll E_P$, $h(u,v)$ should have a
discontinuity along the hyperbola $\{ uv=m^2,\: u>0\}$, be zero
below (i.e.\ for $uv<m^2$) and be nearly constant above. Furthermore, we know
that $h$ decays at infinity on the scale of the Planck energy.
Similar to our discussion of $\hat{P}$ after (\ref{p:2ll}), the
precise form of $h$ for large energy or momentum is completely
arbitrary. The function $h(u,v)$ corresponding to the example of
Figure~\ref{fig1} is shown in Figure~\ref{fig2}.
The two branches of the hyperbola asymptotic to the $u$ and $v$ axes
are labeled by ``A'' and ``B,'' respectively.

It is instructive to consider the energy scales of our system. The
scale for high energies is clearly given by the Planck energy $E_P$. The
relevant low-energy scale, on the other hand, is $m^2/E_P$ (it is zero
for massless fermions). This is because the hyperbola $uv=m^2$ comes as close
to the $v$-axis as as $v \sim m^2/E_P$ before leaving the low-energy region.
These two energy scales are also marked in Figure~\ref{fig2}. Since we want to
analyze the situation close to the light cone, we choose the
``small'' light-cone parameter $s$ on the Planck scale, i.e.
\begin{equation}
s \sim E_P^{-1} \spc{\mbox{or}}\spc s < E_P^{-1} \:.
    \label{p:2z}
\end{equation}
The ``large'' light-cone parameter $l$, on the other hand, is non-zero
and thus yields a third energy scale. We shall always choose this
scale between the two extremal energy scales, more precisely
\begin{equation}
\frac{1}{E_P} \;\ll\; l \;<\; l_{\mbox{\scriptsize{max}}} \;\ll\;
\frac{E_P}{m^2} \:.
    \label{p:2s}
\end{equation}
The parameter $l_{\mbox{\scriptsize{max}}}$ \label{lmax} was introduced here in
order to avoid $l$ being chosen too large. Namely, we will
always regard $l$ as being small compared to the length scales of
macroscopic physics (a reasonable value for
$l_{\mbox{\scriptsize{max}}}$ would e.g.\ be the Fermi length).
One should keep in mind that the quotient of the two fundamental
energy scales is in all physical situations extremely large; namely
$E_P^2/m^2 \gg 10^{35}$. Thus the constraints (\ref{p:2s}) can be
easily satisfied and still leave us the freedom to vary $l$ on many
orders of magnitude.

In the remainder of this section we shall evaluate the Fourier
integral (\ref{p:2r}) using the scales (\ref{p:2z}) and (\ref{p:2s}).
In preparation, we discuss and specify the function $h(u,v)$ for fixed $u$,
also denoted by $h_u(v)$. As one sees in Figure~\ref{fig2}, $h_u$ will in
general not be continuous. More precisely, in the example of Figure~\ref{fig2},
$h_u$ has  a discontinuous ``jump'' from zero to a finite value
on the hyperbola (and its extension to the high-energy region) and
maybe has a second jump to zero for large $v$ (e.g. on line ``a'').
For simplicity, we assume that $h_u$ is always of this general form, i.e.
\begin{equation}
h_u(v) \;=\; \left\{ \begin{array}{cl} 0 & {\mbox{for $v<\alpha_u$ or
$v>\beta_u$}} \\
{\mbox{smooth}} & {\mbox{for $\alpha_u \leq v \leq \beta_u$}}
\end{array} \right.
    \label{p:2t}
\end{equation}
with parameters $\alpha_u<\beta_u$. The case of less than two
discontinuities can be obtained from (\ref{p:2t}) by setting
$h_u(\alpha_u)$ or $h_u(\beta_u)$ equal to zero, or alternatively by
moving the position of the discontinuities $\alpha_u$ or $\beta_u$ to
infinity. We remark that the discontinuity at $v=\beta_u$ will become
irrelevant later; it is here included only to illustrate why the
behavior of the fermionic projector
on the light cone is independent of many regularization details.
Without regularization, $h_u(v)$ is for $v \geq
\alpha_u$ a constant function, (\ref{p:2rr}). Thus the $v$-dependence
of $h_u(v)$ for $\alpha_u \leq v \leq \beta_u$ is merely a
consequence of the regularization, and it is therefore reasonable to
assume that the $v$-derivatives of $h_u(v)$ scale in inverse powers of the
regularization length $E_P$. More precisely, we demand that there is a
constant $c_1 \ll l E_P$ with
\begin{equation}
|h_u^{(n)}(v)| \;\leq\; \left(\frac{c_1}{E_P}\right)^n \:\max
|h_u| \spc {\mbox{for $\alpha_u \leq v \leq \beta_u$}} \:,
    \label{p:2v}
\end{equation}
where the derivatives at $v=\alpha_u$ and $\beta_u$ are understood as
the right- and left-sided limits, respectively. This regularity
condition is typically satisfied for polynomial, exponential and
trigonometric functions, but it excludes small-scale fluctuations of
$h_u$. Clearly, we could also consider a more general ansatz for
$h_u$ with more than two discontinuities or weaker regularity
assumptions. But this does not seem to be the point because all
interesting effects, namely the influence of discontinuities for
small and large $v$, as well as of smooth regions, can already be
studied in the setting (\ref{p:2t}, \ref{p:2v}).

Let us analyze the $v$-integral of the Fourier transform (\ref{p:2r}),
\begin{equation}
P_u(l) \;:=\; \int_{-\infty}^\infty h_u(v) \: e^{-ivl} \;dv \:.
    \label{p:2w}
\end{equation}
According to the first inequality in (\ref{p:2s}), the exponential factor
in (\ref{p:2w}) is highly oscillatory on the scale $v \sim E_P$. Thus we can
expect that the smooth component of $h_u$ gives only a small
contribution to the integral (\ref{p:2w}), so that the discontinuities
at $\alpha_u$ and $\beta_u$ play the dominant role.
In order to make this picture mathematically
precise, we iteratively integrate in (\ref{p:2w}) $K$ times by parts,
\begin{eqnarray}
\lefteqn{ P_u(l) \;=\; \int_{\alpha_u}^{\beta_u} h_u(v) \:e^{-ivl} \:dv \;=\;
-\frac{1}{il} \int_{\alpha_u}^{\beta_u} dv\; h_u(v) \:\frac{d}{dv}
e^{-ivl} } \nonumber \\
&=& -\frac{1}{il} \left. h_u(v) \:e^{-ivl}
\right|_{\alpha_u}^{\beta_u} \:+\: \frac{1}{il}
\int_{\alpha_u}^{\beta_u} h_u^\prime(v) \:e^{-ivl} \:dl
\;=\; \cdots \;=\; \nonumber \\
&=& -\frac{1}{il} \sum_{n=0}^{K-1}
\left(\frac{1}{il}\right)^n \: \left. h_u^{(n)}(v) \:e^{-ivl}
\right|_{\alpha_u}^{\beta_u} \:+\: \left( \frac{1}{il} \right)^K
\int_{\alpha_u}^{\beta_u} h_u^{(K)}(v) \:e^{-ivl} \:dl \:.\spc \label{p:2x}
\end{eqnarray}
If we bound all summands in (\ref{p:2x}) using the first inequality in
(\ref{p:2s})
and the regularity condition (\ref{p:2v}), each
$v$-derivative appears in combination with a power of $l^{-1}$, and
this gives a factor $c_1/(l E_P) \ll 1$. Thus we can in the limit
$K \to \infty$ drop the integral in (\ref{p:2x}) and obtain
\begin{equation}
P_u(l) \;=\; -\frac{1}{il} \sum_{n=0}^\infty
\left(\frac{1}{il}\right)^n \: \left. h_u^{(n)}(v) \:e^{-ivl}
\right|_{\alpha_u}^{\beta_u} \: .
    \label{p:2y}
\end{equation}
This expansion converges, and its summands decay like $(c_1/(l E_P))^n$.

Using (\ref{p:2w}), we can write the Fourier transform (\ref{p:2r}) as
\begin{equation}
P(s,l) \;=\; \int_{-\infty}^\infty P_u(l) \:e^{-ius} \:du \:.
    \label{p:21}
\end{equation}
Notice that, apart from the constraints (\ref{p:2s}), the ``large''
variable $l$ can be freely chosen. We want to study the functional
dependence of (\ref{p:21}) on the parameter $l$. In preparation, we
consider an integral of the general form
\begin{equation}
\int_a^b f(u) \:e^{-i \gamma(u) \:l} \:du \:,
    \label{p:22}
\end{equation}
where we assume that $(u, \gamma(u))$ is a curve in the high-energy
region, more precisely $\gamma \sim E_P$. Assume furthermore that $\gamma$ is
monotone with $|\gamma^\prime| \sim 1$ and that $(b-a) \sim E_P$. By
transforming the integration variable, we can then write (\ref{p:22}) as
the Fourier integral
\begin{equation}
\int_{\gamma(a)}^{\gamma(b)} f \:|\gamma^\prime|^{-1} \:e^{-i \gamma
l} \:d\gamma \:.
    \label{p:23}
\end{equation}
If the function $f \:|\gamma^\prime|^{-1}$ is smooth,
its Fourier transform (\ref{p:23}) has rapid decay in
the variable $l$. Under the stronger assumption that $f \:|\gamma^\prime|^{-1}$
varies on the scale $E_P$, we conclude that the length scale for
this rapid decay is of the order $l \sim E_P^{-1}$. As a consequence, the
rapid decay can be detected even under the constraint
$l<l_{\mbox{\scriptsize{max}}}$ imposed by (\ref{p:2s}), and we say
that (\ref{p:23}) has {\em{rapid decay in $l$}}. The reader who feels
uncomfortable with this informal definition can immediately make this
notion mathematically precise by an integration by parts argument
similar to (\ref{p:2x}) imposing for $f \:|\gamma^\prime|^{-1}$ a
condition of type (\ref{p:2v}). The precise mathematical meaning of
rapid decay in $l$ for the integral (\ref{p:22}) is that for every
integer $k$ there should be constants $c \sim 1$ and
$l_{\mbox{\scriptsize{min}}} \ll l_{\mbox{\scriptsize{max}}}$ such
that for all $l \in (l_{\mbox{\scriptsize{min}}},
l_{\mbox{\scriptsize{max}}})$,
\[ \int_a^b f(u) \:e^{-i \gamma(u) \:l} \:du \;\leq\; c \: (l E_P )^{-k}
\:\int_a^b |f(u)| \:du \:. \]

We return to the analysis of the integral (\ref{p:21}).
The boundary terms of (\ref{p:2y}) at $\beta_u$ yield
contributions to $P(s,l)$ of the form
\begin{equation}
-\left(\frac{1}{il}\right)^{n+1} \int_{-\infty}^\infty
h_u^{(n)}(\beta_u) \:e^{-i \beta_u l - i u s} \:du \:.
    \label{p:24}
\end{equation}
Recall that the points $(u, \beta_u)$ are in the high-energy region
(in the example of Figure~\ref{fig2}, these points lie on curve ``a'').
According to (\ref{p:2z}), the length scale for the oscillations of
the factor $\exp (-ius)$ is $u \sim E_P$. Under the reasonable assumption
that $\beta_u$ is monotone and that the functions $|\beta^\prime(u)|^{-1}$
and $h_u^{(n)}(\beta_u)$ vary on the scale $E_P$, the
integral (\ref{p:24}) is of the form (\ref{p:23}), and the above
consideration yields that (\ref{p:24}) has rapid decay in $l$.
We remark that this argument could be extended to the case where
$\beta_u$ has extremal points (basically because the extrema give
contributions only for isolated momenta $u$ and thus can be shown to
be negligible), but we will not go into this here. Having established
rapid decay in $l$ for (\ref{p:24}), it remains to consider the
boundary terms in (\ref{p:24}) at $\alpha_u$, more precisely
\begin{eqnarray}
P(s,l) & = & \sum_{n=0}^\infty \left(\frac{1}{il}\right)^{n+1}
\int_{-\infty}^\infty h_u^{(n)}(\alpha_u) \:e^{-i \alpha_u l - ius}
\:du \nonumber \\
&&\:+\: {\mbox{(rapid decay in $l$)}} \:. \label{p:25a}
\end{eqnarray}
We cannot again apply our ``oscillation argument'' after (\ref{p:22}),
because $\alpha_u$ tends asymptotically to zero on branch ``A''
of the hyperbola (see Figure~\ref{fig2}), so that the factor $\exp(-i
\alpha_u l)$ is non-oscillating in this region. We expand this factor
in a Taylor series,
\begin{equation}
P(s,l) \;=\; \sum_{n,k=0}^\infty \frac{1}{k!} \:(il)^{k-n-1}
\int_{-\infty}^\infty h_u^{(n)}(\alpha_u) \:(-\alpha_u)^k \: e^{-ius}
\:du \: .
    \label{p:25}
\end{equation}
In the region where $l \alpha_u \not \ll 1$, this expansion might
seem problematic and requires a brief explanation. First of all,
$\alpha_u$ becomes large near $u=0$ (on branch ``B'' of the hyperbola
in Figure~\ref{fig2}). In the case without regularization, the power
expansion of the factor $\exp(-i \alpha_u \:l)$ corresponds to an
expansion in the mass parameter (recall that in this case,
$\alpha_u = m^2/u$ according to (\ref{p:2rr})), and in (\ref{p:25})
it would lead to a singularity of the integrand at the origin.
Indeed, this difficulty is a special case of the logarithmic mass
problem which was mentioned in~{\S}\ref{jsec5} and was resolved by
working with the ``regularized'' distribution~$T^{\mbox{\scriptsize{reg}}}_a$ (\ref{Tadef}). Using these results, the behavior of the unregularized $P(s,l)$ for small momenta $u \ll E_P$ is well understood. Our oscillation argument after (\ref{p:22})
yields that the regularization for $u \ll E_P$ (i.e.\ the form of the
extension of branch ``B'' of the hyperbola to the high-energy region)
affects $P(s,l)$ merely by rapidly decaying terms. \label{pl3}
Thus it is sufficient to consider here the integrand in (\ref{p:25}) away
from the origin $u=0$. When combined with the results in~{\S}\ref{jsec5},
our analysis will immediately yield a complete description of the
regularized fermionic projector near the light cone.
Furthermore, the function $\alpha_u$ might become large for $u \sim
E_P$, and this is a more subtle point. One way of justifying
(\ref{p:25}) would be to simply assume that $l_{\mbox{\scriptsize{max}}}
\alpha_u \ll 1$ along the whole extension of
branch ``A'' to the high-energy region. A more general method would be
to split up the curve $(u, \alpha_u)$ in the high-energy region $u
\sim E_P$ into one branch where the expansion (\ref{p:25}) is justified and
another branch where our oscillation argument after (\ref{p:22}) applies. The
intermediate region $l \alpha_u \sim 1$, where none of the two
methods can be used, is generically so small that it can
be neglected. In order to keep our analysis reasonably simple, we
here assume that $\alpha_u$ is sufficiently small away from the
origin, more precisely
\begin{equation}
\alpha_u \;<\; \alpha_{\mbox{\scriptsize{max}}}
\;\ll\;l_{\mbox{\scriptsize{max}}}^{-1} \spc
{\mbox{for $u \sim E_P$.}} \label{p:25n}
\end{equation} \label{alphamax}

For a fixed value of $k-n$, all summands in (\ref{p:25}) have the
same $l$-dependence. Let us compare the relative size of these terms.
According to our regularity assumption (\ref{p:2v}), the derivatives
of $h$ scale like $h_u^{(n)} \sim E_P^{-n}$. Using the bound
(\ref{p:25n}), we conclude that, for a fixed
power of $l$, the summands in (\ref{p:25}) decrease like
$(\alpha_{\mbox{\scriptsize{max}}}/E_P)^n$.
Thus it is a very good approximation to drop the summands for large $n$.
At first sight, it might seem admissible to take into account only the first
summand $n=0$. But the situation is not quite so simple.
For example, it may happen that, when restricted to the curve $(u, \alpha_u)$,
the function $h(u,v)$ is so small that the summands for $n=0$ in
(\ref{p:25}) are indeed not dominant. More generally, we need to know that
for some $n_0 \geq 0$, the function $h^{(n_0)}_u(\alpha_u)$ is really of the
order given in (\ref{p:2v}), i.e.
\begin{equation}
|h^{(n_0)}_u(\alpha_u)| \;\geq\; c \left( \frac{c_1}{E_P} \right)^{n_0} \:
\max |h_u| \spc {\mbox{for $u \sim E_P$}}
\label{p:27h}
\end{equation}
with a positive constant $c$ which is of the order one.
If this condition is satisfied, we may neglect
all summands for $n>n_0$, and collecting the terms in powers of $l$, we
conclude that
\begin{eqnarray}
\lefteqn{ P(s,l) } \nonumber \\
&=& \frac{1}{(il)^{n_0+1}} \sum_{k=0}^\infty (-il)^k \!\!\!\!\!\!\!
\sum_{n=\max (n_0-k, 0)}^{n_0} \frac{(-1)^{n_0 - n}}{(k-n_0+n)!}
\int_{-\infty}^\infty h_u^{(n)}(\alpha_u) \:\alpha_u^{k-n_0+n}
\:e^{-ius} \:du \nonumber \\
&&+\sum_{n=n_0+1}^\infty \frac{1}{(il)^{n+1}} \int_{-\infty}^\infty
h_u^{(n)}(\alpha_u) \:e^{-ius} \:du
\:+\:{\mbox{(rapid decay in $l$)}} \nonumber \\
&&+\: {\mbox{(higher orders in
$(\alpha_{\mbox{\scriptsize{max}}}/E_P)$}}\:.
    \label{p:26}
\end{eqnarray}
We point out that, according to (\ref{p:25n}),
\[ \alpha_{\mbox{\scriptsize{max}}}/E_P \ll
(l_{\mbox{\scriptsize{max}}}E_P)^{-1}\:, \]
and this explains why we disregard the higher orders in $\alpha_{\mbox{\scriptsize{max}}}/E_P$.
In our case, the function $h_u$ has in the low-energy region
according to (\ref{p:2rr}) the form $h_u(\alpha_u) = m/(32 \pi^3)\:
\Theta(u)$. Hence it is natural to assume that
(\ref{p:27h}) is satisfied for $n_0=0$. Introducing the shorter notation
\begin{equation}
h(u) \;:=\; h_u(\alpha(u)) \:,\;\;\;\;\; h^{[n]}(u) \;:=\;
h_u^{(n)}(\alpha_u) \:,\;\;\;\;\; \alpha(u) \;:=\;
\alpha_u \:,
    \label{p:27x}
\end{equation}
we have thus derived the following result.\\[.5em]
{\bf{Expansion of the scalar component:}} {\em{Close to the light cone
(\ref{p:2z}, \ref{p:2s}), the scalar component (\ref{p:25s}) of
the fermionic projector of the vacuum has the expansion}}
\begin{eqnarray}
\lefteqn{ P(s,l) \;=\; \frac{1}{il} \sum_{k=0}^\infty \frac{(-il)^k}{k!}
\int_{-\infty}^\infty h \:\alpha^k \:e^{-ius} \:du } \label{p:27a} \\
&&+\sum_{n=1}^\infty \frac{1}{(il)^{n+1}} \int_{-\infty}^\infty
h^{[n]} \:e^{-ius} \:du \label{p:27b} \\
&&+\:{\mbox{(rapid decay in $l$)}} \:+\: {\mbox{(higher orders in
$(\alpha_{\mbox{\scriptsize{max}}}/E_P)$}} \spc
    \label{p:27c}
\end{eqnarray}
{\em{with suitable regularization functions $h$, $h^{[n]}$ and $\alpha$.
In the low-energy region $u \ll E_P$, the regularization functions are}}
\begin{equation}
    h(u) \;=\; \frac{m}{32 \pi^3} \:\Theta(u) \:,\spc
    h^{[n]}(u) \;=\; 0 \:,\spc \alpha(u) \;=\; \alpha_u \;=\;
    \frac{m^2}{u} \:.
    \label{p:2rs}
\end{equation}

In this expansion, the $l$-dependence is written out
similar to a Laurent expansion. The main simplification compared to
our earlier Fourier representation (\ref{p:2k}) is that the
dependence on the regularization is now described by functions of only
one variable, denoted by $h$, $h^{[n]}$ and $\alpha$. In composite
expressions in $P(s,l)$, we will typically get convolutions of these
functions; such one-dimensional convolutions are convenient and
can be easily analyzed. The simplification to one-dimensional
regularization functions became possible because many details of the
regularization affect only the contribution with rapid decay in $l$,
which we do not consider here. Notice that the summands in
(\ref{p:27a}) and (\ref{p:27b}) decay like $(l
\:\alpha_{\mbox{\scriptsize{max}}})^k/k! \ll
(l/l_{\mbox{\scriptsize{max}}})^k/k!$ and $(l E_P)^{-n}$,
respectively. In the low-energy limit (\ref{p:2rs}), the expansion
(\ref{p:27a}) goes over to a power series in $m^2$, and we thus refer to
(\ref{p:27a}) as the {\em{mass expansion}}\index{mass expansion}.
In the mass expansion, the regularization is
described by only two functions $h$ and $\alpha$. The series
(\ref{p:27b}), on the other hand, is a pure regularization effect and
is thus called the {\em{regularization
    expansion}}\index{regularization expansion}. It involves an
infinite number of regularization functions $h^{[n]}$. Accordingly, we
will use the notions of mass and regularization expansions also for
other expansions of type (\ref{p:26}).

In the expansion (\ref{p:26}), the fermionic projector is described
exclusively in terms of the function $h(u,v)$ in a neighborhood of the
discontinuity along the curve $(u, \alpha_u)$. Let us go back to the
definition of $h$, (\ref{p:2q}), and consider what this result means
for the regularized fermionic projector in momentum space
(\ref{p:25s}). In the case without regularization (\ref{p:2rr}), we saw that
integrating out the cylindrical coordinates $\rho$ and $\varphi$
yields a discontinuity of $h$ whenever the $2$-plane $(\omega,
k)={\mbox{const}}$ meets and is tangential to the hyperboloid
$\omega^2 - k^2 - \rho^2 = m^2$. This picture is true in the general
case in the sense that the discontinuity of $h$ can be associated
to a contribution to $\hat{P}$ which describes a hypersurface in
four-dimensional momentum space. The simplest way to recover the
discontinuity of $h$ when integrating out the cylindrical coordinates
would be to choose $\hat{P}$ of the form (\ref{p:25s}) with a function
$\phi$ and the spherically symmetric distribution $f=\delta(|\vec{p}| - \omega -
\alpha(-|\vec{p}| - \omega))$. Since spherically symmetric
regularizations seem too restrictive, it is preferable to describe
the discontinuity of $h$ more generally by a contribution to $\hat{P}$
of the form
\begin{equation}
\phi(\vec{p}) \;\delta(\omega - \Omega(\vec{p})) \:,
    \label{p:27e}
\end{equation}
which is singular on the hypersurface $\omega=\Omega(\vec{p})$. For small
momentum $|\vec{p}| \ll E_P$, the surface should clearly go over to
the mass shell given by $\Omega = -\sqrt{|\vec{p}|^2 + m^2}$ and
$\phi=m/|2 \Omega|$; also, it is reasonable to assume that $\phi$
and $\Omega$ are smooth and sufficiently regular. This
consideration shows that for the behavior of the fermionic
projector on the light cone (\ref{p:26}), the essential role is
played by states lying on a hypersurface. We refer to these one-particle
states as the {\em{surface states}} \label{pl4}\index{surface states} of the fermionic projector
of the vacuum. This result seems physically
convincing because the surface states naturally generalize the states
on the lower mass shell known from relativistic quantum mechanics.
By integrating out the cylindrical coordinates for the ansatz
(\ref{p:27e}), one can express the regularization functions
$h_u^{(n)}$ in (\ref{p:26}) in terms of $\phi$ and the geometry of the
hypersurface. But we point out that, in contrast to the just discussed
discontinuity of $h$, the partial derivatives of $h$ depend also on
states other than surface states. For example, a contribution to
$\hat{P}$ of the form $b(\omega, \vec{p}) \:\Theta(\omega -
\Omega(\vec{p}))$ with $\Omega$ as in (\ref{p:27e}) and a smooth
function $b$ has a discontinuity on the surface $\Omega$ and affects
all the regularization functions $h_u^{(n)}$ for
$n \geq 1$ (as one verifies by a short computation).
Thinking of the decomposition of the fermionic projector into the
one-particle states, such non-surface contributions would consist of a
large number of states and would thus make it necessary to introduce
many additional fermions into our system. It does not seem quite
reasonable or appropriate to considerably increase the number of particles
of the system with the only purpose of having more freedom for
the derivative terms of $h$ in (\ref{p:26}). It seems easiest and
physically most convincing to assume that all the regularization
functions in (\ref{p:26}) come about as a consequence of surface
states. We refer to this assumption as the {\em{restriction to surface
states}}\index{surface states!restriction to}. It is of no relevance for the scalar component
(\ref{p:27a}, \ref{p:27c}), but it will yield an important relation
between the regularization functions for the vector component in the
next section. To avoid confusion, we point out that the restriction
to surface states clearly does not imply that $\hat{P}$ is of the
form (\ref{p:27e}). It imposes a condition only on the behavior of
$\hat{P}$ in a neighborhood of our hypersurface; namely that the
only distributional or non-regular contribution to $\hat{P}$ in this
neighborhood should be the hypersurface itself.

For clarity, we finally review our assumptions on the regularization.
Our first assumption was that the function $h(u,v)$ has, for every
fixed $u$, at most two discontinuities at $\alpha(u)$ and $\beta(u)$,
and is sufficiently regular otherwise (\ref{p:2v}). Furthermore,
the function $\beta(u)$ had to be monotone and again sufficiently
regular. For the function $\alpha(u)$, we assumed that (\ref{p:25n})
holds. Since $h$ is obtained
from $\hat{P}$, (\ref{p:25s}), by integrating out the cylindrical
coordinates (\ref{p:2q}), these assumptions implicitly pose
conditions on the fermionic projector of the vacuum. Although they
could clearly be weakened with more mathematical effort, these
conditions seem sufficiently general for the moment. In order to
understand this better, one should realize that integrating out the
cylindrical coordinates does generically (i.e.\ unless
there are singularities parallel to the plane
$(\omega,k)={\mbox{const}}$) improve the regularity. The restriction to
the generic case is in most situations justified by the fact that the direction
$y-x$ and the coordinate system in (\ref{p:2pq}) can be freely chosen.
Using the above assumptions on $h(u,v)$, we showed that the dominant
contribution to the fermionic projector on the light cone comes from
states on a hypersurface in four-dimensional momentum space. With
the ``restriction to surface states'' we assumed finally that the
behavior on the light cone (\ref{p:26}) is completely characterized
by these states.

\section{The Regularized Vacuum on the Light Cone, Vector Component}
\label{psec25} \setcounter{equation}{0}
We shall now extend the previous analysis to the
vector component\index{vector component} in (\ref{p:2f2}). More precisely, we will analyze
the Fourier integral (\ref{p:2k}) for
\begin{equation}
    \hat{P}(p) \;=\; v_j(p) \:\gamma^j \:f(p)
    \label{p:25v}
\end{equation}
close to the light cone. We again choose light-cone coordinates $(s,l,x_2,
x_3)$ with $y-x=(s,l,0,0)$ ($s$ and $l$ are given by (\ref{p:2q1}),
while $x_2$ and $x_3$ are Cartesian coordinates in the orthogonal
complement of the $sl$-plane). The associated momenta are denoted by
$p=(u,v,p_2, p_3)$ with $u$ and $v$ according to (\ref{p:2q2}).
As in (\ref{p:2pp}), we integrate out the coordinates
perpendicular to $u$ and $v$,
\begin{equation}
h_j(u,v) \;:=\; \frac{1}{2 \:(2 \pi)^4} \int_{-\infty}^\infty dp_2
\int_{-\infty}^\infty dp_3 \; (v_j \:f)(u,v,p_2,p_3) \:,
    \label{p:28a}
\end{equation}
and obtain a representation of the fermionic projector involving
two-dimensional Fourier integrals
\[ P(s,l) \;=\; \gamma^j \:P_j(s,l) \]
with
\begin{equation}
P_j(s,l) \;:=\; \int_{-\infty}^\infty du \int_{-\infty}^\infty dv \;
h_j(u,v) \:e^{-i(us + vl)} \:. \label{p:28b}
\end{equation}
The tensor indices in (\ref{p:28a}) and (\ref{p:28b}) refer to the
coordinate system $(s,l,x_2,x_3)$. For clarity, we denote the range of
the indices by $j=s,l,2,3$; thus
\begin{equation}
\gamma^s \;=\; \frac{1}{2}\:(\gamma^0 - \gamma^1) \:,\spc
\gamma^l \;=\; \frac{1}{2}\:(\gamma^0 + \gamma^1) \:,
    \label{p:28c}
\end{equation}
where $\gamma^0,\ldots,\gamma^3$ are the usual Dirac matrices of
Minkowski space. According to the continuum kernel (\ref{p:2b}),
$\hat{P}$ has in the case without regularization the form $\hat{P}=p
\slsh \:\delta(p^2 - m^2) \:\Theta(-p^0)$ and $h_j$ can be computed
similar to (\ref{p:2rr}) to be
\begin{equation}
\gamma^j \:h_j(u,v) \;=\; \frac{1}{32 \pi^3} (-u \gamma^s - v \gamma^l)
\:\Theta(uv - m^2) \:\Theta(u) \:.
    \label{p:29}
\end{equation}
This limiting case specifies the regularized $h_j(u,v)$ for small
energy-momentum $u,v \ll E_P$. In order to keep the form of the
functions $h_j$ in the high-energy region sufficiently general, we
merely assume in what follows that the $h_j$ satisfy all the
conditions we considered for the function $h$ in the previous
section (see the summary in the last paragraph of {\S}\ref{psec24}).
Our main result is the following.\\[.5em]
{\bf{Expansion of the vector component:}} {\em{Close to the light cone
(\ref{p:2z}, \ref{p:2s}), the vector component (\ref{p:25v}) of
the fermionic projector of the vacuum has the expansion $P = \gamma^j
P_j$ with}}

\begin{eqnarray}
\lefteqn{ P_s(s,l) \;=\; \frac{1}{il} \sum_{k=0}^\infty \frac{(-il)^k}{k!}
\int_{-\infty}^\infty -u \:g_s \: \:\alpha^k
\:e^{-ius} \:du } \nonumber \\
&&+\sum_{n=1}^\infty \frac{1}{(il)^{n+1}} \int_{-\infty}^\infty
- u \:g_s^{[n]} \:e^{-ius} \:du \nonumber \\
&&+\:{\mbox{(rapid decay in $l$)}} \:+\: {\mbox{(higher orders in
$(\alpha_{\mbox{\scriptsize{max}}}/E_P)$}} \label{p:210} \\
\lefteqn{ P_l(s,l) \;=\; \frac{1}{(il)^2} \sum_{k=0}^\infty \frac{(-il)^k}{k!}
\int_{-\infty}^\infty \left[ (k-1) \: \alpha^k \:+\: k
\:\frac{b}{u} \: \alpha^{k-1} \right] \:g_l
\:e^{-ius} \:du } \nonumber \\
&&+\sum_{n=1}^\infty \frac{1}{(il)^{n+2}} \int_{-\infty}^\infty
- (n+1)\: g_l^{[n]} \:e^{-ius} \:du \nonumber \\
&&+\:{\mbox{(rapid decay in $l$)}} \:+\: {\mbox{(higher orders in
$(\alpha_{\mbox{\scriptsize{max}}}/E_P)$}}\label{p:211} \\
\lefteqn{ P_{2\!/\!3}(s,l) \;=\; \frac{1}{(il)^2} \sum_{k=0}^\infty \frac{(-il)^k}{k!}
\int_{-\infty}^\infty \left[ \alpha^k \:+\: k
\:\frac{b_{2\!/\!3}}{u} \: \alpha^{k-1} \right] \:g_{2\!/\!3}
\:e^{-ius} \:du } \nonumber \\
&&+\sum_{n=1}^\infty \frac{1}{(il)^{n+2}} \int_{-\infty}^\infty
g_{2\!/\!3}^{[n]} \:e^{-ius} \:du \nonumber \\
&&+\:{\mbox{(rapid decay in $l$)}} \:+\: {\mbox{(higher orders in
$(\alpha_{\mbox{\scriptsize{max}}}/E_P)$}} \spc \label{p:212}
\end{eqnarray}
{\em{and suitable regularization functions $g_j$, $g_j^{[n]}$, $b$,
$b_{2\!/\!3}$ and the mass regularization function $\alpha$ as in
(\ref{p:27a}, \ref{p:2rs}). In the low energy region $u \ll E_P$, the
regularization functions have the form}}
\begin{eqnarray}
g_s(u) &=& \frac{1}{32 \pi^3} \:\Theta(u) \:,\spc
g_s^{[n]}(u) \;=\; 0 \label{p:regs} \\
g_l(u) &=& \frac{1}{32 \pi^3} \:\Theta(u) \:,\spc
g_l^{[n]}(u) \;=\; b(u) \;=\; 0 \label{p:2rv2} \\
g_{2\!/\!3}(u) &=& g_{2\!/\!3}(u) \;=\; b_{2\!/\!3}(u) \;=\; 0 \:.
    \label{p:59c}
\end{eqnarray}

Before entering the derivation, we briefly discuss these formulas.
To this end, we consider the situation where, like in the case without
regularization, the vector $v(p)$ in (\ref{p:25v}) points into the
direction $p$. In this case we can write the vector component as
\begin{equation}
    \hat{P}(p) \;=\; p_j \gamma^j \:(\phi f)(p) \:,
    \label{p:2pr}
\end{equation}
where $(\phi f)$ has the form of the scalar component considered
in {\S}\ref{psec24}. Since multiplication in momentum space
corresponds to differentiation in position space, we obtain for
(\ref{p:28b})
\[ P(s,l) \;=\; -i \left( \gamma^s \frac{\partial}{\partial s} +
\gamma^l \frac{\partial}{\partial l} +
\gamma^2 \frac{\partial}{\partial x^2} +
\gamma^3 \frac{\partial}{\partial x^3} \right)
P_{\mbox{\scriptsize{scalar}}}(s,l) \:, \]
where $P_{\mbox{\scriptsize{scalar}}}$ is the scalar component
(\ref{p:2r}) with $h$ as in (\ref{p:2q}). We now substitute for
$P_{\mbox{\scriptsize{scalar}}}$ the expansion on the light cone
(\ref{p:27a}--\ref{p:27c}) and carry out the partial
derivatives. For the $s$- and $l$-components, this gives
exactly the expansions (\ref{p:210}, \ref{p:211}) with
\begin{equation}
g_s \;=\; g_l \;=\; h \:,\spc g_s^{[n]} \;=\; g_l^{[n]} \;=\; h^{[n]}
\:,\spc b \;=\; 0 \:. \label{p:2M1}
\end{equation}
For the components $j=2,3$, the calculation of the partial
derivatives is not quite so straightforward because the expansion of
the scalar component (\ref{p:27a}--\ref{p:27c}) was carried out
for fixed $x_2$ and $x_3$. Nevertheless, one can deduce also the
expansion (\ref{p:212}) from (\ref{p:27a}--\ref{p:27c}) if one
considers $x_2$ and $x_3$ as parameters of the regularization
functions $h$, $h^{[n]}$ and $\alpha$, and differentiates through,
keeping in mind that differentiation yields a factor $1/l$ (to get the
scaling dimensions right). In this way, the simple example (\ref{p:2pr})
explains the general structure of the expansions
(\ref{p:210}--\ref{p:212}). We point out that the regularization
function $b$ vanishes identically in (\ref{p:2M1}). This means that
$b$ is non-zero only when the direction of the vector field $v$ is modified
by the regularization. Thinking in terms of the decomposition into the
one-particle states, we refer to this regularization effect as the
{\em{shear of the surface states}}\index{surface states!shear of the}.

We shall now derive the expansions (\ref{p:210}--\ref{p:212}).
Since the Fourier integrals in (\ref{p:28b}) are of the form
(\ref{p:2r}), they have the expansion (\ref{p:26}), valid
close to the light cone (\ref{p:2z}, \ref{p:2s}). It remains to determine
the parameter $n_0$ in (\ref{p:26}). We consider the
components $j=s,l,2$ and $3$ separately.
According to (\ref{p:29}), the function $h_s$ in the
low-energy region looks similar to the hyperbola depicted in
Figure~\ref{fig2}. The main difference to the low-energy behavior of
the scalar component (\ref{p:2rr})
is the additional factor $u$ in $h_s$ which grows linearly along
branch ``A'' of the hyperbola. Thus in the low-energy region away from the
origin,
\begin{equation}
(h_s)_u(\alpha_u) \;\sim\; E_P \spc{\mbox{and}}\spc
\max_{v \in (0, E_P)} |(h_s)_u(v)| \;\sim\; E_P \:.
\label{p:29hs}
\end{equation}
Hence it is natural to assume that $h_s$ satisfies the
bound (\ref{p:27h}) with $n_0=0$. Because of
the linearly growing factor $u$ in the low-energy region, it is
convenient to write the regularization functions in the form
\begin{equation}
(h_s)_u(\alpha_u) \;=:\; -u\:g_s(u) \:,\spc
(h_s)_u^{(n)}(\alpha_u) \;=:\; -u\:g_s^{[n]}(u)
    \label{p:213}
\end{equation}
with suitable functions $g_s$ and $g_s^{[n]}$ (this can be done because,
as explained after (\ref{p:25}), close to the origin $u=0$, we can
work with the unregularized fermionic projector). This yields the
expansion (\ref{p:210}). According to (\ref{p:29}) and
(\ref{p:213}), the regularization functions have the low-energy limit
(\ref{p:regs}).
For the $l$-component, the situation is much different. According to
(\ref{p:29}), the function $h_l$ in the low-energy limit has the form
\begin{equation}
h_l(u,v) \;=\; -\frac{1}{32 \pi^3} \:v \:\Theta(uv-m^2) \:.
    \label{p:27f}
\end{equation}
The factor $v$ decreases like $m^2/u$ along branch ``A'' of the hyperbola.
Thus in the low-energy region away from the origin,
\begin{equation}
(h_l)_u(\alpha_u) \;\sim\; m^2/E_P \spc{\mbox{whereas}}\spc
\max_{v \in (0, E_P)} |(h_l)_u(v)| \;\sim\; E_P \:.
\label{p:29hl}
\end{equation}
Therefore, we cannot assume that $h_l$ satisfies the bound (\ref{p:27h})
with $n_0=0$. But $(h_l)^{(1)}_u(\alpha_u) \sim 1$ in the low-energy region,
and thus we may choose $n_0=1$. We conclude that it is necessary to
take into account two inner summands in (\ref{p:26}), more precisely
\begin{eqnarray}
P_l(s,l) &=& \frac{1}{(il)^2} \sum_{k=0}^\infty \frac{(-il)^k}{k!}
\int_{-\infty}^\infty \left[ (h_l)^\prime_u(\alpha_u) \:\alpha_u^k
\:-\: k \:(h_l)_u(\alpha_u) \: \alpha_u^{k-1} \right] e^{-ius} \:du
\nonumber \\
&&+\: \cdots , \label{p:27g}
\end{eqnarray}
where ``$\cdots$'' stands for the regularization expansion and all
terms neglected in (\ref{p:26}). In the low-energy region, we have
according to (\ref{p:27f}, \ref{p:2rs}),
\[ (h_l)_u(\alpha_u) \;=\; -\frac{1}{32 \pi^3}\:\frac{m^2}{u} \;=\;
(h_l)^\prime_u(\alpha_u) \:\alpha_u \: . \]
Thus in this region, the two summands in the square brackets of
(\ref{p:27g}) are of the same order of magnitude, and none of them can be
neglected. In view of the low-energy limit, we introduce the regularization
functions as
\begin{eqnarray}
(h_l)^\prime_u(\alpha_u) &=:& -g_l(u) \nonumber \\
(h_l)^{[1+n]}_u (\alpha_u) &=:& -(n+1) \:g^{[n]}_l(u) \nonumber \\
(h_l)^\prime_u(\alpha_u) \:\alpha_u - (h_l)_u(\alpha_u) &=:&
\frac{b(u)}{u} \:g_l(u) \:;
    \label{p:214}
\end{eqnarray}
this yields the expansion (\ref{p:211}). According to (\ref{p:29}), the
regularization functions have the low-energy limit (\ref{p:2rv2}).
We finally consider the components $j=2$ and $3$. According to (\ref{p:29}),
these components are identically equal to zero in the low-energy limit.
But for $u \sim E_P$, they might behave similar to $P_s$ or $P_l$. To be on
the safe side, we choose $n_0=1$. Denoting the regularization
functions by
\begin{eqnarray}
(h_{2\!/\!3})^\prime_u(\alpha_u) &=:& g_{2\!/\!3}(u) \nonumber \\
(h_{2\!/\!3})^{[1+n]}_u (\alpha_u) &=:& g^{[n]}_{2\!/\!3}(u) \nonumber \\
- (h_{2\!/\!3})_u(\alpha_u)  &=:& \frac{b_{2\!/\!3}(u)}{u}
\:g_{2\!/\!3}(u) \:,
    \label{p:215}
\end{eqnarray}
we obtain the expansion (\ref{p:212}). According to
(\ref{p:29}), the regularization functions $g_{2\!/\!3}$,
$g_{2\!/\!3}^{[h]}$ and $b_{2\!/\!3}$ vanish in the low-energy
region, (\ref{p:59c}).

For clarity, we point out that choosing $n_0=1$ (as in (\ref{p:211}, \ref{p:212})) is a generalization of setting $n_0=0$ (as in (\ref{p:210})),
obtained by taking into account more summands of the expansion
(\ref{p:25}). Nevertheless, the different behavior in the low-energy
region (\ref{p:29hs}, \ref{p:29hl}) suggests that
(\ref{p:211}) and
(\ref{p:212}) should not be merely more general formulas
than (\ref{p:210}), but that the behavior of $P_j(s,l)$, $j=l,2,3$, should
be really different from that of $P_s(s,l)$. We shall now make this difference
precise.
Comparing (\ref{p:29hs}) and (\ref{p:29hl}) (and using that
$h_{2\!/\!3}$ vanishes in the low-energy region), it is reasonable to impose
that there should be a constant $\varepsilon_{\mbox{\scriptsize{shear}}}>0$
\label{varepsilon_shear} with
\begin{equation}
|(h_j)_u(\alpha_u)| \;<\; \varepsilon_{\mbox{\scriptsize{shear}}} \:
|(h_s)_u(\alpha_u)| \spc {\mbox{for $u \sim E_P$ and $j=l,2$, or $3$}}.
\label{p:2Es}
\end{equation}
In view of (\ref{p:29hs}) and (\ref{p:29hl}),
$\varepsilon_{\mbox{\scriptsize{shear}}}$ should be as small as
\begin{equation}
\varepsilon_{\mbox{\scriptsize{shear}}} \;\sim\; \frac{m^2}{E_P^2} \:.
\label{p:2Et}
\end{equation}
However, if the surface states have shear (as defined earlier in this
section), the constant $\varepsilon_{\mbox{\scriptsize{shear}}}$ must
in general be chosen larger. In order to keep our analysis as general as
possible, we will not specify here how
$\varepsilon_{\mbox{\scriptsize{shear}}}$ scales in the Planck energy,
but merely assume that $m^2/E_P^2 < \varepsilon_{\mbox{\scriptsize{shear}}}
\ll 1$. Using (\ref{p:213}), (\ref{p:214}) and (\ref{p:215}), the
condition (\ref{p:2Es}) can be expressed in terms of the regularization
functions $g_j$ and $b_j$ as
\begin{equation}
\left(\frac{b}{u}+\alpha_u \right) g_l ,\; \frac{b_{2\!/\!3}}{u}\: g_{2\!/\!3}
\;<\; \varepsilon_{\mbox{\scriptsize{shear}}} \:u\:g_s
\spc {\mbox{for $u \sim E_P$.}}
\label{p:2Eu}
\end{equation}
It is interesting to discuss what the condition (\ref{p:2Es})
means for the functions $P_j$. We begin with the case without
regularization. In this case, the vector component of $P(x,y)$
points into the direction $y-x$, more precisely $P(x,y) = i
(y-x)_j \gamma^j \:S(x,y)$ with a scalar distribution $S$. In a
composite expression like the closed chain $P(x,y) \:P(y,x)$, one
can contract the tensor indices and obtains in a formal
calculation $P(x,y) \:P(y,x) = (y-x)^2 \: S(x,y) \:S(y,x)$ with a
scalar factor $(y-x)^2$ which vanishes on the light cone. Let us
consider this contraction in our light-cone coordinates. Before
the contraction, each factor $(y-x)_j \gamma^j = 2l \:\gamma^s +
2s \:\gamma^l \approx 2l \gamma^s$ is, if we take only the leading
contribution on the light cone (i.e.\ the lowest order in $s/l$),
proportional to $l$. After the contraction, however, the product
$(y-x)^2 = 4 ls$ is proportional to both $l$ and $s$. Thus the
contraction yields, to leading order on the light cone, a
dimensionless factor $s/l$. While the factor $l^{-1}$ changes the
scaling behavior in the ``large'' variable, the factor $s$ tends
to make the composite expression ``small'' near the light cone.
The analysis of the scaling behavior in $l$ can immediately be
extended to the case with regularization by looking at the
expansions (\ref{p:210}) and (\ref{p:211}). Let us consider as an
example the leading term of the mass expansion. For the expansion
(\ref{p:210}), this is the summand $k=0$, and it scales like
$P_s(s,l) \sim 1/l$. If we assume that (\ref{p:2Es}) holds with
$\varepsilon_{\mbox{\scriptsize{shear}}}$ according to
(\ref{p:2Et}), then (\ref{p:2Eu}) shows that $b(u) \sim 1$, and
the summands in the square bracket in (\ref{p:211}) are of
comparable size. Hence the leading term of the expansion
(\ref{p:211}) is also the summand $k=0$, and it scales in $l$ like
$P_l(s,l) \sim 1/l^2$. Hence the leading term of the sum $\gamma^l
P_l + \gamma^s P_s$ behaves like $P \sim 1/l +
{\mathcal{O}}(1/l^2)$. Since $s$ and $l$ are null directions, a
contraction of the tensor indices in the closed chain leads only
to mixed products of the form $P_s \:P_l$, and this mixed product
scales in $l$ like $P_s \:P_l \sim 1/l^3$. Thus, exactly as in the
case without regularization, the contraction of the tensor indices
yields an additional factor $l^{-1}$. If on the other hand, the
condition (\ref{p:2Es}) were violated, the regularization function
$b$ could be chosen arbitrarily large. But if $b$ becomes large
enough, the cominant contribution to (\ref{p:211}) is the summand
$k=1$ (notice that $b$ does not appear in the summand $k=0$), and
hence $P_l(s,l) \sim 1/l$. This implies that $P_s \:P_l \sim
1/l^2$, and the contraction does no longer yield an additional
factor $l^{-1}$. This consideration is immediately extended to the
components $P_{2\!/\!3}$ by considering the $l$-dependence of the
summands in (\ref{p:212}). We conclude that the condition
(\ref{p:2Es}) with $\varepsilon_{\mbox{\scriptsize{shear}}} \ll 1$
means that the contraction of the tensor indices yields a scalar
factor which is small on the light cone. We refer to this
condition by saying that the {\em{vector component is null on the
light cone}}\index{vector component!is null on the light cone}.
If one wishes, one can simply take this condition as
an additional assumption on the fermionic projector of the vacuum.
However, the property of the vector component being null on the
light cone also arises in the study of composite expressions in
the fermionic projector as a compatibility condition and can thus
be derived from the equations of discrete space-time (see
Remark~\ref{vcnlc})).

The next question is if our regularization functions $\alpha$,
$g_j$, $g_j^{[n]}$ and $b$, which appear in our expansions
(\ref{p:210}--\ref{p:212}), are all independent of each other,
or whether there are some relations between them. Recall that the
regularization functions are given in terms of the boundary values
of the functions $\partial_v^n h_j(u,v)$, $n \geq 0$, on the curve
$(u, \alpha_u)$ (see (\ref{p:213}, \ref{p:214}, \ref{p:215})).
Since the $(h_j)_{j=s,l,2,3}$ were treated in our
two-dimensional Fourier analysis as four independent and (apart
from our regularity assumptions) arbitrary functions, we can
certainly not get relations between the regularization functions
by looking at the situation in the $uv$-plane. But we can hope
that when we consider the surface states in four-dimensional
momentum space (as introduced in {\S}\ref{psec24}), the geometry
of the hypersurface defined by these states might yield
useful restrictions for the regularization functions. First
of all, we mention that our discussion of surface states of the
previous section applies without changes also to the vector
component; we will in what follows make use of the {\em{restriction to
surface states}}. Since in the low-energy region the regularization
is irrelevant and the results of~{\S}\ref{jsec5} apply, we can
furthermore restrict attention to large energy and momentum
$\omega, |\vec{k}| \sim E_P$. We choose polar coordinates
$(\omega, k=|\vec{k}|, \vartheta, \varphi)$ in momentum space and
introduce the ``mass shell coordinates''
\begin{equation}
    U \;=\; -|\vec{k}| - \omega \:,\spc V \;=\; |\vec{k}| - \omega
    \:.
    \label{p:2A}
\end{equation}
Notice that, in contrast to the coordinates $u$ and $v$,
(\ref{p:2q2}), the variables (\ref{p:2A}) are the spherically
symmetric part of a four-dimensional coordinate system $(U, V,
\vartheta, \varphi)$. Extending also the notation (\ref{p:28c}) in a
spherically symmetric way, we introduce the Dirac matrix
\[ \gamma^S \;=\; \frac{1}{2} \left( \gamma^0 \:-\: \frac{\vec{\gamma}
\vec{k}}{k} \right) . \]
Let us consider what the expansions (\ref{p:210}--\ref{p:212}) tell us about the
surface states.  Similar as explained before
(\ref{p:27e}), the discontinuities of $h_j$ come about in (\ref{p:28a}) when
the plane $(u,v)={\mbox{const}}$ meets and is tangential to the hypersurface of
the surface states.  We denote the tangential intersection point of the surface
$(u,v)={\mbox{const}}$ with the hypersurface by $Q=(U,V,\vartheta,\varphi)$.  In
the high-energy region under consideration, the variable $U$ is of the order
$E_P$.  The variable $V$, on the other hand, will be of order $\alpha(U) <
\alpha_{\mbox{\scriptsize{max}}}$.  Thus our hypersurface is close to the mass
cone in the sense that $V/U \sim \alpha_{\mbox{\scriptsize{max}}}/E_P \ll 1$.
As a consequence, the coordinate $\vartheta$ of the intersection point $Q$ must
be small (more precisely, $\vartheta \leq \sqrt{\alpha_{\mbox{\scriptsize{max}}}
/ E_P}$), and we conclude that, to leading order in
$\alpha_{\mbox{\scriptsize{max}}}/E_P$, $V = \alpha(U)$. Hence we
can write the hypersurface as a graph $V = A(U, \vartheta,
\varphi)$ with a function $A$ satisfying the condition
\[ A(U, \vartheta=0) \;=\; \alpha(U) \:+\: {\mbox{(higher orders
in $\alpha_{\mbox{\scriptsize{max}}}/E_P$)}} \:. \]
One can think of the function~$A(u, \vartheta, \varphi)$ as the
extension of $\alpha$ to the four-dimensional setting. In order to
determine the structure of the Dirac matrices, we first recall that
the assumption that the vector component is null on the light cone
implied in our consideration after (\ref{p:210}) that the parameter
$n_0$ corresponding to $P_l$, $P_2$ and $P_3$ was equal to one. This
means that to leading order in
$\alpha_{\mbox{\scriptsize{max}}}/E_P$, only the function $h_s(u,v)$
is discontinuous on the curve $(u, \alpha_u)$, and we conclude that
the distribution $\hat{P}$ is on the hypersurface at the point $Q$
a scalar multiple of $\gamma^s$; we use the short notation
$\hat{P}(Q) \sim \gamma^s$. Using again that $\vartheta$ is small,
we obtain that to leading order in
$\alpha_{\mbox{\scriptsize{max}}}/E_P$, $\hat{P}(U, A(U,\vartheta=0),
\vartheta=0) \sim \gamma^s$. Since the spatial direction of
the vector $y-x$ in (\ref{p:2k}) can be chosen arbitrarily, we can by
rotating our coordinate system immediately extend this result to
general $\vartheta$ and $\varphi$, and obtain that $\hat{P}(U,
\alpha(U, \vartheta, \varphi), \vartheta, \varphi) \sim \gamma^S$.
Hence the surface states are described by a contribution to
$\hat{P}$ of the form
\begin{equation}
-32 \pi^3 \:g(U, \vartheta, \varphi) \:\gamma^S
\;\delta(V-A(U, \vartheta, \varphi)) \:+\:
{\mbox{(higher orders in $\alpha_{\mbox{\scriptsize{max}}}/E_P$)}}
    \label{p:2Aa}
\end{equation}
with some function $g$. It is reasonable to assume that the functions
in (\ref{p:2Aa}) are sufficiently regular. Similar to our regularity
condition (\ref{p:2v}) for $h$, we here assume that the derivatives
of $A$ and $g_S$ have the natural scaling behavior in $E_P$. More
precisely, for all $n_1, n_2, n_3 \geq 0$ there should exist a constant
$c \ll l E_P$ with
\begin{equation}
|\partial_U^{n_1} \partial_\vartheta^{n_2} \partial_\varphi^{n_3}
A(U, \vartheta, \varphi)| \:+\:
|\partial_U^{n_1} \partial_\vartheta^{n_2} \partial_\varphi^{n_3}
g(U, \vartheta, \varphi)| \;\leq\; c \:E_P^{-n_1} \:
\max (|A| + |g|)
    \label{p:2b2}
\end{equation}
for all $U \sim E_P$.

The form of the surface states (\ref{p:2Aa}) allows us to calculate
the regularization functions $g_j$, $g_j^{[n]}$ and $b_j$. For
this, we first represent the matrix $\gamma^S$ in (\ref{p:2Aa}) in
the Dirac basis $(\gamma^j)_{j=s,l,2,3}$; this yields the
contributions of the surface states to the
distributions $(v_j \:f)$. By substituting into (\ref{p:28a}) and
carrying out the integrals over $p_2$ and $p_3$, one obtains the
functions $h_j$. Finally, the regularization functions can be computed
via (\ref{p:213}, \ref{p:214}, \ref{p:215}). This whole
calculation is quite straightforward, and we only state the main
results. To leading order in $v/u$, we can take $A$ and $g$ as constant
functions, and thus the calculation of $\gamma^s h_s + \gamma^l h_l$
reduces to the integral
\begin{eqnarray*}
&& -\frac{1}{\pi} \int_{-\infty}^\infty dp_2 \int_{-\infty}^\infty
dp_3 \; \left( \gamma^s + \frac{v}{u} \:\gamma^l \right) g(u,
\vartheta=0) \: \delta \!\left(\!v-\alpha_u - \frac{p_2^2 + p_3^2}{u}
\!\right) \\
&&+ {\mbox{(higher orders in $v/u$,
$\alpha_{\mbox{\scriptsize{max}}}/E_P$)}} \: .
\end{eqnarray*}
An evaluation in cylindrical coordinates yields that both $g_s(u)$
and $g_l(u)$ are equal to $g(u, \vartheta=0)$, and we thus have the
important relation
\begin{equation}
g_s(u) \;=\; g_l(u) \;=:\; g(u) \:.
    \label{p:2C}
\end{equation}
In the case without shear of the surface states, this relation
was already found in (\ref{p:2M1}); we now see that it holds in a
much more general setting.
The calculation of the angular components $j=2,3$ gives for
$g_{2\!/\!3}$ contributions proportional to $u \:\partial_{2\!/\!3}
A$ and $u \:\partial_{2 \!/\! 3} g$. Unfortunately, this is not
very helpful because we have no information on the derivatives
of $A$ and $g$. The computation of the regularization
functions $g_j^{[n]}$ involves higher derivatives of the functions in
(\ref{p:2Aa}) and becomes quite complicated. We remark that the
above analysis of the surface states can be carried out similarly for
the scalar component of the previous section and gives relations
between the regularization functions $h$ and $h^{[n]}$,
(\ref{p:27x}), but these relations all depend on unknown details of
the geometry of the hypersurface. We thus conclude that (\ref{p:2C}) is
the only relation between the regularization functions which can be
derived with our present knowledge on the surface states.

We finally mention two assumptions on the regularization which, although we will
not use them in the present work, might be worth considering later.  The first
assumption is related to the fact that $P$ should as a projector be idempotent,
$P^2=P$. A formal calculation using (\ref{p:2k}) and (\ref{p:2f2}) yields that
\begin{eqnarray}
(P^2)(x,y) & = & \int \frac{d^4p}{(2 \pi)^4} \:\hat{P}(p)^2 \:e^{-ip(x-y)}
\spc {\mbox{with}}
    \label{p:g1}  \\
\hat{P}(p)^2 & = & \left( 2 \phi(p) \;v_j(p) \:\gamma^j \:+\:
(v_j(p) \:v^j(p) + \phi(p)^2) \right) f(p)^2 \:.
    \label{p:g2}
\end{eqnarray}
In order to make sense out of (\ref{p:g2}), one must regularize in momentum
space, e.g.\ by considering the system in finite 3-volume and take a suitable
limit. Since the results of this analysis depend sensitively on how the
regularization in momentum space is carried out, (\ref{p:g2}) cannot give any detailed
information on the functions $\phi$, $v$, or $f$. The only simple conclusion
independent of the regularization is that the scalars multiplying the factors
$v_j \gamma^j$ in (\ref{p:2f2}) and (\ref{p:g2}) should have the same sign,
and thus $\phi(p) \:f(p)$ should be positive. According to (\ref{p:2q}), this
implies that the regularization function $h$ be positive,
\[ h(u) \;\geq\; 0 \spc {\mbox{for all $u$}}. \]
This assumption is called the {\em{positivity of the scalar
    component}}\index{positivity of the scalar component}.  The
second assumption is obtained by considering the rank of $\hat{P}(p)$.  The $4
\times4$ matrix $(p \slsh + m)$ in the integrand of the unregularized fermionic
projector (\ref{p:2b}) has the special property of being singular of rank two.
This means that the fermionic projector is composed of only two occupied
fermionic states, for every momentum $p$ on the mass shell.  The natural
extension of this property to the case with regularization is that for every $p$
on the hypersurface defined by the surface states, the matrix $\hat{P}(p)$
corresponding to the vector-scalar structure (\ref{p:2f2}) should be of rank
two.  We refer to this property as the assumption of {\em{half occupied surface
states}}\index{surface states!half occupied}.  In terms of the functions $h(u,v)$ and $h_j(u,v)$, it means that
$h_s(u, \alpha(u)) \:h_l(u, \alpha_u) = h(u, \alpha_u)^2$.  Using
(\ref{p:27x}, \ref{p:213}, \ref{p:214}, \ref{p:2C}), the
assumption of half occupied surface states yields the following relation
between the regularization functions of the scalar and vector components,
\begin{equation}
(\alpha(u) \:u + b(u)) \:g(u)^2 \;=\; h(u)^2 \:.
\label{p:hoss}
\end{equation}

\section{The General Formalism}
\label{psec26}  \setcounter{equation}{0}
In this section we shall extend our previous
analysis on the light cone in three ways: to the case with
interaction, to systems of Dirac seas as introduced in~{\S}\ref{jsec3}
and to composite expressions in the fermionic projector. Our first
step is to develop a method which allows us to introduce a
regularization into the formulas of the light-cone expansion
(\ref{fprep}). We here only motivate and describe this method,
the rigorous justification is given in Appendix~\ref{pappB}. Since
the formulas of the light-cone expansion involve the factors
$T^{(n)}$, (\ref{Tldef}, \ref{Tadef}, \ref{Taf}), we
begin by bringing these distributions into a form similar to our
expansion of the regularized scalar component (\ref{p:27a}). By
partly carrying out the Fourier integral (\ref{Taf}) in the
light-cone coordinates introduced in {\S}\ref{psec24} (see
(\ref{p:2q1}, \ref{p:2q2})), we can write the distribution
$T_a$ as
\begin{equation}
T_a(s,l) \;=\; \frac{1}{32 \pi^3} \:\frac{1}{il} \int_0^\infty e^{
-\frac{ial}{u} \:-\: i u s} \:du \:.
    \label{p:2D}
\end{equation}
This formula can be regarded as a special case of the expansion
(\ref{p:25a}) (notice that the function $h(u,v)$ corresponding to
$T_a$ is computed similar to (\ref{p:2rr})), but (\ref{p:2D})
holds also away from the light cone. The distribution $T_a$ is not
differentiable in $a$ at $a=0$, as one sees either directly in
position space (\ref{l:3.1}) or equivalently in (\ref{p:2D}), where formal
differentiation leads to a singularity of the integrand at $u=0$. We
bypassed this problem by working instead of $T_a$
with the distribution $T^{\mbox{\scriptsize{reg}}}_a$ (\ref{Tadef}). Let us
briefly consider what this ``regularization'' means in the integral
representation~(\ref{p:2D}). The formal $a$-derivative of (\ref{p:2D}),
\[ \frac{d}{da} T_a(s,l) \;=\; -\frac{1}{32 \pi^3} \int_0^\infty
\frac{1}{u} \:e^{-\frac{i a l}{u} \:-\: i u s} \:du \:, \]
is well-defined and finite for $a \neq 0$ because of the oscillatory
factor $\exp(-i a l/u)$. However, the limit $a \to 0$ leads to
a logarithmic divergence. Thus one must subtract a logarithmic
counter term before taking the limit; more precisely,
\[ T^{(1)}(s,l) \;=\; -\frac{1}{32 \pi^3}
\:\lim_{a \to 0} \int_{-\infty}^\infty \left[ \frac{1}{u}
\;e^{-\frac{i a l}{u}} \:\Theta(u) \:-\: (1+\log a) \:\delta(u) \right]
e^{-i u s} \:du \:. \]
The higher $a$-derivatives $T^{(n)}$, $n>1$,
are defined similarly using suitable counter\-terms which are localized
at $u=0$. Since we do not need the details in what follows, we simply
write \label{T^(n)}
\begin{equation}
T^{(n)}(s,l) \;=\; -\frac{1}{32 \pi^3} \:
(-il)^{n-1} \int_0^\infty
\left(\frac{1}{u^n}\right)^{\mbox{\scriptsize{reg}}} \:e^{-ius} \:du
\:.
    \label{p:2E}
\end{equation}

Consider a summand of the light-cone expansion (\ref{fprep}),
\begin{equation}
{\mbox{(phase-inserted line integrals)}} \; T^{(n)}(s,l) \:.
    \label{p:2F}
\end{equation}
According to our assumption of macroscopic potentials and wave
functions described in {\S}\ref{psec22}, we shall regularize only
the distribution $T^{(n)}$, keeping the
iterated line integral unchanged. Let us briefly analyze what this
assumption means quantitatively. Not regularizing the iterated
line integral in (\ref{p:2F}), denoted in what follows by $F$,
will be a good approximation if and only if $F$ is nearly constant
on the Planck scale. In other words, not regularizing $F$ is
admissible if we keep in mind that this method can describe the
regularized fermionic projector only modulo contributions of the
order $\partial_jF / E_P$. In the case that this last derivative
acts on the bosonic potentials and fields contained in $F$, we
obtain the limitation already mentioned in {\S}\ref{psec22} that
energy and momentum of the bosonic fields should be small compared
to the Planck energy. More precisely, we can describe the
fermionic projector only to leading order in
$(l_{\mbox{\scriptsize{macro}}} E_P)^{-1}$, where
$l_{\mbox{\scriptsize{macro}}}$ is a typical length scale of
macroscopic physics. A point we did not pay attention to earlier
is that the iterated line integrals also involve factors $(y-x)$
which are contracted with the bosonic potentials and fields.
Thus in light-cone coordinates,
$F$ will in general contain factors of $l$. If the derivative in
$\partial_j F$ acts on a factor $l$, this factor is annihilated.
Hence keeping the iterated line integrals in (\ref{p:2F})
unchanged can describe only the leading order in $(l E_P)^{-1}$ of
the fermionic projector. We conclude that the assumption of
macroscopic potentials and wave functions is justified if and only
if we restrict our analysis to the {\em{leading order in $(l
E_P)^{-1}$ and $(l_{\mbox{\scriptsize{macro}}} E_P)^{-1}$}}. We
remark that going beyond the leading order in $(l E_P)^{-1}$ or
$(l_{\mbox{\scriptsize{macro}}} E_P)^{-1}$ would make it
impossible to describe the interaction by classical fields and is
thus at present out of reach.

The restriction to the leading order in $(l E_P)^{-1}$\index{restriction to the leading order in $(l E_P)^{-1}$} is a
considerable simplification. First of all, we can
neglect all regularization expansions (which are just expansions in
powers of $(l E_P)^{-1}$; see e.g.\ (\ref{p:27b}) and the discussion
thereafter), and thus we do not need to consider the regularization
functions $h^{[n]}$ and $g_j^{[n]}$. Next we compare for given $k$ the
summands in (\ref{p:210}--\ref{p:212}) (the analysis
for fixed $k$ is justified assuming that the vector component is null on the
light cone; see (\ref{p:2Es}) and the discussion thereafter). One sees that
the tensor index $j=s$ gives the leading contribution
in $(l E_P)^{-1}$ to the vector component. This is a great
simplification when tensor indices are contracted in composite
expressions. Namely, when the vector component is contracted
with the bosonic potentials or fields, it suffices to consider the
contribution $P_s$, (\ref{p:210}). If vector components are
contracted with each other, the products of type $P_{2\!/\!3}\:
P_{2\!/\!3}$ are according to (\ref{p:210}--\ref{p:212}) of higher order
in $(l E_P)^{-1}$ or $\varepsilon_{\mbox{\scriptsize{shear}}}$ than
corresponding products of type $P_s \:P_l$. Hence in such contractions,
we must take into account both $P_s$
and $P_l$, but we can again neglect the components $P_2$ and
$P_3$. We conclude that the only
regularization functions which should be of relevance here are those
appearing in (\ref{p:27a}) and in the mass expansions of
(\ref{p:210}) and (\ref{p:211}), i.e.\ the four functions
\begin{equation}
\alpha(u) \:,\;\;\;\;\; g(u) \:,\;\;\;\;\; h(u)\;\;\;\;\;{\mbox{and}}
\;\;\;\;\; b(u)
    \label{p:2G}
\end{equation}
with $g$ given by (\ref{p:2C}).

Under the assumption of macroscopic potentials and wave functions, it suffices
to regularize the factor $T^{(n)}$ in (\ref{p:2F}).
Our method for regularizing $T^{(n)}$ is to
go over to the integral representation (\ref{p:2E}) and to
insert the regularization functions (\ref{p:2G}) into the integrand.
The procedure depends on whether the contribution
to the light-cone expansion is of even or odd order in the mass
parameter $m$. Furthermore, we must treat the factors
$(y-x)_j \gamma^j$ in the light-cone expansion separately. The precise
regularization method is the following.\\[.5em]
{\bf{Regularization of the light-cone expansion:}} {\em{A summand of the
light-cone expansion (\ref{fprep}) which is proportional to $m^p$,}}
\begin{equation}
m^p \:{\mbox{(phase-inserted line integrals)}} \; T^{(n)}(s,l) \:,
    \label{p:2F2}
\end{equation}
{\em{has the regularization}}
\begin{eqnarray}
\lefteqn{ (-1)\: {\mbox{(phase-inserted line integrals)}} } \\
&& \!\!\!\!\!\!\!\times (-il)^{n-1} \int_{-\infty}^\infty du
\left(\frac{1}{u^n}\right)^{\mbox{\scriptsize{reg}}} \:e^{-ius} \times
\left\{ \begin{array}{ll} h(u) \:a(u)^{\frac{p-1}{2}} & {\mbox{for
$p$ odd}} \\
g(u) \:a(u)^{\frac{p}{2}} & {\mbox{for $p$ even}} \end{array} \right.
\nonumber \\
&&\!\!\!\!\!\!\!+\:{\mbox{(rapid decay in $l$)}}
\:+\: {\mbox{(higher orders in $(lE_P)^{-1},
(l_{\mbox{\scriptsize{macro}}} E_P)^{-1},
\varepsilon_{\mbox{\scriptsize{shear}}}$)}} \:.\spc
    \label{p:2F3}
\end{eqnarray}

{\em{A contribution to the light-cone expansion (\ref{fprep}) which is
proportional to $m^p$ and contains a factor $(y-x)_j \gamma^j$,}}
\beq
m^p \;{\mbox{(phase-inserted line integrals)}}
\;(y-x)_j \gamma^j \:T^{(n)}(s,l) \:, \label{p:2M}
\eeq
{\em{is properly regularized according to}}
\begin{eqnarray}
\lefteqn{ (-1) \:
{\mbox{(phase-inserted line integrals)}} } \nonumber \\
&&\!\!\!\!\!\!\times (-il)^{n-1} \int_{-\infty}^\infty du
\left[ 2l \:\gamma^s \left(\frac{1}{u^n}\right)^{\mbox{\scriptsize{reg}}}
\!\!+\: 2i n\:\gamma^l \left(\frac{1}{u^{n+1}}\right)^{\mbox{\scriptsize{reg}}}
\!\!+\: 2l \:b(u) \:\gamma^l
\left(\frac{1}{u^{n+2}}\right)^{\mbox{\scriptsize{reg}}} \right]
\nonumber \\
&&\!\!\!\!\!\!\times \:e^{-ius} \:\times
\left\{ \begin{array}{ll} h(u) \:a(u)^{\frac{p-1}{2}} & {\mbox{for
$p$ odd}} \\
g(u) \:a(u)^{\frac{p}{2}} & {\mbox{for $p$ even}} \end{array} \right.
\:+\: {\mbox{(contributions $\sim \gamma^2, \gamma^3$)}}
\nonumber\\
&&\!\!\!\!\!\!+\:{\mbox{(rapid decay in $l$)}}
\:+\: {\mbox{(higher orders in $(l
E_P)^{-1}, (l_{\mbox{\scriptsize{macro}}} E_P)^{-1},
\varepsilon_{\mbox{\scriptsize{shear}}}$)}} \:.\spc
    \label{p:2N}
\end{eqnarray}
{\em{In these formulas, the regularization function $a$ is given by}}
\begin{equation}
a(u) \;=\; u \:\alpha(u) \:, \label{p:2J}
\end{equation}
{\em{$\varepsilon_{\mbox{\scriptsize{shear}}}$ is defined via (\ref{p:2Es}),
and $l_{\mbox{\scriptsize{macro}}}$ is a macroscopic length scale.}} \\[-.5em]

Let us briefly explain and motivate this regularization method
(see Appendix~\ref{pappB} for the derivation). First of all, we
note that, after writing the factor $(y-x)_j \gamma^j$ together
with the iterated line integrals, the expression (\ref{p:2M}) is
of the form (\ref{p:2F2}), and the regularization rule
(\ref{p:2F3}) applies. Thus (\ref{p:2N}) is an extension of
(\ref{p:2F3}) giving additional information on the $l$-component of
the factor $(y-x)_j \gamma^j$. As we shall see later, this
information is essential when the factor $(y-x)_j$ in (\ref{p:2M})
is to be contracted with another factor $(y-x)^j$ in a composite
expression. To explain the formula (\ref{p:2F3}), we first point
out that the expansions of the scalar and vector components
(\ref{p:27a}--\ref{p:27c}, \ref{p:210}, \ref{p:211}) do
not involve the mass parameter $m$. The reason is that $m$ was
absorbed into the regularization functions $g$, $h$ and $\alpha$,
as one sees by considering the low-energy limit; see
(\ref{p:2rs}, \ref{p:regs}, \ref{p:2rv2}). Furthermore, we
note that each contribution to the mass expansions of the scalar
or vector components contains either a factor $h$ or $g$ (see
(\ref{p:27a}, \ref{p:210}, \ref{p:211})), and it is
therefore reasonable that we should also use exactly one of these
factors here. As a consequence, the power $m^p$ in (\ref{p:2F2})
uniquely determines how many factors of each regularization
function we should take. Namely for even $p$, we must take one
factor $g$ and $p/2$ factors $\alpha$, whereas the case of odd $p$
gives rise to one factor $h$ and $(p-1)/2$ factors $\alpha$. On
the other hand, we know that the insertion of the regularization
functions into (\ref{p:2E}) should modify the behavior of the
integrand only for large $u \sim E_P$; in particular, we should
for small $u$ have a behavior $\sim u^{-n}$. In order to comply
with all these conditions, one must insert the regularization
functions precisely as in (\ref{p:2F3}). In order to motivate
(\ref{p:2N}), we consider the expansion of the vector component
(\ref{p:210}, \ref{p:211}). Recall that the regularization
function $b$ vanishes in the low-energy region (\ref{p:2rv2}) and
describes the shear of the surface states (as explained after
(\ref{p:2M1})). Since this effect is not related to the mass of
the Dirac particle, it is plausible that we should not associate
to $b$ a power of $m$. For the mass expansion of the vector
component, we should thus collect all terms to a given power of
$\alpha$. The contribution $\sim \alpha^k$ to $\gamma^s P_s +
\gamma^l P_l$ takes according to (\ref{p:210}, \ref{p:211})
the form
\[ \frac{1}{il} \:\frac{(-il)^k}{k!} \int_{-\infty}^\infty \left(
-u \:\gamma^s \:+\: \frac{k-1}{il} \:\gamma^l \:-\: \frac{b}{u}
\:\gamma^l \right) g \:\alpha^k \:e^{-ius} \:du \:. \]
In order to obtain the correct behavior in the low-energy region, we
must multiply this formula by $-2l$ and choose $k=n+1$. This
explains the form of the square bracket in (\ref{p:2N}). The
combination of the regularization functions $g$, $h$ and $a$ in
(\ref{p:2N}) can be understood exactly as in (\ref{p:2F3}) using
power counting in $m$.

Our constructions so far were carried out for the case $N=1$ of one
Dirac sea. We will now generalize our regularization method to systems
of Dirac seas as introduced in~(\ref{jsec3}) and will also introduce a
compact notation for the regularization. Exactly as in~{\S}\ref{jsec5}
we only consider the {\em{auxiliary fermionic projector}}, because the
fermionic projector is then obtained simply by taking the partial
trace~(\ref{part}). We first outline how
chiral particles (e.g.\ neutrinos) can be described. Without
regularization, a chiral Dirac sea is obtained by
multiplying the Dirac sea of massless particles with the chiral
projectors $\chi_{L\!/\!R} = \frac{1}{2}\:(\1 \mp \rho)$; for
example in the vacuum and left/right handed particles,
\begin{equation}
\hat{P}(p) \;=\; \chi_{L\!/\!R} \:p \slsh \:\delta(p^2)
\:\Theta(-p^0) \:.
    \label{p:D0}
\end{equation}
The most obvious regularization method is to deduce the regularized
chiral Dirac sea from a Dirac sea regularized with our above methods
again by multiplying from the left with a chiral projector. This simple method
indeed works, under the following assumptions. First, we must
ensure that the regularized fermionic projector of the vacuum is a Hermitian
operator. To this end, we must assume that the scalar component $\phi$
in (\ref{p:2f2}) be identically equal to zero (this generalizes the
requirement of massless particles needed in the case
without regularization). Hence we regularize (\ref{p:D0}) by setting
\[ \hat{P}(p) \;=\; \chi_{L\!/\!R} \:v_j(p) \gamma^j \:f(p)
\:. \]
The expansions near the light cone are then obtained
from (\ref{p:27b}, \ref{p:27c}) and (\ref{p:210}--\ref{p:212})
by setting the scalar regularization functions $h$ and $h^{[n]}$
to zero and by multiplying with $\chi_{L\!/\!R}$. Assuming
furthermore that the bosonic potentials are causality
compatible (see Def.~\ref{ccc}), the formulas of the light-cone
expansion are
regularized likewise by taking the regularizations (\ref{p:2F3},
\ref{p:2N}) with $h$ set identically equal to zero, and by
multiplying from the left by a chiral projector
$\chi_{L\!/\!R}$.

We next consider the generalization to systems
of Dirac seas. In the vacuum, we can describe a system of Dirac
seas by taking a direct sum of regularized Dirac seas
and by using instead of the chiral projectors $\chi_{L\!/\!R}$ the
chiral asymmetry matrix $X$ (see~(\ref{81n})). Since we may choose the
regularization functions for each Dirac sea independently, this
procedure clearly increases the total number of regularization
functions. However, it is natural to impose that the
regularization should respect all symmetries among the Dirac seas.
More precisely, if the fermionic projector of the vacuum contains
identical Dirac seas (e.g.\ corresponding to an underlying color
$SU(3)$ symmetry), then we will always use the same regularization
functions for all of these Dirac seas. Once the regularization has
been specified for the vacuum, we can again apply the rules
(\ref{p:2F2}--\ref{p:2N}) to regularize the light-cone
expansion. In the special case that the bosonic potentials are
diagonal in the Dirac sea index, we can simply take the direct sum
of the contributions (\ref{p:2F3}, \ref{p:2N}), using in each
summand the regularization functions of the corresponding vacuum
Dirac sea. In the general case of a non-diagonal bosonic field,
the regularization functions can be inserted uniquely if one uses
that, according to the assumption of macroscopic potentials and
wave functions of {\S}\ref{psec22}, the fermionic projector is
modified by the bosonic fields only on the macroscopic scale, so
that its microscopic structure is the same as in the vacuum. For
example, one can in the case of a gravitational and Yang-Mills field make
the bosonic potential locally to zero by transforming to a suitable
coordinate sytem and gauge, can in this system insert the
regularization functions as in the vacuum and can finally
transform back to the original system. We conclude that the
generalization of our regularization method to systems of Dirac
seas is quite straightforward and canonical. Therefore we can
introduce a short notation for the regularizations of the factors
$T^{(n)}$ in the light-cone expansion by
simply adding a label for the order in the mass parameter. More
precisely, we introduce in the case $N=1$ of one Dirac sea the
following abbreviations for the Fourier integrals in (\ref{p:2F3})
and (\ref{p:2N}), \label{T^(n)_[p]}
\begin{eqnarray}
T^{(n)}_{[p]} &\equiv& -(-il)^{n-1} \int_{-\infty}^\infty du
\left(\frac{1}{u^n}\right)^{\mbox{\scriptsize{reg}}} \:e^{-ius} \nonumber \\
&& \hspace*{3.87cm} \times
\left\{ \begin{array}{ll} h(u) \:a(u)^{\frac{p-1}{2}} & {\mbox{for
$p$ odd}} \\
g(u) \:a(u)^{\frac{p}{2}} & {\mbox{for $p$ even}} \end{array} \right.\;\;\;\;\;
\label{p:D1} \\
(\xi\slsh \:T^{(n)}_{[p]}) &\equiv& -(-il)^{n-1} \int_{-\infty}^\infty
du \:e^{-ius} \:\times
\left\{ \begin{array}{ll} h(u) \:a(u)^{\frac{p-1}{2}} & {\mbox{for $p$ odd}} \\
g(u) \:a(u)^{\frac{p}{2}} & {\mbox{for $p$ even}} \end{array} \right.
\nonumber \\
&& \hspace*{-1.4cm} \times \left[ 2l \:\gamma^s \left(\frac{1}{u^n}\right)^{\mbox{\scriptsize{reg}}}
\:+\: 2i n\:\gamma^l \left(\frac{1}{u^{n+1}}\right)^{\mbox{\scriptsize{reg}}}
\:+\: 2l \:b(u) \:\gamma^l
\left(\frac{1}{u^{n+2}}\right)^{\mbox{\scriptsize{reg}}} \right]
\spc \label{p:D2} \\
T^{(n)}_{\{p\}} &\equiv& -(-il)^{n-1} \int_{-\infty}^\infty du
\left(\frac{1}{u^n}\right)^{\mbox{\scriptsize{reg}}} \:e^{-ius}
\:b(u) \nonumber \\
&& \hspace*{3.87cm} \times \left\{ \begin{array}{ll} h(u) \:a(u)^{\frac{p-1}{2}}
& {\mbox{for $p$ odd}} \\
g(u) \:a(u)^{\frac{p}{2}} & {\mbox{for $p$ even .}} \end{array}
\right. \label{p:D3} \label{T^(n)_{p}}
\end{eqnarray}
In the case of a system of Dirac seas~(\ref{vafp1}), we use the same
notation for the corresponding direct sum.
With this notation, the regularization of the light-cone expansion is
carried out (modulo all the contributions neglected in (\ref{p:2F3})
and (\ref{p:2N})) merely by the replacement $m^p \:T^{(n)}(x,y)
\to T^{(n)}_{[p]}$ and by marking with brackets that the
factors $(y-x)_j \gamma^j$ and $T^{(n)}_{[p]}$ belong together (where
we use the abbreviation $\xi \equiv y-x$) \label{xi}. We call a factor
$\xi \slsh$ inside the brackets $(\xi\slsh T^{(n)}_{[p]})$ an
{\em{inner factor}} $\xi\slsh$\index{inner factor $\xi\slsh$}.
Notice that the functions $T^{(n)}_{\{p\}}$ in (\ref{p:D3})
involve the regularization
function $b$; they will be needed below to handle contractions between
the inner factors.

We finally come to the analysis of composite expressions in the
fermionic projector. In {\S}\ref{psec23} we already discussed
the simplest composite expression, the closed chain $P(x,y)\:
P(y,x)$ in the vacuum. In order to analyze the closed chain near
the light cone, we substitute for $P(x,y)$ and $P(y,x)$ the
regularized formulas of the light-cone expansion and multiply out.
It is convenient to use that the fermionic projector is Hermitian
and thus $P(y,x)=P(x,y)^*$; hence the light-cone
expansion of $P(y,x)$ is obtained from that for $P(x,y)$ by taking
the adjoint (with respect to the spin scalar product). The
iterated line integrals can be multiplied with each other giving
smooth functions; also we can simplify the resulting product of
Dirac matrices using their anti-commutation relations. Denoting
the adjoints of (\ref{p:D1}) and (\ref{p:D2}) by
$\overline{T^{(n)}_{[p]}}$ and $(\overline{\xi\slsh
T^{(n)}_{[p]}})$, respectively, we thus obtain for the closed
chain a sum of terms of the following forms,
\begin{equation}
\begin{array}{clc}
F \:T^{(n_1)}_{[r_1]} \: \overline{T^{(n_2)}_{[r_2]}} &,\quad&
F \:(\xi_{j_1} T^{(n_1)}_{[r_1]}) \: \overline{T^{(n_2)}_{[r_2]}} \\
F \:T^{(n_1)}_{[r_1]} \: (\overline{\xi_{j_2} T^{(n_2)}_{[r_2]}}) &,\quad&
F \:(\xi_{j_1} T^{(n_1)}_{[r_1]})\: (\overline{\xi_{j_2} T^{(n_2)}_{[r_2]}})
\;, \end{array} \label{p:Dm1}
\end{equation}
where $F$ is a smooth function in $x$ and $y$ and
$n_j$, $r_j$ are integer parameters. Here the tensor indices of the
inner factors $\xi$ are contracted either with each other or with tensor
indices in the smooth prefactor $F$. In order to analyze Euler-Lagrange
equations like for example~(\ref{p:52}, \ref{p:54}), we need to consider
more general expressions. More precisely, all Euler-Lagrange equations
in this book can be written in terms of expressions being a product of a
smooth function with a quotient of two monomials in~$T^{(n)}_{[r]}$
and~~$\overline{T^{(n)}_{[r]}}$, possibly with inner factors~$\xi$
in the numerator. Thus our key problem is to mathematically handle
expressions of the form
\begin{eqnarray}
\lefteqn{ {\mbox{(smooth function)}} \times
\left[ T^{(l_{1})}_{[s_{1}]} \cdots T^{(l_f)}_{[s_f]} \;
\overline{ T^{(l_{f+1})}_{[s_{f+1}]} \cdots T^{(l_g)}_{[s_g]} }
\right]^{-1} } \nonumber \\
&&\times (\xi_{j_1} T^{(n_1)}_{[r_1]}) \cdots
(\xi_{j_a} T^{(\na)}_{[r_a]})\; T^{(n_{a+1})}_{[r_{a+1}]} \cdots
T^{(n_b)}_{[r_b]} \nonumber \\
&&\hspace*{2.15cm} \times \:
\overline{ ( \xi_{j_{b+1}} T^{(n_{b+1})}_{[r_{b+1}]}) \cdots
(\xi_{j_c} T^{(n_c)}_{[r_c]}) \;
T^{(n_{c+1})}_{[r_{c+1}]} \cdots T^{(n_d)}_{[r_d]} }
\label{p:D4}
\end{eqnarray}
with $0 \leq f \leq g$, $0 \leq a \leq b \leq c \leq d$,
parameters $l_j, s_i, n_i, p_i$ and tensor indices $j_i$.
Here the tensor indices of the inner factors $\xi_i$ are again contracted
either with other inner factors or with tensor indices in the smooth
prefactor. We mention for clarity that, since the factors in (\ref{p:D4})
are complex functions or, in the case $N>1$ of systems of Dirac seas, direct
sums of complex functions, the product (\ref{p:D4}) clearly is commutative.

The inner factors in (\ref{p:D4}) can be simplified using the particular form
(\ref{p:D1}, \ref{p:D2}) of $T^{(n)}_{[r]}$
and $(\xi_j T^{(n)}_{[r]})$. We begin with the case
of an inner factor which is contracted with a tensor index in the
smooth prefactor, i.e.\ with products of the form
\[ \cdots F^j \:(\xi_j T^{(n)}_{[r]}) \cdots \spc{\mbox{or}}\spc
\cdots F^j \:(\overline{ \xi_j T^{(n)}_{[r]} }) \cdots \]
and a smooth vector field $F$, where ``$\cdots$'' stands for any other
factors of the form as in (\ref{p:D4}). According to (\ref{p:D2}),
to leading order in $(l E_P)^{-1}$ it suffices to take into account
the $s$-component, and thus (\ref{p:D1}) yields that
$(\xi\slsh T^{(n)}_{[r]}) \approx 2l \:\gamma^s \:T^{(n)}_{[r]}$.
Since $2l \:\gamma^s$ coincides on the light cone with $\xi\slsh$, we
conclude that, to leading order in $(l E_P)^{-1}$,
\begin{equation}
F^j \:( \xi_j T^{(n)}_{[r]}) \;=\; F^j \xi_j \:T^{(n)}_{[r]}
\spc{\mbox{and}}\spc F^j \:(\overline{ \xi_j \:T^{(n)}_{[r]} }) \;=\;
F^j \xi_j \overline{T^{(n)}_{[r]}} \:. \label{p:D9}
\end{equation}
These relations coincide with what one would have expected naively.
We next consider the case of two inner factors
which are contracted with each other, i.e.\ products of the
following form,
\begin{equation}
\begin{array}{cll}
\cdots (\xi_j T^{(n_1)}_{[r_1]})(\xi^j T^{(n_2)}_{[r_2]})\cdots &,\quad&
\cdots (\xi_j T^{(n_1)}_{[r_1]})(\overline{\xi^j T^{(n_2)}_{[r_2]}})\cdots \\
{\mbox{or}} &&
\cdots (\overline{\xi_j T^{(n_1)}_{[r_1]}})(\overline{\xi^j T^{(n_2)}_{[r_2]}}) \cdots \;. \end{array} \label{p:D5}
\end{equation}
In this case, the product cannot be calculated naively because the
factor $\xi_j \xi^j=\xi^2$ vanishes on the light cone. But we can still
compute the product using the Fourier representation (\ref{p:D2}). Since the
$s$- and $l$-directions are null, only the mixed products of
the $s$- and $l$-components in (\ref{p:D2}) contribute, and we obtain
\begin{eqnarray*}
\lefteqn{ (\xi_j T^{(n_1)}_{[r_1]})(\xi^j T^{(n_2)}_{[r_2]}) } \\
&=& (-il)^{n_1-1} \:l \int_{-\infty}^\infty du_1 \:
\frac{1}{u_1^{n_1}} \: e^{-iu_1s} \:\times
\left\{ \begin{array}{ll} h(u_1) \:a(u_1)^{\frac{r_1-1}{2}}
& {\mbox{for $r_1$ odd}} \\
g(u_1) \:a(u_1)^{\frac{r_1}{2}} & {\mbox{for $r_1$ even}} \end{array} \right. \\
&&\quad \times (-il)^{n_2-1} \int_{-\infty}^\infty du_2 \left[
\frac{2i n_2}{u_2^{n_2+1}} + \frac{2l \:b(u_2)}{u_2^{n_2+2}}
\right] e^{-iu_2s} \nonumber \\
&& \hspace*{4.9cm} \times
\left\{ \begin{array}{ll} h(u_2) \:a(u_2)^{\frac{r_2-1}{2}} & {\mbox{for $r_2$ odd}} \\
g(u_2) \:a(u_2)^{\frac{r_2}{2}} & {\mbox{for $r_2$ even}} \end{array}
\right. \\
&&+ (-il)^{n_1-1} \int_{-\infty}^\infty du_1 \left[
\frac{2i n_1}{u_1^{n_1+1}} + \frac{2l \:b(u_1)}{u_1^{n_1+2}}
\right] e^{-iu_1s} \nonumber \\
&& \hspace*{4.9cm} \times
\left\{ \begin{array}{ll} h(u_1) \:a(u_1)^{\frac{r_1-1}{2}}
& {\mbox{for $r_1$ odd}} \\
g(u_1) \:a(u_1)^{\frac{r_1}{2}} & {\mbox{for $r_1$ even}} \end{array}
\right. \\
&&\quad \times (-il)^{n_2-1} \:l \int_{-\infty}^\infty du_2
\:\frac{1}{u_2^{n_2}}\: e^{-iu_2s} \:\times
\left\{ \begin{array}{ll} h(u_2) \:a(u_2)^{\frac{r_2-1}{2}}
& {\mbox{for $r_2$ odd}} \\
g(u_2) \:a(u_2)^{\frac{r_2}{2}} & {\mbox{for $r_2$ even ,}} \end{array} \right.
\end{eqnarray*}
and similarly for the two other
products in~(\ref{p:D5}). In the case of systems of Dirac seas, this
calculation can be done for each summand of the direct sum
separately. Rewriting
the Fourier integrals using the notation (\ref{p:D1}) and
(\ref{p:D3}), we get the following result.\\[.5em]
{\bf{Contraction rules:}} {\em{To leading order in $(l E_P)^{-1}$,}}
\begin{eqnarray}
\lefteqn{(\xi_j T^{(n_1)}_{[r_1]}) (\xi^j T^{(n_2)}_{[r_2]}) }
\nonumber \\
&=& -2 \:T^{(n_1)}_{[r_1]} \:(n_2 \:T^{(n_2+1)}_{[r_2]}
+ T^{(n_2+2)}_{\{r_2\}})
\:-\: 2 \:(n_1 \:T^{(n_1+1)}_{[r_1]} + T^{(n_1+2)}_{\{r_1\}})
\:T^{(n_2)}_{[r_2]} \spc\label{p:D6} %\\
\end{eqnarray}
\begin{eqnarray}
\lefteqn{ (\xi_j T^{(n_1)}_{[r_1]}) (\overline{\xi^j T^{(n_2)}_{[r_2]}}) }
\nonumber \\
&=& -2 \:T^{(n_1)}_{[r_1]} \:(n_2 \:\overline{T^{(n_2+1)}_{[r_2]}}
+ \overline{T^{(n_2+2)}_{\{r_2\}}})
\:-\: 2 \:(n_1 \:T^{(n_1+1)}_{[r_1]} + T^{(n_1+2)}_{\{r_1\}})
\:\overline{T^{(n_2)}_{[r_2]}}\spc \label{p:D7} \\
\lefteqn{ (\overline{\xi_j T^{(n_1)}_{[r_1]}})
(\overline{\xi^j T^{(n_2)}_{[r_2]}}) }
\nonumber \\
&=& -2 \:\overline{T^{(n_1)}_{[r_1]}} \:(n_2 \:\overline{T^{(n_2+1)}_{[r_2]}}
+ \overline{T^{(n_2+2)}_{\{r_2\}}})
\:-\: 2 \:(n_1 \:\overline{T^{(n_1+1)}_{[r_1]}} +
\overline{T^{(n_1+2)}_{\{r_1\}}}) \:\overline{T^{(n_2)}_{[r_2]}}
\:. \spc\;\label{p:D8}
\end{eqnarray}
By iteratively applying (\ref{p:D9}) and the contraction rules
(\ref{p:D6}--\ref{p:D8}), we can in (\ref{p:D4}) eliminate all
inner factors $\xi$ to end up with products of the form
\begin{equation}
{\mbox{(smooth function)}} \:
\frac{ T^{(a_1)}_\circ \cdots T^{(a_\alpha)}_\circ \:
\overline{T^{(b_1)}_\circ \cdots T^{(b_\beta)}_\circ} }
{ T^{(c_1)}_\circ \cdots T^{(c_\gamma)}_\circ \:
\overline{T^{(d_1)}_\circ \cdots T^{(d_\delta)}_\circ} }
    \label{p:D11}
\end{equation}
with parameters~$\alpha, \beta \geq 1$, $\gamma, \delta \geq 0$
and $a_i, b_i, c_i, d_i$ (if~$\gamma=0=\delta$ the denominator
clearly is equal to one). Here each subscript~``$_\circ$''
stands for an index $[r]$ or $\{r\}$. The quotient of the
two monomials in~(\ref{p:D11}) is called a {\em{simple
    fraction}}\index{simple fraction}.

We point out that the above transformation rules for the inner factors
(\ref{p:D9}) and (\ref{p:D6}--\ref{p:D8}) are identities valid
pointwise (i.e.\ for fixed $x$ and $y$) close to the light cone.
We anticipate that Euler-Lagrange equations like
(\ref{p:52}, \ref{p:54}) do not lead us to evaluate the products of the
form (\ref{p:D4}) pointwise, but merely in the weak sense.
Therefore, we now go over to a weak analysis of the
simple fraction. In the case of a continuous regularization, we thus
consider the integral
\begin{equation}
\int d^4x \; \eta(x) \:
\frac{ T^{(a_1)}_\circ \cdots T^{(a_\alpha)}_\circ \:
\overline{T^{(b_1)}_\circ \cdots T^{(b_\beta)}_\circ} }
{ T^{(c_1)}_\circ \cdots T^{(c_\gamma)}_\circ \:
\overline{T^{(d_1)}_\circ \cdots T^{(d_\delta)}_\circ} }
    \label{p:D12}
\end{equation}
with a test function~$\eta$. Before coming to the derivation of
calculation rules for the integrand in (\ref{p:D12}), we must
think about how the test function $\eta$ is to be chosen. As
explained in {\S}\ref{psec23} in the example of the closed chain
(\ref{p:2p}), a weak integral in general depends essentially on
the unknown high-energy behavior of the fermionic projector and is
therefore undetermined. To avoid this problem, we must evaluate
(\ref{p:D12}) in such a way that our expansions near the light
cone become applicable. To this end, we assume that $\eta$ has its
{\em{support near the light cone}}\index{support near the light cone}, meaning that in light-cone
coordinates $(s, l, x_2, x_3)$, the ``large'' variable $l$
satisfies on the support of $\eta$ the conditions (\ref{p:2s}).
For clarity, we remark that this definition does not state that
the support of $\eta$ should be in a small neighborhood of the
light cone, but merely in a strip away from the origin. This is
sufficient because we shall extract information on the behavior
near the light cone by considering the singularities of the
integral for $E_P \to \infty$ (see (\ref{p:Dwe}) below).
Furthermore, we assume that $\eta$ is {\em{macroscopic}} in the
sense that its partial derivatives scale in powers of $l^{-1}$ or
$l^{-1}_{\mbox{\scriptsize{macro}}}$. Under these assumptions, the
integrand in (\ref{p:D12}) is macroscopic in $l$, and carrying out
the $s$- and $l$-integrals in (\ref{p:D12}) gives a function which
is macroscopic in the ``transversal'' variables $x_2$ and $x_3$.
Therefore, in the three variables $(l, x_2, x_3)$, a weak analysis
is equivalent to a pointwise analysis, and thus it suffices to
consider the $s$-integral in (\ref{p:D12}), i.e. the expression
\begin{equation}
\int_{-\infty}^\infty ds \; \eta \:
\frac{ T^{(a_1)}_\circ \cdots T^{(a_\alpha)}_\circ \:
\overline{T^{(b_1)}_\circ \cdots T^{(b_\beta)}_\circ} }
{ T^{(c_1)}_\circ \cdots T^{(c_\gamma)}_\circ \:
\overline{T^{(d_1)}_\circ \cdots T^{(d_\delta)}_\circ} }
    \label{p:D13}
\end{equation}
for fixed $l$, $x_2$ and $x_3$. In the case of a discrete
regularization, the integral in (\ref{p:D12}) must be replaced
by a sum over all space-time points, i.e.\ we must consider instead of
(\ref{p:D12}) the weak sum
\begin{equation}
\sum_{x \in M} \eta(x) \:\frac{ T^{(a_1)}_\circ \cdots T^{(a_\alpha)}_\circ \:
\overline{T^{(b_1)}_\circ \cdots T^{(b_\beta)}_\circ} }
{ T^{(c_1)}_\circ \cdots T^{(c_\gamma)}_\circ \:
\overline{T^{(d_1)}_\circ \cdots T^{(d_\delta)}_\circ} } \:,
    \label{p:Dsum}
\end{equation}
where $M \subset \R^4$ are the discrete space-time points and $\eta$ is
a macroscopic function in $\R^4$ with support near the light cone. Up to
a normalization factor, (\ref{p:Dsum}) can be regarded as a Riemann sum
which approximates the integral (\ref{p:D12}). Assuming that the space-time
points have a generic position in $\R^4$ and keeping in mind that the function
inside the sum (\ref{p:Dsum}) is macroscopic in the variables $l$, $x_2$,
and $x_3$, the Riemann sum and the integral indeed coincide to leading order
in $(l E_P)^{-1}$ and $(l_{\mbox{\scriptsize{macro}}} E_P)^{-1}$. Hence
it is admissible to work also in the discrete case with the
one-dimensional integral (\ref{p:D13}).

Let us analyze the integral (\ref{p:D13}) in more detail. We first consider
how (\ref{p:D13}) scales in the Planck energy. In the limit $E_P \to
\infty$, the factors $T^{(n)}_\circ$ go over to distributions which are in
general singular on the light cone. Hence their product in (\ref{p:D13})
becomes ill-defined for $E_P \to \infty$ even in the distributional
sense, and thus we expect that the integral (\ref{p:D13}) should diverge
for $E_P \to \infty$. The order of this divergence can be determined
using the following power counting argument. Keeping in mind that the
regularization functions decay on the Planck scale $u \sim E_P$, the Fourier
integrals (\ref{p:D1}) and (\ref{p:D3}) behave on the light cone (i.e.\ for
$s=0$) like
\[ T^{(n)}_\circ \;\sim\; \log^g(E_P) \: E_P^{-n+1} \]
with $g=1$ in the case $n=1$ and $g=0$ otherwise. Hence the product in the
integrand of (\ref{p:D13}) scales on the light cone as
\begin{equation}
\frac{ T^{(a_1)}_\circ \cdots T^{(a_\alpha)}_\circ \:
\overline{T^{(b_1)}_\circ \cdots T^{(b_\beta)}_\circ} }
{ T^{(c_1)}_\circ \cdots T^{(c_\gamma)}_\circ \:
\overline{T^{(d_1)}_\circ \cdots T^{(d_\delta)}_\circ} }
\;\sim\; \log^g(E_P) \:E_P^L
\label{p:Dp}
\end{equation}
with $g \in \Z$ and \label{L}
\begin{equation}
L \;=\; \alpha + \beta - \gamma - \delta
-\sum_{j=1}^\alpha a_j -\sum_{j=1}^\beta b_j
+\sum_{j=1}^\gamma c_j +\sum_{j=1}^\delta d_j\:.
\label{p:DL}
\end{equation}
We call $L$ the {\em{degree}} of the simple fraction\index{degree!of
  the simple fraction}.
We will here restrict attention to the case $L>1$. In this case, the
simple fraction~(\ref{p:Dp}) diverges in the limit~$E_P \to \infty$ at least
quadratically. If $s$ is not zero, the oscillations of the factor
$\exp(-ius)$ in (\ref{p:D1}, \ref{p:D3}) lead to a decay of
$T^{(n)}_\circ$ on the scale $s \sim E_P^{-1}$. This consideration
shows that the dominant
contribution to the integral (\ref{p:D13}) when $E_P \to \infty$
is obtained by evaluating $\eta$ on the light cone, and the scaling behavior
of this contribution is computed by multiplying (\ref{p:Dp}) with
a factor $E_P^{-1}$. We conclude that (\ref{p:D13}) diverges in the limit
$E_P \to \infty$, and its leading divergence scales in~$E_p$ like
\begin{equation}
\eta(s=0) \:\log^g(E_P) \:E_P^{L-1}\:.
\label{p:Dsing}
\end{equation}
Collecting the logarithmic terms in the light-cone expansion, one can
easily compute the parameter~$g$.
We remark that due to possible zeros in the denominator, the
integral~(\ref{p:D13}) might diverge even for finite~$E_P$. In this
case we can still use~(\ref{p:Dsing}) if we set the proportionality
factor equal to plus or minus infinity. We also note that,
by substituting the Fourier representations (\ref{p:D1}, \ref{p:D3})
into~(\ref{p:D13}), one can rewrite the products in
(\ref{p:D13}) in terms of the regularization functions
(this is explained in detail in Appendix~\ref{pappC} for a
particular choice of regularization functions).
Collecting the factors of~$l$ in (\ref{p:D1}) and (\ref{p:D3}), we
end up with the following result.\\[.5em]
{\bf{Weak evaluation on the light cone:}}
{\em{Consider the integral (\ref{p:D13}) for a simple fraction of degree $L>1$.
Then there is an integer $g \geq 0$ and a real coefficient
$c_{\mbox{\scriptsize{reg}}}$ independent of~$s$ and~$l$ such that for every
macroscopic test function $\eta$,}}
\begin{eqnarray}
\lefteqn{ \int_{-\infty}^\infty ds \; \eta \:
\frac{ T^{(a_1)}_\circ \cdots T^{(a_\alpha)}_\circ \:
\overline{T^{(b_1)}_\circ \cdots T^{(b_\beta)}_\circ} }
{ T^{(c_1)}_\circ \cdots T^{(c_\gamma)}_\circ \:
\overline{T^{(d_1)}_\circ \cdots T^{(d_\delta)}_\circ} }
\;=\; \frac{c_{\mbox{\scriptsize{reg}}}}{(il)^L}
\:\eta(s=0) \:\log^g(E_P) \:E_P^{L-1} } \nonumber \\
&&\:+\: {\mbox{(higher orders in $(l
E_P)^{-1}$ and  $(l_{\mbox{\scriptsize{macro}}} E_P)^{-1}$)}} \:.
\spc\spc\spc
\label{p:Dwe}
\end{eqnarray}
The coefficient $c_{\mbox{\scriptsize{reg}}}$ clearly depends on the
indices of the simple fraction and on the
details of the regularization. We call $c_{\mbox{\scriptsize{reg}}}$ \label{c_reg}
a {\em{regularization parameter}}\index{regularization parameter}.

Integrals of type (\ref{p:D13}) can be transformed using integration by parts.
For clarity we begin with the special case of a monomial,
\begin{eqnarray}
\lefteqn{ \int_{-\infty}^\infty ds \: \left(\frac{d}{ds} \eta \right)
 \:T^{(a_1)}_\circ \cdots \overline{T^{(b_q)}_\circ} \;=\;
-\int_{-\infty}^\infty ds \; \eta \;\frac{d}{ds} \left(
T^{(a_1)}_\circ \cdots \overline{T^{(b_q)}_\circ} \right) }
\label{p:D14} \\
&=&-\int_{-\infty}^\infty \!\!\!ds \: \eta \left[
\left(\frac{d}{ds} T^{(a_1)}_\circ \right) T^{(a_2)}_\circ
\cdots \overline{T^{(b_q)}_\circ} \right. \nonumber \\
&& \hspace*{3cm} \left. \;+\; \cdots \;+\;
T^{(a_1)}_\circ \cdots \overline{T^{(b_{q-1})}_\circ}
\left( \frac{d}{ds} \overline{T^{(b_{q})}_\circ} \right) \right] , \;\;
\;\;\;\;\;\; \label{p:D16}
\end{eqnarray}
where in the last step we applied the Leibniz rule. Differentiating
(\ref{p:D1}) and (\ref{p:D3}) with respect to $s$ yields that

\begin{equation}
\frac{d}{ds} T^{(n)}_\circ \;=\; -l \: T^{(n-1)}_\circ \spc
{\mbox{and}} \spc
\frac{d}{ds} \overline{T^{(n)}_\circ} \;=\; -l \:
\overline{T^{(n-1)}_\circ} \:. \label{p:D15}
\end{equation}
With these relations, we can carry out the derivatives in (\ref{p:D16}).
Notice that the differentiation rules (\ref{p:D15}) decrease the index $n$
by one. According to (\ref{p:DL}) and (\ref{p:Dwe}), decrementing the
upper index of a factor $T^{(a_j)}_\circ$ or $\overline{T^{(b_k)}_\circ}$
increments the degree of the monomial and yields in the
weak integral a factor of the order $E_P/l$. Using
furthermore that $\eta$ is macroscopic (as defined after (\ref{p:D12})),
we conclude that each summand in (\ref{p:D16}) dominates the left side
of (\ref{p:D14}) by one order in $l E_P$ or
$l_{\mbox{\scriptsize{macro}}} E_P$. We
have thus derived the following result.\\[.5em]
{\bf{Integration-by-parts rule for monomials:}}
{\em{Consider a monomial of degree $L>1$. In a weak analysis near the light cone, we
have to leading order in $(l E_P)^{-1}$ and
$(l_{\mbox{\scriptsize{macro}}} E_P)^{-1}$,}}
\begin{eqnarray}
\lefteqn{ 0 \;=\; T^{(a_1-1)}_\circ \cdots T^{(a_p)}_\circ
\: \overline{T^{(b_1)}_\circ} \cdots \overline{T^{(b_q)}_\circ}
+ \cdots + T^{(a_1)}_\circ \cdots T^{(a_p-1)}_\circ
\: \overline{T^{(b_1)}_\circ} \cdots \overline{T^{(b_q)}_\circ} } \nonumber \\
&& \hspace*{-.5cm}
+ \:T^{(a_1)}_\circ \cdots T^{(a_p)}_\circ
\: \overline{T^{(b_1-1)}_\circ} \cdots \overline{T^{(b_q)}_\circ}
+ \cdots +
T^{(a_1)}_\circ \cdots T^{(a_p)}_\circ
\: \overline{T^{(b_1)}_\circ} \cdots \overline{T^{(b_q-1)}_\circ} \:.\spc\;
\label{p:D17}
\end{eqnarray}
The integration-by-parts method works similarly for simple fractions.
For ease in notation we state it more symbolically. \\[.5em]
{\bf{Integration-by-parts rule for simple fractions:}}\index{integration-by-parts rule}
{\em{Consider a simple fraction of degree $L>1$. In a weak analysis near the light cone
and to leading order in $(l E_P)^{-1}$ and
$(l_{\mbox{\scriptsize{macro}}} E_P)^{-1}$,}}
\beq \label{sfinte}
\nabla \left(
\frac{ T^{(a_1)}_\circ \cdots T^{(a_\alpha)}_\circ \:
\overline{T^{(b_1)}_\circ \cdots T^{(b_\beta)}_\circ} }
{ T^{(c_1)}_\circ \cdots T^{(c_\gamma)}_\circ \:
\overline{T^{(d_1)}_\circ \cdots T^{(d_\delta)}_\circ} }
\right) \;=\;  0 \:.
\eeq
{\em{Here~$\nabla$ acts on all factors like a derivation (i.e.\ it is
linear and satisfies the Leibniz rule), commutes with complex conjugations}}
and
\[ \nabla T^{(n)}_\circ \;=\; T^{(n-1)}_\circ \:,\spc
\nabla \frac{1}{T^{(n)}_\circ} \;=\; -\frac{T^{(n-1)}_\circ}
{(T^{(n)}_\circ)^2}\:. \]
The integration-by-parts rule gives us relations between simple fractions.
We say that simple fractions are {\em{independent}}\index{simple fraction!independence of} if the integration-by-parts
rules gives no relations between them.
More systematically, we consider the vector space of linear combinations
of simple fractions. We say that two such linear combinations are {\em{equivalent}}
if they can be transformed into each other with the integration-by-parts rules.
We refer to the equivalence classes as the {\em{basic
fractions}}\index{basic fraction}.
Taking the linear combination of the corresponding regularization
parameters~$c_{\mbox{\scriptsize{reg}}}$, we can associate to every
basic fraction a so-called {\em{basic regularization
parameter}}\index{regularization parameter!basic}.
In Appendix~\ref{pappC} it is shown for all
simple fractions which will appear in this book that the
corresponding basic fractions are linearly independent in the sense
that there are no further identities between them. Therefore it seems
a reasonable method to take the basic regularization parameters as
empirical parameters modeling the unknown microscopic structure of space-time.

We remark that the notion of the basic fraction
can be made more concrete by choosing from each equivalence class
one representative. Then one can identify every basic fraction
with the distinguished simple fraction in its equivalence class.
For simplicity we give this construction in the special case that
the simple fractions are monomials of the form
\[ T^{(a_1)}_\circ \cdots T^{(a_p)}_\circ
\: \overline{T^{(b_1)}_\circ} \cdots \overline{T^{(b_q)}_\circ} \]
(the construction can immediately be extended to simple fractions, but
it becomes a bit complicated and we do not need it here). If
only one factor~$T^{(a)}_\circ$ appears ($p=1$), one can by applying the
integration-by-parts rule iteratively increment the parameter $a_1$; this
clearly decreases the other parameters $b_1,\ldots,b_q$. In order to avoid that
any of the parameters $b_1,\ldots,b_q$ becomes smaller than $-1$, we stop the
integration-by-parts procedure as soon as one of the $b_j$ equals $-1$. In this
way, we can express every monomial as a unique linear combination of monomials
of the form
\begin{equation}
T^{(a_1)}_\circ \:\overline{T^{(b_1)}_\circ} \cdots \overline{T^{(b_q)}_\circ}
\spc{\mbox{with}}\spc -1=b_1 \leq \cdots \leq b_q.
\label{p:Dbm1}
\end{equation}
Similarly for $p>1$, the integration-by-parts rule allows us to
increment the smallest of the parameters $a_j$, unless either one of the
parameters $b_j$ equals $-1$ or there are two factors $T^{(a_j)}$ with
$a_j=\min (a_1,\ldots,a_p)$. By iteration, we can thus transform any monomial
into a linear combination of monomials of the following type,
\begin{eqnarray}
\lefteqn{ \hspace*{-2cm} T^{(a_1)}_\circ \cdots T^{(a_p)}_\circ
\: \overline{T^{(b_1)}_\circ} \cdots \overline{T^{(b_q)}_\circ}
\;\;\;\;\;{\mbox{with}}\;\;\;\;\; a_1 \leq \cdots \leq a_p ,\;\;
b_1 \leq \cdots \leq b_q } \nonumber \\
&&\hspace*{2.5cm} {\mbox{and}} \;\;\;\;\; a_1=a_2 {\mbox{ or }} b_1=-1 \:.
\label{p:Dbm}
\end{eqnarray}
We can now consider~(\ref{p:Dbm1}) and~(\ref{p:Dbm}) as the basic
monomials.

With the above constructions we have developed the mathematical
framework for analyzing composite expressions in the fermionic
projector in the continuum. Our procedure is outlined as follows.
We first substitute for the fermionic projector the regularized
formulas of the light-cone expansion; this yields sums of products
of the form (\ref{p:D4}), where the smooth prefactor involves the
bosonic potentials and fields as well as the wave functions of the
Dirac particles and anti-particles of the system. Applying our
contraction rules, we then eliminate all inner factors and obtain
terms of the form (\ref{p:D11}). When evaluated in the weak sense
(\ref{p:Dwe}), the $l$-dependence determines the degree $L$ of the
simple fraction, and the dependence on the regularization is described
for each simple fraction by the corresponding regularization parameters
$c_{\mbox{\scriptsize{reg}}}$. Using our integration-by-parts
rule, we can furthermore restrict attention to the basic fractions
and the corresponding basic regularization parameters.
Taking the basic regularization parameters as free empirical
parameters, the composite expressions in the fermionic projector
reduce to expressions in the bosonic fields and fermionic wave
functions, involving a small number of free parameters. This
procedure for analyzing composite expressions in the fermionic
projector is called the {\em{continuum limit}}\index{continuum limit}.

\mainmatter
\setcounter{page}{123}
\include{pfp2}
\begin{appendix}
\include{pfp3}
\end{appendix}

\backmatter
%!TEX root = finsterpfp.tex
\bibliographystyle{amsalpha}

\vspace*{.3cm}%
\noindent
{\footnotesize{%
{{NWF I - Mathematik, Universit\"at Regensburg, D-93040 Regensburg, Germany}} \\
Email: {\tt{Felix.Finster@mathematik.uni-regensburg.de}}}

\printindex
% Index of Symbols
\cleardoublepage
\markboth{NOTATION INDEX}{}
\addcontentsline{toc}{chapter}{Notation Index}
\twocolumn[\vspace*{2.8cm} \centerline{\bf{\Large{Notation Index}}} \vspace*{1.9cm}]

\begin{itemize}
\item[] $(M,\langle .,.\rangle)$ \,\pageref{Mrangle}
\item[] $\mathrm{E}$ \,\pageref{E}
\item[] $J_x$ \,\pageref{Jx}
\item[] $J$, $J^\lor_x$, $J^\land_x$ \,\pageref{Jlorx}
\item[] $I$, $I^\lor_x$, $I^\land_x$ \,\pageref{Ilorx}
\item[] $L$, $L^\lor_x$, $L^\land_x$ \,\pageref{Elorx}
\item[] $T_pM$ \,\pageref{Tpm}
\item[] $g_{jk}$ \,\pageref{gjk}
\item[] $\nabla$ \,\pageref{nabla}
\item[] $R^i_{jkl}$ \,\pageref{Rijkl}
\item[] $T_{jk}$ \,\pageref{Tjk}
\item[] $S$ \,\pageref{S}, \pageref{S2}, \pageref{e:2a}
\item[] $\Box$ \,\pageref{Box}
\item[] $\{.,.\}$ \,\pageref{{.,.}}
\item[] $\gamma^j$ \,\pageref{gammaj}
\item[] $\vec{\sigma}$ \,\pageref{vecsigma}
\item[] $\Pdd$ \,\pageref{Pdd}
\item[] $\Sl .|. \Sr$ \,\pageref{Sl.|.Sr}
\item[] $A^\ast$ \,\pageref{Aast}
\item[] $J^k$ \,\pageref{Jk}
\item[] $\rho$ \,\pageref{rho}
\item[] $\epsilon_{jklm}$ \,\pageref{epsilonjklm}
\item[] $\chi_L$, $\chi_R$ \,\pageref{chiL}
\item[] $(.|.)$ \,\pageref{(.|.)}, \pageref{(.|.)b}
\item[] $h$ \,\pageref{h}
\item[] $\Psi_{\vec{p} s \epsilon}$ \,\pageref{Psivecpsepsilon}
\item[] $d\mu_{\vec{p}}$ \,\pageref{dmuvecp}
\item[] $\land$ \,\pageref{land}
\item[] $\mathcal{F}$, $\mathcal{F}^n$ \,\pageref{mathcalFn}
\item[] $\hat{\Psi}_{\vec{p} s\epsilon}$, $\hat{\Psi}^\dagger_{\vec{p} s \epsilon}$ \,\pageref{hatPsidagger}
\item[] $\mathcal{G}$ \,\pageref{mathcalG}
\item[] $\mathcal{D}$ \,\pageref{mathcalD}, \pageref{mathcalD2}
\item[] $\sigma^{jk}$ \,\pageref{sigmajk}
\item[] $\Tr$ \,\pageref{Tr}, \pageref{Tr2}
\item[] $D$ \,\pageref{gcd}, \pageref{D}
\item[] $\bra .|. \ket$ \,\pageref{langle.|.rangle}
\item[] $\mathcal{B}$ \,\pageref{mathcalB}
\item[] $s_m$ \,\pageref{sm}
\item[] $s^\lor_m$, $s^\land_m$ \,\pageref{slorm}
\item[] $s^F_m$ \,\pageref{sFm}
\item[] $P^{\mbox{\scriptsize{sea}}}$ \,\pageref{Psea}, \pageref{Psea2}, \pageref{84}
\item[] $p_m$ \,\pageref{pm}
\item[] $k_m$ \,\pageref{km}
\item[] $\tilde{k}_m$ \,\pageref{tildekm}
\item[] $\tilde{p}_m$ \,\pageref{tildepm}
\item[] $\tilde{s}_m$ \,\pageref{sm2}
\item[] $p$ \,\pageref{p}
\item[] $k$ \,\pageref{k}
\item[] $X$ \,\pageref{X}
\item[] $Y$ \,\pageref{Y}
%\item[] $P^{\mbox{\scriptsize{sea}}}(x,y)$ \,\pageref{Pvac}, \pageref{vs1},
%\pageref{84}
\item[] $P(x,y)$ \,\pageref{1g}, \pageref{Pfut}, \pageref{P(x,y)}
\item[] $c_{\mbox{\scriptsize{norm}}}$ \,\pageref{1g}, \pageref{Pfut}, \pageref{cnorm}
\item[] $\Psi_{\mbox{\scriptsize{in}}}$, $\Psi_{\mbox{\scriptsize{out}}}$ \,\pageref{Psiin}
\item[] $\Aslsh_L$, $\Aslsh_R$ \,\pageref{AslshR}
\item[] $\Phi$ \,\pageref{Phi}
\item[] $\Xi$ \,\pageref{Xi}
\item[] $S^\lor_{m^2}$ \,\pageref{Slorm2}
\item[] $Y_L$, $Y_R$ \,\pageref{YL}
\item[] $S^{(l)}$ \,\pageref{Sl}
\item[] $\Pexp$ \,\pageref{Pexp}
\item[] $T_{m^2}$ \,\pageref{Tm2}
\item[] $T_{m^2}^{\mbox{\scriptsize{reg}}}$ \,\pageref{Tregm2}
\item[] $T^{(l)}$ \,\pageref{T(l)}, \pageref{T^(n)}
\item[] $P^{\mbox{\scriptsize{he}}}$ \,\pageref{Phe}
\item[] $P^{\mbox{\scriptsize{le}}}$ \,\pageref{Ple}
\item[] $X^i$ \,\pageref{X^i}, \pageref{X^i2}
\item[] $\vert x\alpha \ket$ \,\pageref{vertxalpha}
\item[] $\Sl \Psi \:|\: \Phi \Sr$ \,\pageref{SlPsi|PhiSr}
\item[] $M$ \,\pageref{M}
\item[] $E_x$ \,\pageref{E_x}
\item[] $\mathcal{L}$ \,\pageref{mathcalL}
\item[] $A$ \,\pageref{A}
\item[] $|.|$ \,\pageref{|.|}
\item[] $\overline{A}$ \,\pageref{overlineA}
\item[] ${\tr}$ \,\pageref{tr}
\item[] $Q(x,y)$ \,\pageref{Q}, \pageref{e:2c}
\item[] $\mathcal{B}$ \,\pageref{mathcalB2}
\item[] $\hat{P}$ \,\pageref{hatP}
\item[] $E_P$ \,\pageref{E_P}
\item[] $s$ \,\pageref{s}
\item[] $l$ \,\pageref{l}
\item[] $u$ \,\pageref{u}
\item[] $v$ \,\pageref{v}
\item[] $l_{\mbox{\scriptsize{max}}}$ \,\pageref{lmax}
\item[] $\alpha_{\mbox{\scriptsize{max}}}$ \,\pageref{alphamax}
\item[] $\varepsilon_{\mbox{\scriptsize{shear}}}$
  \,\pageref{varepsilon_shear}
\item[] $T^{(n)}_{[p]}$ \,\pageref{T^(n)_[p]}
\item[] $T^{(n)}_{\{p\}}$ \,\pageref{T^(n)_{p}}
\item[] $\xi$ \,\pageref{xi}
\item[] $L$ \,\pageref{L}
\item[] $c_{\mbox{\scriptsize{reg}}}$ \,\pageref{c_reg}
\item[] $\mathcal{M}$ \,\pageref{mathcalM}, \pageref{mathcalM2}
\item[] $\lambda_+$, $\lambda_-$ \,\pageref{lambda_pm}
\item[] $F_+$, $F_-$ \,\pageref{F_pm}
\item[] $z^{(n)}_{[r]}$ \,\pageref{z^(n)_[r]}
\item[] $\deg$ \,\pageref{deg}
%\item[] $\mathcal{M}$ \,\pageref{mathcalM2}
\item[] $\hat{Q}(p)$ \,\pageref{hatQ(p)}
\item[] $\mathcal{C}$ \,\pageref{mathcalC}
\item[] $\mathcal{C}^\lor$, $\mathcal{C}^\land$ \,\pageref{mathcalClor}
\item[] $A_L$, $A_R$ \,\pageref{A_L/R}
\item[] $Y_L$, $Y_R$ \,\pageref{Y_L/R}
\item[] $W_L$, $W_R$ \,\pageref{W_c}
\item[] $\nu_{nc}$ \,\pageref{nu_nc}
\item[] $I_{nc}$ \,\pageref{I_nc}
\item[] $B_p$ \,\pageref{e:3f1}
\item[] $F_p$ \,\pageref{e:3f1}
\item[] $\mathcal{B}_p$ \,\pageref{mathcalB_p}
\item[] $\mathcal{F}_p$ \,\pageref{mathcalF_p}
\item[] $\mathcal{G}_p$ \,\pageref{e:3A}
\item[] $\acute{Y}_L$, $\acute{Y}_R$ \,\pageref{acuteY_LR}
\item[] $\grave{Y}_L$, $\grave{Y}_R$ \,\pageref{graveY_LR}
\item[] $\hat{Y}_L$, $\hat{Y}_R$ \,\pageref{hatY_LR}
\item[] $I_\uparrow$, $I_\downarrow$ \,\pageref{I_arrows}
\item[] $P^q$ \,\pageref{P^q}
\item[] $P^l$ \,\pageref{P^l}
\item[] $\overline{V}$ \,\pageref{overlineV}
\item[] $Y^\eff$ \,\pageref{e:5A}
\item[] $A_L^\eff$, $A_R^\eff$ \,\pageref{e:5D}
\item[] $K_a$ \,\pageref{K_a}
\item[] $H_a$ \,\pageref{H_a}
\item[] $S^\Join_a$ \,\pageref{SJoin_a}
\item[] $K^{(n)}$ \,\pageref{K^(n)}
\item[] $S^{(n)}_\Join$ \,\pageref{S^(n)Join}
\item[] $H^{(n)}$ \,\pageref{H^(n)}

\end{itemize}

%%% Local Variables: 
%%% mode: latex
%%% TeX-master: t
%%% End: 

\end{document}